%% file: 00main.tex
\documentclass[
    letterpaper,
    10pt,
    english,
    aps,
    showpacs,
    showkeys,
    preprintnumbers,
    prd,
    floatfix,
    nofootinbib,
    superscriptaddress,
    notitlepage,
    twocolumn]{revtex4-1}

\usepackage{bbm}
\usepackage{comment}
\usepackage{graphicx}
\usepackage{nicefrac}
\usepackage{lineno}
\usepackage{amsmath}
\usepackage{amsfonts}
\usepackage[colorlinks = true,linkcolor = blue, urlcolor  = blue,citecolor = red,anchorcolor = blue]{hyperref}
\usepackage[normalem]{ulem} %
\usepackage{color}
\usepackage{enumitem}
\usepackage{multirow}
\usepackage{array}
\usepackage{lipsum}
\usepackage{siunitx}
\usepackage{placeins}
\usepackage[capitalise]{cleveref}
\crefname{section}{Sec.}{Secs.}
\Crefname{section}{Section}{Sections}
\crefname{subsection}{Sec.}{Secs.}
\Crefname{subsection}{Section}{Sections}
\usepackage{booktabs}
\usepackage[labelfont=bf]{caption}
\usepackage{subcaption}
\def\ncteqpp{{\tt \textbf{nCTEQ++}}}

\def\chidof{\chi^2/N_{\text{dof}}}
\def\chidata{\chi^2/N_{\text{data}}}

\definecolor{purple}{RGB}{128,0,128}
\definecolor{orange}{RGB}{255,102,0}
\definecolor{darkteal}{RGB}{3,102,97}

\let\oldcite\cite
\renewcommand{\cite}[1]{\mbox{\oldcite{#1}}}

\newcolumntype{L}[1]{>{\raggedright\arraybackslash}p{#1}}
\newcolumntype{R}[1]{>{\raggedleft\arraybackslash}p{#1}}
\newcolumntype{C}[1]{>{\centering\arraybackslash}p{#1}}
\newcommand{\rowpad}{\rule{0pt}{2.6ex}}
\newcommand\setrowstyle[1]{\gdef\rowmacro{#1}#1\ignorespaces}
\newcommand\clearrowstyle{\global\let\rowmacro\relax}
\clearrowstyle

\newcommand{\tocSpace}{\protect\vspace{-1em}}

\begin{document}

\input{0_Front.tex}
\begin{abstract}
We present the new release of nuclear parton distribution functions (nPDFs) from the nCTEQ collaboration: nCTEQ26. In this global analysis, we combine our recent, more selective studies of different experimental measurements and extend them to determine a modern set of nPDFs based on the largest selection of data considered to date. This selection includes neutral- and charged-current deep-inelastic scattering (DIS), fixed-target Drell-Yan, and a large variety of collider data from RHIC and LHC Runs I and II. On the theoretical side, we match this considerably extended kinematic range with a new nPDF parametrization that combines a flexible functional form of the $x$-dependence with a quadratic logarithmic nuclear $A$-dependence as well as target-mass and deuteron corrections for the reliable inclusion of high-$x$ DIS data. Following our previous successful work, we partially employ a data-driven approach that allows us to not only include open heavy-quark data, but also quarkonium data. In addition, we establish the consistency of this approach by comparing it to an alternative fit based on perturbative predictions of open heavy-quark data alone. We determine the uncertainties of our nCTEQ26 nPDFs with a Hessian error analysis, test their reliability by comparing our nPDFs to other, similar recent analyses, and make predictions for current and future experimental measurements that have not been included in the analysis.
\end{abstract}

\date{\today}

\maketitle
\tableofcontents{}

\clearpage
\section{Introduction}

With an unprecedented wealth of precision data from experiments at the CERN LHC, JLab, and BNL RHIC, nuclear and particle physics have entered a new era in the investigation of nuclear matter at high energies. Precise experimental information on nuclear structure is now available across a broad range of energy scales from those near the nucleon mass to hundreds of GeV. At such high energies, nuclei are described in terms of nuclear parton distribution functions (nPDFs) in the context of collinear QCD factorization \cite{Bodwin:1984hc,Collins:1985ue,Collins:1998rz,Qiu:2003cg}. This theoretical framework allows to separate the universal, non-perturbative PDFs from the perturbatively calculable short-distance scattering, thereby enabling systematic predictions for high-energy collider processes with PDFs serving as essential inputs. At high density and temperatures, phenomena such as gluon saturation and the creation of a deconfined state of hot nuclear matter, the so-called quark-gluon plasma, can lead to higher-twist nuclear effects beyond those encoded in leading-twist nPDFs. A precise determination of nPDFs is therefore essential to establish a robust baseline for isolating and quantifying hot as opposed to so-called cold nuclear effects. The complexity of QCD implies that (n)PDFs must currently be determined through global analyses of experimental data~\cite{Gao:2017yyd,Kovarik:2019xvh,Ethier:2020way,Klasen:2023uqj}, despite considerable progress in lattice determinations of PDFs \cite{Lin:2017snn,Cichy:2018mum,Lin:2025hka,Detmold:2020snb}.

Beyond serving as a baseline for thermal QCD phenomena, nPDFs are of fundamental interest in their own right, as they provide a quantitative description of the partonic structure in nuclei compared to the one in a free nucleon and encode phenomena such as shadowing, antishadowing, the EMC effect and Fermi motion \cite{Seely:2009gt,HallC:2022utd,JeffersonLabHallATritium:2024las,EuropeanMuon:1988lbf}. Global analyses then make it possible to study the partonic structure of nuclei for each individual parton flavor in dependence of the nuclear mass number $A$ and allow to test microscopic models for the nuclear structure. Prominent examples are the connection between the EMC effect and short-range correlated nucleon pairs~\cite{nCTEQ:2023cpo,CLAS:2019vsb} or isovector nuclear effects~\cite{Cocuzza:2021rfn}. Nuclear PDFs are equally relevant for neutrino physics, where nuclear targets are needed to compensate for the small interaction cross sections. The interpretation of neutrino deep-inelastic scattering (DIS) measurements on nuclear targets~\cite{NuTeV:2005wsg,CCFRNuTeV:2000qwc,CHORUS:2005cpn,Berge:1989hr,MINERvA:2016oql} requires a quantitative understanding of nuclear modifications in certain weak-interaction processes~\cite{Schienbein:2007fs,Kovarik:2010uv,Morfin:2021ujm,Schienbein:2009kk}. Finally, global nPDF analyses provide a nontrivial test of QCD factorization in the nuclear environment. By simultaneously describing diverse processes, including DIS, Drell–Yan, and hadron-collider measurements, with a single set of universal nPDFs, these analyses directly test whether the factorization framework and nuclear parton distributions can consistently account for different classes of experimental data. Reported tensions, such as those between charged-lepton and neutrino DIS data~\cite{Muzakka:2022wey,Paukkunen:2010hb,Paukkunen:2013grz,Kalantarians:2017mkj}, may point to limitations of this picture. Global nPDF analyses thus offer a framework for determining the kinematic domain in which leading-twist collinear factorization holds and to identify regions and processes where effects such as higher-twist contributions~\cite{Blumlein:2008kz,Alekhin:2024dqu,Cerutti:2025yji}, parton energy loss~\cite{Brodsky:1989ex,Gavin:1991qk,Gavin:1990gm,Arleo:2012hn}, or gluon saturation~\cite{vanHameren:2023oiq,Armesto:2022mxy} may become relevant.

Recent precision measurements provide an opportunity to capitalize on the increasingly rich datasets by combining advanced theoretical tools and computational techniques to perform comprehensive global analyses and obtain nPDFs with unprecedented precision. Since discoveries often appear as subtle tensions or deviations from the Standard Model (SM) predictions, highly accurate (n)PDFs are essential for establishing robust SM baselines. Precise determinations of nPDFs therefore also provide a critical foundation for interpreting future measurements at the High-Luminosity LHC (HL-LHC) and the Electron-Ion Collider (EIC). Beyond colliders, precise nPDFs will be equally important for the next generation of neutrino experiments. For instance, DUNE~\cite{DUNE:2024wvj,DUNE:2025zsi,DUNE:2025lvs} will employ argon targets at energies where shallow- and deep-inelastic scattering contribute significantly to the event rate~\cite{Pandey:2025vpa,SajjadAthar:2020nvy}. At higher energies, the emerging neutrino program has already enabled measurements of high-energy neutrino interactions by FASER$\nu$~\cite{FASER:2024hoe}, with substantially larger datasets anticipated at the Forward Physics Facility~\cite{Feng:2022inv}, while neutrino-nucleus cross sections also enter measurements at neutrino telescopes~\cite{IceCube:2017roe,IceCube:2020rnc,Weigel:2024gzh}. Together, these facilities will provide unprecedented opportunities to address fundamental, long-standing questions about hadronic and nuclear structure and to deepen our understanding of the underlying dynamics of the strong interaction.

Beyond our own program, nPDFs are determined by several international collaborations, whose analyses differ in data selection, parametrization, theoretical inputs, and treatment of uncertainties, most extensively by the EPPS \cite{Eskola:2021nhw} and nNNPDF \cite{AbdulKhalek:2022fyi} collaborations, with further global sets provided by the TUJU \cite{Helenius:2021tof}, KSASG \cite{Khanpour:2020zyu}, and DSSZ \cite{deFlorian:2011fp} analyses (for overviews see Refs.~\cite{Klasen:2023uqj, Arleo:2025oos}). Recent methodological developments across nPDF analyses include the propagation of uncertainties of the baseline proton PDFs into nPDFs \cite{Eskola:2022rlm,Eskola:2021nhw,AbdulKhalek:2022fyi} and NNLO determinations based on a subclass of available experimental data \cite{Helenius:2021tof,Khanpour:2020zyu}. Further advances are provided by an improved uncertainty quantification using Bayesian sampling based on Markov Chain Monte Carlo methods, removing the need for the Gaussian assumptions underlying current (n)PDF determinations \cite{Derakhshanian:2026zkx}, and a novel parametrization of nuclear corrections motivated by short-range correlated nucleon pairs \cite{nCTEQ:2023cpo}. Beyond these advances, several tools established for proton PDFs have, however, not yet been adopted in nuclear analyses, including the treatment of theory uncertainties from missing higher orders \cite{NNPDF:2024dpb,NNPDF:2019ubu} and a range of statistical and computational techniques \cite{Hunt-Smith:2022ugn,Kotz:2023pbu,Kotz:2025une,Capel:2024qkm,Candido:2024hjt,Risse:2025qlo,Barontini:2025lnl,Costantini:2025wxp,Costantini:2025agd}.

Our previous  nCTEQ15~\cite{Kovarik:2015cma} nPDF set has served as a valuable pre-LHC reference for numerous studies. In the following, we provide an overview of the principal features introduced with nCTEQ15 and the accompanying series of follow-up analyses~\cite{Kusina:2020lyz,Duwentaster:2021ioo,Muzakka:2022wey,Duwentaster:2022kpv,Segarra:2020gtj,nCTEQ:2023cpo}, all of which are fully incorporated into and extended upon in the new nCTEQ26 nPDF set introduced here. The nCTEQ15 nPDFs included data on neutral-current (NC) DIS,
Drell-Yan (DY) lepton-pair production, and pion production at RHIC. This was the last nCTEQ global analysis not to include LHC data, and it provided an unbiased reference for subsequent experimental and phenomenological studies such as the ones in~\cite{ATLAS:2019maq,CMS:2019leu,LHCb:2018jry,ALICE:2022wpn}. Its primary focus was to obtain a reliable estimate of the uncertainties, a particularly challenging task at the time given the limited experimental constraints. In a following study, the nCTEQ15WZ~\cite{Kusina:2020lyz} set included $W$ and $Z$ boson production data from the LHC. This analysis yielded improved bounds on the gluon PDF and provided strange-quark PDF constraints independent of neutrino-nucleon DIS in a newly accessible kinematic region~\cite{ATLAS:2012sjl,ATLAS:2016nqi}. To further constrain the nuclear gluon distribution, the nCTEQ15WZSIH~\cite{Duwentaster:2021ioo} set
extended that analysis by adding single inclusive hadron (SIH) production data from RHIC and LHC. The gluon distribution was further investigated and constrained in the nCTEQ15HQ analysis~\cite{Duwentaster:2022kpv}, which incorporated heavy-quark and quarkonium production data from the LHC as a novel probe of the nuclear gluon distribution. These new data provided significant constraints in an extended kinematic range down to $x\,{\lesssim}\,10^{-5}$, but required a specialized theoretical treatment to obtain reliable predictions. Finally, the nCTEQ15HIX~\cite{Segarra:2020gtj} analysis used relaxed $\{x,Q^2\}$ kinematic cuts to investigate the high-$x$ region by incorporating DIS data from the JLab experiments, and the nCTEQ15$\nu$~\cite{Muzakka:2022wey} set revisited the compatibility between existing charged-lepton and neutrino DIS data.

In the current nCTEQ26 analysis, we incorporate all the developments described above and build upon them with several additional improvements, the most notable of which are summarized here:

\begin{itemize}
\item 
We significantly expand the datasets used in the analysis by both relaxing the kinematic selection cuts, thereby retaining more of the previously used data, and incorporating new measurements that were previously unavailable. Together, these improvements substantially increase the dataset size to more than 3500 data points, compared with 740 in nCTEQ15, while also providing much broader kinematic coverage.
\item 
To accurately model data across the expanded kinematic domain, we incorporate essential theoretical refinements, including target mass corrections (TMCs)~\cite{Schienbein:2007gr,Ruiz:2023ozv} and deuteron corrections. Furthermore, we adopt an updated baseline proton PDF~\cite{Accardi:2016qay} that applies a consistent treatment of these theoretical effects while explicitly eliminating the double counting of nuclear data.
\item 
With the goal to obtain a refined understanding of how the nuclear PDFs vary with the atomic mass number $A$, we introduce a new, more flexible parameterization of the nuclear $A$-dependence. This $A$-dependence also reduces correlations among the fit parameters and avoids imposing an artificially smooth or monotonic evolution of the PDFs and their uncertainties with $A$, particularly in regions where direct experimental constraints are limited or absent. We also introduce a new PDF parametrization that provides greater flexibility in the $x$-dependence inherited from the proton baseline. As a result, the parametrization provides more realistic and reliable estimates of the nPDF uncertainties across the full range of nuclei.
\end{itemize}

The resulting nCTEQ26 set represents a new generation of nuclear PDFs, seamlessly incorporating the full range of theoretical and methodological advancements developed since the nCTEQ15 series. This next-generation set establishes a critical benchmark for future theoretical and experimental studies. It marks a significant step forward in our ability to precisely describe and compute QCD phenomena across a broad kinematic range. These capabilities are especially vital for major programs such as the HL-LHC, the EIC, DUNE, and other forthcoming experiments. Together, these efforts will lay the foundation for the next generation of precision QCD studies and deepen our understanding of the fundamental structure of matter.

The remainder of the paper is organized as follows: In \cref{sec:method}, we present the methodology used to determine the nCTEQ26 nuclear PDFs, including the PDF parametrization, QCD input parameters, minimization procedure, and our approach for estimating nPDF uncertainties. In \cref{sec:data}, we provide a detailed description of the experimental datasets and the corresponding theoretical calculations used in the fit. \cref{sec:res} discusses the results of the analysis. We first assess the quality of the fit by examining its description of the experimental data and verifying the validity of the Hessian uncertainty propagation. We then examine the resulting nPDFs, with particular attention to their nuclear $A$-dependence. In \cref{subsec:comp}, we compare our results with the nPDFs obtained by other groups. \cref{sec:discussion} contains a discussion of two particularly interesting aspects of the analysis, i.e., the strange-quark and gluon distributions. In \cref{sec:pred}, we discuss experimental datasets not included in our analysis and provide theoretical predictions for them using our new nCTEQ26 nPDFs. In \cref{sec:sum}, we summarize the results obtained. Finally, \cref{app:usage} provides instructions on the correct use of the nCTEQ26 nPDF set, while the remaining appendices contain technical details of the fit.

\section{Methodology}
\label{sec:method}

We begin by outlining the methodology used in our nCTEQ26 analysis in order to obtain an optimal set of nPDFs and their associated uncertainties. We first discuss the employed nPDF parametrization, its $A$-dependence, and the implementation of number and momentum sum rules. We then fix the QCD input parameters such as the strong coupling constant and the heavy-quark masses. Third, we describe the minimization procedure, and finally we discuss the details of our implementation of the Hessian method that we use for determining nPDF uncertainties.

\subsection{Nuclear PDF parametrization}
\label{subsec:param}

In order to preserve the conceptual similarity with free proton PDFs, $f_k^p(x,Q)$, as well as fundamental sum rules on a per-nucleon basis, it is convenient, though not necessary, to reformulate the full nuclear PDFs $f_k^A(x,Q)$ of a parton $k$ in a nucleus with mass number $A$ and electric charge $Z$ in terms of effective bound proton ($f_k^{p/A}$) and neutron ($f_k^{n/A}$) PDFs, i.e.
\begin{equation}
  \begin{split}
    f_k^A(x,Q) &= \frac{Z}{A}f_k^{p/A}(x,Q) + \frac{A-Z}{A}f_k^{n/A}(x,Q)\ .
  \end{split}
  \label{eq:nPDFA}
\end{equation}
Similarly to isospin symmetry for free nucleons, the down-quark $(d/\bar{d})$ distributions in the bound neutron can then be identified with the up-quark $(u/\bar{u})$ distributions in the bound proton and vice versa. Note that this does not take into account isovector effects as discussed, e.g., in Ref.~\cite{Cocuzza:2021rfn}.

In nCTEQ26, we use a new parametrization of the bound proton PDFs at the initial factorization scale $Q_0=m_c=1.3$ GeV,
\begin{equation}
    xf_k^{p/A}(x,Q_0) = c^k_0 x^{c^k_1} (1-x)^{c^k_2} 
    (1+c^k_3\sqrt{x}+c^k_4x)e^{c^{k}_5\sqrt{x}}\,,
    \label{eq:parametrization1}
\end{equation}
with $k= \{u_v,\bar{u}+\bar{d},g,s+\bar{s}\}$, which offers more flexibility in the high-$x$ region to accommodate the considerably enlarged analyzed kinematic region and experimental datasets than our previous parametrization \cite{Kovarik:2015cma,Kusina:2016fxy,Muzakka:2022wey,Duwentaster:2022kpv,Duwentaster:2021ioo,Segarra:2020gtj,Kusina:2020lyz,nCTEQ:2023cpo}. The $d$-valence quark distribution receives an additional admixture of $u$-valence quarks,
\begin{equation}
  \begin{split}
    xd_v^{p/A}(x,Q_0) &= c^{k}_0 \Big[
      x^{c^{k}_1} (1-x)^{c^{k}_2}
      (1+c^{k}_3\sqrt{x}+c^{k}_4x)e^{c^{k}_5\sqrt{x}}
        \\ & \quad + c^{k}_6x^{c^{k}_7} \,xu_v^{p/A}(x,Q_0)
      \Big]\,,\quad k = d_v \,, \raisetag{16pt}
  \end{split}
  \label{eq:parametrization2}
\end{equation}
the difference between light sea quarks is parametrized in the form of the ratio
\begin{equation}
  \frac{\bar{d}}{\bar{u}}(x,Q_0) = c_0^{k}x^{c_1^{k}}(1-x)^{c_2^{k}} + 1 + c_3^{k}x(1-x)^{c_4^{k}}\,,\ k=\bar{d}/\bar{u},
\end{equation}
and we assume that there is no strange-quark asymmetry,
\begin{equation}
  s(x) - \bar{s}(x) = 0.
\end{equation}

This new parametrization in $x$ has been adopted, and extended by an exponential factor for the strange quark (i.e.\ $c_5^k=0$ otherwise), from the recent series of free-proton analyses by the CJ collaboration \cite{Accardi:2016qay,Accardi:2023gyr,Accardi:2026hdv}, which offer several advantages for the analysis of nuclear PDFs. First, they are free of nuclear data, thereby avoiding potential nuclear contamination or double counting, whereas it is standard practice in many free-proton PDF fits to use nuclear data like neutrino DIS and/or dimuon data from the NuTeV and CCFR experiments~\cite{NuTeV:2001dfo,NuTeV:2005wsg,CCFRNuTeV:2000qwc} taken on an iron target to obtain better sensitivity to less constrained flavors. Second, as demonstrated in our previous nCTEQ15HIX analysis \cite{Segarra:2020gtj}, it is essential for the extension of the kinematic range into the large-$x$, low-$Q^2$ and low-$W^2$ regions and the inclusion of the high-precision 6~GeV JLab data \cite{CLAS:2019vsb,Seely:2009gt} to account for deuteron and target-mass corrections, which have been carefully studied in the CJ analyses. Third, it is important to build on a parametrization that allows to consistently include higher-twist corrections, either additively or multiplicatively~\cite{Blumlein:2008kz,Alekhin:2024dqu,Cerutti:2025yji}, and that has been thoroughly scrutinized. Although newer CJ fits are now available, we have decided to use the well-tested and stable (in particular in the high-$x$ region) CJ15 \cite{Accardi:2016qay} release as our proton baseline. Note that, unlike ratio-based nuclear PDF approaches, we always parametrize the full nuclear PDFs, so that they depend only indirectly on the free-proton baseline through the boundary condition $A\to1$ and datasets with a proton as an incoming beam.\footnote{A recent study investigated the impact of proton PDF uncertainties on $W^\pm$-boson production in $p$Pb collisions and found that they have only a minimal effect on the extraction of nuclear modifications \cite{Eskola:2022rlm}.} Proton PDF uncertainties are in principle available from the CJ15 analysis \cite{Accardi:2016qay} and can be combined with the nuclear uncertainties determined here, e.g.~by adding them in quadrature~\cite{Eskola:2016oht} or by using the procedure outlined in Ref.~\cite{Eskola:2021nhw}. The impact of the nuclear-free proton fit CT25pd currently in preparation~\cite{Ablat:2025gbp} and of a simultaneous proton and nuclear PDF fit is left for study in future work.

In our previous analyses, we directly incorporated the dependence of our nuclear PDFs on the mass number $A$ in our free parameters with ${c_j(A) = p_j + a_j(1-A^{-b_j})}$. However, this functional form induces correlations between the parameters $a_j$ and $b_j$ that are difficult to control. Inspired by early experimental observations of a logarithmic $A$-dependence of nuclear cross section ratios \cite{Arnold:1983mw,Gomez:1993ri}, we therefore now rather adopt the form
\begin{equation}
 c_j(A) = p_j + a_j\ln A + b_j\ln^2A.
 \label{eq:Amod2}
\end{equation}
Note that in the relevant region up to $A\sim208$, the terms $\ln A$ and $\ln^2 A$ agree within $\sim\!10\%$ and $\sim\!25\%$ with the terms $A^{1/3}$ and $A^{2/3}$ reminiscent of nuclear modeling. This new functional form still ensures that for $A=1$ we retrieve the free-proton parametrization given by the set of $p_j$ parameters (cf.\ \cref{tab:params}), which are kept fixed during the fitting, so that the nuclear dependence is fitted only with the parameters $a_j$ and $b_j$. Similarly to other nuclear PDF analyses, we do not include a dependence on the nuclear charge $Z$ beyond the leading dependence in \cref{eq:nPDFA}, since with the current precision of experimental data it would be insufficiently constrained. However, in the future it may become interesting to include nuclei with the same $A$, but different $Z$, and isospin-breaking effects, as well as to go beyond the mean-field single-nucleon approximation \cite{nCTEQ:2023cpo}. 

As is well known, the parameters introduced above are not completely independent, but constrained by number and momentum sum rules. There are two number sum rules and one momentum sum rule for each nucleus, which we use to constrain three normalization parameters for each value of $A$. We fix the valence normalizations $c_0^{u_v}$ and $c_0^{d_v}$ such that the number sum rules
\begin{align}
    \int_0^1 u_v(x, Q_0) \, \mathrm{d}x = 2 \;\; \text{and} \;\; \int_0^1 d_v(x, Q_0) \, \mathrm{d}x = 1
\end{align}
are satisfied. We then define the momentum contribution of each flavor $k$ as
\begin{equation}
    M_k = \int_0^1 x f_k(x, Q_0) \mathrm{d}x ,
\end{equation}
so that the momentum sum rule becomes
\begin{equation}
    M_{u_v} + M_{d_v} + 2 M_{\bar{u}+\bar{d}} + M_{s+\bar{s}} + M_g = 1,
\end{equation}
and use this sum rule to fix the normalization parameter $c_0^{\bar{u}+\bar{d}}$. The remaining normalizations $c_0^g$ and $c_0^{s+\bar{s}}$ are then in principle free parameters. However, we fit instead the related parameters
\begin{equation}
    c_\mathrm{sum}^g \equiv M_g \;\; \text{and} \;\; c_\mathrm{sum}^{s+\bar{s}} \equiv \frac{3 M_{s+\bar{s}}}{M_{s+\bar{s}} + 2 M_{\bar{u}+\bar{d}}} ,
\end{equation}
where $c_\mathrm{sum}^g$ encodes the gluon momentum contribution and $c_\mathrm{sum}^{s+\bar{s}}$ measures the fraction of the strange-quark momentum relative to the momentum of all sea quarks consisting of the sum of $\bar{u}$, $\bar{d}$ and $s+\bar{s}$. The $c_\mathrm{sum}^g$ and $c_\mathrm{sum}^{s+\bar{s}}$ parameters are equipped with an $A$-dependence as described in \cref{eq:Amod2}. Note that in the proton baseline $M_{s+\bar{s}}/M_{\bar{u}+\bar{d}}=0.4$, which entails $p_\mathrm{sum}^{s+\bar{s}} = 0.5$.

\subsection{QCD input parameters}
\label{sec:dglap}

We fix the value of the strong coupling constant at the $Z$-boson mass $M_Z=91.188$ GeV to ${\alpha_S(M_Z) = 0.118}$, as it was done in the employed CJ15 proton baseline. However, we evolve it with the renormalization scale $\mu$ at NLO QCD with the Runge--Kutta method, as opposed to the analytic solution in the original CJ15 analysis. This can in practice result in small numerical differences. We therefore recommend using \texttt{nCTEQ26\_1\_1} instead of \texttt{CJ15nlo} as the proton PDF in observables involving free protons to avoid artifacts due to this difference. The heavy-quark masses, where the charm and bottom thresholds are crossed, are set to the same values \mbox{$m_c=1.3$~GeV} and \mbox{$m_b=4.5$~GeV} as in the CJ15 analysis. The DGLAP evolution of the PDFs is performed at NLO QCD in the variable-flavor number scheme (VFNS). We use a modified version of the HOPPET package~\cite{Salam:2008qg} to handle multiple grids for different nuclei simultaneously. We restrict our analysis to five active flavors with the same mass thresholds as in the $\alpha_S$ evolution, i.e. we do not introduce a PDF for the top quark, but only the light quarks, charm and bottom heavy quarks, and the gluon.

\subsection{Minimization procedure}
\label{subsec:min}

For normally distributed uncorrelated data points, the goodness of fit can be estimated with a loss function
\begin{align} \label{eq:chi2-naive}
    \chi^2(\textbf{a})=\sum_i\frac{[D_i-T_i(\textbf{a})]^2}{\sigma_i^2},
\end{align}
where $D_i$ are the measured experimental values, $T_i$ are the corresponding theoretical predictions, which depend on the parameters $\textbf{a}$ defined at the input scale $Q_0$ and discussed in \crefrange{eq:parametrization1}{eq:Amod2}, and \mbox{$\sigma_i^2\equiv \hat\sigma_{i}^2+\bar\sigma_{i}^2$} adds the statistical uncertainty $\hat\sigma_{i}^2$ and uncorrelated systematic uncertainty $\bar\sigma_{i}^2$ for each data point $D_i$ in quadrature.

The above naive $\chi^2$ definition is insufficient to incorporate the full knowledge of correlated experimental uncertainties, in particular normalization uncertainties, which affect not only the central values of the data, but also the sizes of the other uncertainties. One prescription to take correlations and normalizations into account is to modify the definition of the loss function to
\begin{align} \label{eq:chi2-correlated}
    \chi^2 = (\textbf{D}-\textbf{T})^TC_T^{-1}(\textbf{D}-\textbf{T}),
\end{align}
where without loss of generality we now assume a single dataset. The correlation matrix is calculated as
\begin{align} \label{eq:chi2-correlation-matrix}
    C_{T,ij}(\textbf{a}) = \sigma_i^2\delta_{ij} + \sigma_{\text{norm}}^2T_i(\textbf{a})T_j(\textbf{a}) + \sum_\alpha \bar{\sigma}_{i\alpha}\bar{\sigma}_{j\alpha}\ ,
\end{align}
where $\sigma^2_{\text{norm}}$ accounts for the normalization uncertainty of this dataset and $\bar\sigma_{i\alpha}$ is the correlated systematic uncertainty on data point $D_i$ from the $\alpha$-th source of uncertainty. 
The normalization uncertainty is accounted for as a multiplicative correction to the theoretical prediction, which introduces additional nuisance parameters. The usual treatment is to integrate them out, but one can also eliminate them by minimizing the $\chi^2$-function with respect to them~\cite{Muzakka:2022wey}. This leads to
\begin{align} 
\label{eq:chi2-with-nuisance-params}
    \chi^2 = \sum_i \frac{\left(D_i - r_{\mathrm{norm}} T_i - \sum_\alpha \bar{\sigma}_{i \alpha} \bar{r}_\alpha\right)^2}{\sigma_i^2} & \nonumber \\
    + \sum_\alpha \bar{r}_\alpha^2 + \frac{(1 - r_{\mathrm{norm}})^2}{\sigma_{\mathrm{norm}}^2} &,
\end{align}
where the nuisance parameters are set to their minimal values
\begin{align}
    \bar{r}_\alpha &= \sum_\beta (A^{-1})_{\alpha \beta} B_{\beta}, \\
    r_{\mathrm{norm}} &= \frac{1 + \sigma_{\mathrm{norm}}^2 \mathbf{D}^T C^{-1} \mathbf{T}}{1 + \sigma_{\mathrm{norm}}^2 \mathbf{T}^T C^{-1} \mathbf{T}}
\end{align}
with
\begin{align}
    A_{\alpha \beta} &= \delta_{\alpha \beta} + \sum_i \frac{\bar{\sigma}_{i \alpha} \bar{\sigma}_{i \beta}}{\sigma_i^2}, \\
    B_{\alpha} &= \sum_i \frac{\bar{\sigma}_{i \alpha} (D_i - r_{\rm norm} T_i)}{\sigma_i^2}.
\end{align}
While other prescriptions for the treatment of normalization uncertainties exist, this choice has been shown to avoid the d'Agostini bias \cite{DAgostini:1993arp} without introducing additional nuisance parameters to the fit. 
For a more thorough comparison of different normalization treatments, we refer the reader to Ref.~\cite{Muzakka:2022wey}.

The $\chi^2$-function of \cref{eq:chi2-with-nuisance-params} is not directly suited for analyzing the fit quality data point by data point, as due to the correlation matrix no single value of $\chi^2$ can be associated with a given data point and its theory prediction. However, this definition has the advantage that it can be manifestly divided into an uncorrelated contribution $\tilde{\chi}^2_i$ per data point $D_i$ and a correlated penalty contribution. In the former, we can absorb the nuisance parameters into the theory values,
\begin{equation}
    \hat{T}_i = r_{\mathrm{norm}} T_i + \sum_\alpha \bar{\sigma}_{i \alpha} \bar{r}_\alpha,
\end{equation}
such that the uncorrelated contribution to $\chi^2$ becomes
\begin{equation}
    \tilde{\chi}^2 = \sum_i \tilde{\chi}_i^2 = \sum_i \frac{\bigl(D_i - \hat{T}_i\bigr)^2}{\sigma_i^2} \,.
\end{equation}
The penalty contribution
\begin{equation} \label{fullCorrelated}
    \chi^2 - \tilde{\chi}^2 = 
     \sum_\alpha \bar{r}_\alpha^2 + \frac{(1 - r_{\mathrm{norm}})^2}{\sigma_{\mathrm{norm}}^2} \, ,
\end{equation}
potentially correlated over multiple datasets, is then distributed among the individual data points with relative weights corresponding to their uncorrelated $\chi^2$ contributions, allowing us to meaningfully assign a $\chi^2$ value to each point. We rely on this prescription throughout the manuscript.%
\footnote{In particular, when comparing our theoretical predictions to the fitted data we shift the theory prediction to account for the $r_{\text{norm}}$ correction. By construction, this correction is different for each eigenvector, which are then rescaled independently. This rescaling can result in effective reduction of the uncertainty band as compared to the unrescaled prediction.}
While this procedure takes into account all available experimental uncertainties, the study of theoretical and model uncertainties (e.g.~missing higher order corrections \cite{Harland-Lang:2018bxd,Kassabov:2022orn,NNPDF:2024dpb} and the choice of parametrization \cite{H1:2015ubc,Courtoy:2022ocu,Kotz:2023pbu}) is left for future work.

Finally, to minimize the $\chi^2$-function, we use a sequence of gradient descent~\cite {Levenberg1944,Marquardt1963,Nocedal:2006jvv} and Simplex~\cite{Nelder:1965zz} algorithms. This typically allows to find a minimum more reliably than employing only one of them.

\subsection{Uncertainty determination}
\label{subsec:errors}

Apart from determining the best fit of nuclear PDFs to a given dataset and within the chosen theoretical approach, one would also like to obtain a quantitative estimate of the PDF uncertainties. Various approaches to determine these uncertainties have been proposed in the literature \cite{Giele:1998gw,Stump:2001gu,Pumplin:2001ct,NNPDF:2019ubu,Costantini:2024wby} including our own studies based on a Markov Chain Monte Carlo framework \cite{Risse:2025qlo,Derakhshanian:2026zkx}. As in our previous analyses, we choose the Hessian method \cite{Pumplin:2001ct,Pumplin:2000vx}, which probes the vicinity of the minimal $\chi^2$ in a quadratic approximation and transforms the original parameter basis into normalized orthogonal eigenvectors of the Hessian matrix. In the following brief description of the Hessian method, we employ the notation of Ref.~\cite{Kovarik:2015cma}. 

Given a set of parameters $\{a_{i,0}\}$ that correspond to the minimum value of the $\chi^2$ function $\chi^2_0\equiv\chi^2(\{a_{i,0}\})$, the region near the minimum can be approximated as 
\begin{align}
    \chi^2 \approx \chi^2_0 + \sum_{i,j} H_{ij}y_iy_j, 
\end{align}
where $y_i\equiv a_i-a_{i,0}$ are the parameter shifts from the minimum and $H_{ij}$ is the Hessian matrix defined as 
\begin{align}
    H_{ij} = \frac12 \left(\frac{\partial^2 \chi^2}{\partial y_i \partial y_j}\right)_{a_i=a_{i,0}}.
    \label{eq:hessian}
\end{align}
The Hessian is a symmetric $n\times n$ matrix (where $n$ is the number of parameters $a_i$) with $n$ orthogonal eigenvectors $V_i^{(k)}$ in $\{y_i\}$-space, which are determined by the characteristic equation 
\begin{align}
    \sum_j H_{ij} V_j^{(k)} = \lambda_k V_i^{(k)}
\end{align}
and the normalization condition
\begin{align}
    \sum_i V_i^{(j)} V_i^{(k)} = \delta_{jk}.
\end{align}
We calculate the partial double derivatives in \cref{eq:hessian} numerically using a noise-reducing finite-difference prescription \cite{lanczos}.

The orthogonal eigenvectors $V_i^{(k)}$ can be used to define a transformation matrix 
\begin{align}
    D \equiv (V^{(1)},V^{(2)},...,V^{(n)}) = 
    \begin{pmatrix}
        V_1^{(1)} & V_1^{(2)} & ... & V_1^{(n)}\\
        V_2^{(1)} & V_2^{(2)} & ... & V_2^{(n)} \\
        \vdots & \vdots & \ddots & \vdots \\
        V_n^{(1)} & V_n^{(2)} & ... & V_n^{(n)}
    \end{pmatrix} 
\end{align}
with $D^{-1}=D^T$ leading to a new set of orthogonal basis parameters
\begin{align}
    \textbf{z} \equiv D^{-1}\textbf{y},\label{eq:hessianParameterDefinition1}
\end{align}
for which the Hessian matrix is diagonal,
\begin{align}
    \sum_{i,j} H_{ij}y_iy_j &= \sum_{i,j} H_{ij}^Dz_iz_j = \textbf{z}^T D^T H D \, \textbf{z}\\
    &= \textbf{z}^T
    \begin{pmatrix}
        \lambda_1 & 0 & ... & 0\\
        0 & \lambda_2 &  & \vdots \\
        \vdots &  & \ddots & 0\\
        0 & ... & 0 & \lambda_n
    \end{pmatrix} 
    \textbf{z} \ .
\end{align}

Eq.~\eqref{eq:hessianParameterDefinition1} can be expressed in index notation as
\begin{align}
    z_k = \sum_j y_jV_j^{(k)},\label{eq:hessianParameterDefinition2}
\end{align}
or conversely
\begin{align}
    y_i = \sum_k V_i^{(k)}z_k = \sum_k \tilde{V}_i^{(k)}\tilde{z}_k ,
\end{align}
where in the last step we have rescaled the basis vectors $\tilde{z}_k \equiv s_kz_k $ and the eigenvectors $\tilde{V}_i^{(k)} \equiv V_i^{(k)}/s_k$, so that
\begin{align}
    \Delta \chi^2 \equiv \chi^2 - \chi^2_0 = \sum_k \tilde{z}_k^2.\label{eq:deltaChiZ}
\end{align}
If the quadratic approximation exactly matched the true $\chi^2$ function, these rescaling factors would be symmetric and given by ${s_k = \sqrt{\lambda_k}}$. However, since the quadratic approximation does not hold exactly in any real PDF analysis, we choose to determine the scaling factors $s^\pm_k$ separately in both directions, such that for each eigenvector direction the data are described within the chosen tolerance $\Delta \chi^2$. 

The uncertainty of the PDFs is estimated by varying the $n$ orthogonal basis parameters $\tilde{z}_k$ in both directions within the tolerance $\Delta\chi^2=T$. While a number of prescriptions for the choice of the value of $T$ has been proposed in the literature \cite{Stump:2001gu,Martin:2009iq,Eskola:2009uj,Kovarik:2015cma,Eskola:2021nhw}, there is currently no consensus. In the current analysis we deviate from our earlier studies \cite{Kovarik:2015cma} and rely on a prescription based on the Wilks' theorem \cite{Wilks:1938dza,Cowan:1998ji} described in \cref{app:tolerance}, which aligns with the one used in EPPS21 \cite{Eskola:2021nhw}. With this prescription we obtain the value $T=50$. The shift corresponding to $\Delta\chi^2=T$ induces $2n$ changes in the original parameters $a_i$ and therefore also in the PDFs,
\begin{align}
    f_k^\pm = f\left(a_{i,0}\pm \frac{\sqrt{\Delta\chi^2}}{s^\pm_k}V_i^{(k)}\right), \quad \mathrm{for}\ k=1,2,...,n\ .
    \label{eq:errorPDFs}
\end{align}
The PDF uncertainty of an observable $X$ can be obtained by evaluating it with these error PDFs, ${X_k^\pm=X(f_k^\pm)}$, and Gaussian error propagation,
\begin{align}
    \Delta X = \frac12 \sqrt{\sum_{k=1}^{n} (X_k^+-X_k^-)^2},
    \label{eq:ObservableUncertainty}
\end{align}
i.e.\ by adding the uncertainties corresponding to each individual direction in quadrature. This also holds for the uncertainties of the PDFs themselves or ratios of physical observables. Ratios of observables for different nuclei should of course be evaluated by taking the same error PDFs for both numerator and denominator.

Instead of the symmetric definition of uncertainties in \cref{eq:ObservableUncertainty}, it is possible to introduce asymmetric PDF uncertainties in an observable $X$ with
\begin{align}    \label{eq:ObservableUncertaintyAsym}
\begin{split}
        \Delta X^+ &= \sqrt{\sum_{k=1}^{n}
        \left[ \max\left( X_k^+-X_0, X_k^--X_0, 0 \right) \right]^2}, \\
        \Delta X^- &= \sqrt{\sum_{k=1}^{n}
        \left[ \max\left( X_0-X_k^+, X_0-X_k^-, 0 \right) \right]^2},
\end{split}
\end{align}
where $X_0=X(f_0)$. For most of the PDFs, we find that in practice the two prescriptions agree very well. However, as will be discussed later, for the strange quark the asymmetric prescription leads to a significantly smaller lower variation at small $x$ despite the fact that the strange PDF is not directly constrained in this region by any experimental data. We therefore recommend using the symmetric uncertainties defined in \cref{eq:ObservableUncertainty}.

\section{Experimental data and theoretical modeling}\label{sec:data}

\begin{figure*}[ht!]
    \centering
    \includegraphics[width=0.85\linewidth]{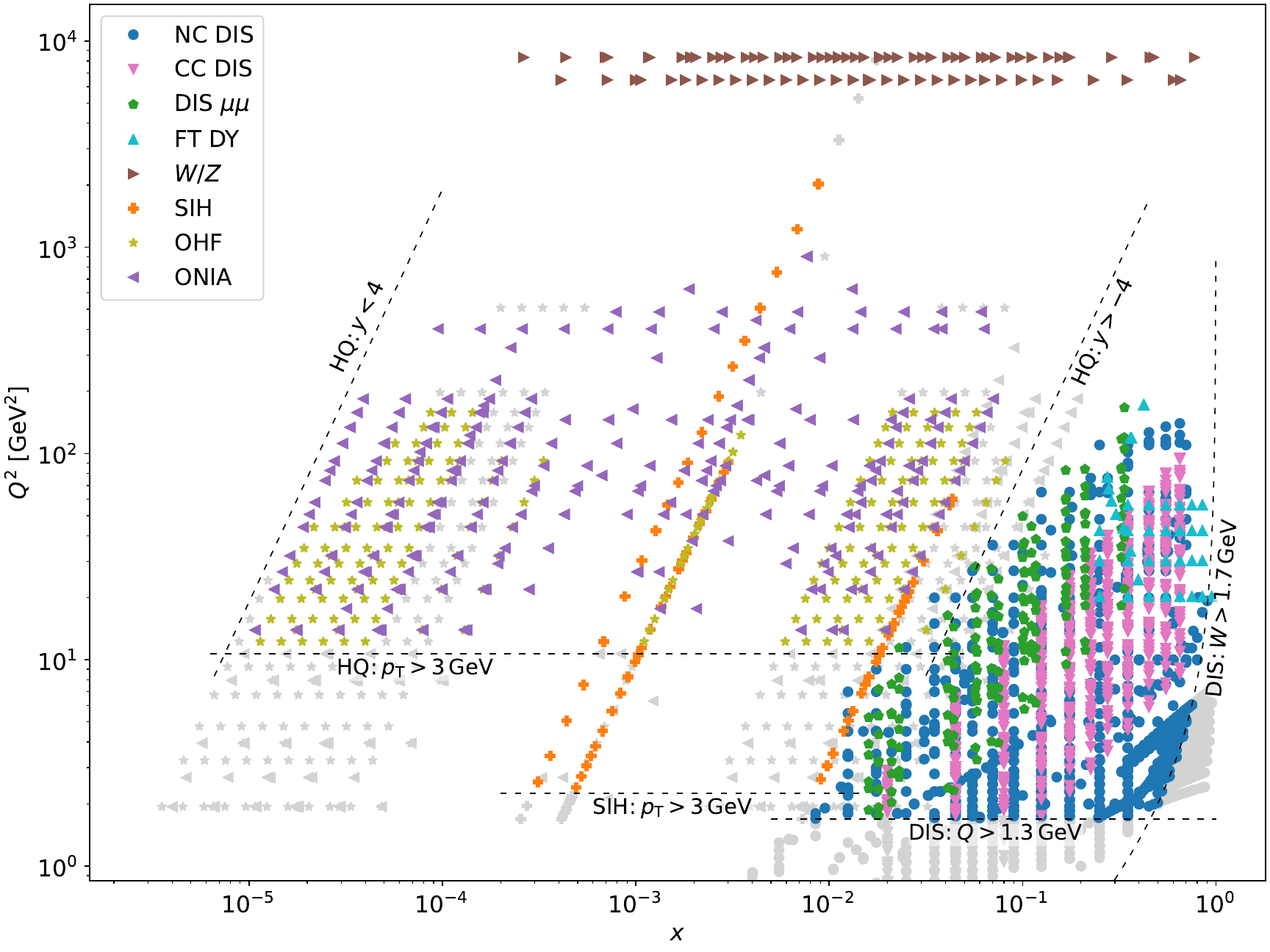}
    \caption{Experimental data points in the $(x, Q^2)$ plane colored by process. The dashed lines and their labels indicate additional process-specific cuts. Data points that do not enter the nCTEQ26 analysis are displayed in gray. The location of the points is computed using leading-order kinematics as described in \cref{app:approximate-kinematical-variables}.
    }
    \label{fig:x-Q2-plane}
\end{figure*}

In this section, we describe the experimental data included in the nCTEQ26 analysis together with the corresponding theoretical calculations. We start with general aspects and then describe the different data types individually. 

To provide a first general overview of the analyzed experimental data in terms of the kinematic reach, we show in \cref{fig:x-Q2-plane} all considered data points in the $(x,Q^2)$ plane together with additional selection cuts for specific data types. The precise meaning of the variables $x$ and $Q^2$ depends on the underlying scattering process and is often only approximate, as it is derived from leading-order kinematics (cf.\ \cref{app:approximate-kinematical-variables}). In addition, the incoming parton flavors and associated production channels are also not specified, so that this figure only provides a rough impression of where to expect constraints on nPDFs from the various datasets. Furthermore, the actual constraining power depends strongly on the relative precision of the data. Nevertheless, there are a few aspects that are readily visible from \cref{fig:x-Q2-plane}. Most prominent is the coverage of the low-$x$ region ($x\lesssim10^{-4}$), where the only constraints originate from heavy-quark(onium) data. Moving up in $x$, in addition to heavy-quark(onium) data we observe constraints from $W/Z$-boson and single-inclusive hadron (SIH) production starting between \mbox{$x\sim 10^{-4}$} and \mbox{$x\sim 10^{-3}$}. The region above \mbox{$x\sim 10^{-2}$} is covered by neutral-current (NC) and charged-current (CC) DIS, and the highest $x$ region receives constraints from NC DIS and fixed-target (FT) Drell-Yan (DY) data. In total, our analysis includes 3598 data points spread across different processes. NC and CC DIS comprise 1277 and 1242 points, respectively. The DY process contributes 92 plus 180 data points from FT and collider experiments. Furthermore, we have 112 SIH data points from RHIC and LHC. Finally, we include 695 data points from the heavy-quark(onium) production process, out of which 209 correspond to open heavy-flavor (OHF) production (i.e.\ of $D^0$ mesons), whereas the remaining data points originate from prompt and non-prompt quarkonium (ONIA) production. It is worth noting that the proton-lead data from the LHC contributes a total of 953 data points. Note that as a matter of principle we do not consider data involving collisions of more than one nucleus (with the exception when one of them is a deuteron). This is because data from $AA$ collisions are likely to be more susceptible to final state effects. That includes also the ultraperipheral collision (UPC) data~\cite{ATLAS:2024mvt,CMS:2025jjx,CMS:2023snh}, in which case we believe that theoretical uncertainties related to modeling of the corresponding predictions are not sufficiently advanced at the moment to keep uncertainties under control.

We perform our analysis at NLO QCD. The theoretical predictions are either calculated directly at this order or by using interpolating grid techniques to reduce the computational effort. Additionally, for heavy-quark(onium) production a data-driven approach is partially used. The details are described in the following sections, each dedicated to a specific data type.

\subsection{Charged-lepton DIS data}
\label{subsec:dis}

The electron and muon NC DIS data are quite numerous. They provide information on a wide range of nuclear targets and thus form the backbone of any nPDF global analysis. In our new nCTEQ26 analysis, we use the same DIS datasets, listed in \cref{tab:dis1data,tab:dis2data}, that we included in our previous global release of nCTEQ nPDFs~\cite{Kovarik:2015cma}. Most of these datasets are also used in analyses performed by other groups, cf.\ Refs.~\cite{AbdulKhalek:2022fyi,Eskola:2021nhw,Helenius:2021tof}. In addition, we include the JLab DIS data from the CLAS and Hall~C experiments that we first analyzed in Ref.~\cite{Segarra:2020gtj} and further add the E12-10-008 data from Hall C~\cite{HallC:2022utd} that were not included in that study. These data provide particularly precise constraints in the high-$x$ EMC region and are listed in \cref{tab:jlabdata}. All the NC DIS data are reported as ratios of observables measured on one nucleus to the corresponding observable measured on a different nucleus (the deuteron in the majority of cases). For the earlier datasets listed in \cref{tab:dis1data,tab:dis2data}, the used observables are either $F_2$ structure functions or DIS cross sections, whereas in the case of the JLab data they are always cross sections. In many of the earlier experiments tabulated in \cref{tab:dis1data,tab:dis2data}, the experimental data were originally corrected to be isoscalar, which limits their potential to distinguish between up- and down-quark distributions. To use the full potential of these data, the isoscalar corrections were undone following Ref.~\cite{Segarra:2020gtj}. There is also an interesting measurement of the mirror nuclei ${}^3$He and tritium from the MARATHON collaboration \cite{JeffersonLabHallATritium:2024las}. These data could in principle be included in our analysis but, to fully exploit their potential, would require an explicit $Z$-dependence, since the two nuclei have the same $A$.

\input{tables/data_new/DIS_NC_AoD}
\input{tables/data_new/DIS_NC_FAoFAprime}
\input{tables/data_new/DIS_NC_sigmaAosigmaD_JLAB}

In order to be able to constrain nPDFs at high $x$ (${x\gtrsim0.65}$), we relax the conservative kinematic cuts $Q > 2$ GeV and $W > 3.5$ GeV used in most of our earlier analyses \cite{Kovarik:2015cma,Muzakka:2022wey,Duwentaster:2022kpv} and rather follow our study of Ref.~\cite{Segarra:2020gtj} by employing the selection cuts
\begin{equation}
  Q > 1.3 \ \text{GeV}
  \qquad \text{and} \qquad
  W > 1.7 \ \text{GeV}.
\end{equation}
Here, $Q^2=-q^2$ is the four-momentum transfer of the virtual photon and $W^2=Q^2(1-x)/x + M_N^2$ is the mass of the photon-quark system with $M_N$ being the nucleon mass. These relaxed cuts in $Q$ and $W$ allow us to include many more NC DIS data points compared to the more conservative treatment in our earlier analyses. Specifically, the number of included data points is now 1277 and includes 416 points from the JLab experiments, whereas in earlier analyses only 415 points were included and no JLab data survived the conservative cuts. Such an increase not only allows for a more precise determination of the nPDFs in the high-$x$ region, but also provides additional indirect information on the quark PDFs in general. 

Theoretical calculations for the DIS process are performed at NLO QCD with our code \ncteqpp\ in the ACOT general-mass variable-flavor-number scheme~\cite{Aivazis:1993kh,Aivazis:1993pi,Kretzer:1998ju,Collins:1998rz}. Since we relax the kinematic cuts to allow for the inclusion of the low-$Q^2$ and high-$x$ data beyond the conventional DIS regime toward the shallow-inelastic scattering (SIS) region, additional corrections need to be considered. In particular, we include target-mass corrections (TMCs) as discussed in our recent review~\cite{Ruiz:2023ozv} and in earlier works concentrating on nucleons~\cite{Schienbein:2007gr,Kretzer:2003iu}. Also, since most of the NC DIS data are reported as ratios of structure functions or cross sections for a given nucleus to the same observable for the deuteron, which we do not fit, deuteron corrections are included as well. They have been extensively studied in the context of proton PDFs, where the deuteron data are used to improve constraints for the $d$-quark distribution~\cite{Ball:2020xqw,Owens:2012bv,Alekhin:2017fpf}. The treatment of deuteron corrections in the current analysis follows our recent work concentrating on high-$x$ nPDFs~\cite{Segarra:2020gtj} and employs the corrections determined in the CJ15~\cite{Accardi:2016qay} fit. 
We also investigated the importance of higher-twist (HT) effects following the approach of Refs.~\cite{Accardi:2016qay,Accardi:2023gyr,Segarra:2020gtj}. However, we found that they did not improve the description of the considered data and therefore did not include them in the final analysis. 
Nevertheless, further investigation is needed to disentangle the HT and TMC contributions and to more precisely characterize the transition between the DIS and SIS regimes.
In the past~\cite{Segarra:2020gtj}, we have also explored the possibility of increasing the flexibility of our input parametrization in the extreme high-$x$ region predominantly affected by Fermi motion. This was achieved by a small shift in the variable $x$, which in turn relaxed the requirement that the nPDFs vanish at \mbox{$x\to1$}. In the current study, we do not enable this \mbox{so-called} \mbox{$x$-rescaling}, as most of the data that would have benefited from this treatment are removed by the cut on $W$.

\subsection{Neutrino DIS data}
\label{subsec:nudis}

The neutrino or CC DIS data have a long history in global analyses for both proton and nuclear PDFs. Initially, they were used in Ref.~\cite{Olness:2003wz} to constrain the strange-quark distributions in the proton, which at that time lacked reliable constraints from other proton data~\cite{NuTeV:2001dfo,Kusina:2012vh,Faura:2020oom}.  More recently, questions have been raised about their compatibility with NC DIS data, as reflected by the fact that PDF groups now limit their use to selected datasets. The compatibility issues became even more evident when neutrino data were included in nPDF analyses and treated alongside NC DIS data on iron and lead targets. The first analysis to clearly demonstrate this problem came from the nCTEQ collaboration~\cite{Schienbein:2007fs} and subsequently prompted several follow-up studies from other groups~\cite{Kovarik:2010uv,Paukkunen:2010hb,Paukkunen:2013grz,Paukkunen:2013grz,Kalantarians:2017mkj}.

In the current paper, we follow the results of our recent reanalysis of the available neutrino data~\cite{Muzakka:2022wey}. This means we include only the inclusive data from the CHORUS collaboration taken on lead~\cite{CHORUS:2005cpn} and the semi-inclusive dimuon data from the NuTeV and CCFR collaborations~\cite{NuTeV:2001dfo}, which were taken on iron. We exclude the inclusive data from NuTeV~\cite{Tzanov:2005kr}, CCFR~\cite{Yang:2001rm}, and CDHSW~\cite{Berge:1989hr}. We also do not include the dimuon data from the NOMAD experiment~\cite{NOMAD:2013hbk}. However, we will later compare them to predictions obtained with our nCTEQ26 nPDFs, in \cref{subsec:nomad}. Another CC DIS dataset that could in principle be included comes from ratios of inclusive cross sections measured by MINER$\nu$A~\cite{MINERvA:2016oql}, which, however, extend down to ${Q=1}$~GeV and would therefore require both less stringent cuts and a lower input scale. The datasets included in our current analysis are summarized in
\cref{tab:nudata}, and the justification for this selection is detailed in Ref.~\cite{Muzakka:2022wey}. Note that even though we include the same CC DIS datasets as in Ref.~\cite{Muzakka:2022wey}, we use different, i.e.\ more relaxed, kinematic cuts. Specifically, we use the same cuts as for the NC DIS data,
\begin{equation}
  Q > 1.3 \ \text{GeV}
  \qquad \text{and} \qquad
  W > 1.7 \ \text{GeV}.
\end{equation}
As a result, we include many more data points in the large-$x$ region. To be precise, we now have 1068 data points for the CHORUS experiment and 174 data points for the dimuon data from NuTeV and CCFR as compared to the 824 and 150 points used in Ref.~\cite{Muzakka:2022wey}.

\input{tables/data_new/neutrino_DIS}

Theoretical calculations for the neutrino DIS cross sections are performed with the same code as for NC DIS using the ACOT heavy-quark scheme. The dimuon cross section is obtained from the neutrino DIS charm-production cross section, supplemented by an acceptance correction that accounts for charm fragmentation, semileptonic decay, detector geometry, and the applied kinematic cuts, with the correction determined from a Monte Carlo simulation. Further details of this correction are provided in \cref{app:dimu}. Note, however, that an improved treatment based on an NLO calculation for dimuons in the SACOT-$\chi$ scheme has recently been published~\cite{Helenius:2024fow,Paukkunen:2025kjb}. As in the NC DIS case, we include TMCs~\cite{Ruiz:2023ozv}. We also investigated the impact of HT corrections following our study in Ref.~\cite{Segarra:2020gtj}, but found that they do not improve, or even slightly degrade, the description of the neutrino data. This may be related to the relatively simple multiplicative form of the HT corrections employed in Refs.~\cite{Accardi:2016qay,Segarra:2020gtj}. However, a more realistic accounting for nuclear dependence could change this situation. Furthermore, HT corrections partially overlap with the TMCs and will be investigated in more detail in future work, e.g., by also considering additive HT corrections. Note that more detailed studies of TMCs, HT and other non-perturbative QCD effects are particularly important for lower neutrino-energy oscillation experiments such as DUNE, in particular in view of experimental indications~\cite{Morfin:2021ujm,Schienbein:2007fs,Schienbein:2009kk,Kovarik:2010uv,Muzakka:2022wey,Eskola:2016oht,Paukkunen:2010hb,Eskola:2021nhw,Kalantarians:2017mkj} that HT effects for neutrino weak interactions could differ from those for electron or muon interactions.

\subsection{Fixed-target Drell-Yan data}
\label{subsec:dy}

In the current analysis, we include the standard set of the FT DY lepton-pair production data from the FNAL-E772-90 and FNAL-E866-99 experiments, which are summarized in \cref{tab:DYdata}. The measurements are reported as ratios of DY differential cross sections taken on different nuclei compared to either deuteron or beryllium. The calculations for these data are performed using our \ncteqpp\ code at NLO QCD.

\input{tables/data_new/Drell_Yan}

\subsection{LHC \texorpdfstring{$W/Z$}{W/Z}-boson data}
\label{subsec:wz}

Among the proton-lead LHC data, the production of leptonically decaying vector bosons provides the cleanest probe of nuclear structure, as the colorless final state ensures that other nuclear effects such as final-state energy loss~\cite{Brodsky:1989ex,Gavin:1991qk,Gavin:1990gm,Arleo:2012hn} are not present, so that the leading-twist approximation is well justified~\cite{Qiu:2003cg}. The production of $W^{\pm}$- and $Z$-bosons as well as the DY process then not only provide important constraints on the valence-quark distributions, but also on the distributions of the strange and light sea quarks~\cite{Kusina:2012vh,Kusina:2016fxy,Kusina:2020lyz,ATLAS:2012sjl,ATLAS:2016nqi,Cooper-Sarkar:2018ufj,Faura:2020oom}. Moreover, for a significant epoch in nPDF fits these processes provided the most important constraints on the gluon distribution. Even now, as the nPDF groups have begun to incorporate other LHC data, e.g.\ from jet or heavy-quark(onium) production, that are directly sensitive to the gluon PDF, electroweak probes remain important, as they allow us to test the influence of other nuclear-matter effects on colored final states and thus provide a valuable cross-check.

The $W$- and $Z$-boson data considered in this analysis have all been obtained in proton-lead collisions at Runs I and II of the LHC. We use the absolute cross sections depending on the rapidity of the $Z/\gamma^*$ boson or on the pseudorapidity of the charged lepton originating from the leptonic decay of the $W$ boson. The individual datasets used in the analysis are summarized in \cref{tab:WZdata}. New data that were not used in our previous analyses are indicated by bold IDs.

Theoretical calculations for these processes are performed at NLO QCD using MCFM 6.8~\cite{Campbell:2015qma} interfaced with APPLgrid~\cite{Carli:2010rw}, which allows us to produce fast and reliable predictions without resorting to approximations. Since the data lie entirely in a perturbative region and no resummations are needed, we do not apply any kinematic selection cuts.

\input{tables/data_new/WZ}

\subsection{RHIC/LHC single-inclusive hadron data}
\label{subsec:sih}

Inclusive production of light hadrons provides valuable insights into the PDFs, in particular of the gluon, at low values of $x$. To obtain the cross section for the single-inclusive production of a hadron $h$, the PDFs of the two initial state particles $f^A_i$ and $f^B_j$ are convolved with the cross section of the partonic subprocess $\hat{\sigma}_{ij\to k}$ and the final state fragmentation function (FF) $D^{h}_k$,
\begin{align}
    \sigma_{A{+}\mathrm{B}\to h+X} = \sum_{i,j,k} f_{i}^{A} \otimes f_{j}^{\mathrm{B}} \otimes \hat{\sigma}_{ij\to k}  \otimes D_k^{h} \ . \label{eq:xsec}
\end{align}
Here, the sum runs over all possible subprocesses with two incoming partons $i$ and $j$ producing an outgoing parton $k$.

This expression introduces technical and theoretical challenges. On the technical side, the nested convolutions are computationally expensive in a PDF fit, where cross sections for hundreds of data points must be evaluated thousands of times each. We perform these calculations with INCNLO~\cite{INCNLO} and precompute the convolution of the proton (or deuteron in the case of RHIC) PDF using our updated proton baseline with the partonic cross section and the FF, such that during fitting only the convolution with the nuclear PDF has to be computed. Additionally, we have modified the INCNLO code to improve the numerical stability of the computation, as outlined in Ref.~\cite{Helenius:2014jza}. The theoretical challenge arises because the FFs are non-perturbative objects that must be determined from data, much like PDFs. We have performed a comprehensive analysis of the impact of the choices of FF and perturbative scales on inclusive hadron production in Ref.~\cite{Duwentaster:2021ioo}. In that analysis, we have determined that setting the renormalization scale $\mu_r$, the initial factorization scale $\mu_i$, and the fragmentation scale $\mu_f$ all equal to $p_T/2$ yields the most accurate description of the measured cross sections for transverse momenta $p_T>3$\,GeV. Regarding FFs, we found that using DSS14~\cite{deFlorian:2014xna} for pions and DSS17~\cite{deFlorian:2017lwf} for kaons is a good choice, since these FFs agree well with other analyses and their Hessian uncertainties span a range similar to the spread of other FF determinations. To account for the FF uncertainty in our PDF fit, we have added it in quadrature with the experimental systematic uncertainties. Note that in the meantime more recent fits of FFs have become available~\cite{Soleymaninia:2026edd,Soleymaninia:2026xjq,Soleymaninia:2023dds,Borsa:2022vvp,Soleymaninia:2022alt,AbdulKhalek:2022laj,Borsa:2021ran,Abdolmaleki:2021yjf,Soleymaninia:2018uiv,Bertone:2018ecm,Hirai:2016loo}. Their impact will be explored in a follow-up study, but is expected to be small given the subleading importance of the SIH data in our global dataset.

In the current analysis, we consider data for the production of pions and kaons from RHIC (PHENIX and STAR) collected at $\sqrt{s_{NN}}=200$ GeV per nucleon and from the LHC (ALICE) collected at $\sqrt{s_{NN}}$ of 5.02 and 8.16 TeV. The specific datasets included in our fit are listed in Tab.~\ref{tab:SIHdata}. For all of these data, we use a kinematic selection cut 
\begin{equation}
p_T > 3\ \mathrm{GeV}
\end{equation}
excluding data with small transverse momenta, which can not be described well with our perturbative calculations. In some cases, the minimal values of the scaling variable $z$ used in the FFs further limit the kinematics of the used data. For details, we refer the reader to Sec.~III.A in Ref.~\cite{Duwentaster:2021ioo}. After these cuts, we end up with a total of 112 SIH data points. Note that there is also a new dataset for $\pi^0$ production from LHCb~\cite{LHCb:2022tjh}, which we do not consider here. The reason is that we do not expect this dataset to have an appreciable impact on the gluon distribution in the presence of heavy-quark(onium) and $W/Z$-boson production data, but we may include it in future studies.

\input{tables/data_new/SIH}

\subsection{LHC heavy-quark(onium) data}
\label{subsec:hq}

\input{tables/data_new/HQ_pPb}

Measurements of open heavy-flavor (OHF) meson and quarkonium (ONIA) production, hereafter collectively referred to as heavy-quark (HQ) production, at the LHC have resulted in a large dataset that is particularly sensitive to the gluon PDF due to the dominance of the gluon-gluon channel in HQ production at high energies. This was initially shown in the case of proton-proton collisions in Ref.~\cite{PROSA:2015yid}. Later, the reweighting study presented in Refs.~\cite{Kusina:2017gkz, Kusina:2020dki} demonstrated that the LHC HQ data can constrain the nuclear gluon PDF at very small $x\approx10^{-5}$. This observation was then followed by a full analysis from the nCTEQ collaboration~\cite{Duwentaster:2022kpv}. More recently, the $D$-meson production data have been used in nuclear PDF analyses by the EPPS and NNPDF groups to reduce the uncertainty on the gluon PDF~\cite{Eskola:2019bgf,Eskola:2021nhw,Khalek:2022zqe}. In addition, the influence of OHF data on the PDFs of lighter nuclei in SMOG2~\cite{BoenteGarcia:2024kba} was recently investigated in Ref.~\cite{Flore:2025uhv}.

As in our dedicated nCTEQ15HQ analysis \cite{Duwentaster:2022kpv}, we use a wide range of available HQ data from proton-lead collisions at the LHC and include data from $D^0$-meson, $J/\psi$, $\psi(2S)$, and $\Upsilon(1S)$ production. Table~\ref{tab:HQdatapPb} lists all the included datasets, indicating also with bold IDs the new datasets that were not used in Ref.~\cite{Duwentaster:2022kpv} and are included here for the first time. In order to stay in the perturbative region and to be able to obtain a good description of the data, we use the kinematic selection cuts
\begin{equation}
p_T > 3\ \mathrm{GeV}
\quad {\rm and } \quad
-4.1 < y < 4.1
\end{equation}
excluding data with small transverse momentum and extreme rapidities. For certain datasets, we applied additional cuts ensuring they can be described by our calculations.\footnote{Specifically, these cuts are \mbox{$p_T^{\mathrm{ID}\,=\,3101,3122,3123} {<} 16$ GeV}, \mbox{$p_T^{\mathrm{ID}\,=\,3133,3134} {<} 15$ GeV}, 
and 
\mbox{$2 < y^{\mathrm{ID}\,=\,3133} {<} 4$}, 
\mbox{$-4 < y^{\mathrm{ID}\,=\,3134} {<} -2$}, where IDs 3133 and 3134 correspond to the same Run-II $D^0$ measurement from LHCb in the forward and backward regions, respectively.} With these kinematic selections, we are left with 695 HQ data points, which includes 209 OHF and 486 quarkonium points and increases the number of fitted HQ data points in Ref.~\cite{Duwentaster:2022kpv} by almost 150.

In the case of OHF production, there are a number of established perturbative QCD approaches to provide NLO theoretical predictions, cf.~Refs.~\cite{Olness:1997yc,Kniehl:2004fy,Kniehl:2005mk,Kniehl:2012ti,Helenius:2018uul,Cacciari:1998it,Cacciari:2012ny}. However, the situation is different in the case of quarkonium production, see e.g.~Ref~\cite{Lansberg:2019adr}. For this reason, we decided to analyze the HQ data in two different ways. In our main nCTEQ26 fit, we employ the data-driven approach that we have already tested in Ref.~\cite{Duwentaster:2022kpv}. In addition, we also consider here an alternative fit, nCTEQ26OHF, that includes only the OHF data, relies on NLO QCD calculations using the GM-VFNS scheme~\cite{Wissmann:2024bca} and will be discussed in detail in \cref{subsec:gluon}. The data-driven method used in the main fit was originally described in Ref.~\cite{Kom:2011bd} before being applied to nuclear PDFs in the reweighting study in Refs.~\cite{Kusina:2017gkz, Kusina:2020dki}. Under the assumptions that the gluon-gluon channel contributes the dominant fraction of the total cross section and that it is sufficient to consider subprocesses with 2~$\rightarrow$~2 kinematics, the cross section can be expressed using the Crystal-Ball (CB) parametrization as
\begin{align*}
    &\sigma(A B \rightarrow \mathcal{Q}+X)= \\
    &\int \mathrm{d} x_{1} \mathrm{~d} x_{2} f_{1,g}\left(x_{1}, Q\right) f_{2,g}\left(x_{2},Q\right) \frac{1}{2 \hat{s}} \overline{\left|\mathcal{A}_{g g \rightarrow \mathcal{Q}+X}\right|^{2}} \mathrm{dPS}, 
\end{align*}
where
\begin{align} \label{eq:crystal-ball-matrix-element}
    \begin{split}
        &\overline{\left|\mathcal{A}_{g g \rightarrow \mathcal{Q}+X}\right|^{2}}= \frac{\lambda^2\kappa\hat{s}}{M_\mathcal{Q}^2} e^{a|y|}\\ 
        &\times
        \begin{cases} 
            e ^{ -\kappa \frac{p_{T}^{2}}{M_{\mathcal{Q}}^{2}}} & \text { if } p_{T} \leq\left\langle p_{T}\right\rangle \\ e ^ {-\kappa \frac{\left\langle p_{T}\right\rangle^{2}}{M_{\mathcal{Q}}^{2}}}\left(1+\frac{\kappa}{n} \frac{p_{T}^{2}-\left\langle p_{T}\right\rangle^{2}}{M_{\mathcal{Q}}^{2}}\right)^{-n} & \text { if } p_{T}>\left\langle p_{T}\right\rangle
        \end{cases}
    \end{split}
    \raisetag{65pt}
\end{align}
and the five parameters $\lambda$, $\kappa$, $\left<p_T \right>$, $n$ and $a$ are fitted to $pp \rightarrow \mathcal{Q}+X$ data for each final state $\mathcal{Q}$ independently. As described in Ref.~\cite{Duwentaster:2022kpv}, we extended the CB parametrization by introducing the fifth parameter $a$, which was not present in the original parametrization~\cite{Gaiser:1982yw}, to allow for a more accurate reproduction of the rapidity dependence. The data we use in the proton CB baseline fit are listed in \cref{tab:HQdatapp}. Their experimental uncertainties are propagated into the extracted CB parameters using the Hessian approach, and the resulting CB-parameter uncertainties are then added in quadrature to the experimental uncertainties of the proton-lead data used in the nuclear analysis. Since our previous analysis~\cite{Duwentaster:2022kpv}, the ALICE, CMS, and LHCb collaborations have released new data from both proton-proton and proton-lead collisions, which are highlighted in Tabs.\ \ref{tab:HQdatapPb} and \ref{tab:HQdatapp} with bold IDs. Our new fit of the CB parameters to the proton-proton data has been performed as described in Ref.~\cite{Duwentaster:2022kpv}, but with our updated proton PDF baseline and these new datasets, but still applying the same kinematic cuts of $p_T>3$ GeV and $-4<y_{cms}<4$. The resulting best-fit parameters and their $\chidof$ values are presented in \cref{tab:CrystalBallBaseline}. As in nCTEQ15HQ, the agreement in the proton-baseline fit between data and theory is very good with the $\chidof$ for the different final states being either around or below one. Comparing the different final states, one observes that the $\chidof$ increases with the number of data points. This can be explained by the fact that the CB parametrization has to compromise more as the number of points increases. Note also that the $\langle p_T \rangle$ fit parameter becomes negative for prompt $D^0$ and prompt $\Upsilon(1S)$ production, which implies that only the second case in \cref{eq:crystal-ball-matrix-element} is used in the predictions for these final states.

Another potentially interesting HQ dataset comes from the SMOG experiment at LHCb~\cite{LHCb:2014vhh}. It is, however, challenging to include in a global analysis because of the involved low transverse momenta. Available are currently first charm ($D$-meson and $J/\psi$) measurements~\cite{LHCb:2018jry} in $p$He and $p$Ar collisions at $\sqrt{s_{NN}}=86.6$ GeV and $\sqrt{s_{NN}}=110.4$ GeV, respectively, as well as the corresponding measurements in $p$Ne collisions at $\sqrt{s_{NN}}=68.5$ GeV for $J/\psi$~\cite{LHCb:2022sxs} and $D$-mesons~\cite{LHCb:2022cul}. In the meantime, data with much larger statistics have been taken by the SMOG2 experiment~\cite{2707819} and can potentially be included in future global analyses~\cite{Flore:2025uhv}.

\input{tables/data/HQ_pp}
\input{tables/crystal_ball}
\section{Results}\label{sec:res}

We now turn to the main results of the global analysis. 
First, we assess the overall fit quality and reliability, discussing the description of each dataset and any notable tensions. 
We then examine how the experimental information constrains the fit by analyzing data sensitivities and correlations, which allows us to identify the key processes and kinematic domains that drive the individual parton flavors. 
Finally, we present the extracted nuclear PDFs, emphasizing their flavor dependence, uncertainties, behavior in the relevant $(x,Q)$ regions, and their dependence on mass number $A$, tracing how the nuclear modifications evolve from heavy to light nuclei. 
A comparison to results obtained in other nPDF analyses and a detailed discussion focusing on strange and gluon distributions is postponed to the following sections.

\subsection{Fit quality}
\label{subsec:fit}

We begin by discussing the statistical quality of the fit as measured by the total $\chi^2$ defined in \cref{eq:chi2-correlated} and how it is distributed across different types of scattering processes, nuclei and individual experiments. We begin with the total $\chi^2$ for the nCTEQ26 analysis, which is 3357.97 for 3598 data points and 36 free parameters, resulting in $\chidof=0.943$. This indicates a good fit of the data overall.

\begin{figure}[tb]
    \centering
    \includegraphics[width=0.49\textwidth]{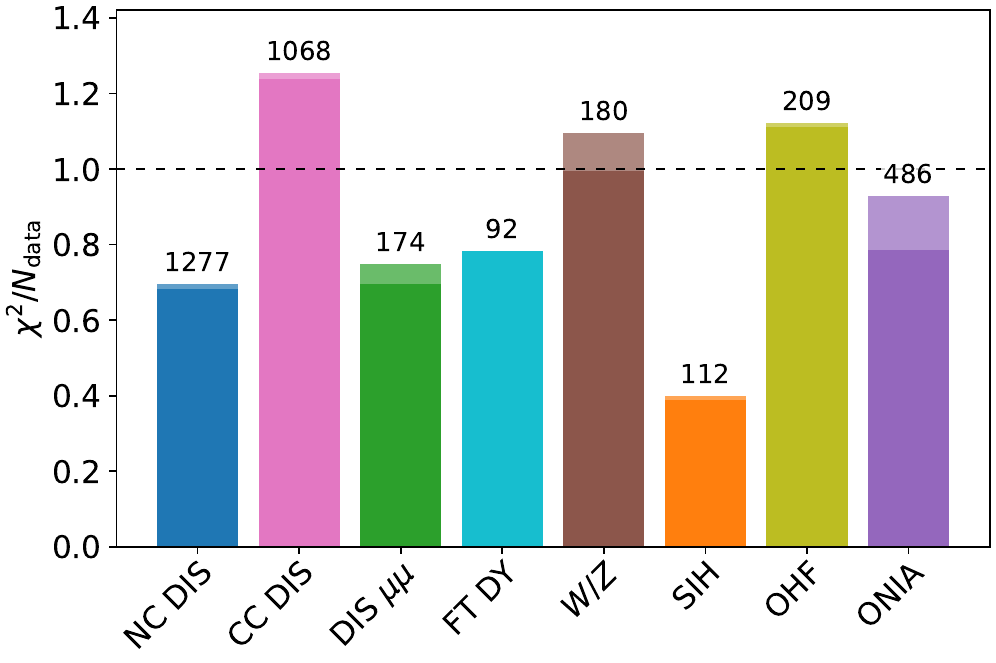}
    \caption{$\chidata$ breakdown per type of scattering process. The number of data points for each process is indicated above the bars. Additionally, the lighter shaded parts of the bars indicate the $\chi^2$-contributions originating from the normalization penalty.
    }
    \label{fig:chi2DATAtype}
\end{figure}
\begin{figure*}[ht!]
    \centering
    \includegraphics[width=\textwidth]{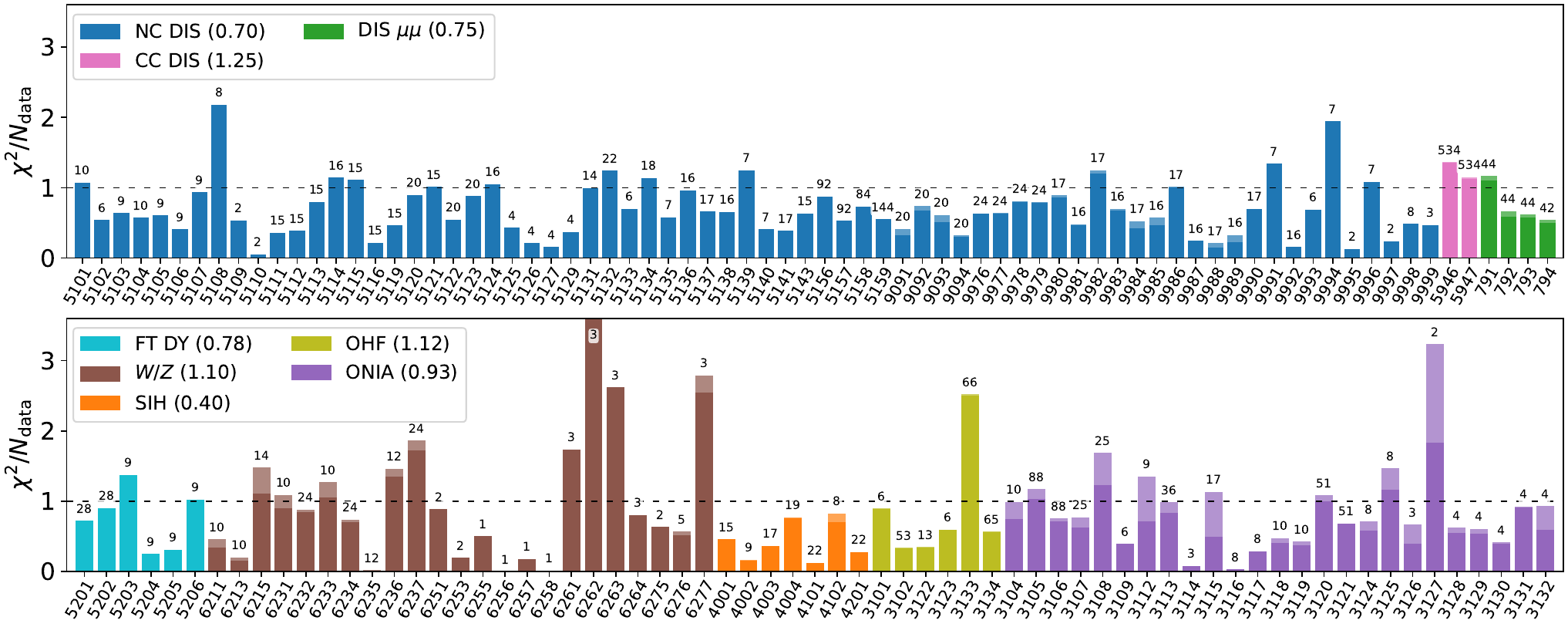}
    \caption{$\chidata$ for individual experiments used in the fit. The number of data points for each dataset is indicated above the bars. The experiment IDs indicated on the $x$-axis are defined in Tabs.~\ref{tab:dis1data}-\ref{tab:HQdatapPb}.}
    \label{fig:chi2dataset}
\end{figure*}

Next, we show in \cref{fig:chi2DATAtype} the $\chidata$ for the different types of scattering processes used in the fit. As one can see, most data types are described with $\chidata\sim1$ as expected for a good fit. The most significant deviation from this behavior appears in the single inclusive hadron (SIH) data from RHIC and LHC, which yields $\chidata\sim0.4$, i.e.\ much smaller than one and potentially indicating overfitting. This phenomenon was already observed in our dedicated analysis of the SIH data~\cite{Duwentaster:2021ioo}. The two main reasons are (i) large uncertainties inherent in these data and (ii) the necessary exclusion of low-transverse momentum data (cf.\ \cref{subsec:sih}), which are typically measured most precisely. Furthermore, theoretical uncertainties associated with the FFs have been added to the experimental uncertainties in quadrature, and they are often significant and sometimes even dominate the total uncertainty. Note that SIH production is mostly induced by the gluon-gluon channel in a region of $x$ that overlaps with other data (cf.\ \cref{fig:x-Q2-plane}). As a result, the impact of these data, which contribute only about one percent of the total $\chi^2$, on the fit is rather limited. A deviation from $\chidata\sim1$ towards larger values is observed for the charged-current (CC) neutrino DIS data from the CHORUS experiment with $\chidata\sim1.3$, indicating that the description of the data is not perfect, but still acceptable. The question of the compatibility of neutrino DIS data with NC DIS data has already been extensively discussed elsewhere~\cite{Schienbein:2007fs,Kovarik:2010uv,Paukkunen:2010hb,Paukkunen:2013grz,Muzakka:2022wey}. In this work, we follow our recent analysis~\cite{Muzakka:2022wey}, in which we concluded that the CHORUS inclusive CC DIS data and the DIS dimuon data from the NuTeV and CCFR experiments can be fitted together with other data without introducing large tensions. This observation still holds despite the fact that we have extended our kinematic cuts and included more CC DIS data in the high-$x$ and low-$Q$ regions and many more points from other processes. In \cref{fig:chi2DATAtype}, we also show separately the contribution to the $\chi^2$ originating from the normalization penalty, given by the last term in \cref{eq:chi2-with-nuisance-params}. Non-zero values indicate that the normalization of a given dataset was adjusted for a better fit quality. We can see that these contributions are typically small with the exception of $W/Z$-boson and quarkonium (ONIA) data, where they can reach 5\% and 10\% of the total $\chi^2$, respectively.

\begin{figure}
    \centering
    \includegraphics[width=0.45\textwidth]{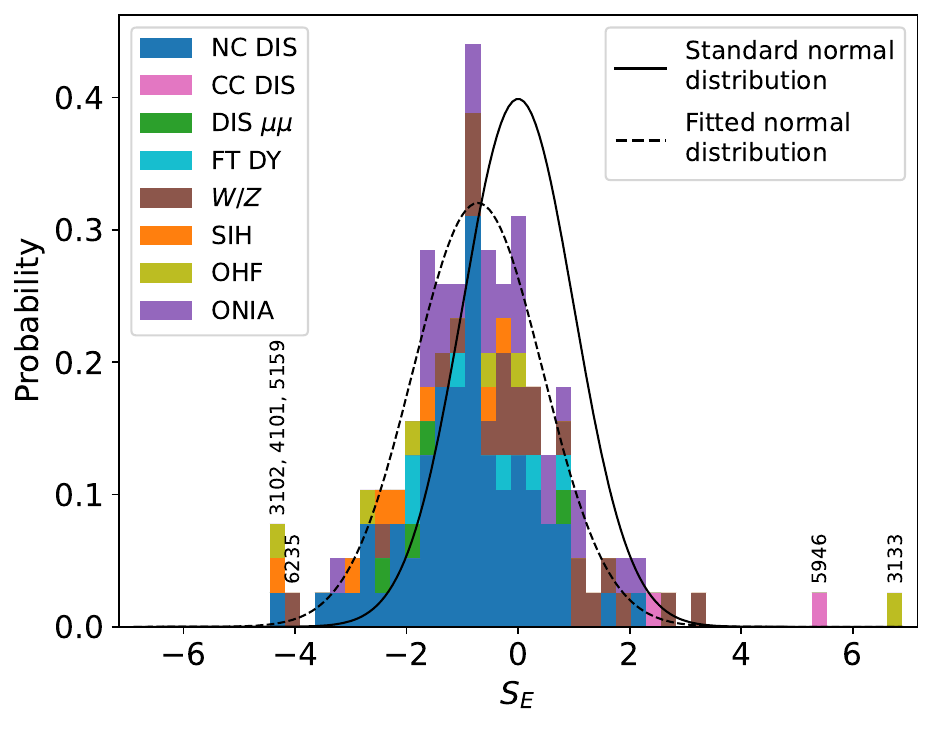}
    \caption{$S_E$ distribution stacked by data type. Outliers with $|S_E| \geq 4$ are annotated by their dataset IDs.
    }
    \label{fig:SE}
\end{figure}

To examine how well each experimental dataset is described individually, we show their $\chidata$ in Fig.~\ref{fig:chi2dataset}. There are only a few outliers deviating appreciably from $\chidata\sim1$ towards higher or smaller values, which will be discussed in more detail below. In \cref{fig:SE}, we then assess their distribution in the effective Gaussian variable $S_E = \sqrt{2 \chi^2_E} - \sqrt{2 N_E - 1}$, where $\chi^2_E$ is associated with an experimental dataset $E$ with $N_E$ points. For an ideal fit, $S_E$ is distributed like a standard normal distribution. We see that the $S_E$ distribution of the nCTEQ26 fit is consistent with the normal distribution within $1\sigma$, but slightly wider and shifted towards the negative direction. Highlighted in \cref{fig:SE} are the dataset IDs of outliers at the edges of the distribution, i.e.\ datasets with ${|S_E| > 4}$. In general, NC DIS, DIS $\mu \mu$ and OHF datasets tend to be fitted better than CC DIS and $W/Z$-boson production, and the FT DY and quarkonium datasets are distributed approximately evenly around $S_E=0$. Datasets that are fitted unusually well come from the NC DIS, SIH and OHF categories, while datasets that are difficult to describe include the newest LHCb Run-II $D^0$ measurement and the CC DIS CHORUS data. It is particularly interesting that datasets 3102 (LHCb Run-I prompt $D^0$ \cite{LHCb:2017yua}) and 3133 (LHCb Run-II prompt $D^0$ \cite{LHCb:2022dmh} at forward rapidity) are at opposite ends of the distribution.

\begin{figure}
    \centering
    \includegraphics[width=0.45\textwidth]{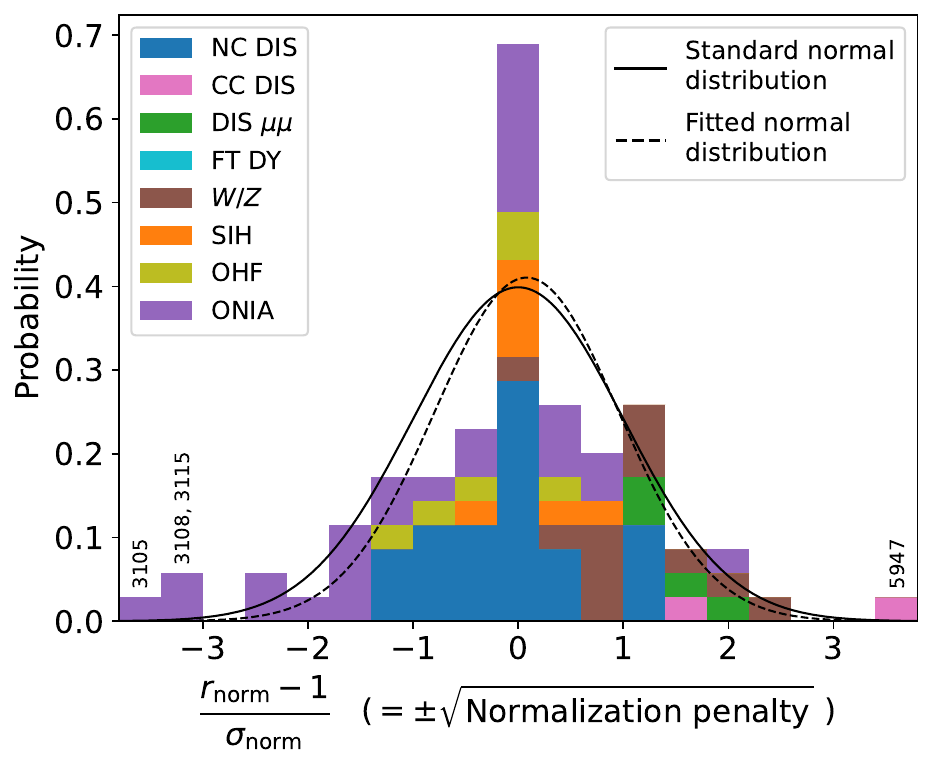}
    \caption{Distribution of normalization penalties \mbox{$(r_{\rm norm}-1)/\sigma_{\rm norm}$} stacked by data type. Datasets requiring a large normalization correction are annotated by their dataset IDs.}
    \label{fig:normalization-histogram}
\end{figure}

Shifting our focus to the normalization penalty breakdown per experiment, we observe in \cref{fig:chi2dataset} that the large normalization penalties in the $W/Z$-boson and quarkonium data noticed in \cref{fig:chi2DATAtype} do not originate from a single dataset, but rather from a combination of contributions from several datasets. The distribution of the square root of the applied normalization penalties as defined by the last term in \cref{eq:chi2-with-nuisance-params} is shown in \cref{fig:normalization-histogram}. In an ideal case, these values would follow again a normal distribution. The reason why we chose this metric instead of the normalization coefficient itself is that each coefficient follows its own normal distribution with the width corresponding to the normalization error. The first two moments are consistent with the Gaussian distribution, which is indicated by the close alignment of the fitted analytic curve (dashed line) and the ideal normal distribution (solid line). 
The central peak is more pronounced than expected, but the tails are symmetric. Most normalization adjustments are well within expectation, with the exception of a few datasets. 
Most notably the CHORUS anti-neutrino data (ID~5947) require a large positive multiplicative correction $r_{\text{norm}}$ to the theory prediction, which corresponds effectively to a reduction of the data normalization. Conversely, three quarkonium datasets (LHCb $J/\psi$ IDs~3105, 3108 and CMS $\psi(2S)$ ID~3115) require large opposite adjustments.

To summarize this section, we have obtained a good description of the fitted data in the nCTEQ26 global nPDF analysis with a total value of $\chi^2$ only slightly below one. We also found that the distribution of the individual values of $\chi_E^2$ per experimental dataset $E$ was close to the ideal case of Gaussian random fluctuations between experimental data and theoretical predictions. We did observe slight tensions for a small number of datasets, but this is of course expected in global analyses of PDFs~\cite{Kovarik:2019xvh}.

\begin{figure*}
    \centering
    \subfloat[\centering NMC Sn/C data]{
        \label{fig:dataTH_DIS_NMC}
        \includegraphics[height=0.28\textwidth]{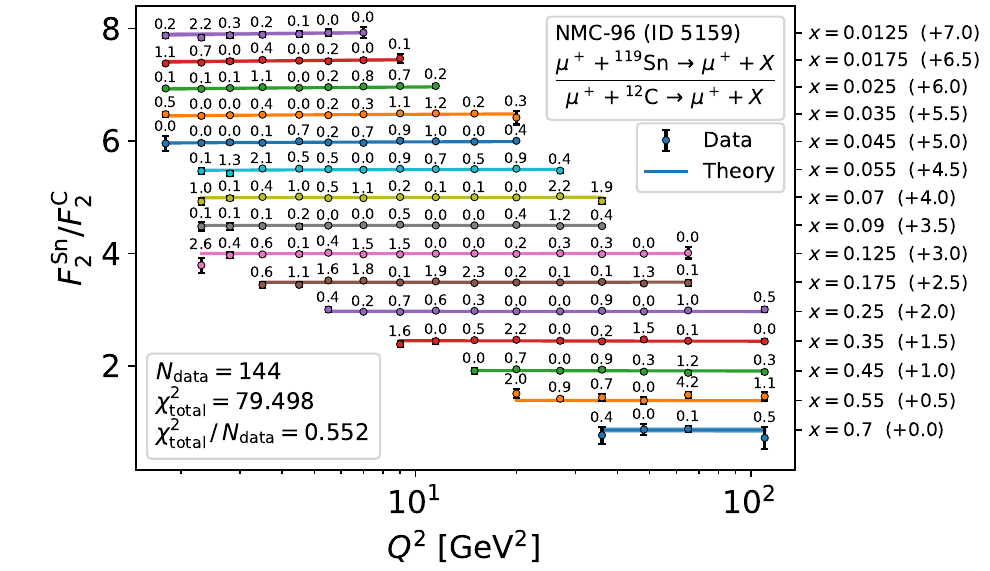}
    }
    \subfloat[\centering Hermes Kr/D data]{
        \label{fig:dataTH_DIS_Hermes}
        \includegraphics[height=0.28\textwidth]{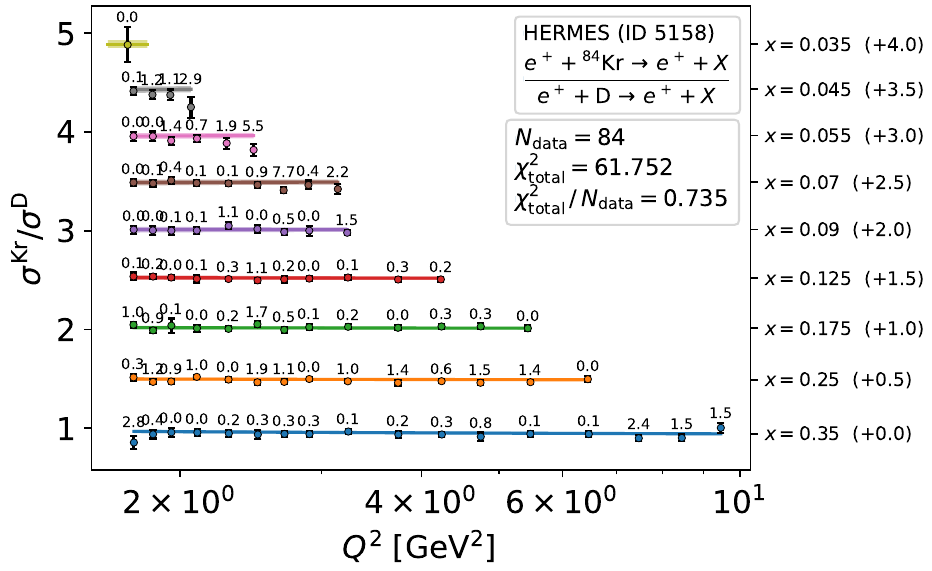}}
\\
    \subfloat[\centering JLab Al/D data]{
        \label{fig:data-vs-theory-jlab-A-D}
        \includegraphics[height=0.28\textwidth]{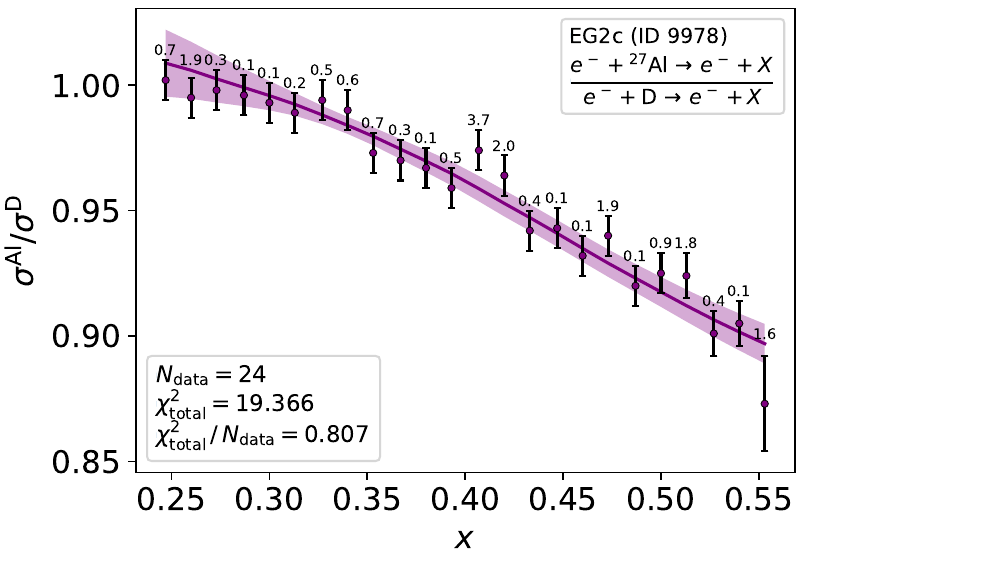}
    }
    \subfloat[\centering JLab C/D data]{
        \label{fig:data-vs-theory-jlab-C-D}
        \includegraphics[height=0.28\textwidth]{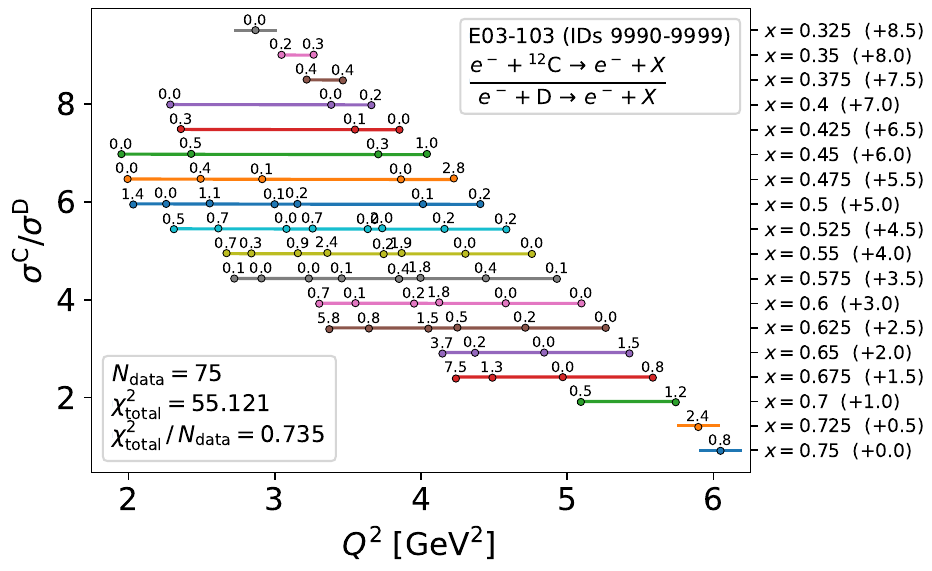}
    }
    \caption{Comparison of nCTEQ26 theory predictions with NC DIS data from (a) NMC Sn/C, (b) Hermes Kr/D, (c) JLab Al/D, and (d) JLab C/D, including the corresponding values of $\chi^2$ per data point. In panels (a), (b), and (d) the ratios corresponding to different values of $x$ are separated vertically in steps of 0.5.}
    \label{fig:dataTH_DIS} 
\end{figure*}

\subsection{Description of selected datasets}
\label{subsec:datavstheory}

In order to assess the fit quality in more detail, we examine in the following the agreement between the theoretical predictions obtained with our nCTEQ26 PDFs and a representative selection of measurements, ordered by scattering process.

We start with NC DIS data, which are in general very well described. This can already be seen in Fig.~\ref{fig:chi2dataset}, where the only significant exception to this rule comes from the EMC-88 data for the ratio of $F_2$ structure functions on tin and deuteron (ID 5108). It is well known that this dataset exhibits deviations from other NC DIS data~\cite{Kovarik:2015cma,deFlorian:2011fp}, so that we do not discuss this here again. Instead, we compare in Fig.~\ref{fig:dataTH_DIS} a representative selection of nCTEQ26 theory predictions to other NC DIS data. Specifically, we show the NMC data for Sn/C (ID 5159), Hermes data for Kr/D (ID 5158), JLab data for Al/D (ID 9978), and JLab data for C/D \mbox{(IDs 9990-9999)}. These data range from heavy to light nuclei and provide numerous data points over a relatively large kinematic range. In panels (a), (b), and (d), we plot ratios of $F_2$ structure functions and cross sections as a function of $Q^2$ for a number of different values of $x$ separated vertically in steps of 0.5. Due to this condensed representation, PDF uncertainties are not clearly visible, but in the majority of cases the theory predictions agree with the experimental data within the uncertainties. In panel (c), we plot instead the ratio of cross sections as a function of $x$ without differentiating the dependence on $Q^2$, which is very mild in the considered measurement ($Q^2\in[1.72,3.37]$~GeV$^2$). It then becomes clear that the agreement between data and theory is well within the PDF and/or data uncertainties. Similar observations hold for all NC DIS data. The pull of each individual data point on these plots is indicated by its contribution to the total $\chi^2$ printed above the data points. Note, that these per-point $\chi^2$ values are calculated following the prescription detailed in \cref{subsec:min}. Furthermore, the theory predictions are shifted accordingly, to facilitate the visual comparison. 

\begin{figure*}
    \setcounter{subfigure}{0}
    \subfloat[JLab Hall C (2022) $^{9}$Be/D]{
        \centering
        \hspace*{-2em}
        \includegraphics[width=0.33\linewidth, page=2]{figs/data_vs_theory/selected_datasets_DIS.pdf}
        \label{subfig:dvt_9092}
    }
    \hspace{0.5em}
    \setcounter{subfigure}{2}
    \subfloat[JLab Hall C (2022) $^{11}$B/D]{
        \centering
        \hspace*{-3em}
        \includegraphics[width=0.33\linewidth, page=4]{figs/data_vs_theory/selected_datasets_DIS.pdf}
        \label{subfig:dvt_9094}
    }
    \hspace{0.5em}
    \setcounter{subfigure}{4}
    \subfloat[JLAB Hall C (2009) $^{12}$C/D, 32 deg]{
        \centering
        \hspace*{-3em}
        \includegraphics[width=0.33\linewidth, page=6]{figs/data_vs_theory/selected_datasets_DIS.pdf}
        \label{subfig:dvt_9994}
    }\\
    \setcounter{subfigure}{1}
    \subfloat[JLab Hall C (2022) $^{10}$B/D]{
         \centering
         \hspace*{-2em}
         \includegraphics[width=0.33\linewidth, page=3]{figs/data_vs_theory/selected_datasets_DIS.pdf}
         \label{subfig:dvt_9093}
    }
    \hspace{0.5em}
    \setcounter{subfigure}{3}
    \subfloat[JLab Hall C (2022) $^{12}$C/D]{
        \centering
        \hspace*{-3em}
        \includegraphics[width=0.33\linewidth, page=1]{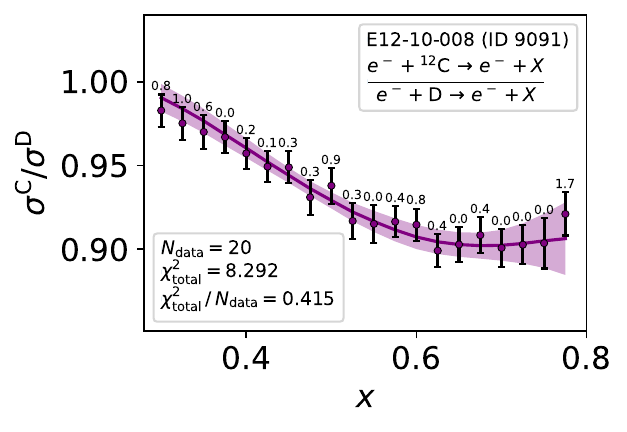}
        \label{subfig:dvt_9091}
    }
    \hspace{0.5em}
    \setcounter{subfigure}{5}
    \subfloat[JLab Hall C (2009) $^{12}$C/D, 40 deg]{
        \centering
        \hspace*{-3em}
        \includegraphics[width=0.33\linewidth, page=5]{figs/data_vs_theory/selected_datasets_DIS.pdf}
        \label{subfig:dvt_9992}
    }
    \caption{Comparison of nCTEQ26 theory predictions with NC DIS data from JLab Hall C on (a) $^{9}$Be, (b) $^{10}$B, (c) $^{11}$B and (d-f) $^{12}$C targets collected at beam energies of (a-d) 11 GeV~\cite{HallC:2022utd} and (e,f) 6 GeV~\cite{Seely:2009gt} with the corresponding values of $\chi^2$ per data point.}
    \label{fig:proplematic_datasets2}
\end{figure*}

In \cref{fig:proplematic_datasets2}, we compare our nCTEQ26 theory predictions with the recent 11 GeV NC DIS data on $^{10}$B, $^{11}$B, $^{11}$Be, and $^{12}$C targets from JLab Hall C~\cite{HallC:2022utd}, which have up to now not been included in any nPDF analysis. We also compare our predictions to the older 6 GeV JLab Hall C data taken on $^{12}$C~\cite{Seely:2009gt}. As before, the JLab measurements are provided as ratios of cross sections to those on the deuteron. From \cref{fig:proplematic_datasets2} (a-d), we observe that all of the new JLab data are very well described by the nCTEQ26 nPDFs yielding $\chidata<1$. Thanks to the higher beam energy, these data reach very large values of ${x\sim0.8}$. Our theoretical predictions describe the flattening of the spectrum there very well. In particular, we do not observe any degradation of the fit quality despite the fact that we include only deuteron and target-mass corrections (which largely cancel in the ratio~\cite{Ruiz:2023ozv,Segarra:2020gtj}), but no other higher-twist effects. The earlier JLab Hall C data on carbon were taken at an electron-beam energy of 6 GeV and different electron scattering angles (in \cref{subfig:dvt_9994,subfig:dvt_9992} we show results for of 32 and 40 degrees, respectively). As one can see in \cref{subfig:dvt_9994,subfig:dvt_9992}, these lower-energy data have a more limited reach in $x\leq0.7$. While the 40-degree data exhibit the same plateau starting at around $x\sim 0.65$ as the newer data and the theory predictions, the \mbox{32-degree} data still decrease monotonically and are part of the worst described JLab data in our fit (cf.\ also \cref{fig:data-vs-theory-jlab-C-D} where data for other scattering angles are shown). The absence of a plateau in the 32-degree data is somewhat unexpected, given that the values of $Q^2$ at these data points agree with those in the 40-degree data within 10\%.

\begin{figure*}[htbp]
    \centering
    \includegraphics[width=0.84\textwidth]{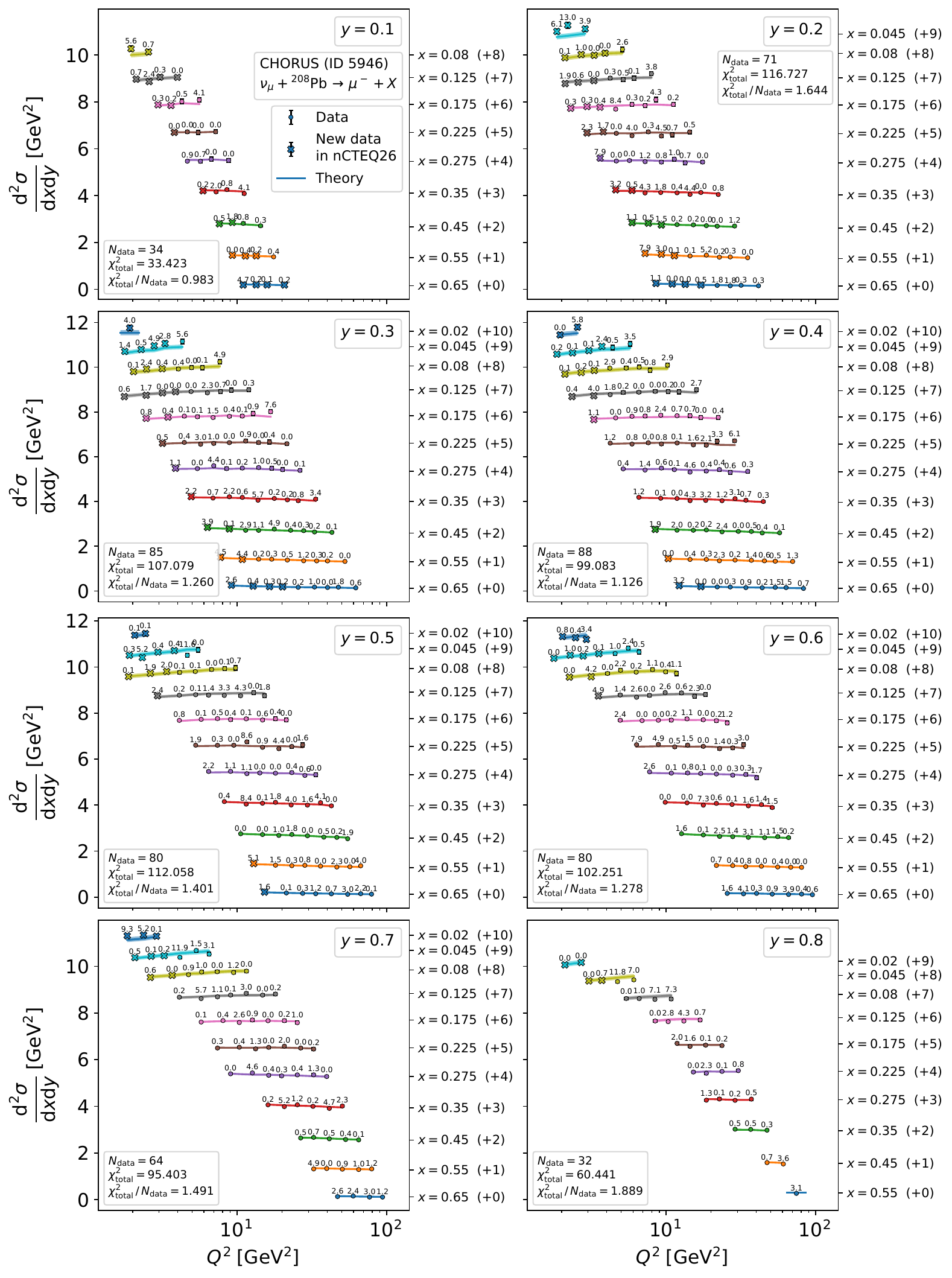}
    \caption{Comparison of nCTEQ26 theory predictions with CC DIS data from CHORUS as a function of $Q^2$ for different values of the inelasticity $y$ and with the corresponding values of $\chi^2$ per data point. Cross sections corresponding to different values of $x$ are separated vertically in steps of 0.5. Newly added points due to relaxed kinematic cuts as compared to Ref.~\cite{Muzakka:2022wey} are indicated by crosses.}    
    \label{fig:data-vs-theory:CHORUS}
\end{figure*}

In the case of neutrino DIS data, the situation is more complex. It has long been known that there are tensions between NC and CC DIS data, which is why the selection of CC DIS data for the current analysis was informed by our dedicated study of Ref.~\cite{Muzakka:2022wey} on this matter. Based on this study, we included only the inclusive CHORUS cross section data and the dimuon data from the NuTeV and CCFR experiments. However, the current analysis includes more data points, as we have extended the kinematic range of the neutrino data used by relaxing the selection cuts (cf.\ Sec.~\ref{subsec:nudis}). First, we look at the CHORUS data. In Fig.~\ref{fig:data-vs-theory:CHORUS} we present the inclusive cross section for neutrinos (ID 5946) as a function of $Q^2$. The eight panels in the figure correspond to different values of the inelasticity ($y$). In each panel, we plot the cross section for several slices of $x$. To make the plots more readable, we vertically offset the lines corresponding to different $x$ slices by 0.5. This allows us to display all 534 neutrino data points. 
The figure does not show a clear pattern of which kinematic region is described better or worse. For instance, some of the bins in inelasticity (e.g.,~${y=\{0.1,0.4\}}$) are very well described with $\chidata \sim 1$, and some (e.g.,~$y=\{0.2, 0.8\}$) yield $\chidata \gtrsim 1.5$. There may be a tendency of points at low $Q^2$ to result in larger $\chi^2$ values, but no clear pattern emerges. Also a combination of these low-$Q^2$ points with different values of $x$ and $y$ does not exhibit clear systematic features. In Fig.\ \ref{fig:data-vs-theory:CHORUS}, data points newly introduced by the relaxed kinematic cuts relative to our previous analysis~\cite{Muzakka:2022wey} are marked by crosses instead of circles. It then becomes clear that they also do not introduce any systematic effects. In addition to the neutrino CHORUS data shown in Fig.~\ref{fig:data-vs-theory:CHORUS}, we also fit the analogous set of anti-neutrino data (ID 5947). We do not show them here because the experimental uncertainties for the anti-neutrino data are larger, which results in a better data description (see Tab.~\ref{tab:nudata} and Figs.~\ref{fig:chi2dataset} and~\ref{fig:SE}).

\begin{figure}
    \centering
    \includegraphics[width=0.47\textwidth]{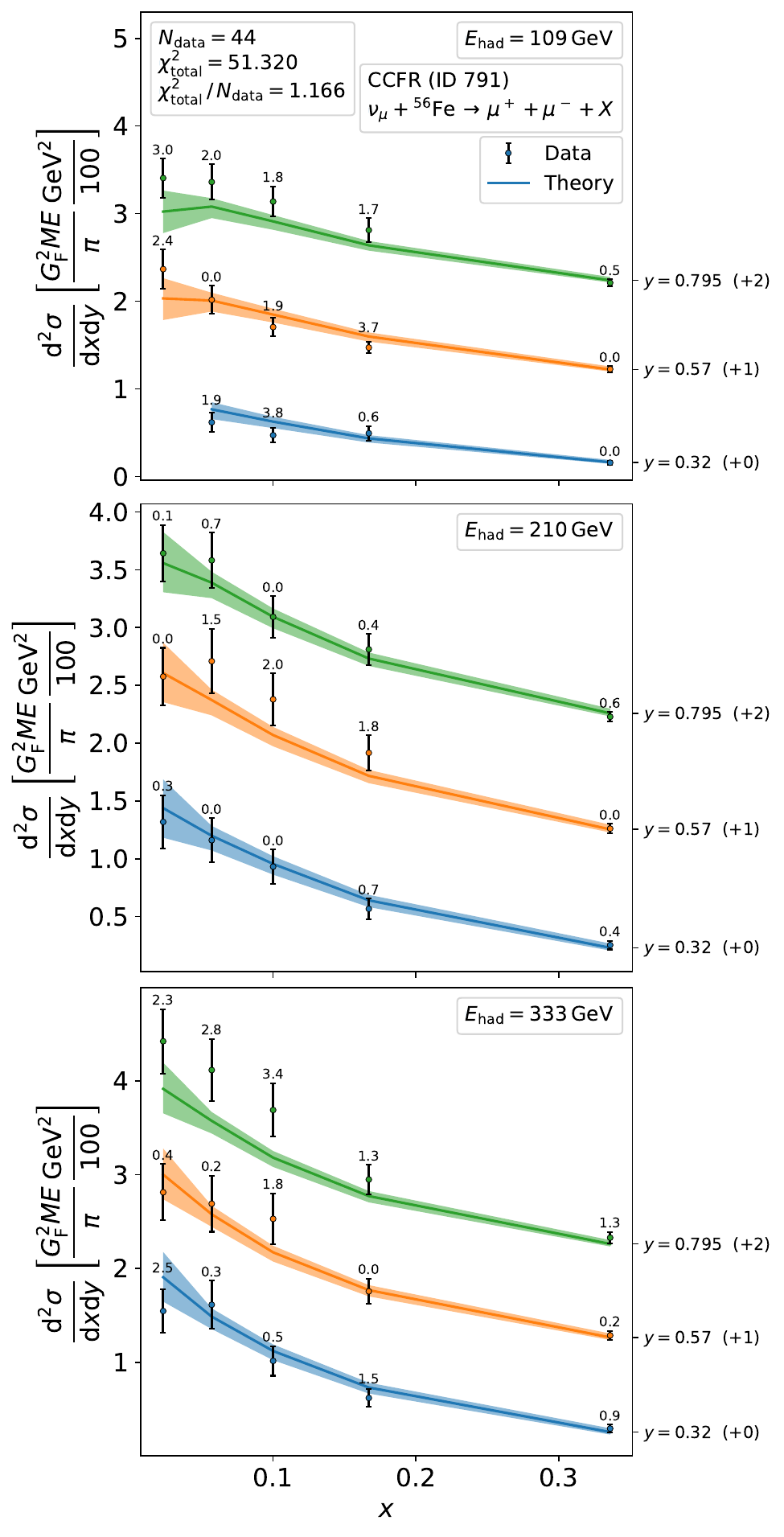}
    \caption{Comparison of nCTEQ26 theory predictions with CC DIS dimuon data from CCFR as a function of $x$ for different values of the hadronic energy $E_{\mathrm{had}}$, including the corresponding values of $\chi^2$ per data point. Cross sections corresponding to different values of $y$ are separated vertically in steps of 1.}
    \label{fig:data-vs-theory:CCFR}
\end{figure}

Moving on to the dimuon data, we find that all the dimuon datasets are well described by our fit. 
In Fig.~\ref{fig:data-vs-theory:CCFR} we compare the CCFR neutrino data (ID 791) with the theory predictions, since it yields the largest $\chi^2$ among these datasets (cf.\ Fig.~\ref{fig:chi2dataset}). 
The figure shows absolute cross sections plotted as a function of $x$ for different values of inelasticity $y$ separated vertically in steps of 1 and for three different values of the hadronic energy $E_{\mathrm{had}}$. 
From Fig.~\ref{fig:data-vs-theory:CCFR} we observe that the bins with ${E_{\mathrm{had}}=333}$~GeV, the largest values of $y$, and the lowest ${x~\sim0.01}$ are described the worst. 
A similar observation holds for low $x$ and large $y$ for the lowest value of ${E_{\mathrm{had}}=109}$~GeV, but not for the intermediate value ${E_{\mathrm{had}}=210}$~GeV. 
We note that the data included by relaxing the kinematic cuts, which lie predominantly at low $y$ and ${E_{\mathrm{had}}=\{109,210\}}$~GeV, do not increase the global $\chidata$. 
The new region introduced by relaxing the cuts on $Q$ and $W$ is believed to be sensitive to higher-twist (HT) effects. 
To test this, we followed our own work in Ref.~\cite{Segarra:2020gtj}, where HT effects were included according to their treatment in the CJ15 analysis~\cite{Accardi:2016qay}. 
Employing HT corrections extracted from there, we were unable to obtain an improved description of the data. 
In fact, including these corrections yields a very similar quality of description for most of the data, apart from the neutrino data from the CHORUS experiment, for which $\chidata$ degraded from 1.26 to 1.29. 
One should keep in mind that these HT corrections were obtained in an analysis of proton and deuteron data. An appropriate accounting for these effects in the case of nuclear targets needs to be more general. 
This, however, goes beyond the current analysis and is left for future studies. 
To refine theoretical predictions for neutrino cross sections in low-energy oscillation experiments such as DUNE, future investigations would benefit from examining the interplay between TMC and HT effects with the increased statistics in the expanded low-$Q$ region. 
Such a study would also allow independent comparison with the neutrino community’s Bodek-Yang model in this region.

\begin{figure*} %
    \centering  
    \subfloat[\centering $W^+$-boson]{
        \label{fig:dataTH_Wp}
        \includegraphics[width=0.48\textwidth]{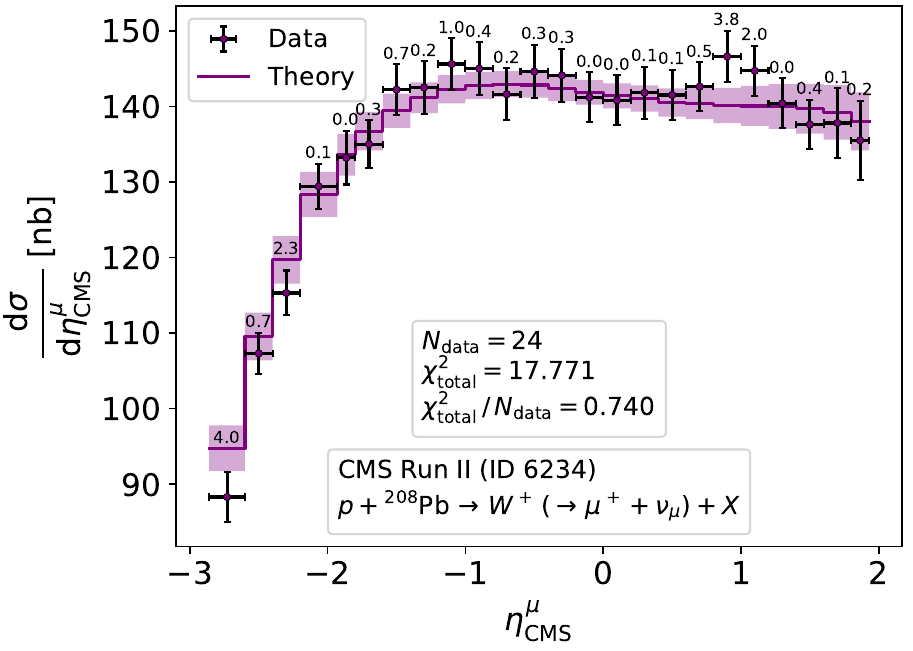}
    }
    \subfloat[\centering $W^-$-boson]{\label{fig:dataTH_Wm}
        \includegraphics[width=0.48\textwidth]{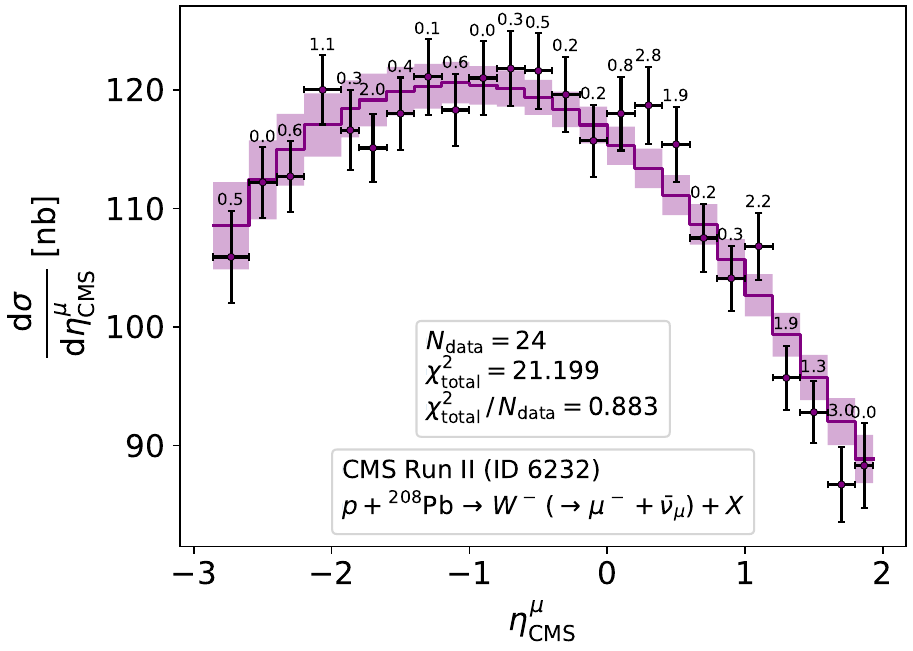}
    }
\\
    \subfloat[\centering $Z$-boson on peak ($60 < M_{ll} < 120$ GeV)]{ 
        \label{fig:dataTH_ZhM}
        \includegraphics[width=0.48\textwidth]{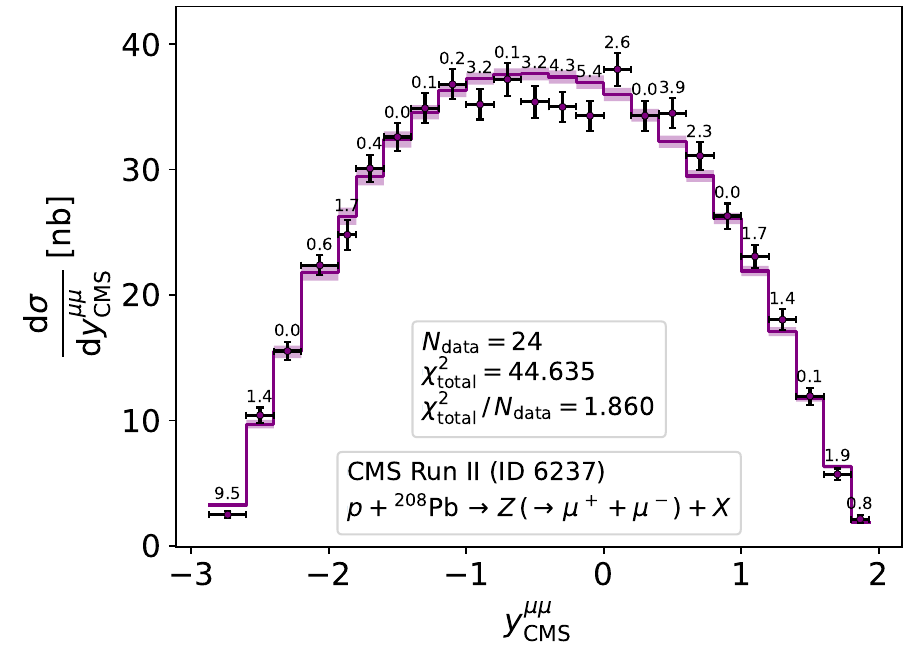}
    }
    \subfloat[\centering $Z$-boson off peak ($15 < M_{ll} < 60$ GeV)]{
        \label{fig:dataTH_ZlM}
        \includegraphics[width=0.48\textwidth]{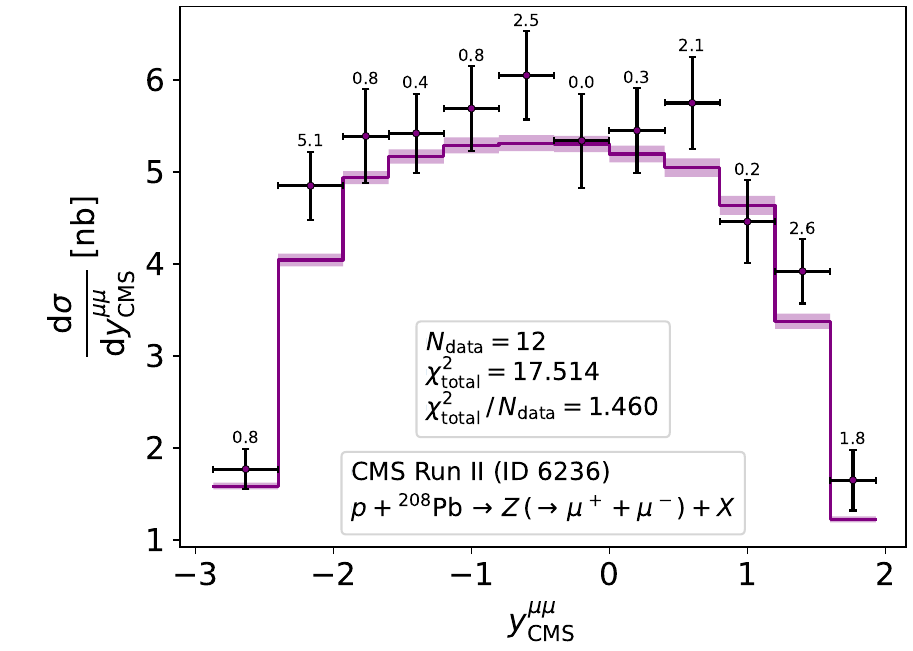}
    }
    \caption{Comparison of nCTEQ26 theory predictions with (a) $W^+$, (b) $W^-$, (c) on-peak, and (d) off-peak $Z$-boson data from CMS Run-II in proton-lead collisions at the LHC with the corresponding values of $\chi^2$ per data point.}
    \label{fig:dataTH_WZ} 
\end{figure*}

Next, we consider $W/Z$-boson production from proton-lead collisions at the LHC. There, the most precise data were collected by the CMS experiment in Run II (IDs 6232–6237). In \cref{fig:dataTH_WZ} we compare these data with predictions using the fitted lead PDFs. We can see that for both $W^+$ and $W^-$ bosons the descriptions are very good with values of $\chidata$ below one. In the case of the $Z$ boson, the obtained description is worse with ${\chidata=1.46}$ in the low-mass region and ${\chidata=1.86}$ in the peak region. Looking at the more precisely measured peak region, shown in \cref{fig:dataTH_ZhM}, one can see that for central rapidities the data feature a curious dip, which is unlikely to be described properly even if more precise NNLO or NNLO+NLO electroweak calculations were used. Conversely, in the low-mass region (\cref{fig:dataTH_ZlM}), we observe that the theory predictions are consistently below the data. This difference could potentially be reduced by choosing a different scale or by including higher-order corrections. The impact of the NNLO QCD corrections and the scale choice are currently under investigation~\cite{ChenDY}.
\begin{figure*}[hbt]
    \setcounter{subfigure}{0}
    \subfloat[ALICE $W^-$ (backward rapidity)]{
        \centering
        \hspace*{-1.5em}
        \includegraphics[width=0.33\linewidth, page=4]{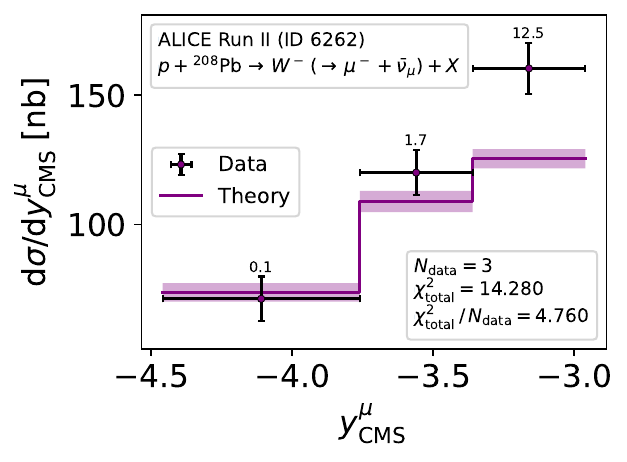}
        \label{subfig:dvt_6236}
    }
    \setcounter{subfigure}{2}
    \subfloat[ALICE $W^+$ (backward rapidity)]{
        \centering
        \hspace*{-1.5em}
        \includegraphics[width=0.33\linewidth, page=6]{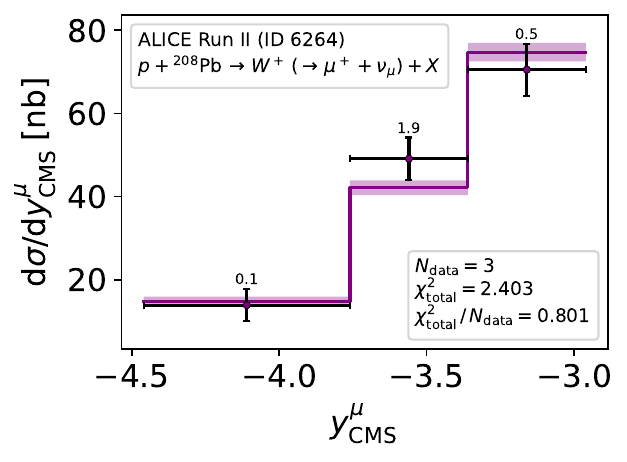}
        \label{subfig:dvt_6262}
    }
    \setcounter{subfigure}{4}
        \subfloat[LHCb $Z$ (backward rapidity)]{
        \centering
        \hspace*{-1.5em}
        \includegraphics[width=0.33\linewidth, page=8]{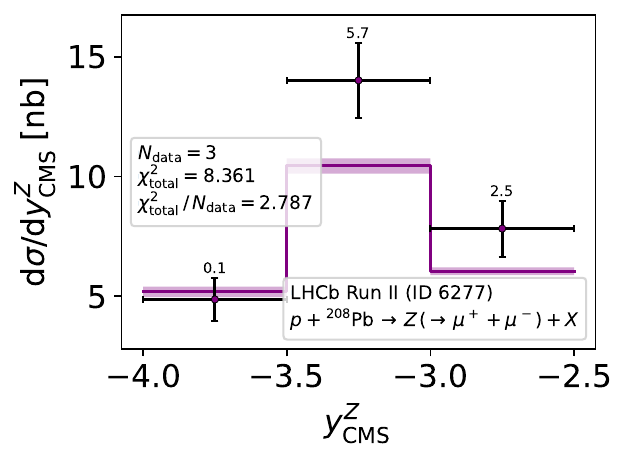}
        \label{subfig:dvt_6277}
    }
    \\
    \setcounter{subfigure}{1}
    \subfloat[ALICE $W^-$ (forward rapidity)]{
        \centering
        \hspace*{-1.5em}
        \includegraphics[width=0.33\linewidth, page=3]{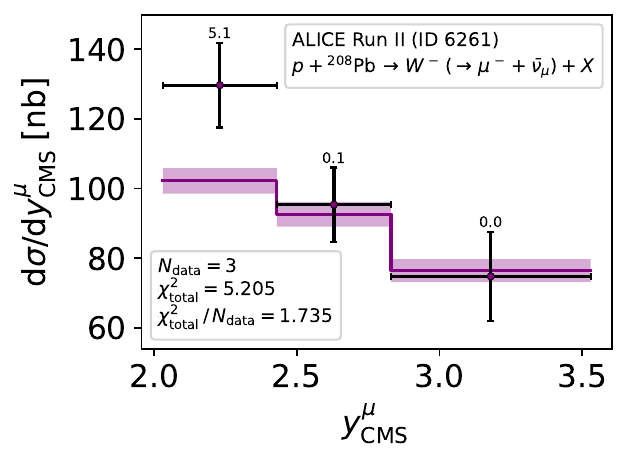}
        \label{subfig:dvt_6261}
    }
    \setcounter{subfigure}{3}
    \subfloat[ALICE $W^+$ (forward rapidity)]{
        \centering
        \hspace*{-1.5em}
        \includegraphics[width=0.33\linewidth, page=5]{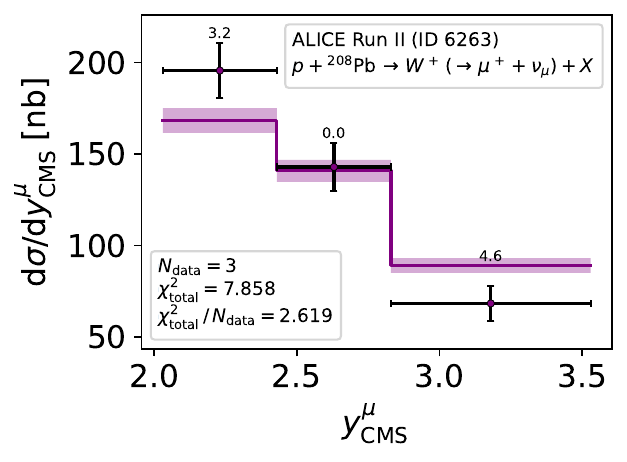}
        \label{subfig:dvt_6263}
    }
    \setcounter{subfigure}{5}
    \subfloat[LHCb $Z$ (forward rapidity)]{
        \centering
        \hspace*{-1.5em}
        \includegraphics[width=0.33\linewidth, page=7]{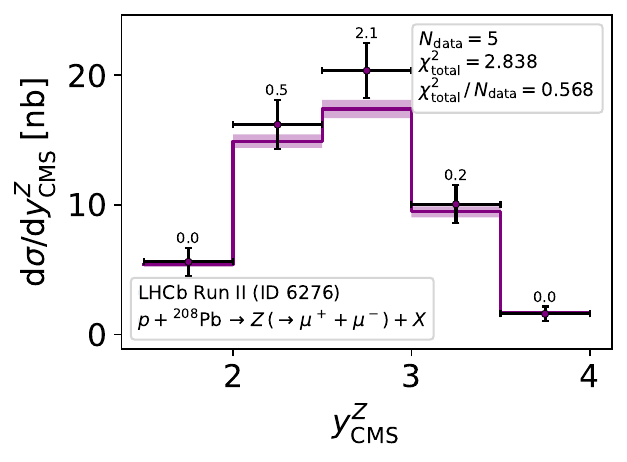}
        \label{subfig:dvt_6276}
    }
    \caption{Comparison of nCTEQ26 theory predictions with (a-d) $W^\pm$-boson production data from ALICE~\cite{ALICE:2022cxs} and (e,f) $Z$-boson production data from LHCb~\cite{LHCb:2022kph}.}
    \label{fig:proplematic_datasets1}
\end{figure*}
Besides the CMS Run-II $Z$-boson data, there are also a few other weak-boson datasets that are not particularly well described, specifically the Run-II data for $W^{\pm}$ bosons from ALICE (IDs 6261–6263)~\cite{ALICE:2022cxs} and $Z$ bosons from \mbox{LHCb (IDs 6276, 6277)~\cite{LHCb:2022kph}} in the backward ($y\geq-4.5$) and forward ($y\leq4$) regions. They contain only a very small number of data points, so that their impact on the overall global $\chi^2$ is negligible. The corresponding data to theory comparisons are shown in \cref{fig:proplematic_datasets1}. At first sight, the average $\chi^2/N_{\mathrm{data}}\sim 2$ for these datasets seems to suggest a poor overall description of the data. However, an inspection of the distributions shown in panels (a-d) for ALICE and (e,f) for LHCb shows that the large $\chi^2$ is dominated by a few data points with $\chi^2 \gtrsim 3$. In ALICE, the majority of the outliers are the most central data points, e.g.~$y^\mu_{\rm CMS}\in[-3.36;-2.96]$ in \cref{subfig:dvt_6236} with ${\chi^2 = 12.5}$. In LHCb, the shape of the spectra in the data are more concave than the theoretical prediction, see e.g.~the central bin in \cref{subfig:dvt_6277} with ${\chi^2=5.7}$. These predictions are generally consistent with the theoretical predictions quoted in the experimental papers and obtained using earlier nPDFs. As this is the first time that these datasets are included in a global nuclear PDF analysis, no comparison with or insights from other analyses are available. In principle the LHCb $Z$-boson data could be compared to the CMS $Z$-boson data~\cite{CMS:2021ynu}, which only exhibits a monotonically increasing spectrum in the negative rapidity region. However, the overlap region is rather small and the binning and selection cuts are not the same. This makes a direct comparison difficult.

\begin{figure*}
    \centering  
    \subfloat[\centering LHCb $D^0$ at $\sqrt{s_\text{NN}} = 5.02\,\text{TeV}$]{
        \label{fig:data-vs-theory:HQ3102}
        \includegraphics[width=0.48\textwidth]{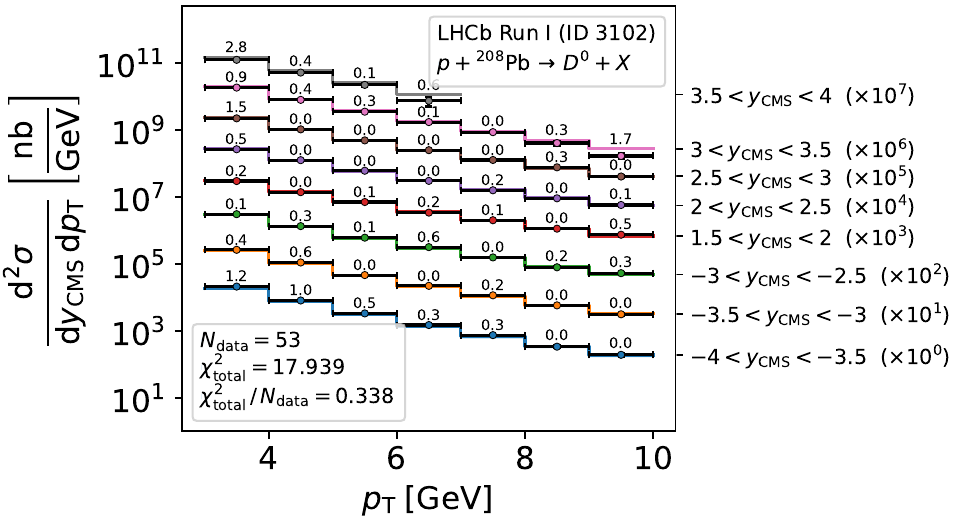}
    }
    \subfloat[\centering LHCb $J/\psi$ at $\sqrt{s_\text{NN}} = 5.02\,\text{TeV}$]{
        \label{fig:data-vs-theory:HQ3108}
        \includegraphics[width=0.48\textwidth]{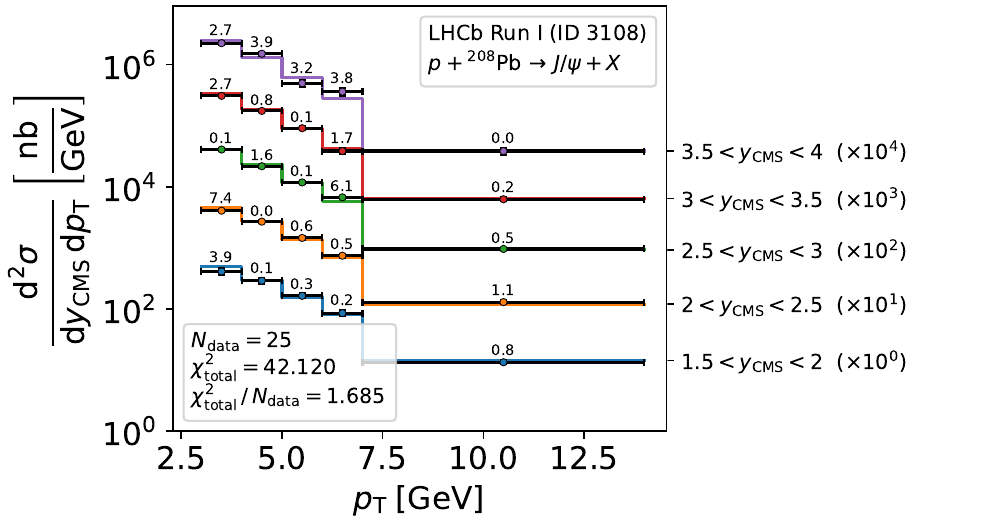}
    }
\\
    \subfloat[\centering LHCb $D^0$ at $\sqrt{s_\text{NN}} = 8.16\,\text{TeV}$ (forward rapidity)]{ 
        \label{fig:data-vs-theory:HQ3133}
        \includegraphics[width=0.48\textwidth]{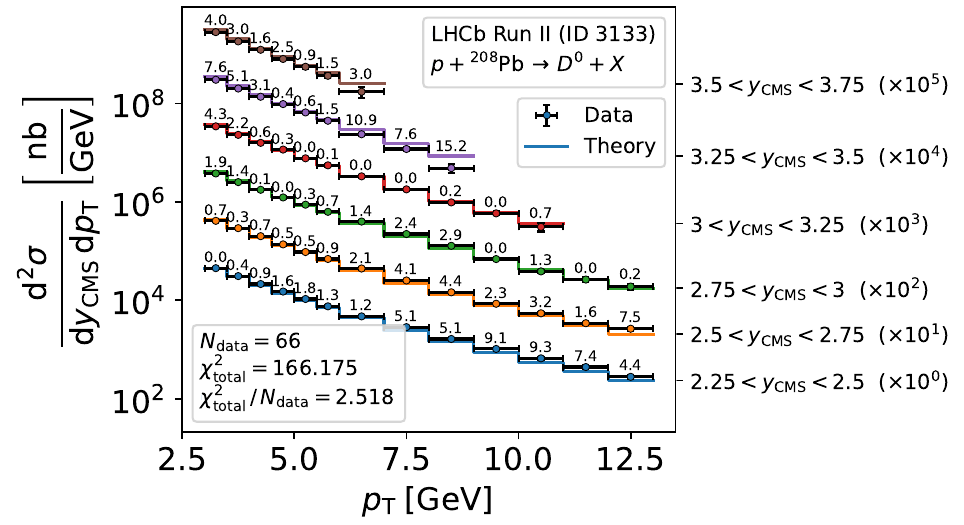}
    }
    \subfloat[LHCb $D^0$ at $\sqrt{s_\text{NN}} = 8.16\,\text{TeV}$ (backward rapidity)]{
        \label{fig:data-vs-theory:HQ3134}
        \includegraphics[width=0.48\textwidth]{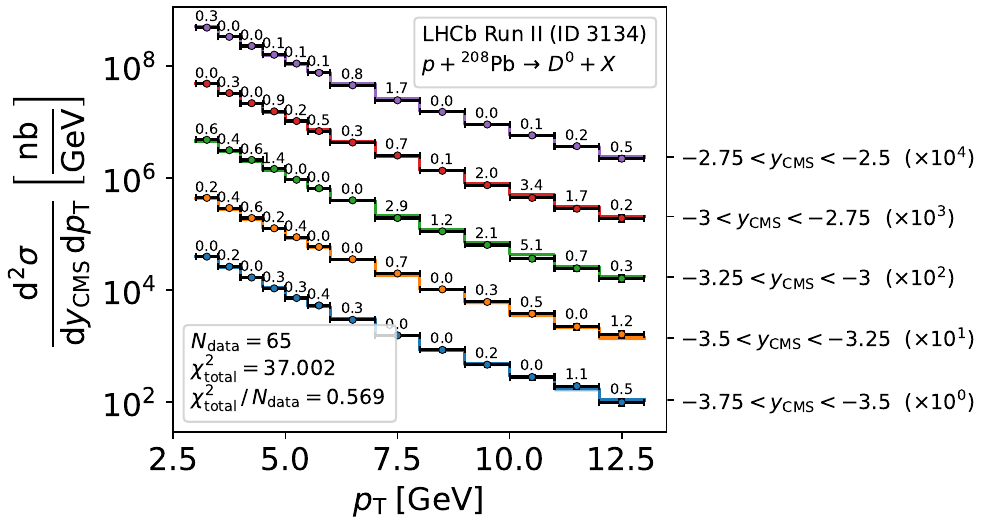}
    }
    \caption{Comparison of nCTEQ26 theory predictions with selected (a,c,d) $D^0$ and (b) $J/\psi$ cross section data from LHCb Run-I (a,b) and Run-II (c,d) in proton-lead collisions at the LHC as a function of the transverse momentum $p_T$ with the corresponding values of $\chi^2$ per data point. Cross sections corresponding to different values of rapidity are separated vertically by factors of 10.
    }
    \label{fig:data-vs-theory:HQ} 
\end{figure*}

The description of the heavy-quark production data, both for open heavy flavor and quarkonia, is generally very good. The few exceptions, exhibiting larger deviations, are the Run-II $D^0$-meson data from LHCb in the forward region (ID 3133), the Run-I $J/\psi$ data from LHCb (ID 3108), and the Run-I $\psi(2S)$ data from ALICE (ID 3127) with only two data points, which leave little imprint on the fit. On the other hand, the $J/\psi$ and $D^0$ data are quite numerous, providing 25 and 66 data points, respectively. In case of the $J/\psi$ data shown in Fig.~\ref{fig:data-vs-theory:HQ3108}, we observe that the regions that are difficult to describe originate primarily from low transverse momentum (mostly the lowest $p_T$ bin) and from the most forward rapidity bin. A more stringent cut on $p_T$ could potentially reduce the $\chi^2$ for this dataset. However, the cuts are based on the quality of $pp$ data description with the Crystal-Ball parametrization, and keeping them identical to those for the $p$Pb data allows us to assess whether we encounter any additional effects in $pA$ collisions. The situation is quite intriguing for the $D^0$-meson data from LHCb. The Run-I data (ID 3102) taken at ${\sqrt{s}=5.02}$ TeV is very well described by the fit throughout the measured region (cf.\ Fig.~\ref{fig:data-vs-theory:HQ3102}). On the other hand, the Run-II data in the forward region (cf.\ Fig.~\ref{fig:data-vs-theory:HQ3133}) are poorly described, with no clear indication of what might account for this discrepancy. The uncertainties in the Run~I data are larger, but not to an extent that could explain such a difference in $\chi^2$. Conversely, the same $D^0$ data in the backward region (cf.~Fig.~\ref{fig:data-vs-theory:HQ3134}) is well described throughout the phase space.

\begin{figure}[!tbp]
    \centering  
        \includegraphics[width=0.95\linewidth]{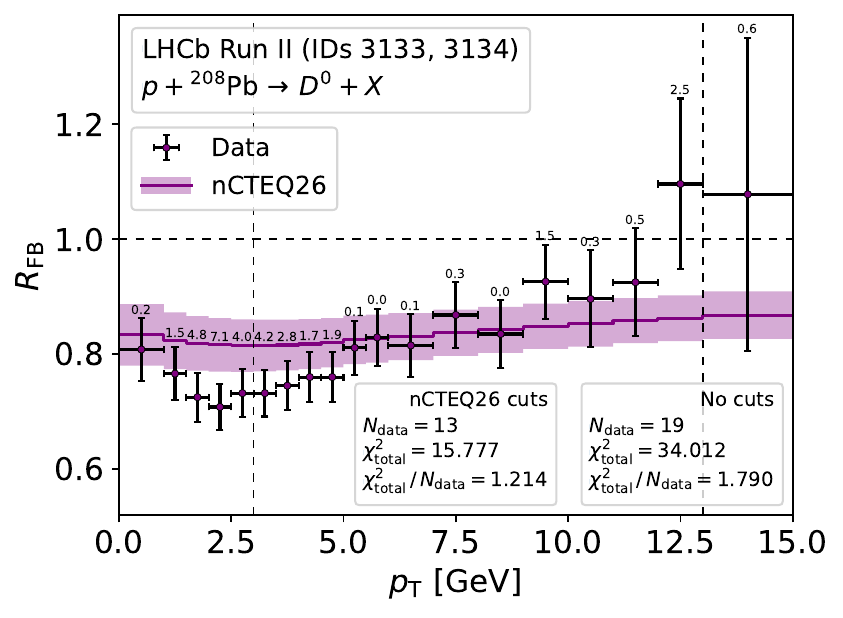}
    \caption{Comparison of the nCTEQ26 theory prediction with $D^0$ forward-backward asymmetry data from LHCb Run-II in proton-lead collisions at the LHC integrated over rapidity as a function of the transverse momentum $p_T$ with the corresponding values of $\chi^2$ per data point. Note that only the region with $p_T\in[3,13]$~GeV is fitted.}
    \label{fig:data-vs-theory:D0AFB} 
\end{figure}

The $D^0$ LHCb Run-II datasets have not been included in global analyses so far. However, Ref.~\cite{Klasen:2023uqj} reports partially diverging predictions for the rapidity-integrated forward/backward ratio from EPPS21, nNNPDF3.0 and nCTEQ15HQ. To verify this, we also show in \cref{fig:data-vs-theory:D0AFB} predictions for this ratio obtained with nCTEQ26 nPDFs. In the region covered by the fit (${p_T>3}$~GeV), the theoretical prediction describes the data well within the experimental uncertainties, but yields a flatter ratio slightly undershooting the data at larger $p_T$ and overestimating the data at lower $p_T$. The largest discrepancy is observed in the region around $p_T\sim 2$~GeV, which is not fitted. Overall, the insufficient description of the data in the forward region observed in \cref{fig:data-vs-theory:HQ3133} is washed out in the rapidity-integrated ratio.

An effect that could potentially interfere with the good description of heavy-quark data in the analysis is saturation~\cite{Ducloue:2015gfa,Ma:2015sia,Qiu:2013qka,vanHameren:2023oiq}. The larger values of $\chi^2$ reported above all originate from the forward-rapidity region, which corresponds to low values of $x\sim10^{-5}$. This is the region where saturation effects could appear and change the evolution of the PDFs~\cite{Armesto:2022mxy}. Furthermore, saturation is expected to be enhanced by the nuclear mass (in case of lead $A=208$), which could explain why the same rapidity region in $pp$ baseline data used for the CB fit is well described. Note also that the heavy-quark and in particular the quarkonium data can be potentially affected by other nuclear effects going beyond nuclear shadowing and saturation, such as energy loss in a deconfined medium or comovers~\cite{Arleo:2021bpv,Arleo:2020hat,Lansberg:2019adr,Ferreiro:2014bia,Vogt:1999cu}. However, these kinds of effects can be expected to be somewhat different for quarkonia and OHF production, while we observe a very consistent picture coming from both types of data leading to nearly the same gluon PDFs. We investigate this in more detail in \cref{subsec:gluon} with an alternative fit, nCTEQ26OHF, in which we exclude the quarkonium data and use perturbative QCD GM-VFNS theory predictions for the open heavy-flavor production data rather than the data-driven approach. As we will see, in this fit we obtain a less satisfactory description 
of the OHF data overall and in particular in the forward region of the $D^0$ LHCb Run-II data, while the backward region remains similar. Alternative approaches to reduce other nuclear effects include more stringent kinematic selection cuts or using only color-neutral probes in the same kinematic region as the HQ data. Unfortunately, at present such probes are either not available, do not cover the same kinematics, or are not precise enough. Data from $W/Z$-boson or direct-photon production in $p$Pb collisions at the LHC could potentially help, provided they can be measured precisely enough in the forward region corresponding to low $x$. In the future, unambiguous probes will be provided by measurements at the EIC~\cite{AbdulKhalek:2021gbh}. 

\subsection{Hessian diagnostics} 
\label{app:numericalPrecisionHessian}

\begin{figure*}
    \centering
    \includegraphics[width=0.99\textwidth]{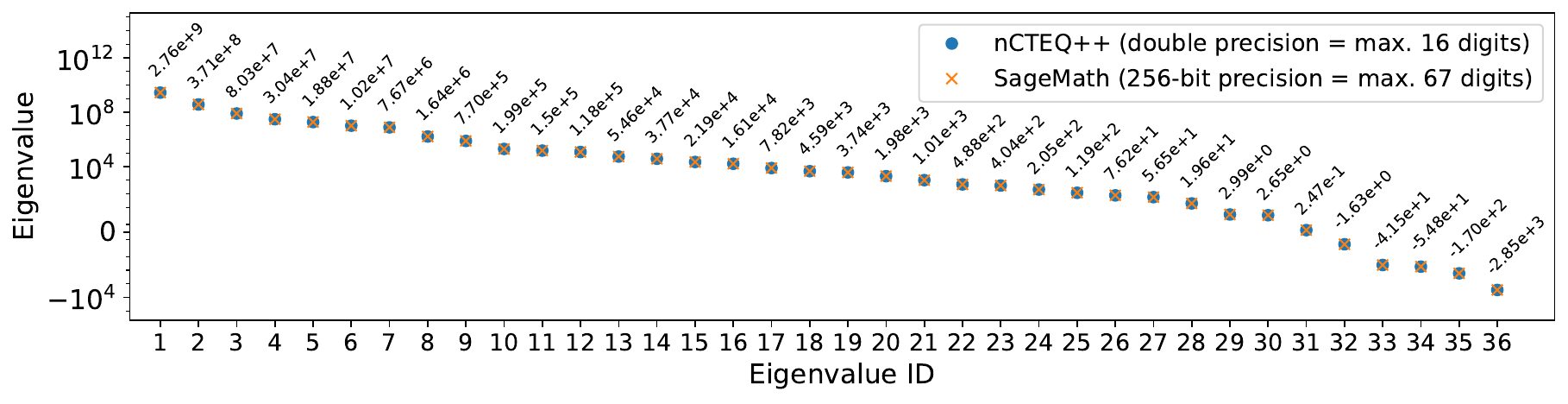}
    \caption{Eigenvalues of the Hessian matrix calculated with \ncteqpp\ (double precision) and SageMath (256-bit precision).}
    \label{fig:hessian-precision-eigenvalues}
\end{figure*}

We now examine the quality of the quadratic approximation inherent in our use of the Hessian method of uncertainty determination. 
We remind the reader that the calculation of the Hessian matrix, its diagonalization, the construction of the basis of eigenvectors, and the estimation of the tolerance were described in \cref{subsec:errors} and \cref{app:tolerance}. 
The eigenvalues of the Hessian matrix of our nCTEQ26 fit are shown in  \cref{fig:hessian-precision-eigenvalues} ordered by numerical value. 
Since they span over ten orders of magnitude, we checked explicitly whether our numerical diagonalization of the Hessian matrix is reliable. 
In \ncteqpp\, this diagonalization is performed with the C++ library Armadillo~\cite{armadillo} at double precision. 
We have verified that all eigenvalues agree to better than $10^{-7}$ (relative precision) with those obtained at 256-bit precision with SageMath~\cite{sagemath} and the acb module in FLINT~\cite{flint,flint-arb}, a library that can perform calculations with a user-defined number of bits and track the numerical uncertainty. 

\begin{figure}
    \centering
    \includegraphics[width=0.99\linewidth]{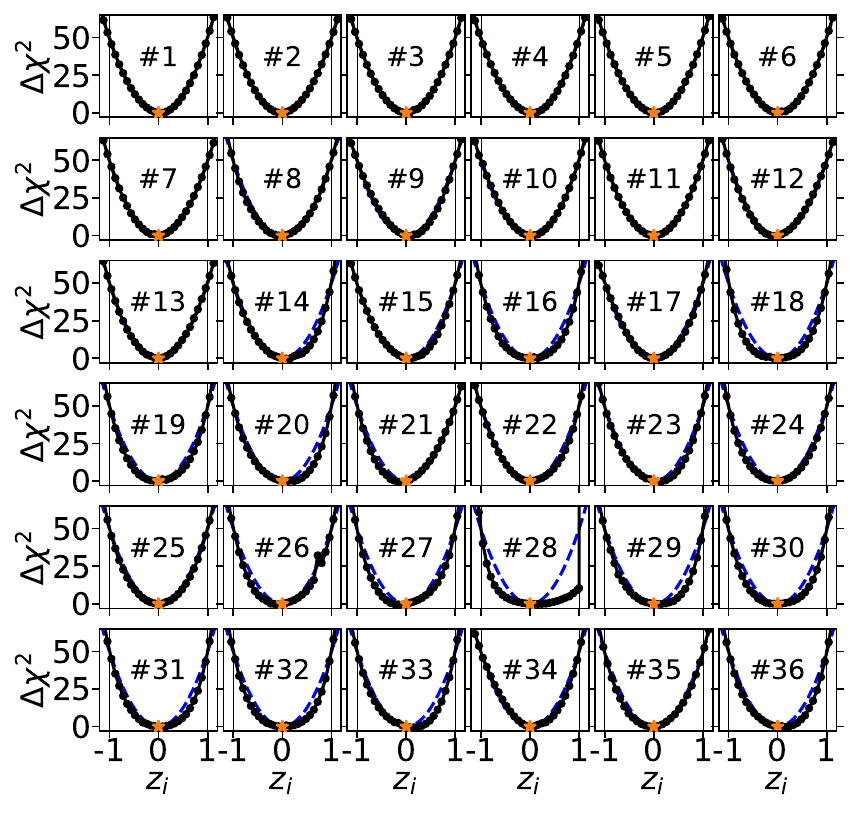}
    \caption{$\Delta\chi^2=\chi^2-\chi_0^2$ plotted along the $z_i$ directions in the Hessian eigenvector space. The full black lines show the actual $\chi^2$ function, and the dashed blue lines represent the quadratic approximation.
    }
    \label{fig:ev1d_all}
\end{figure}

In \cref{fig:ev1d_all} we show the scans of the $\chi^2$ function around its minimal value along the eigendirections $z_i$ of the Hessian matrix with their normalization indicated on the $x$-axis. Each of the 36 panels shows a scan in one direction (full black line) sampled at regular intervals, indicated by the black circles. The panels are ordered from left to right and top to bottom in descending order of eigenvalue magnitude. For large eigenvalues within four orders of magnitude of the largest one, we find a nearly perfect quadratic behavior. For lower eigenvalues, we find notable deviations from the quadratic approximation 
(dashed blue lines) manifested through asymmetric features and a broadening of the profile. However, only in one case does the quadratic approximation fail to capture the true profile of the $\chi^2$ function, namely for eigenvector \#28, whose eigenvalue is, however, suppressed by eight orders of magnitude compared to eigenvector \#1.%
\footnote{For positive direction of eigenvector \#28 we obtain an unphysical configuration which leads to sum rule violation before reaching the $\Delta\chi^2=T$ condition.}
Only the smallest five eigenvalues are negative, which is a consequence of the local $\chi^2$ behavior very close to the minimum. The corresponding $\chi^2$ profiles at the scale of the tolerance $\Delta \chi^2 = T$, however, clearly feature a minimum (rather than a maximum).

\begin{figure}
    \centering
    \includegraphics[width=0.99\linewidth]{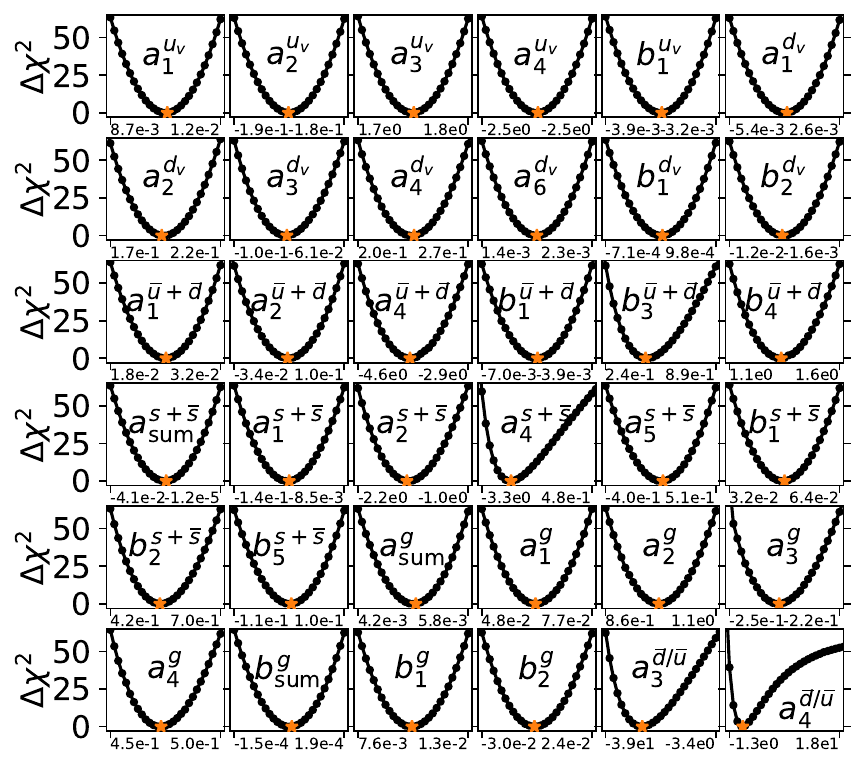}
    \caption{$\Delta\chi^2=\chi^2-\chi_0^2$ plotted along the fitted PDF parameters.}
    \label{fig:param1d_all}
\end{figure}

The one-dimensional scans along the open original parameters are shown in \cref{fig:param1d_all}, now ordered by flavor. All parameters are well constrained, with most exhibiting the expected symmetric profile. The most notable exceptions are the $a_4$ parameters in the strange and $\bar{d}/\bar{u}$ parametrizations.

\begin{figure}
    \centering
    \begin{subfigure}[b]{0.4\textwidth}
        \centering
        \includegraphics[width=\textwidth]{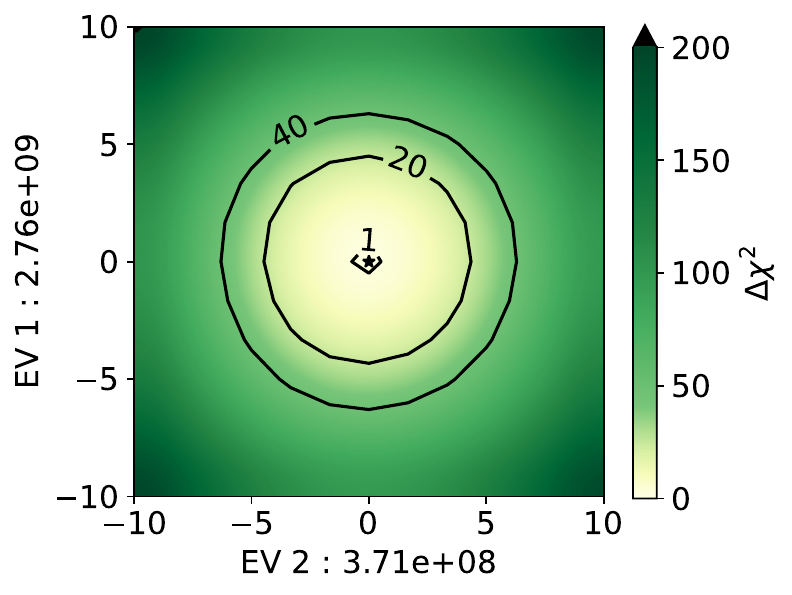}
        \caption{}
        \label{subfig:EV34_EV33}
    \end{subfigure}
    \hfill
    \begin{subfigure}[b]{0.4\textwidth}
        \centering
        \includegraphics[width=\textwidth]{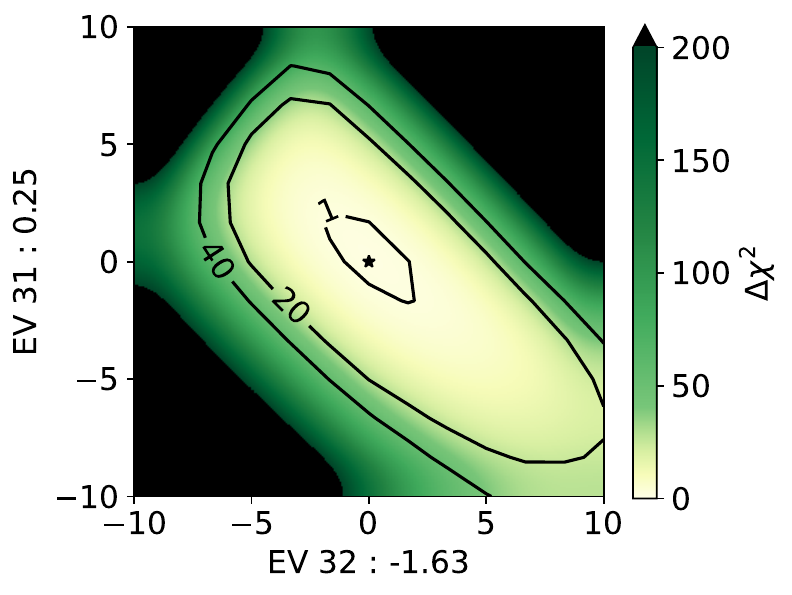}
        \caption{}
        \label{subfig:EV6_EV2}
    \end{subfigure}
    \\
    \caption{Two-dimensional ${\Delta \chi^2=\chi^2-\chi^2_0}$ scans in the Hessian eigenvector space. Shown is representative behavior for directions corresponding to large eigenvalues (a) and small eigenvalues (b).
    }
    \label{fig:2d_EVscan}
\end{figure}

To further investigate the quality of the Hessian approximation, we show in \cref{fig:2d_EVscan} two representative two-dimensional $\chi^2$ scans along the Hessian eigenvector directions. In \cref{subfig:EV34_EV33} for the two largest eigenvalues, the obtained profile is nearly circular, indicating independence of the two eigenvector directions, which largely align with the unrelated parameters $b^g_{\rm sum}$ and $a_6^{d_v}$. In \cref{subfig:EV6_EV2} corresponding to the two smallest eigenvalues in absolute numbers, the profile departs noticeably from the ideal circular form  and clearly indicates a correlation of the two eigenvector directions, which in this case are dominated by the $s+\bar{s}$ large-$x$ parameters $a_2$ and $a_5$.

Overall, we find that for our selection of open parameters the quadratic approximation performs sufficiently well and conclude that using the Hessian error analysis in our global fit is appropriate. We also note that the eigenvalue rescaling we perform is crucial for adjusting the Hessian approximation to match the true value of the $\chi^2$ profile at $\Delta\chi^2=T$. We also examined the possibility of extending the parameter space by fitting with more open parameters and analyzed the corresponding one- and two-dimensional $\chi^2$ scans. However, we found that opening additional parameters does not improve the quality of the fit, but can rather lead to features inconsistent with Hessian error propagation, including flat directions and strong correlations between pairs of parameters.\footnote{Recent studies suggest that these difficulties could be overcome using Bayesian methods, e.g.~\cite{Hunt-Smith:2022ugn,Capel:2024qkm,Candido:2024hjt,Costantini:2024wby,Barontini:2025lnl,Costantini:2025wxp,Costantini:2025agd,Risse:2025qlo,Derakhshanian:2026zkx}.} To minimize the bias from special values of parameters kept fixed, we set all of them to zero.
 
\subsection{Data constraints on parameters and PDFs}
\label{subsec:constraints}

\begin{figure*}
    \centering
        \subfloat[Scan along the $a_1^{u_v}$ parameter.]{
        \includegraphics[width=0.33\linewidth,page=1]{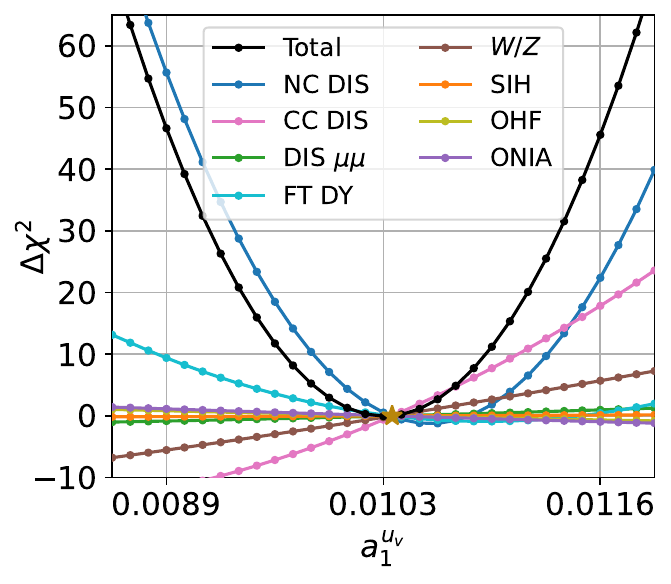}
        \label{subfig:par-uv_a1}
    }
    \qquad
        \subfloat[Scan along the $a_3^{d_v}$ parameter.]{
        \includegraphics[width=0.33\linewidth,page=8]{figs/scans/param1d_all_newDimuon_May26.pdf}
        \label{subfig:par-dv_a3}
    }
    \\
    \vspace{0.05cm}
    \subfloat[Scan along the $a_{\mathrm{sum}}^{g}$ parameter.]{
        \includegraphics[width=0.33\linewidth,page=27]{figs/scans/param1d_all_newDimuon_May26.pdf}
        \label{subfig:par-g_a0}
    }
    \qquad
    \subfloat[Scan along the $a_{\mathrm{sum}}^{s+\bar{s}}$ parameter.]{
        \includegraphics[width=0.33\linewidth, page=19]{figs/scans/param1d_all_newDimuon_May26.pdf}
        \label{subfig:par-ssb_a0}
    }
    \\
    \vspace{0.05cm}
    \subfloat[Scan along the $a_1^{g}$ parameter.]{
        \includegraphics[width=0.33\linewidth,page=28]{figs/scans/param1d_all_newDimuon_May26.pdf}
        \label{subfig:par-g_a1}
    }
    \qquad
    \subfloat[Scan along the $a_2^{s+\bar{s}}$ parameter.]{
        \includegraphics[width=0.33\linewidth, page=21]{figs/scans/param1d_all_newDimuon_May26.pdf}
        \label{subfig:par-ssb_a2}
    }
    \\
    \vspace{0.05cm}
    \subfloat[Scan along the $a_2^{\bar{u}+\bar{d}}$ parameter.]{
        \includegraphics[width=0.33\linewidth,page=14]{figs/scans/param1d_all_newDimuon_May26.pdf}
        \label{subfig:par-a2_ubdb}
    }
    \qquad
    \vspace{0.05cm}
    \subfloat[Scan along the $a_4^{\bar{d}/\bar{u}}$ parameter.]{
        \centering
        \includegraphics[width=0.33\linewidth, page=36]{figs/scans/param1d_all_newDimuon_May26.pdf}
        \label{subfig:par-a4_dboud}
    }
    \caption{One-dimensional $\chi^2$ scans along selected PDF parameters. Plotted is ${\Delta \chi^2=\chi^2-\chi^2_0}$ together with a decomposition into contributions from different data types. The position of the minimum is indicated by the yellow star.}
    \label{fig:1Dscan_par}
\end{figure*}

In this section, we investigate the origin of the constraints on different parton flavors and track them back to the specific datasets. To this end, we examine two complementary diagnostics: (i)~one-dimensional parameter scans, and (ii)~$L_2$ sensitivities. Together, they provide a comprehensive understanding of how individual datasets constrain the fit.

Scans of the $\chi^2$ function along a specific parameter with all other parameters held fixed allow for an exploration of the region surrounding the identified minimum. Note that these scans differ from Lagrange multiplier scans~\cite{Stump:2001gu}, where all parameters are adjusted. The total $\chi^2$ function can then be decomposed further into contributions from individual datasets, data types, or nuclei. Here, we focus on the decomposition based on data types, which directly illustrate their impact on a given parameter. While these scans are relatively easy to interpret, they do not allow for the isolation of constraints on different regions in $x$, which are in general influenced by several parameters.

\begin{figure*}
    \includegraphics[width=0.95\textwidth]{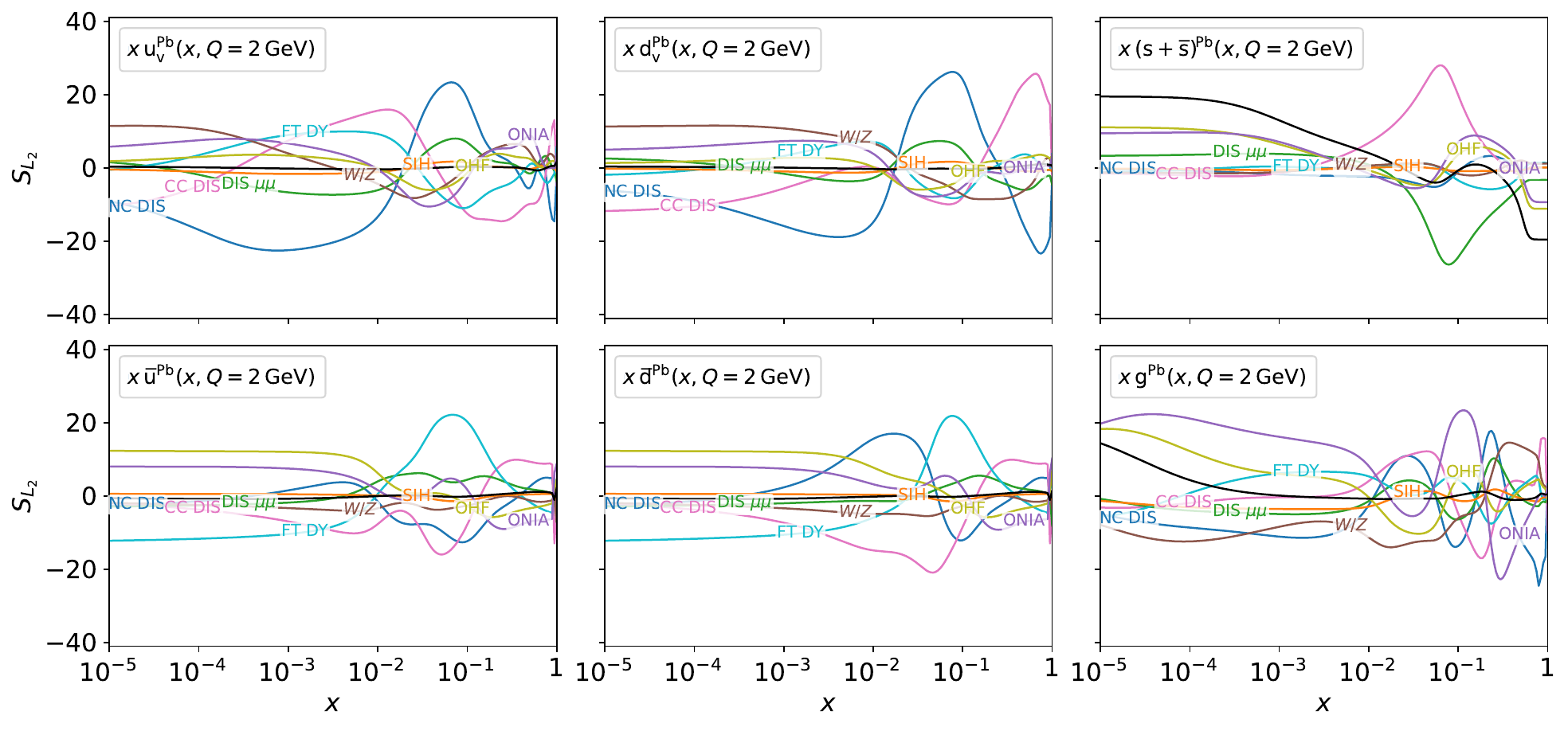}
    \caption{$L_2$ sensitivities for different flavors computed at $Q=2$ GeV. The sum of all experiments is shown in black.}
    \label{fig:L2}
\end{figure*}

Such a separation can instead be achieved with the so-called $L_2$ sensitivity, which correlates each PDF flavor with the $\chi^2$ function and has an explicit dependence on $x$ and $Q^2$ \cite{Jing:2023isu}. It is related to similar measures such as $\cos \phi$ or $\Delta \chi^2_\text{eff}$~\cite{Wang:2018heo} used in our previous nCTEQ15 analysis~\cite{Kovarik:2015cma} and defined as
\begin{eqnarray}
    S_{f,j_{\mathrm{exp}}}^{L_2} (x, Q^2)&=& 
   \frac{1}{4 \, \delta f} \sum_{i_{\mathrm{ev}} = 1}^D \left[ f_{i_{\mathrm{ev}}}^{(+)}(x, Q^2) - f_{i_{\mathrm{ev}}}^{(-)}(x, Q^2) \right]\nonumber\\
   &\times& \left[ \chi^{2(+)}_{j_{\mathrm{exp}},i_{\mathrm{ev}}} - \chi^{2(-)}_{j_{\mathrm{exp}},i_{\mathrm{ev}}} \right] \ ,
\end{eqnarray}
where
\begin{equation}
    \delta \hspace{-0.08em} f = \frac{1}{2} \sqrt{\sum_{i_{\mathrm{ev}} = 1}^D\left[ f_{i_{\mathrm{ev}}}^{(+)}(x, Q^2) - f_{i_{\mathrm{ev}}}^{(-)}(x, Q^2) \right]^2}
\end{equation}
and
\begin{equation}
    \delta \hspace{-0.05em} \chi^2_{j_{\mathrm{exp}}} = \frac{1}{2} \sqrt{\sum_{i_{\mathrm{ev}} = 1}^D \left[ \chi^{2(+)}_{j_{\mathrm{exp}},i_{\mathrm{ev}}} - \chi^{2(-)}_{j_{\mathrm{exp}},i_{\mathrm{ev}}} \right]^2}
\end{equation}
are variations of the PDFs and the $\chi^2$ function in the tolerance hypersphere. The $L_2$ sensitivity indicates the direction in which an experiment or group of experiments with index $j_\text{exp}$ pulls a PDF flavor $f$ at each point in $x$ and $Q^2$. By construction, all contributions add up to approximately zero.%
\footnote{If the total $\chi^2$ function is perfectly symmetric in the positive and negative eigenvector direction, then the $L_2$ sensitivity summed over all experiments is zero. However, in \cref{fig:L2}, we observe that for the strange distribution, and also partially for the gluon, this is not the case. This departure from zero is a consequence of the asymmetry of the total $\chi^2$ function in eigenvector direction \#28 (see \cref{fig:ev1d_all}) which almost exclusively contributes to \mbox{$s + \bar{s}$}. The gluon does not deviate from zero at the initial scale suggesting that the deviation at $Q = 2\,\mathrm{GeV}$ is caused by the evolution.}
Deviations from zero indicate that a given group of experiments prefers a different value of the PDF at a given point in $x$ and $Q^2$. It is important to note that large values of the $L_2$ sensitivity do not necessarily imply large overall pulls in the fit. It is therefore crucial to correlate the information from the $L_2$ sensitivity with the information from the $\chi^2$ scans. Moreover, while a positive $L_2$ sensitivity generally indicates that an experiment favors a smaller PDF value and vice versa, this interpretation is valid only when the PDF changes monotonically along the eigenvector direction \cite{Jing:2023isu}. When this is not the case, the local slope of the PDF, approximated by the finite difference \mbox{$(f^{(+)}_{i_\text{ev}} -f^{(-)}_{i_\text{ev}}) / 2$}, changes sign near an extremum, and the sign of the $L_2$ sensitivity also changes. As a result, the absolute sign of the $L_2$ sensitivity should be interpreted with care. However, because this sign ambiguity originates from the PDF itself rather than from the experimental data, comparing the relative signs of $S_{f,j_{\mathrm{exp}}}$ remains meaningful: experiments with opposite signs can still be regarded as pulling the PDF in opposite directions. Note also that only values comparable to or larger than the value of the employed Hessian tolerance criterion indicate significant tensions.

One-dimensional $\chi^2$ scans along selected PDF parameters are shown in \cref{fig:1Dscan_par}. For the valence parameters $a_1^{u_v}$ and $a_3^{d_v}$, which are representative of the behavior of the other valence parameters, it is clear that the total $\chi^2$ is dominated by the contribution from the NC DIS data (blue lines). Additional constraints come from the CC DIS data (pink lines) and $W/Z$-boson production data from the LHC (brown lines). However, the last two datasets prefer slightly different values of the parameters as compared to NC DIS, lower or higher depending on the parameter under consideration. Even weaker constraints are provided by the fixed-target Drell-Yan data (cyan lines), which, however, aligns with the minimum preferred by the NC DIS data.

The corresponding $L_2$ sensitivities are shown in \cref{fig:L2}. For the valence distributions (two upper left panels), we 
observe that the NC and CC DIS data also prefer different shapes with a crossing at around $x{\sim}2\times10^{-2}$, however they converge at very small $x$. On the other hand, the $W/Z$ data is pulling in opposite direction than the NC DIS also at the smallest $x$ values, which is consistent with the observations from the $\chi^2$ scans. It should be however noted, that values of $L_2$-sensitivity for valence distributions (but also all other flavors) are always considerably smaller that the Hessian tolerance, which means that the opposite pulls shouldn't be interpreted as tensions and that the data are compatible within the tolerance.

Moving on to the other flavors, it is interesting to first investigate the normalization parameters separately. We remind the reader that the normalization parameters for the valence and light-sea \mbox{($\bar{u}+\bar{d}$)} distributions are not free parameters, but fixed by the number and the momentum sum rules. In \cref{subfig:par-g_a0}, we therefore present the scans for the $a_{\mathrm{sum}}^{g}$ parameter, while \cref{subfig:par-ssb_a0} displays the scans for the $a_{\mathrm{sum}}^{s+\bar{s}}$ parameter. These parameters capture the leading coefficient of the $A$ dependence of the gluon and the strange quark normalizations, respectively. For both the $a_{\mathrm{sum}}^{g}$ and $a_{\mathrm{sum}}^{s+\bar{s}}$ parameters, we observe a dominant contribution coming from the NC~DIS data. This is at first sight surprising, as these data are not typically expected to provide significant constraints on the gluon and strange distributions, but can be explained by the momentum sum rule, which relates all PDF flavors. The remaining constraints on the normalization parameters come from CC~DIS data, FT DY data, and, for the strange quark, neutrino dimuon data. In addition, we note that the CC~DIS data prefer a lower normalization for both gluon and strange distributions as compared to the NC~DIS data.

When examining the shape parameters, the situation is different. We start with the gluon, where we present in \cref{subfig:par-g_a1} a scan along the $a_1^g$ parameter, associated with the low-$x$ gluon behavior. We observe that the main contribution to the total $\chi^2$ originates from three datasets: quarkonium (yellow line), open heavy-flavor (violet line), and $W/Z$-boson production (brown line) data from the LHC. This situation persists for all other gluon shape parameters. Furthermore, we observe that the location of the minimum is a compromise between the HQ data (both ONIA and OHF) pulling towards lower values of the parameter $a_1^g$ and the $W/Z$-boson data pulling towards higher values. Such a compromise is also seen in other gluon parameters, yielding a consistent pattern of behavior throughout the whole range in $x$. 
The $L_2$ sensitivity (lower right panel in \cref{fig:L2}) is in line with this observation as the HQ and the $W/Z$-boson data are pulling in opposite directions for most of the $x$ range, especially at low $x$. However, the absolute values of the corresponding $L_2$ sensitivities are much smaller than the tolerance, which indicates that the variations observed in the $\chi^2$ scans across gluon parameters are insufficient to be interpreted as notable tensions between these datasets.

Regarding the shape parameters of the strange distribution, \cref{subfig:par-ssb_a2} displays a profile for the \mbox{high-$x$} $a_2^{s+\bar{s}}$ parameter. As expected, the dimuon neutrino data provide the primary constraints, as they constitute the dominant contribution to the total $\chi^2$~\cite{Rabinowitz:1993xx,CCFR:1994ikl,Kretzer:2001tc,Olness:2003wz,Mason:2004yf,Mason:2006qa,NuTeV:2007uwm}. Interestingly, the second largest contribution to $\chi^2$ comes from the NC~DIS data, while the CC~DIS data provides only the third but comparable contribution. The sensitivity to the NC~DIS data possibly originates from the relatively large number of data points. The same behavior is also observed for other strange shape parameters with the two exceptions of $a_4^{s+\bar{s}}$ and $b_2^{s+\bar{s}}$. For these parameters, the contribution from the CC~DIS data exceeds that of the NC~DIS data for some of their values. The corresponding $L_2$ sensitivity (upper right panel in \cref{fig:L2}) shows a contrasting preference between the CC~DIS and the dimuon data throughout intermediate to large $x$ range.

Finally, the constraints for the light sea-quark parameters are typically dominated by the NC~DIS data, with smaller contributions from the FT DY, CC~DIS, and $W/Z$-boson data. This hierarchy changes slightly for the parameters of the light-sea ratio, $\bar{d}/\bar{u}$,
for which the FT DY data become more constraining. These observations are illustrated for selected parameters in \cref{fig:1Dscan_par}. Furthermore, as shown by the $L_2$ sensitivity in \cref{fig:L2}, even though the NC~DIS data drives the minimum for a dominant part of the $x$ range, it remains in slight tension with the other experiments. Nevertheless, the $L_2$ sensitivity values remain small relative to the tolerance.

The data constraints relevant for the strange and gluon distributions will be discussed in more detail in sections~\cref{subsec:strange,subsec:gluon}.

\subsection{PDFs}
\label{subsec:pdfs}

For the nCTEQ26 analysis, we parametrize the nPDFs at the initial scale $Q_0=1.3$ GeV using the parametrization introduced in \cref{subsec:param}. As a result of the analysis, we obtain the fitted PDF parameters that fully define the nPDFs at this scale. The values of the fitted and fixed parameters are summarized in \cref{tab:params}, which provides the necessary information to fully reconstruct the nPDFs at the initial scale. The nPDFs are then evolved with the DGLAP evolution equations to higher scales and saved as grids in the LHAPDF format~\cite{Buckley:2014ana}, thereby allowing easy access (cf.\ \cref{app:usage}).

\input{tables/parameter_nCTEQ26}

\begin{figure}[!tbp]
    \centering
    \subfloat[nPDFs for various flavors as a function of $x$.]{
        \label{subfig:ncteq26_Pb_Q2_abs}
        \includegraphics[width=0.95\linewidth]{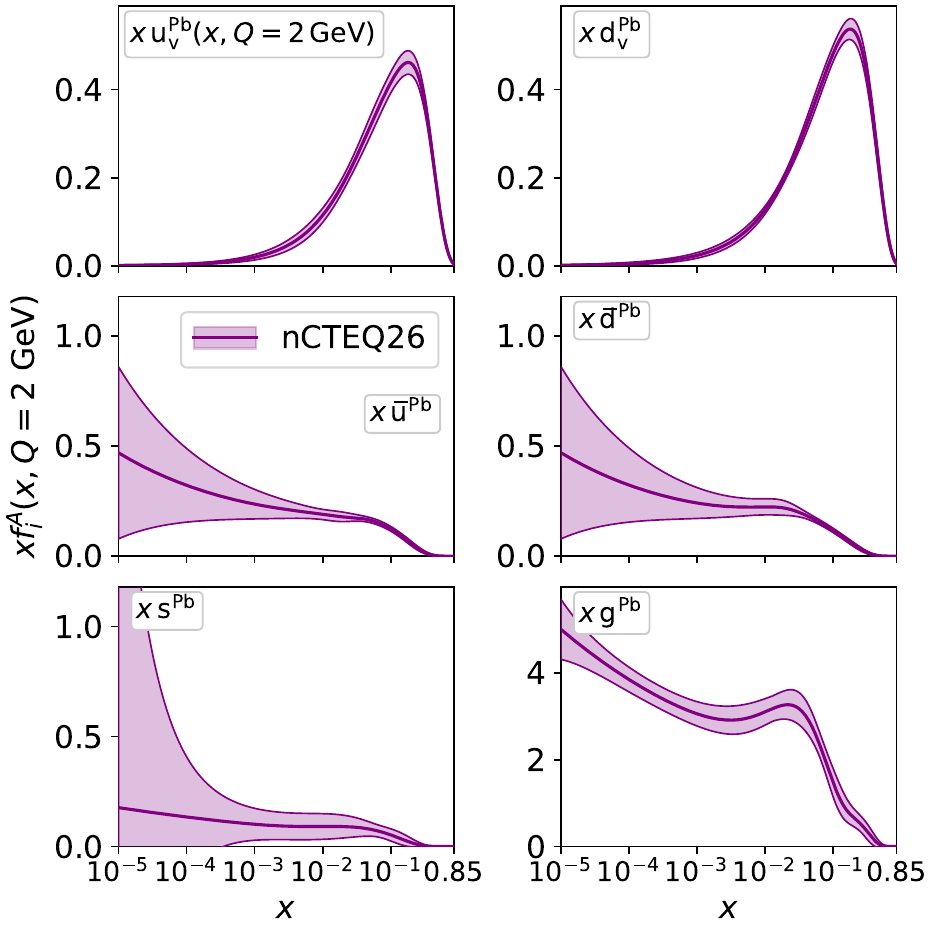}
    }
    \\
    \subfloat[nPDF ratios for various flavors as a function of $x$.]{
        \label{subfig:ncteq26_Pb_Q2_ratio}
        \includegraphics[width=0.95\linewidth]{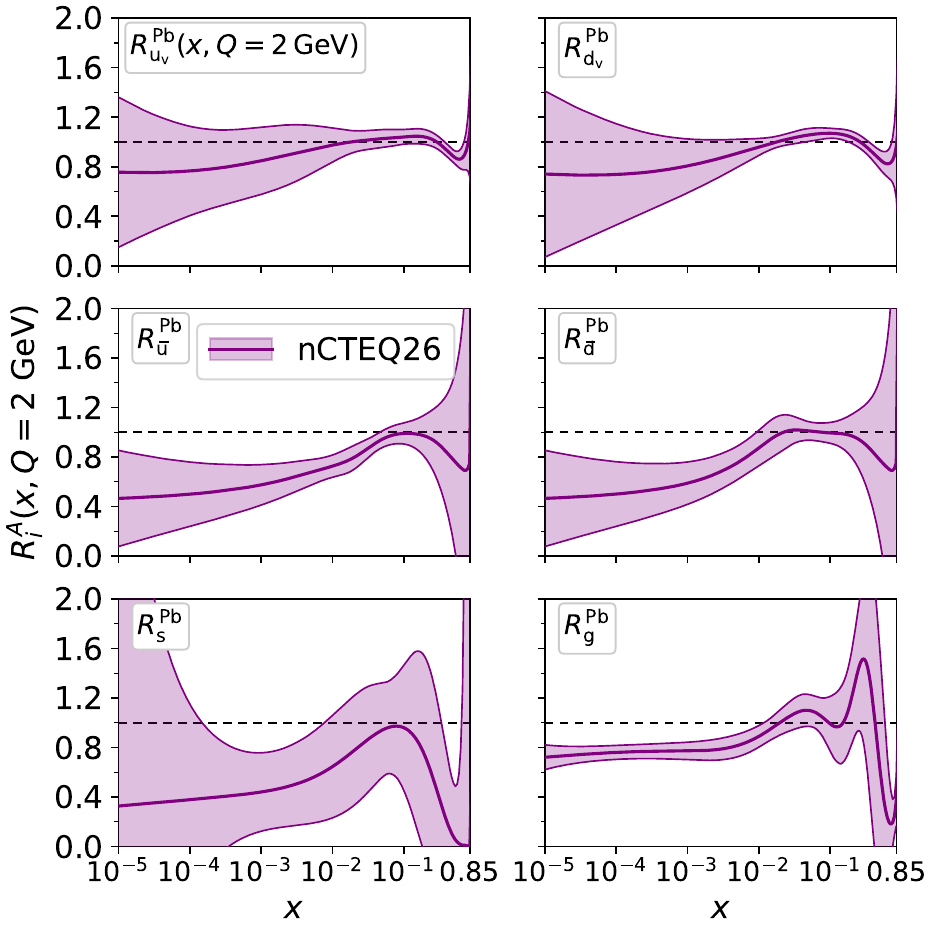}
    }
    \caption{(a) nCTEQ26 nPDFs for lead nuclei at the scale $Q=2$ GeV. (b) Corresponding ratios of the full nPDFs for lead nuclei to those constructed from the free-proton baseline using Eq.~\eqref{eq:nPDFA}.}
    \label{fig:ncteq26_Pb_Q2} 
\end{figure}

\begin{figure}[!tbp]
    \centering
    \subfloat[nPDFs for various flavors as a function of $x$.]{
        \label{fig:ncteq26_Pb_Q10_abs}
        \includegraphics[width=0.95\linewidth]{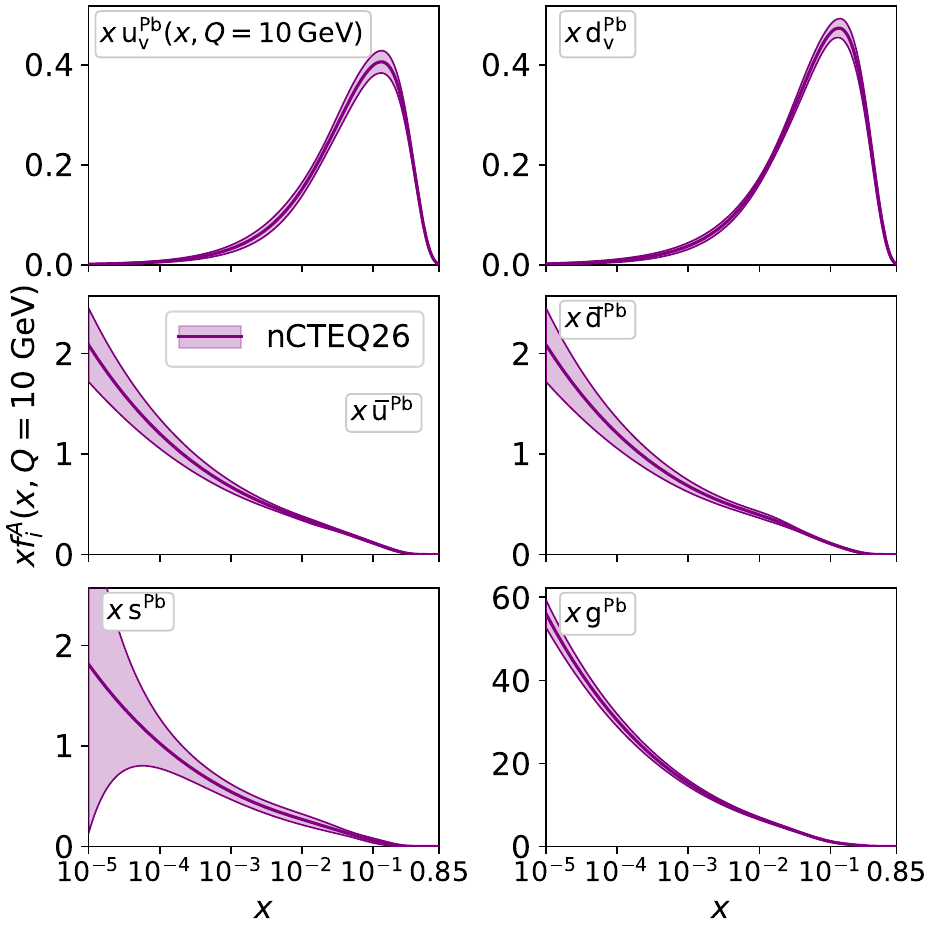}
    }
    \\
    \subfloat[nPDF ratios for various flavors as a function of $x$.]{
        \label{fig:ncteq26_Pb_Q10_ratio}
        \includegraphics[width=0.95\linewidth]{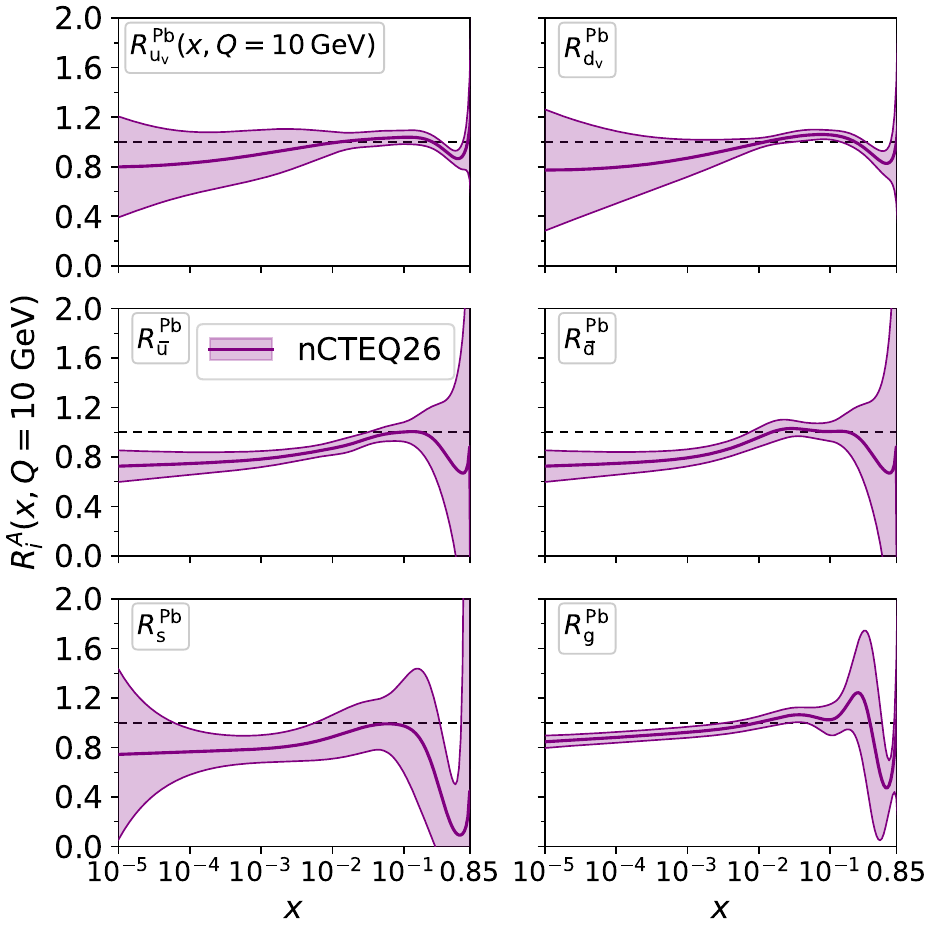}
    }
    \caption{Same as \cref{fig:ncteq26_Pb_Q2}, but at $Q=10$ GeV.}
    \label{fig:ncteq26_Pb_Q10} 
\end{figure}

In \cref{fig:ncteq26_Pb_Q2} and \cref{fig:ncteq26_Pb_Q10} we display the new nCTEQ26 nPDFs in lead nuclei at scales $Q=2$ and $10$ GeV. 
These are the ``full'' nuclear distributions as defined by \cref{eq:nPDFA}. 
The upper panels of these figures show the absolute distributions, and the lower panels show the ratios to the free-proton distributions, which are also constructed according to \cref{eq:nPDFA} using the baseline proton PDFs:
\begin{equation}
\begin{split}
\label{eq:RA}
    &R_i^A
    = \frac{A f_k^{A}}{Zf_k^{\mathrm{free}\,p} + (A-Z)f_k^{\mathrm{free}\,n}}\,.
    \end{split}
\end{equation}
We can see the characteristic valence peak, which, due to the excess of neutrons in lead, is slightly higher in the $d$-valence distribution compared to the $u$-valence. 
The $\bar{u}$ and $\bar{d}$ distributions are rather similar, with some differences visible at a low scale of $2$ GeV, but they mostly vanish already at the scale of $10$ GeV. 
It is also evident that at low-$x$ ($\lesssim10^{-3}$), the relative uncertainties begin to grow substantially across all flavors except the gluon. 
Looking at the strange distribution, we see much larger uncertainties than for the $\bar{u}$ and $\bar{d}$ sea quarks, indicating limited constraints for strange. 
It should be noted that for low $x$, the strange distribution is smaller than the $\bar{d}$ or $\bar{u}$ distribution. This is consistent with the expected suppression of $g\to s\bar{s}$ splittings due to the mass of the strange quark.
Finally, the gluon distribution at low scales has a characteristic valence-like shape that evolves rapidly into a sea-like shape with increasing scale.
At low scales, we also observe a feature in the gluon distribution at $x\sim0.25$, which is visible mostly in the shape of the error band. This is probably related to the shape of the proton baseline and the interplay with sum rules. 
Nevertheless, it is not expected to have a significant impact on the phenomenological predictions, as the gluon PDF at these values of $x$ is very small and vanishes rapidly at larger scales.


Now we turn our attention to the lower panels in \cref{fig:ncteq26_Pb_Q2,fig:ncteq26_Pb_Q10}.
We observe a standard pattern of nuclear effects across all flavors: shadowing at small $x$ (roughly ${x\lesssim 10^{-2}}$), a transition through the intermediate-$x$ region where anti-shadowing may occur ($x\,{\sim}\,0.1$), an EMC-type depletion at larger $x$ (approximately ${x\sim 0.3-0.7}$), and finally a steep rise as $x\to 1$ of distributions as well as their uncertainties associated with Fermi-motion effects.\footnote{The Fermi-motion rise typically starts above $x=0.8$, for some flavors above the upper $x$-axis limit, which we don't show due to lack of reliable data constraints.}
For the light sea, $R^{\text{Pb}}_{\bar{u}}$ and $R^{\text{Pb}}_{\bar{d}}$ are very similar in shape and uncertainty, and they exhibit pronounced shadowing and mild to negligible anti-shadowing. At $Q = 2$ GeV, the ratios drop to $0.5$ at the low end and remain close to one in the $x\sim 10^{-2}$--$10^{-1}$ range before entering the EMC region at higher $x$. 
The valence ratios $R^{\rm Pb}_{u_v}$ and $R^{\rm Pb}_{d_v}$ have a similar shape but appreciably shallower shadowing, consistent with no effect within uncertainties, while the antishadowing is more prominent compared to the up- and down-type sea. Overall, the valence distributions are better constrained, except at low $x$ where the distributions approach zero. 
The strange sector, $R^{\rm Pb}_{s}$, resembles the up and down sea at low $x$ and the gluon in high $x$ in the size and shape of its shadowing and EMC depletion. 
The gluon modification $R^{\rm Pb}_g$ is the most structured: it shows a significant shadowing at small $x$, a nontrivial anti-shadowing feature in the intermediate-$x$ region, and a notably pronounced EMC depletion at larger $x$. Note that the structure visible in the antishadowing region in the ratio appears to correlate with the feature in the absolute distribution.

The fitted lead PDFs exhibit a clear hierarchy in their uncertainties. The valence sector is comparatively well constrained, especially in the high- to mid-$x$ region. The light sea remains broader but still reasonably controlled over much of the relevant $x$ range. The strange distribution features the largest uncertainties, with substantial widening of error bands at the lowest $x$ values, $\lesssim10^{-4}$. Finally, the gluon distribution shows a different pattern with the low-$x$ region displaying the smallest uncertainties that grow at mid/high $x$, especially at the initial scale.

The DGLAP evolution from ${Q=2}$~GeV to ${Q=10}$~GeV generally weakens the magnitude of shadowing and reduces the spread of the uncertainty bands at small $x$, particularly for the sea quarks and the gluon.
As discussed in the following sections, the strange and the gluon sectors provide the most direct window into the remaining limitations and structure of the fit.

\begin{figure}[tb]
     \includegraphics[width=\linewidth]{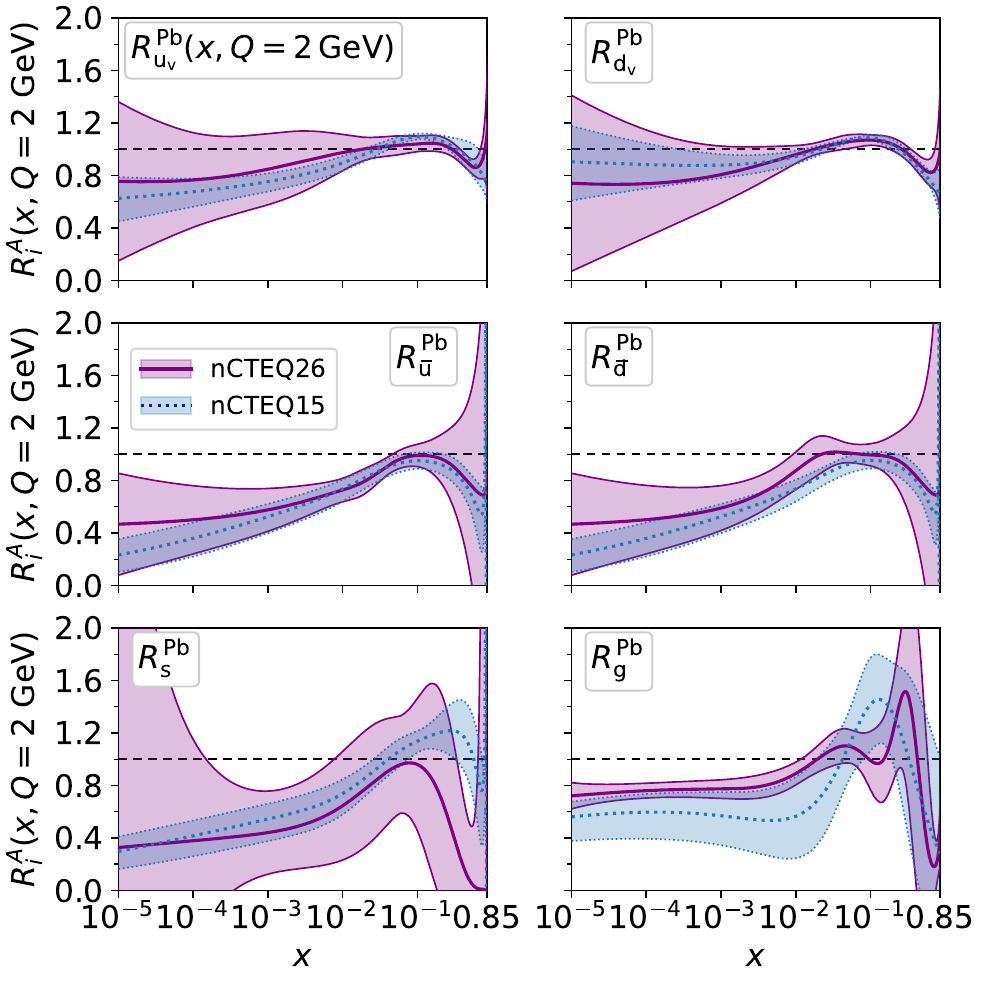}
    \caption{Same as \cref{fig:ncteq26_Pb_Q2} (b), but compared to the corresponding ratios from the nCTEQ15~\cite{Kovarik:2015cma} analysis.}
    \label{fig:ncteq26-vs-ncteq15_Pb_Q2} 
\end{figure}

\begin{figure}[tb]
    \centering
    \includegraphics[width=\linewidth]{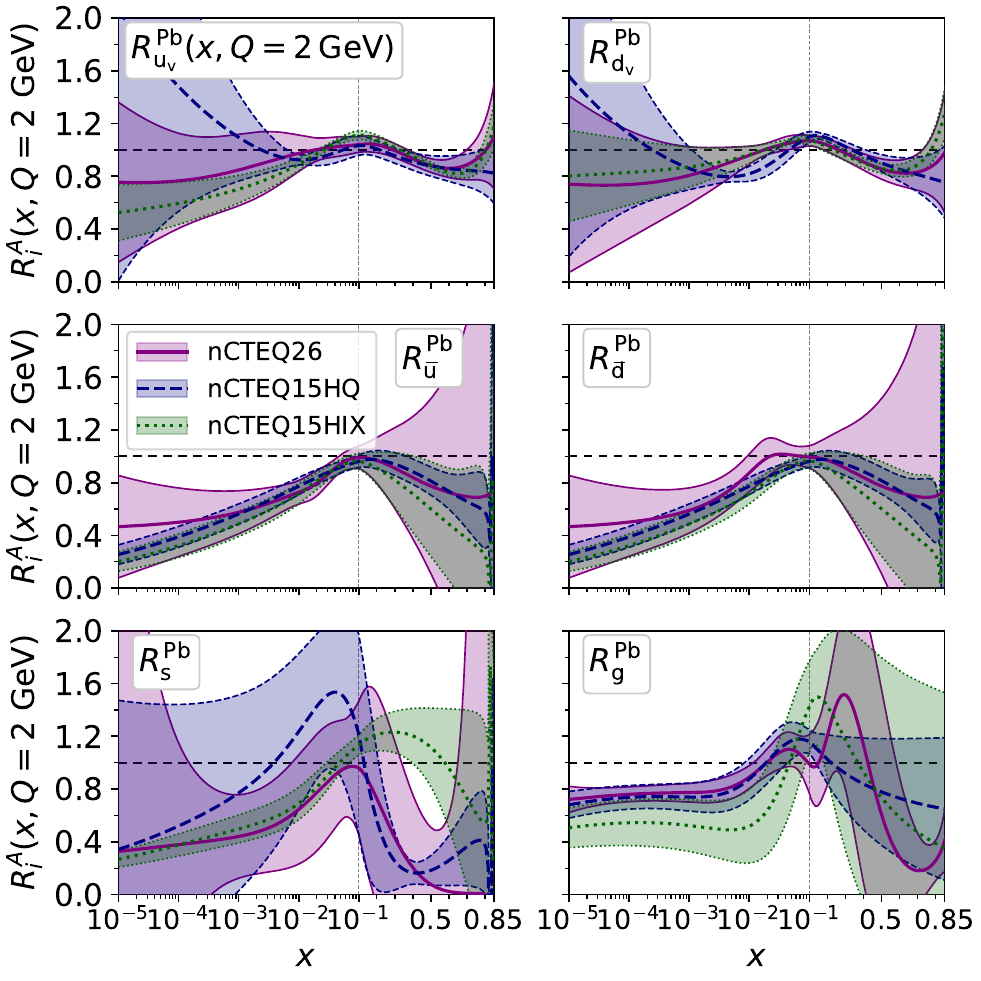}
    \caption{Same as \cref{fig:ncteq26_Pb_Q2} (b), but compared to the corresponding ratios from the nCTEQ15HIX~\cite{Segarra:2020gtj} and nCTEQ15HQ~\cite{Duwentaster:2022kpv} analyses.}
    \label{fig:ncteq26_ncteq15hix_ncteq15hq_Pb_Q2} 
\end{figure}

To provide more context for the obtained lead PDFs, in \cref{fig:ncteq26-vs-ncteq15_Pb_Q2,fig:ncteq26_ncteq15hix_ncteq15hq_Pb_Q2} we compare them with results from our previous analyses~\cite{Duwentaster:2022kpv,Segarra:2020gtj}.
In \cref{fig:ncteq26-vs-ncteq15_Pb_Q2}, we show comparisons with our 10-year-old pre-LHC nCTEQ15 nPDFs. We observe here a result that, at first sight, is rather counterintuitive. For all flavors, apart from the gluon, the uncertainties in the new nCTEQ26 nPDFs are larger than in the case of nCTEQ15. 
This can be attributed to the limited amount of data available in the nCTEQ15 analysis, which restricted our ability to open a sufficient number of parameters.
This, in turn, resulted in significantly underestimated uncertainties due to parametrization bias. 
(Similar results were observed in other nPDF analyses from that time, e.g.~\cite{Eskola:2009uj}.)  
The only exception is the gluon distribution, for which the amount of HQ data from proton-lead collisions at the LHC brings sufficiently stringent constraints that the current gluon uncertainties are still smaller than the underestimated uncertainties present in the nCTEQ15 analysis. In this context, it is rather exceptional that the nCTEQ15 and nCTEQ26 nPDFs agree so well within the provided uncertainties.
With the introduction of new data from the LHC, this situation is improving, allowing more open fitting parameters and yielding more realistic uncertainties. This is reflected in \cref{fig:ncteq26_ncteq15hix_ncteq15hq_Pb_Q2}, where we compare the new PDFs with the results of nCTEQ15HIX and nCTEQ15HQ, which already include more data than nCTEQ15.

Several observations are worth noting in this comparison. At low $x$, we find the reduction of the gluon uncertainties from nCTEQ15HIX to nCTEQ15HQ, and then to the current nCTEQ26 analysis. This is what we expect, as in this progression of analyses we included more and more data providing constraints on gluons at low $x$, particularly the heavy-quark(onium) data.
Another point to note in
\cref{fig:ncteq26_ncteq15hix_ncteq15hq_Pb_Q2} is the behavior at large $x$. In the nCTEQ15HIX analysis, we have focused on the high-$x$ region by extending the $\{Q,W\}$ kinematic cuts and including JLAB data, thereby providing precise information on the high-$x$ EMC region. The same treatment is also employed in the nCTEQ26 fit, but we now include much more data for many different processes, including the CC DIS data from the CHORUS experiment and additional NC DIS data from the Hall~C experiment at JLAB. 
For the valence distributions and their uncertainties we find that nCTEQ26 agrees well with nCTEQ15HIX up to $x \,{\sim}\, 0.6$. At even higher $x$ values, the central predictions align, but the nCTEQ26  uncertainties are significantly larger than in nCTEQ15HIX as a consequence of the greater flexibility introduced by the expanded parametrization.

Furthermore, compared with our previous analyses, we observe a significant evolution in the uncertainties on the strange distribution. Both the strange and the gluon distributions, including their uncertainties, will be discussed in more detail in \cref{subsec:strange,subsec:gluon}.

\subsection{\texorpdfstring{$A$}{A} dependence}
\label{subsec:Adep}

The nCTEQ26 analysis provides predictions for PDFs for a broad range of nuclei.
As discussed in \cref{subsec:param}, a quadratic function in $\ln A$ is used to parametrize the nuclear $A$ dependence.
This allows for rapid changes in light nuclei and slower changes in heavier nuclei, a behavior already preferred in the nCTEQ15 fit and now well represented by the new parametrization.
The initial expectation was that the linear terms proportional to $\ln A$ would provide the leading component of the $A$ dependence with only a small correction from the quadratic term. 
As we will see, this is indeed the case for a majority of parameters.

An analytic $A$ dependence of input parametrization coefficients allows us to obtain predictions for nuclei not included in the fit. 
Note that this analytic prescription also provides a recipe for extrapolating our predictions to nuclei heavier than any nucleus with available data.
It also has the effect of pooling and combining the constraining power of data taken on different nuclei, i.e.~data on iron will have an indirect impact on copper nPDFs. 
The impact of such data pooling was investigated for the first time in a recent analysis~\cite{Derakhshanian:2026zkx} comparing two fits: one with multiple nuclei and an analytic $A$ dependence, and another relying on data for a single nucleus. 
The size of the nuclear effects tends to grow with the mass of the target, and so one may naively expect that the inclusion of data for lighter nuclei in the analysis weakens this trend.
This expectation is confirmed in Ref.~\cite{Derakhshanian:2026zkx}, but the size of the effect is typically smaller than the uncertainty of the extracted nPDFs. 
In the future, it could be considered as an additional source of uncertainty, but it is not accounted for in this study.

\begin{figure}[t]
    \centering
    \includegraphics[width=\linewidth]{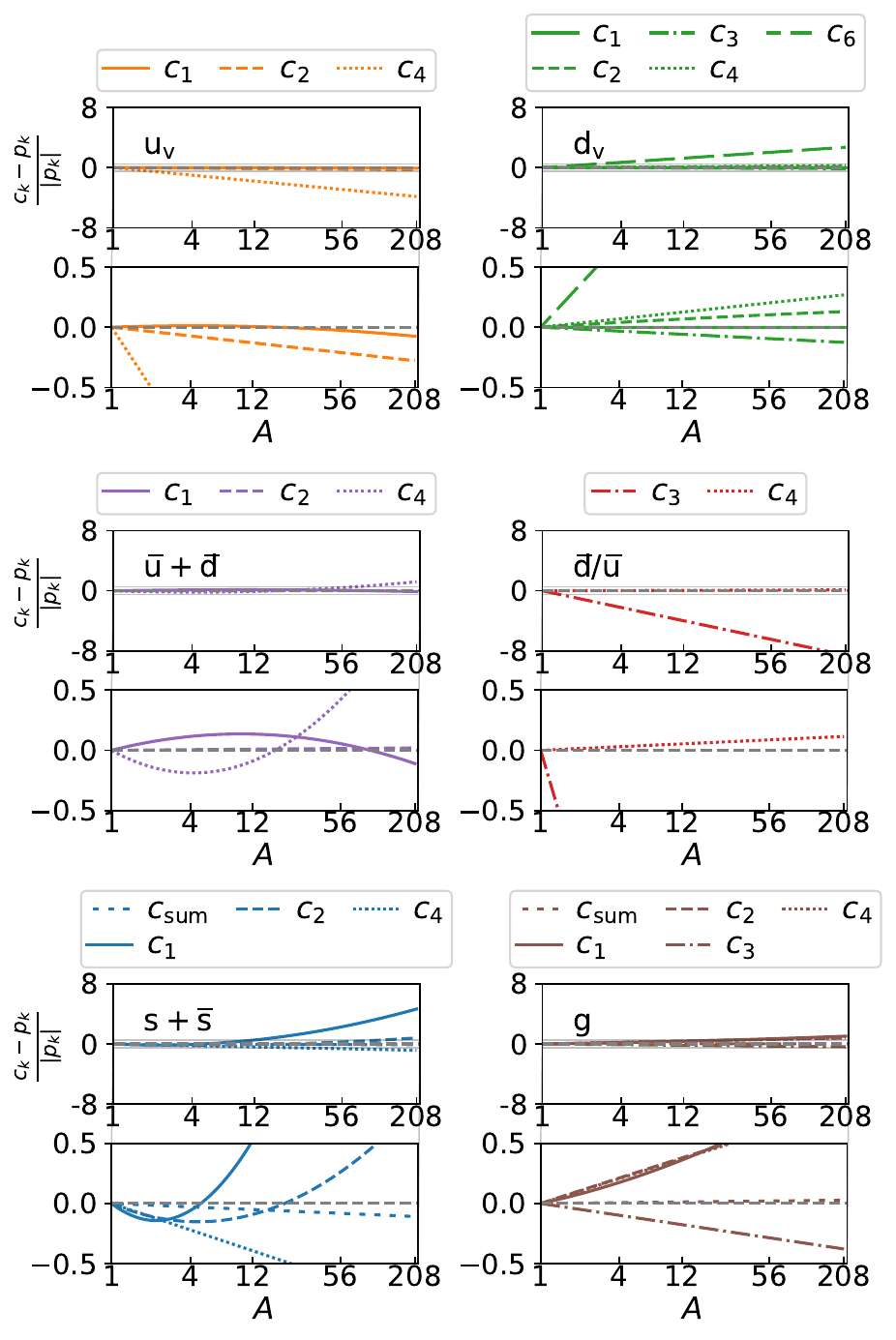}
    \caption{$A$-dependence of nuclear $c_i$ parameters as given in \cref{eq:Amod2}, normalized to its proton component, $p_i$. For each flavor we show two panels with the same content but different $y$-axis ranges, facilitating an easier comparison.
    There are parameters that have a zero $p$ component but a non-zero $a$ and $b$ components, namely $c_3^{u_v}$, $c_3^{\bar{u}+\bar{d}}$, and $c_5^{s+\bar{s}}$. We do not plot those parameters. We also omit the $c_i$ parameters that only have a $p$ component thus and have no $A$-dependence and are fixed to their proton value.
    }
    \label{fig:c-A}
\end{figure}
In \cref{fig:c-A} we show the dependence of the nuclear parameters on $A$. 
We plot the nuclear part of the total $c_k$ parameter of \cref{eq:Amod2} normalized to its free proton component $p_k$.
To facilitate comparison of the magnitude of the coefficients for different flavors, we set all $y$-axes limits of the plots to be identical.
The strongest dependence on $A$, in the range $A \in [1,208]$, is found in the following parameters: $c_3^{\bar{d}/\bar{u}}$, $c_1^{s+\bar{s}}$, $c_4^{u_v}$, and $c_3^{d_v}$.
Since the $d_v$ parametrization is linked to the $u_v$ one, the $A$ dependence of the $u_v$ parameters also influences the $A$ dependence of the $d_v$ distribution.
The dependence on $A$ for the $\bar{u}+\bar{d}$ and gluon parameters is a bit milder.
The plot also shows secondary panels with reduced $y$ limits to resolve such milder dependencies. 
Due to the quadratic parametrization, either straight lines or parabolas can be observed, depending on which parameters were included in the fit and their best-fit values.
For the majority of the parameters, the dependence on $\ln A$ is mostly linear, with the most notable exceptions for the low-$x$ parameter ($c_1$) for the strange and $\bar{u}+\bar{d}$ distributions, the high-$x$ parameter ($c_2$) of the strange distribution, and the $c_4$ parameter for $\bar{u}+\bar{d}$ distribution. The strong quadratic behavior appears to correlate with the size of the uncertainty in the corresponding regions of the relevant distributions.
Those parameters also feature non-monotonic behavior over the plotted range.
The strange $c_1$ parameter plays a special role here, since for small~$x$ the strange distribution is dominated by the term $x^{c_1}$;
hence, $x f_s\sim x^{c_1}$.
If $c_1<0$, the strange distribution, $xf_s$, rises rapidly as $x\to0$, producing a sea-like shape. If $c_1>0$, the distribution instead reaches a maximum and decreases as $x\to0$, resulting in a valence-like shape at the input scale.
However, it should be noted that in this kinematic range, there is almost no data available that could directly constrain the strange distribution. 
This results in large uncertainties in the strange PDFs, since even a modest shift in $c_1$ from negative to positive can dramatically alter the behavior of the $x^{c_1}$ term at small $x$.
It should also be noted that the nonzero $c_5$ parameter in the strange distribution is crucial to capturing this uncertainty. Even though it is not visible from \cref{fig:c-A}, the extra freedom provided by this parameter allows for a more complete exploration of the nPDF uncertainty, something which will be further discussed in \cref{subsec:strange}.

\begin{figure*}[tbh]
    \centering
    \includegraphics[width=0.95\linewidth]{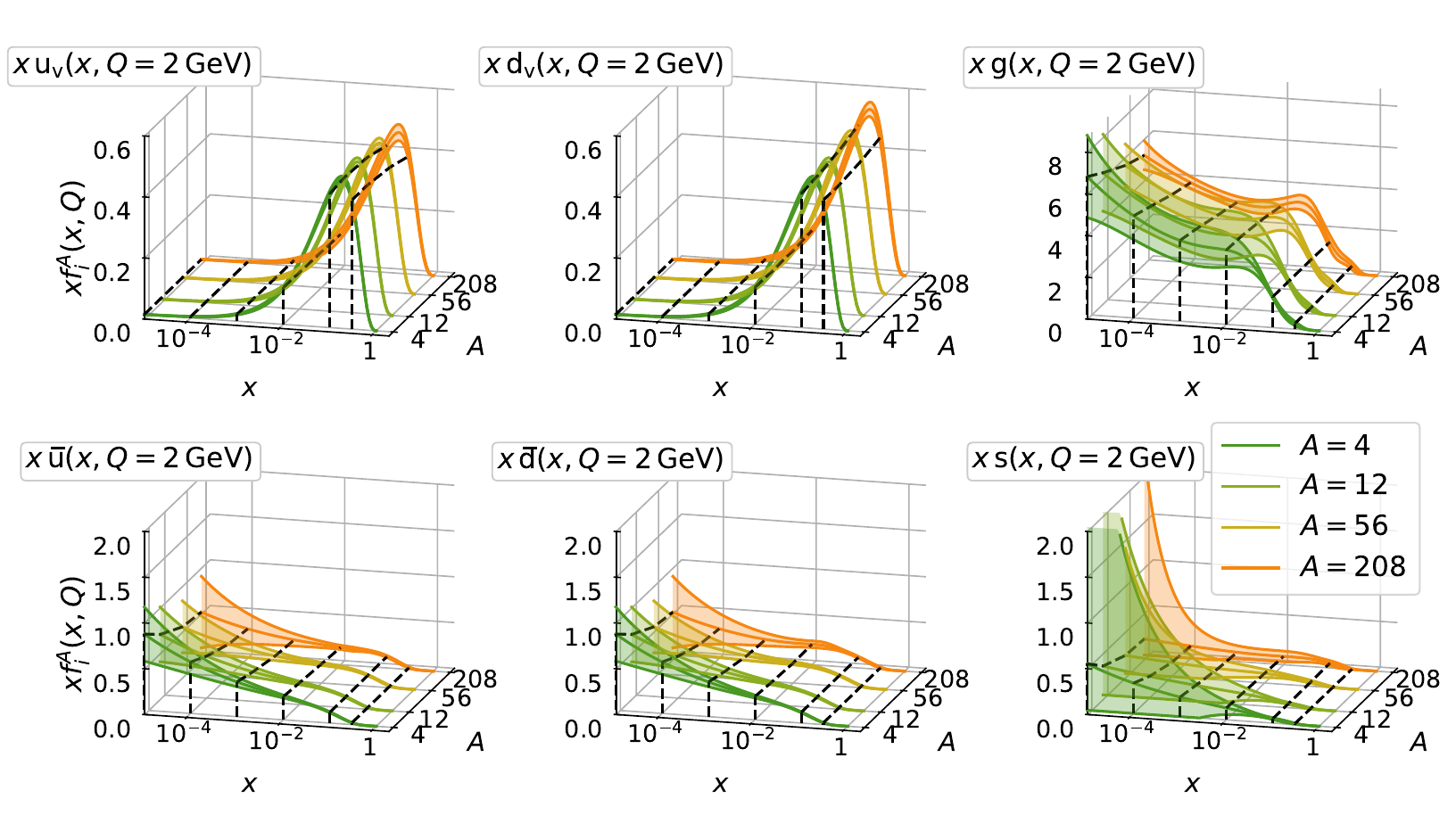}
    \caption{Overview of the $A$ dependence of the nCTEQ26 PDFs shown in three dimensions in dependence on momentum fraction $x$ and nuclear mass number $A$. Plotted are PDFs for helium, carbon, iron, and lead at scale $Q=2$ GeV.}
    \label{fig:a-dependence-true-3d}
\end{figure*}

\begin{figure}[tbh]
    \centering
    \includegraphics[width=\linewidth]{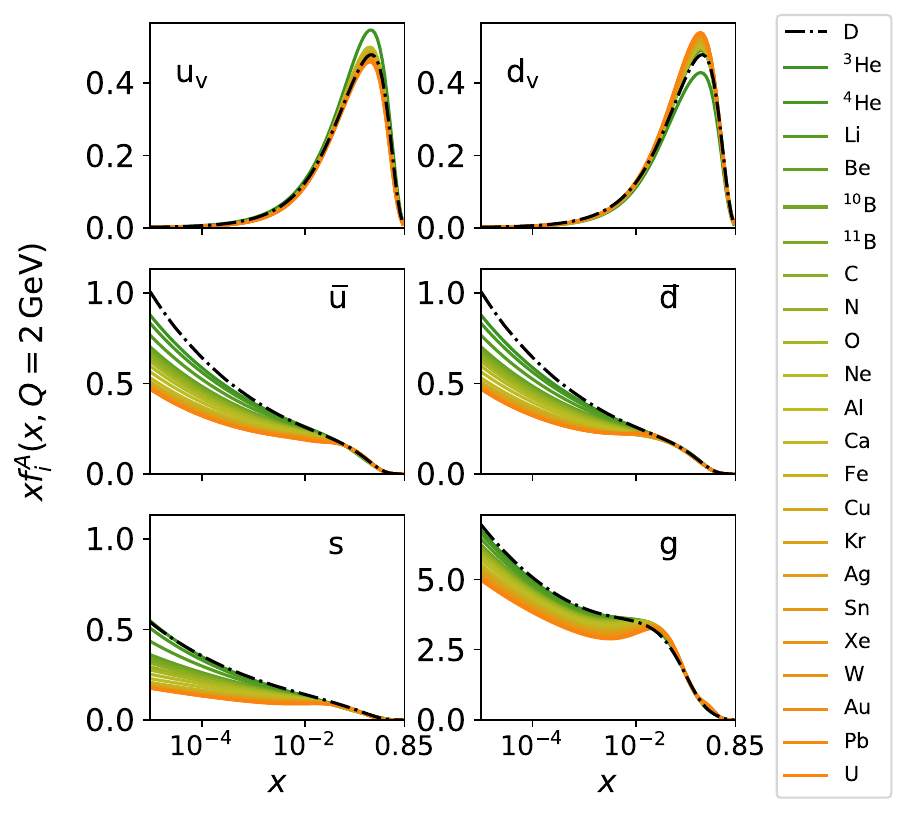}
    \caption{Central nCTEQ26 nPDFs, $f_i^A$, plotted as a function of $x$ for 23 different nuclei ranging from deuteron to uranium.}
    \label{fig:a-dependence-wou}
\end{figure}

\begin{figure}[ht]
    \centering
    \includegraphics[width=\linewidth]{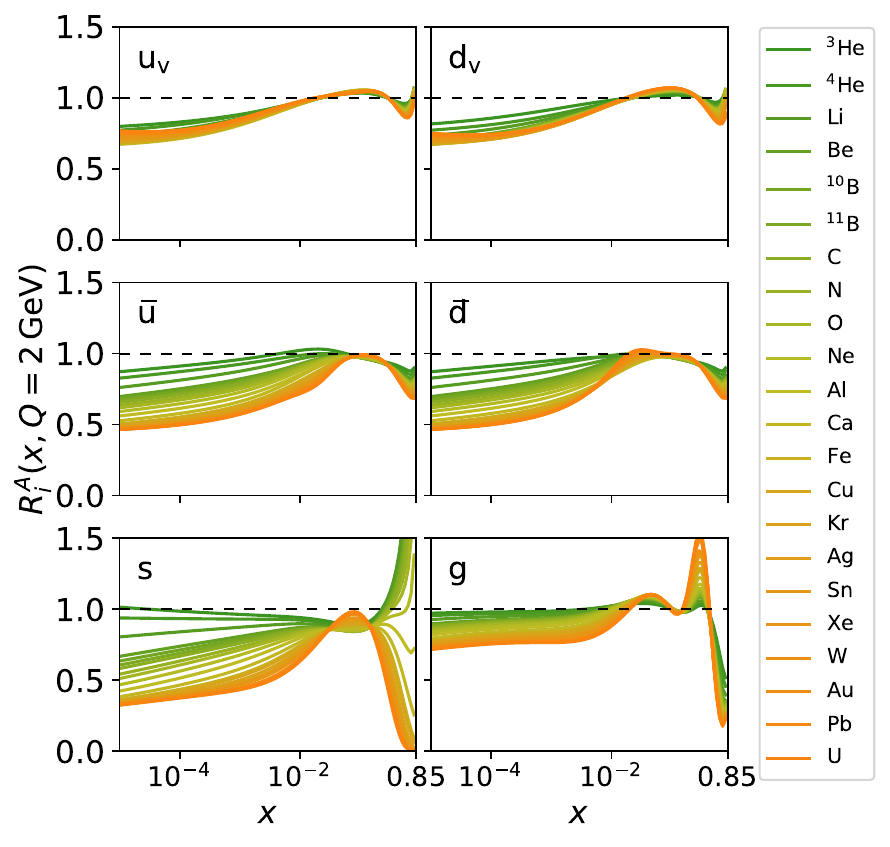}
    \caption{Central values of nuclear modification ratio of \cref{eq:RA} plotted as a function of $x$ for a range of nuclei from helium to uranium.}
    \label{fig:a-dependence-wou-r}
\end{figure}

To examine how the behavior of the $c_k$ parameters with respect to $A$ translates into nuclear dependence in PDFs, we plot them directly. 
In \cref{fig:a-dependence-true-3d}, we first present a general overview of the $A$ dependence where selected nuclei with ${A \in [4,12,56,208]}$ are plotted in three-dimensions displaying the PDFs ${xf_i^A(x, Q= 2~\text{GeV})}$, together with their uncertainties. We observe that the valence distributions are well constrained for light, medium, and heavy nuclei. The light-sea quark distributions also show a rather stable behavior of uncertainties with increasing $A$; however, in this case, the uncertainties grow appreciably at the lowest $x$ values. In case of the strange nPDFs, the behavior is similar to the light sea but with an even more pronounced increase in the uncertainties at low $x$, and a rather clear reduction in uncertainties for the medium $A$ nuclei represented here by the iron nucleus. Finally, for gluons, we observe that uncertainties are the smallest for heavy nuclei (Pb) and increase monotonically with decreasing $A$.
Interestingly, as pointed out above, the trends of increased/reduced uncertainties with the rising mass of nuclei are different between the light-sea quarks and the gluon. 
This happens in spite of the close connection between these distributions 
through the DGLAP evolution where the gluon splitting ($g\to q \bar{q}$) contributes to the quark distributions.
The reason for this is the different origin of data constraints: in the case of gluons, most of the constraints originate from the proton-lead LHC data, whereas in the case of light-sea quarks, an important sensitivity is provided by the fixed target DIS and DY data spanning a broad range of nuclei and, in particular, the lighter ones, e.g., carbon and iron. 
To increase the level of detail, we show the central PDFs, ${xf^A(x, Q = 2~\text{GeV})}$, for a wide selection of nuclei as a function of $x$ in \cref{fig:a-dependence-wou}.
Here, one can see that at low $x$ the central PDFs of light nuclei (green shades) differ significantly from each other, with a relatively large spread, whereas for heavy nuclei (orange shades) only minor differences are observed.
It is also worth noting the results for the two lightest nuclei: deuteron and $^3$He. We can see that the results for the deuteron do not always follow the monotonic changes observed for other nuclei. 
One of the reasons for this is the fact that deuteron PDFs are not fitted in our analysis, and the presented distributions are the isoscalar combinations of free proton and neutron PDFs.\footnote{Deuteron corrections in our analysis are applied on the level of structure functions, but the PDFs themselves are not altered.} 
The $^3$He nucleus is treated as any other nucleus, but because it has the largest possible non-isoscalarity in our analysis (2~protons, 1~neutron), the correction coming from the difference in the number of up and down quarks is maximal. This leads to larger differences in the nuclear distributions, $f_i^A$, which is especially evident for the valence distributions in the peak region.

\begin{figure*}[!ptbh]
    \centering
    \subfloat[$R^A_i$ for $u$- and $d$-valence distributions.]{
        \includegraphics[width=0.99\linewidth]{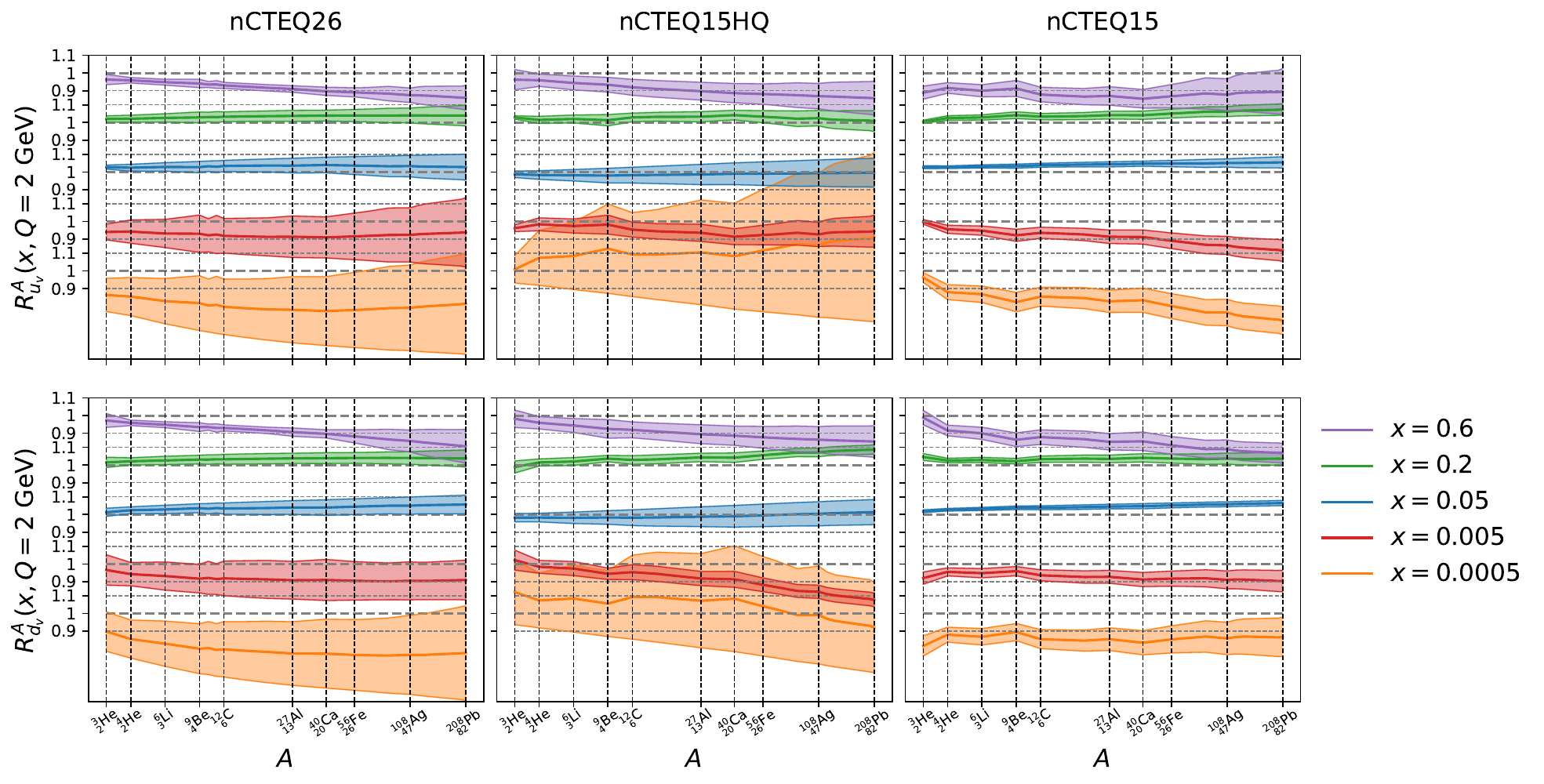}
        \label{subfig:Af-vs-ncteq15-ud}
    }
    \\
    \subfloat[$R^A_i$ for strange and gluon distributions.]{
        \includegraphics[width=0.99\linewidth]{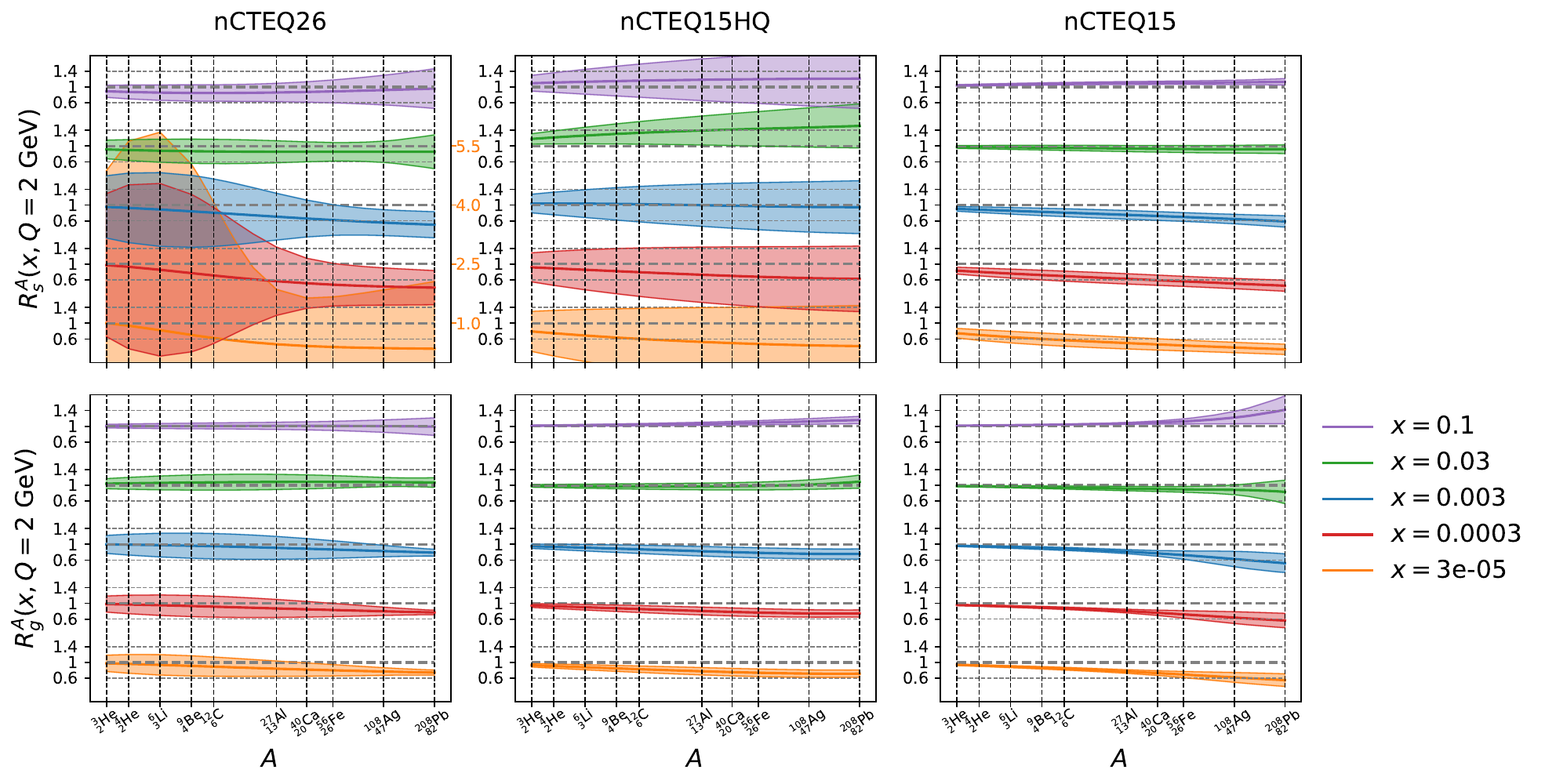}
        \label{subfig:Af-vs-ncteq15-sg}
    }
    \caption{Comparison of nuclear modification ratios $R^A_i$ as defined in \cref{eq:RA} from nCTEQ26, nCTEQ15HQ and nCTEQ15 at $Q = 2$ GeV as a function of nuclear mass number $A$ and for selected values of $x$. We plot a wide selection of nuclei but the tick labels are only shown for a small subset.}
    \label{fig:Af-vs-ncteq15}
\end{figure*}

In order to study the $A$ dependence on the level of nuclear modifications, in \cref{fig:a-dependence-wou-r}, the respective nuclear ratios, as defined in \cref{eq:RA}, are shown. Starting at low~$x$, we see that shadowing is present for all flavors and all nuclei, with the effect weakening for lighter nuclei and nearly vanishing (ratios close to one) for $^3$He in the strange and gluon channels.
Moving up in $x$ to the antishadowing region, we observe a clear antishadowing rise for valence quarks approaching $x\,{\sim}\,0.1$, bringing the ratio above one. A similar rise is visible for the light-sea quarks, but in this case, the ratio stays mostly below one. In both cases, the $A$ dependence in this region is reduced, resulting in only small differences among the distributions for different nuclei.
The situation is more complex for the strange quark in this region. For heavier nuclei (starting at around $A=40$), we observe behavior which is qualitatively similar to the light-sea quarks; however, for lighter nuclei, we observe a very different shape leading to flattening of the ratio or, in a few cases, even a dip.
The case of the gluon is even more involved, where instead of the usual antishadowing rise with a maximum slightly below ${x\sim0.1}$, we observe a double hump structure in this region with an additional minimum at $x\sim0.1$, which also features a pinch in the $A$ dependence.
Finally, at large $x$, we enter the EMC region, where, for both valence and light-sea quarks, we observe the expected drop in the ratio, with a monotonic strengthening of this effect as the nuclear mass increases. A similar suppression is present for the gluon, but in this case, the drop is much sharper.  
For the strange distribution, the picture is more complex. 
For heavier nuclei, we observe the expected drop of the ratio (resembling the behavior observed in the gluon), but for lighter nuclei, we see a transition from the suppression to an enhancement, with the ratio going above one, moving directly to the Fermi motion without explicit EMC suppression. The transition occurs around aluminum and can be linked to the behavior observed in the antishadowing region. The reason for such unexpected behavior in strange is probably the lack of data constraints for lighter nuclei and the need for compensation between antishadowing and the EMC region to satisfy the sum rules. However, as we will see later, the associated uncertainties for light nuclei in this region are large enough to cover the spread.

We now turn to a different representation of the nuclear PDF $A$ dependence that will reveal a complementary perspective, in particular on the dependence of the size of the uncertainty bands on $A$. 
In \cref{fig:Af-vs-ncteq15} we show nuclear modification ratios of full nPDFs, $R^A_i(x,Q)$ as calculated using \cref{eq:RA}, as a function of $A$ for a selection of $x$ slices at $Q = 2$ GeV.
The nuclear modifications for different $x$ slices are stacked, with colors indicating a progression from small $x$ at the bottom to large $x$ at the top; the $x$ values are specified in the legend. 
Additionally, we compare the obtained $A$ dependence of nCTEQ26 PDFs against a selection of previous nCTEQ global analyses: nCTEQ15 and nCTEQ15HQ.

We start by investigating the $A$ dependence of the up- and down-valence distributions shown in \cref{subfig:Af-vs-ncteq15-ud}.\footnote{Note that the small kinks in the $A$ dependence of valence distributions are expected. This is because the $f^A_i$ depends on the $N/Z$ ratio. This dependence in $R^A_i$ cancels only partially, and as a consequence the valence $A$-dependence curves are expected to be perfectly smooth only if nuclei are chosen to be isoscalar (${A/Z=2}$).} 
Firstly, we can see that all nuclei are well-determined, with small uncertainties across the $A$ range. 
Also, the different $x$ slices, especially large and medium $x$, show the distributions are very well constrained with some increase in the size of error bands for lower $x$ values (${x=10^{-4}}$ in the figure). 
This behavior is rather compatible with what was observed in our previous nCTEQ15 and nCTEQ15HQ analyses. Worth noting is the reduction of uncertainties in nCTEQ26 at the highest-$x$ slices for lighter nuclei. 
This is related to the fact that in nCTEQ26 we have loosened the kinematic cuts for DIS data, allowing us to introduce data directly sensitive to this region (416 NC DIS data points from JLAB, as well as $\sim400$ other NC DIS data points, and additional $\sim300$ CC DIS points).

The gluon distribution uncertainties are generally small and typically decrease slowly with increasing $A$, with the exception of the $x=0.1$ slice; see the lower panel of \cref{subfig:Af-vs-ncteq15-sg}. 
This is natural, as most of the gluon constraints originate from proton-lead LHC data, but we also expect DIS data to constrain light-to-intermediate nuclei. 
As we already observed in \cref{fig:a-dependence-wou-r} when inspecting the central predictions, we find shadowing suppression for all nuclei starting from $x \sim 10^{-2}$, but only for heavy nuclei is the nuclear modification smaller than one outside the margin of uncertainty. 
Conversely, the nuclear modification for light nuclei is consistent with one within the uncertainties, indicating a lack of evidence for shadowing.
Compared to our previous analyses, a particularly interesting finding is the opposite pattern in how uncertainty scales with nuclear mass: in earlier results, the uncertainty increased with increasing $A$.
This is especially pronounced in nCTEQ15, due to a lack of gluon-sensitive data, but it is also visible in nCTEQ15HQ. 

As can be seen in the upper panel of \cref{subfig:Af-vs-ncteq15-sg}, the strange PDFs show a similar pattern but with much larger uncertainties, due to their dependence on the gluon via scale evolution.
However, this overlaps with another effect, namely that the dimuon data (at $A=56$) provide a direct constraint on strange.
This results in a minimal strange PDF uncertainty band around $A=56$ for some $x$ slices.
Compared to nCTEQ15 PDFs, the strange uncertainties in nCTEQ15HQ and nCTEQ26 are much larger in most regions of the $(A,x)$  space, which can be attributed to more flexibility in the strange distribution (in nCTEQ15HQ, 3 strange parameters were opened, while in nCTEQ26, 8 strange parameters are open).
However, we observe that in some regions, e.g., ${x=0.0003,~ A>56}$, the strange uncertainty is reduced in nCTEQ26 compared to nCTEQ15HQ.
This reduction can be attributed to the increased constraints provided by the data, in particular, the impact of the CHORUS neutrino DIS measurements.

For all the flavors, a general pattern can be observed. In our previous analyses, which relied on our older model for $A$ dependence, we often found uncertainties shrinking as $A$ decreased, even in the absence of obvious data constraints. 
Our new $A$ dependence used in the nCTEQ26 analysis instead allows sufficient flexibility to reflect the strength of the data constraint as $A$ changes. 
This is particularly noticeable for the gluon in lead and for the strange in iron. 
We also note that some reduction in uncertainty as $A$ decreases may be expected due to the boundary condition of recovering the proton PDF at $A=1$, regardless of the model's analytic form.

Finally, we note that current constraints on the $A$-dependence vary significantly between different flavors. 
For up- and down-quarks we have constraints from data taken on a broad range of nuclei. 
Conversely, for strange and or gluon, the direct constraints originate only from one or two nuclei. 
Specifically, in case of strange, there is data from Pb and Fe, whereas gluon is mostly directly constrained by Pb data. 
This means, that especially for the gluon we have almost no constraints on the $A$-dependence and the only way this can change is through new measurements. 
Particularly valuable would be gluon sensitive measurements on light and medium nuclei. 
Partly can those be provided by the LHC $p$O and $p$Ne runs and SMOG2 if sufficient statistics is collected. 

\section{Comparison with other recent nPDF analyses}
\label{subsec:comp}

In this section, we compare the nuclear PDFs obtained from the nCTEQ26 fit with the results from other recent nPDF analyses. We focus on comparisons with EPPS21 and nNNPDF3.0, as they are the most comprehensive in terms of the used LHC data and therefore most closely aligned with this work.

\subsection{\texorpdfstring{$x$}{x}-dependence}

\begin{figure*}[!ptb]
    \centering
    \subfloat[Absolute nuclear PDFs and their uncertainties for lead.]{
        \includegraphics[width=0.75\linewidth]{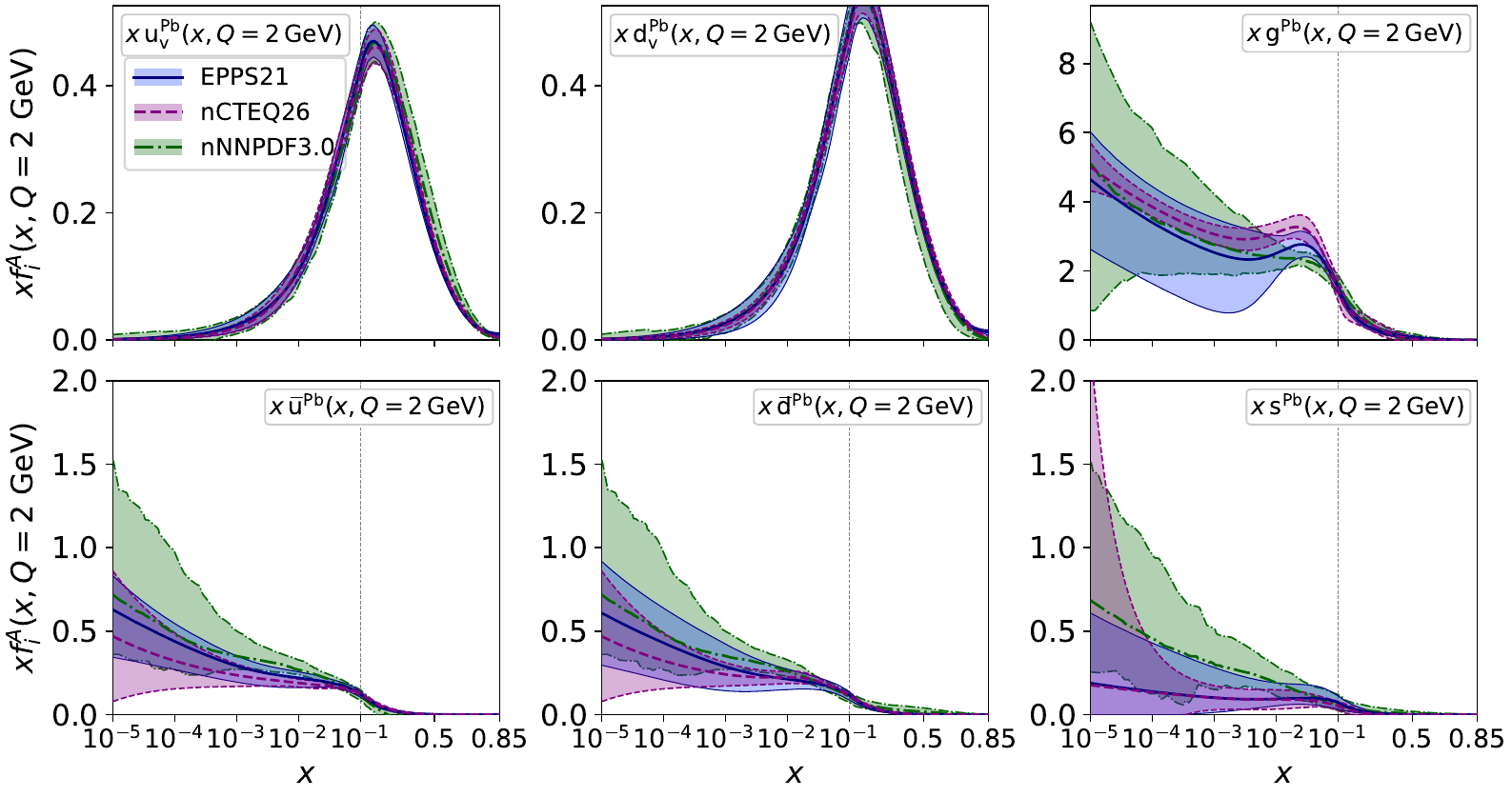}    \label{subfig:ncteq26_epps21_nnnpdf30_Pb_Qs10_abs_loglin}
    }
    \\
    \subfloat[Ratios of nuclear PDFs to the central nCTEQ26 fit and their uncertainties for lead.]{
        \includegraphics[width=0.75\linewidth]{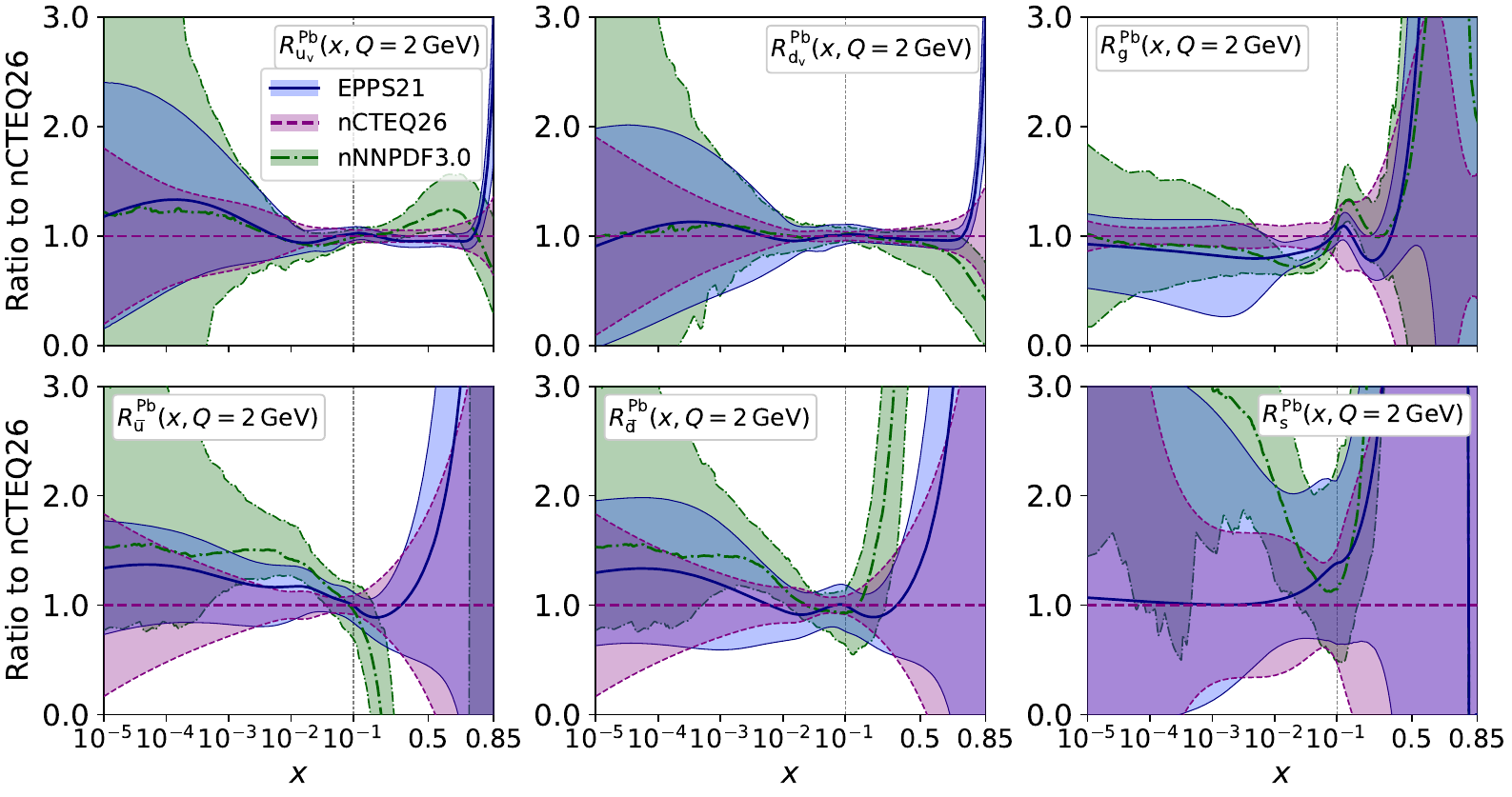}
        \label{subfig:ncteq26_epps21_nnnpdf30_Pb_Qs10_ratio_to_ncteq26_loglin}
    }
    \\
    \subfloat[Nuclear modification ratios as defined in \cref{eq:RA} and their uncertainties for lead. Note that the denominators are computed from the proton baselines of the corresponding analyses.]{
        \includegraphics[width=0.75\linewidth]{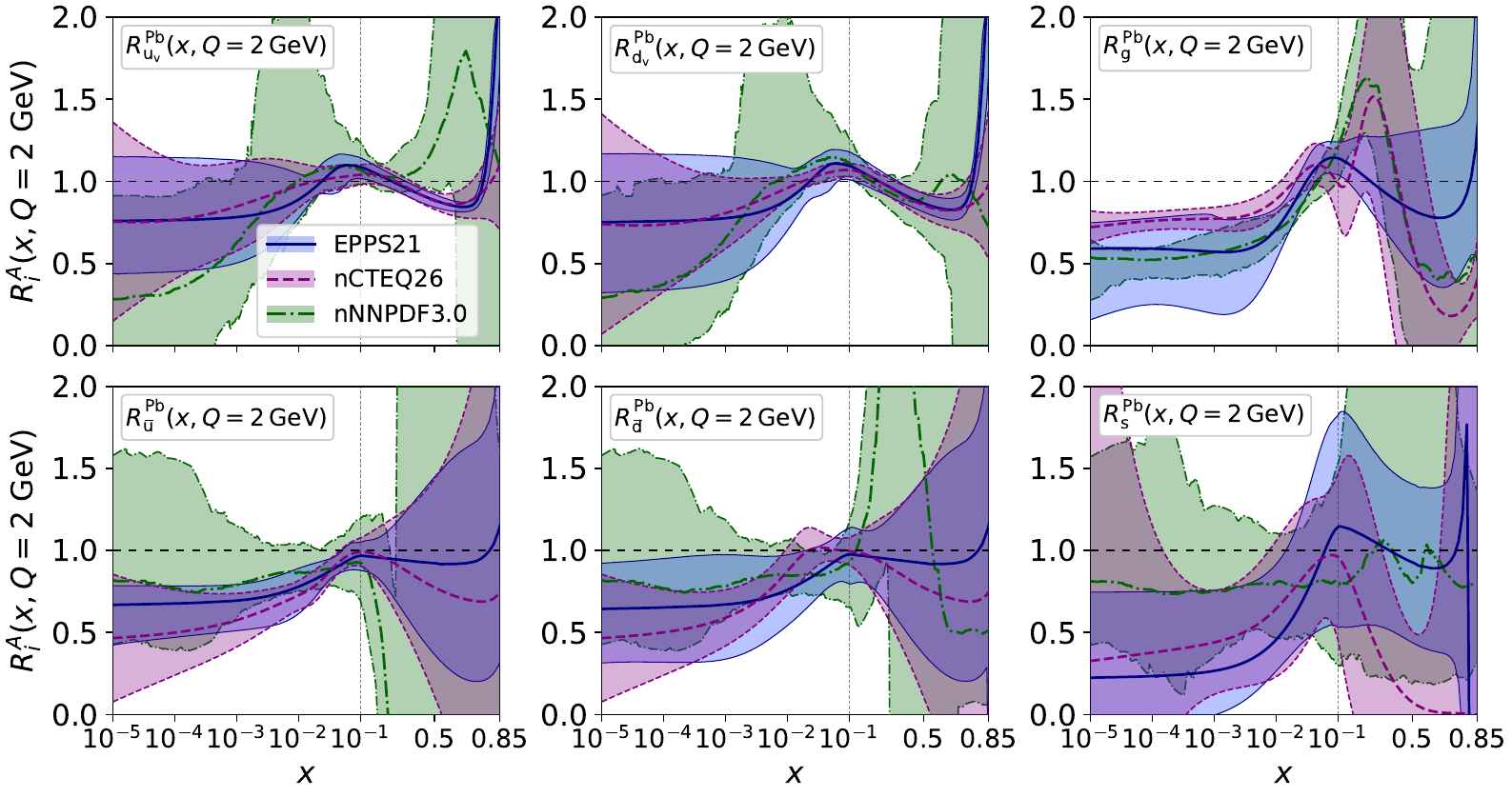}
        \label{subfig:ncteq26_epps21_nnnpdf30_Pb_Qs10_R_loglin}
    }
    \caption{Comparison of nuclear PDFs for lead from the nCTEQ26, EPPS21, and nNNPDF3.0 global analyses. Note the log-linear scaling on the $x$-axis separated by a vertical line.}
    \label{fig:ncteq26_epps21_nnnpdf30_Pb_Qs10_loglin}
\end{figure*}

We start with a comparison of the $x$-dependence in the three global analyses and compare in  \cref{fig:ncteq26_epps21_nnnpdf30_Pb_Qs10_loglin} the PDFs for lead at the scale of $Q=2$~GeV. \cref{subfig:ncteq26_epps21_nnnpdf30_Pb_Qs10_abs_loglin} displays the absolute PDFs, $xf^{\text{Pb}}_i(x,Q)$, for flavors $i \in \{u_v, d_v, g, \bar{u}, \bar{d}, s\}$ in separate panels and in this order, which allows for direct comparisons of the larger $d_v$ peak compared to $u_v$ due to the larger number of neutrons in lead and of the light and strange sea distributions. We immediately notice that on a general level the absolute PDFs from the three different groups agree well. This general agreement is non-trivial, as these three nPDF sets are based on different methodologies and input datasets. Their overall consistency therefore provides confidence that the extracted features are robust and reflect the underlying nuclear structure rather than artifacts of a particular fitting procedure or dataset. Looking at the individual flavors in more detail, we see that the central predictions from nCTEQ26, nNNPDF3.0 and EPPS21 for the valence PDFs agree extremely well across the whole $x$-range up to $x \sim 0.75$, beyond which the available data points are typically excluded by kinematic cuts. 

The differences can be resolved better in \cref{subfig:ncteq26_epps21_nnnpdf30_Pb_Qs10_ratio_to_ncteq26_loglin}, which shows the ratios of the corresponding PDFs to the nCTEQ26 central fit. There, the valence-quark uncertainty bands are also in good agreement in the range ${x \in [0.01, 0.2]}$ with the nCTEQ26 uncertainty band being the smallest, particularly in the $d_v$ distribution. Otherwise, the valence-quark uncertainties vary moderately with the largest band from nNNPDF3.0 throughout most of the remaining $x$-range. The light sea-quark distributions of all groups agree well in the low- and mid-$x$ regions with notable differences between nNNPDF3.0 and nCTEQ26 at ${x\sim5\times10^{-3}}$, where the error bands of the $\bar{u}$ distribution do not overlap. The largest differences in the light sea-quark sector appear at large $x\geq0.2$, where EPPS21 and nCTEQ26 agree within uncertainties, but nNNPDF3.0 features substantially different behavior with appreciably smaller uncertainties, a steeply falling $\bar{u}$ distribution and a steeply rising $\bar{d}$ distribution compared to nCTEQ26 and EPPS21. The strange-quark distribution obtained by all three groups agrees within the quoted uncertainties, but the sizes of the uncertainties vary notably over the entire $x$-range. Among the three predictions, the nCTEQ26 uncertainty is the smallest in the ${x \in [0.001,0.1]}$ range and is the largest outside, where there are no data constraints, indicating increased flexibility in the nCTEQ26 parametrization. This will be discussed in more detail in \cref{subsec:strange}. Looking again at the absolute distributions, \cref{subfig:ncteq26_epps21_nnnpdf30_Pb_Qs10_abs_loglin}, we can see that the central values of EPPS21 and nCTEQ26 are nearly identical below $x\sim0.01$, but the nNNPDF3.0 central value is much larger, especially at low $x$, and is also closer to the light sea-quark PDFs, i.e.~it predicts a more ``democratic'' sea as compared to EPPS21 and nCTEQ26. Also for the gluon distribution, all groups agree within the quoted uncertainties. This agreement is much more significant than in the case of the strange-quark distribution, since the uncertainties are much smaller across nearly the whole $x$-range. Specifically, we observe very good agreement between the EPPS21 and nNNPDF3.0 central values, whereas the nCTEQ26 gluon is slightly larger in the low and mid-$x$ range. The gluon uncertainty in nCTEQ26 is much smaller, especially in the low-$x$ region. This can be traced back to the large amount of heavy-quark data included in our analysis.

We now turn our attention to \cref{subfig:ncteq26_epps21_nnnpdf30_Pb_Qs10_R_loglin}, where we show a comparison of the nuclear modification factors defined according to \cref{eq:RA} for different groups. The general features are the same for all groups, but there are a number of differences worth noting. We emphasize that the nuclear modifications in this comparison depend on the underlying proton PDF baselines used for constructing the denominators, which in \cref{subfig:ncteq26_epps21_nnnpdf30_Pb_Qs10_R_loglin} are different for each of the considered analyses. Note also that in nNNPDF3.0 the uncertainty band in the absolute PDF and the nuclear modification ratio behave differently because of the correlated treatment of nuclear and proton replicas, cf.\ Sec.~7.2 of Ref.~\cite{AbdulKhalek:2022fyi}. As a consequence, the size of the error bands in subfigures (b) and (c) can differ. The first clearly noticeable difference is the much stronger suppression in the nuclear modification of valence distributions at low $x$ in nNNPDF3.0. The central values of the valence nuclear modifications of all groups agree very well down to $x\sim10^{-3}$, but below this value nNNPDF3.0 shows much stronger shadowing with a ratio reaching 0.3, whereas EPPS21 and nCTEQ26 central values stay at $\sim0.8$. Within uncertainties, the results are, however, still consistent. The second notable aspect is the gluon nuclear modification. Here we see a difference between nNNPDF3.0 and nCTEQ26, which goes beyond their respective uncertainties. At the same time, if we look at absolute distributions in \cref{subfig:ncteq26_epps21_nnnpdf30_Pb_Qs10_abs_loglin,subfig:ncteq26_epps21_nnnpdf30_Pb_Qs10_ratio_to_ncteq26_loglin}, we still see relatively good agreement between the two distributions within the quoted uncertainties. This difference is therefore a result of the dependence of the nuclear modification factors on the underlying proton baseline PDFs. 
In principle, one would like to assume that proton PDFs are well-determined and that their differences are subdominant in the extraction of nuclear PDFs. However, in practice, we know that even though they are much better constrained than nPDFs they can differ appreciably. This highlights a question of principle in the global determination of nuclear PDFs on whether one extracts absolute PDFs or rather the nuclear modification factors. 
The answer depends in part on the type of data used in the analysis and on the assumptions about the input scale parametrization of the $x$-dependence. 
In nCTEQ26, and more generally in nCTEQ analyses, we use whenever possible absolute cross section data instead of ratios of cross sections measured on nuclei and free protons. EPPS adopts instead an alternative strategy favoring ratios. Both approaches offer distinct advantages. 
The use of ratios can reduce certain experimental and theoretical uncertainties through partial cancellations, while using absolute cross sections retains additional information, particularly regarding the overall normalization of the data. A related difference concerns the parametrization itself. The nCTEQ and nNNPDF analyses parametrize the absolute nPDFs, whereas EPPS directly parametrizes nuclear modification factors. These are alternative, but complementary frameworks for extracting nPDFs from the available data. Since the absolute nPDFs enter directly into cross section predictions, we treat them as the primary quantities of interest. Consequently, we consider the observed differences in the nuclear modification factors largely methodological rather than fundamental, provided that the resulting nPDFs yield compatible descriptions of the data and comparable predictions for physical observables.

\begin{figure*}[!ptb]
    \centering
    \subfloat[Absolute nuclear PDFs and their uncertainties for carbon.]{
        \includegraphics[width=0.75\linewidth]{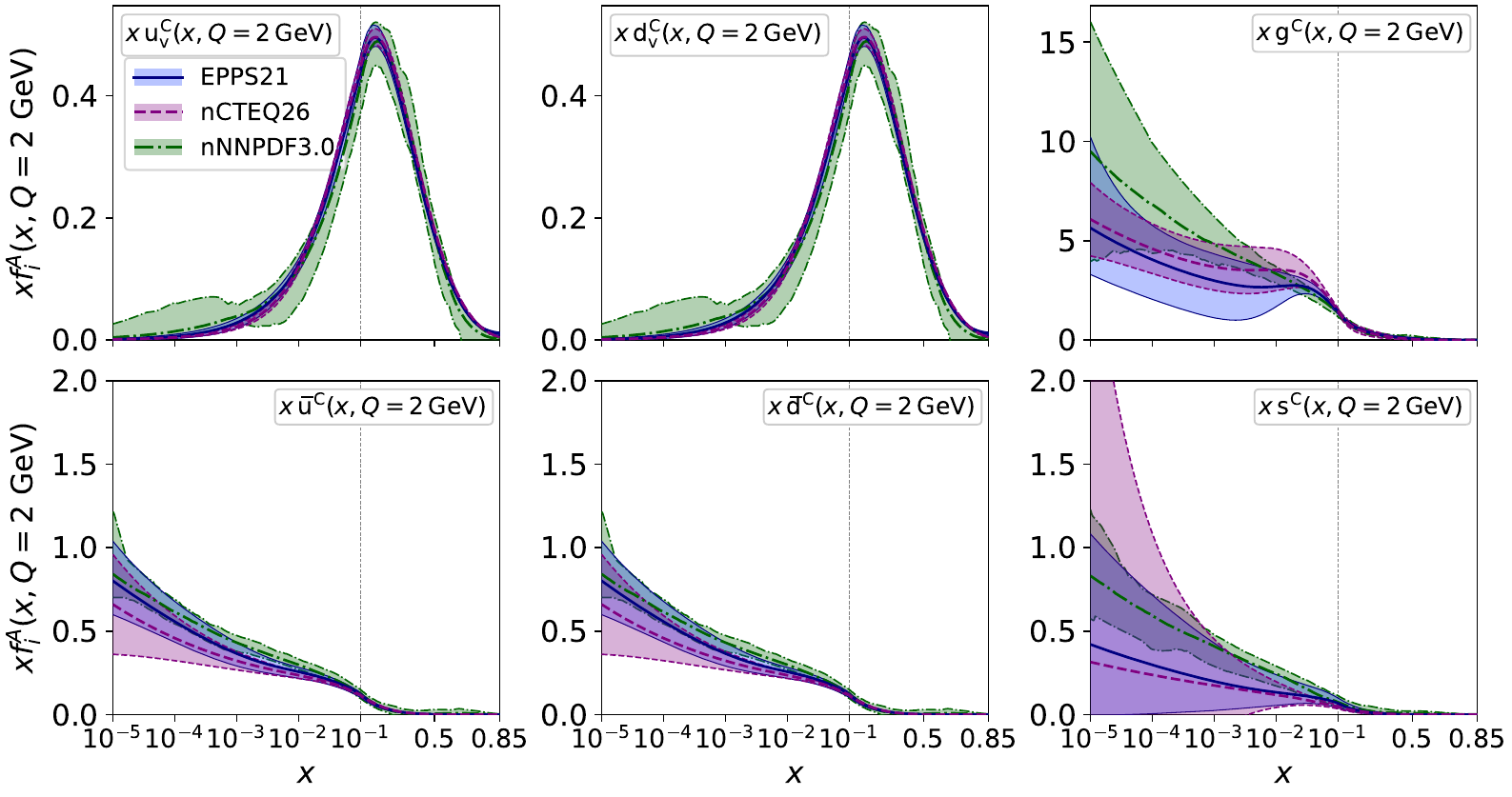}    \label{subfig:ncteq26_epps21_nnnpdf30_C_Qs10_abs_loglin}
    }
    \\
    \subfloat[Ratios of nuclear PDFs to the central nCTEQ26 fit and their uncertainties for carbon.]{
        \includegraphics[width=0.75\linewidth]{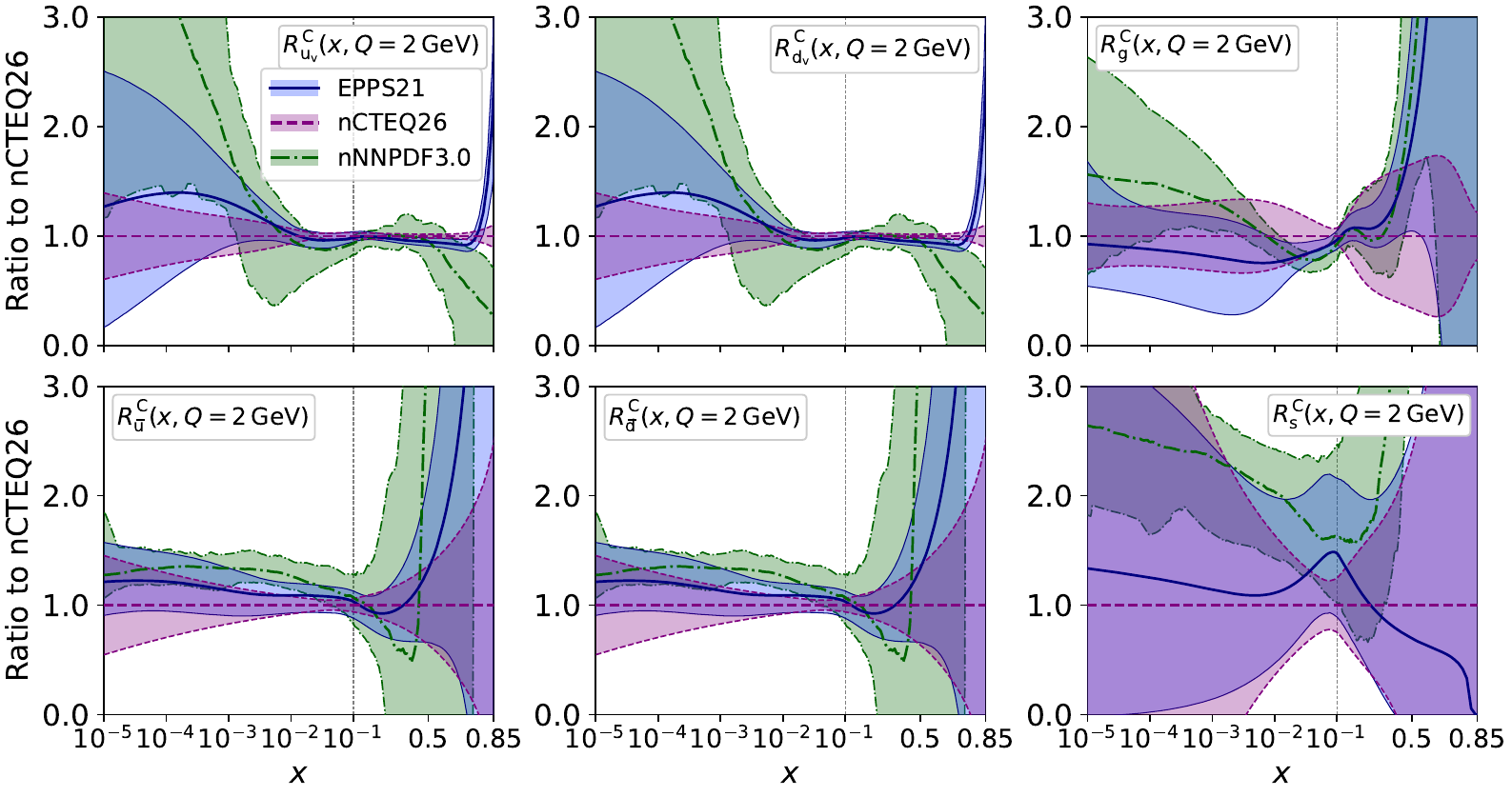}
        \label{subfig:ncteq26_epps21_nnnpdf30_C_Qs10_ratio_to_ncteq26_loglin}
    }
    \\
    \subfloat[Nuclear modification ratios as defined in \cref{eq:RA} and their uncertainties for carbon. Note that the denominators are computed from the proton baselines of the corresponding analyses.]{
        \includegraphics[width=0.75\linewidth]{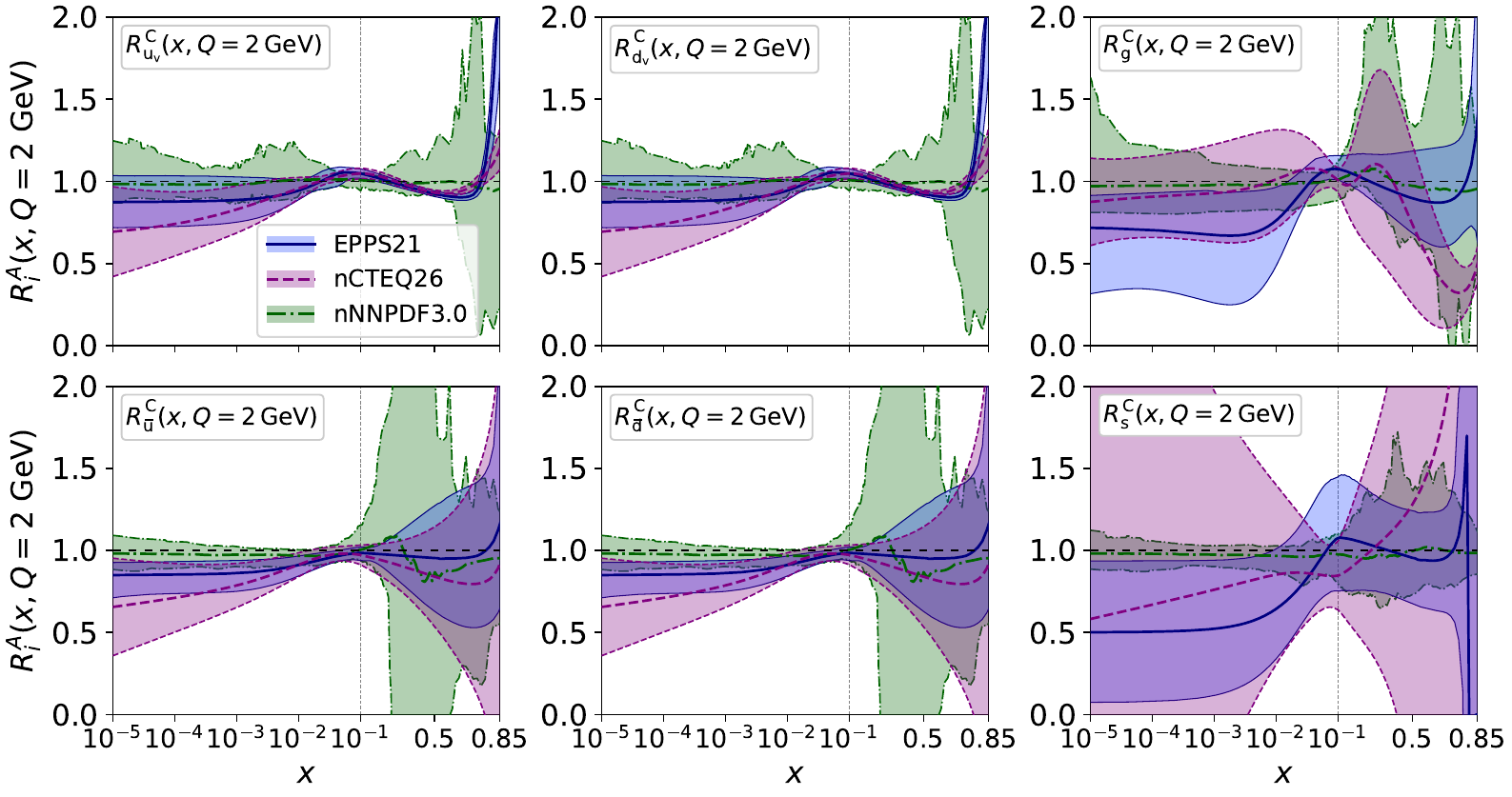}
        \label{subfig:ncteq26_epps21_nnnpdf30_C_Qs10_R_loglin}
    }
    \caption{Comparison of nuclear PDFs for carbon from the nCTEQ26, EPPS21, and nNNPDF3.0 global analyses. Note the log-linear scaling on the $x$-axis separated by a vertical line.}
    \label{fig:ncteq26_epps21_nnnpdf30_C_Qs10_loglin}
\end{figure*}

We now turn our attention to \cref{fig:ncteq26_epps21_nnnpdf30_C_Qs10_loglin}, where we compare the nPDFs from the three groups, but now for the carbon nucleus. We again present the absolute PDFs, $xf^{\text{C}}_i(x,Q)$, the corresponding ratios to the central nCTEQ26 carbon PDFs, and the nuclear modification factors for different flavors at \mbox{$Q=2$ GeV}. We focus here on the features that stand out relative to our observations above regarding lead PDFs. The first observation is the difference in the low-$x$ valence distributions among nNNPDF3.0, EPPS21, and nCTEQ26. For $x\lesssim10^{-3}$, the nNNPDF3.0 valence distributions are much larger than those of nCTEQ26 and EPPS21, and for nCTEQ26, which features smaller errors, this difference goes beyond the corresponding error bands. However, this effect is much smaller for the nuclear modification factors of the valence distribution, see \cref{subfig:ncteq26_epps21_nnnpdf30_C_Qs10_R_loglin}, where nNNPDF3.0 remains larger than nCTEQ26 and EPPS21, but is compatible with both within uncertainties. The second noticeable feature is the difference in the size of the strange uncertainties between the different groups. nCTEQ26 generally has much larger errors for the whole $x$-range apart from the region close to ${x\sim0.1}$, whereas the uncertainties of nNNPDF3.0 are the smallest. We also notice here the dependence on the proton baseline PDFs, where in \cref{subfig:ncteq26_epps21_nnnpdf30_C_Qs10_ratio_to_ncteq26_loglin} the EPPS21 and nNNPDF3.0 strange distributions agree well, whereas the corresponding nuclear modifications of \cref{subfig:ncteq26_epps21_nnnpdf30_C_Qs10_R_loglin} are only marginally compatible. Finally, it is worth noting that the gluon distributions in carbon are compatible across all three groups, both in absolute distributions and in nuclear modification ratios.

\subsection{\texorpdfstring{$A$}{A}-dependence}

Although the $A$-dependence is a central ingredient of global nPDF analyses, its role has traditionally been primarily functional, as it allowed data from many nuclear targets to be described within a common parametrization and enabled constraints from well-measured nuclei to inform those for sparse measurements. Comparatively less attention has been paid to the detailed $A$-dependence that emerges from the fit as an object of study in its own right. Consequently, existing analyses have generally employed relatively economical forms, whose main purpose was to provide a smooth and coherent description across nuclei.

\begin{figure*}[!p]
    \centering
    \subfloat[$R^A_i$ for $u$- and $d$-quark distributions.]{
        \includegraphics[width=0.99\linewidth]{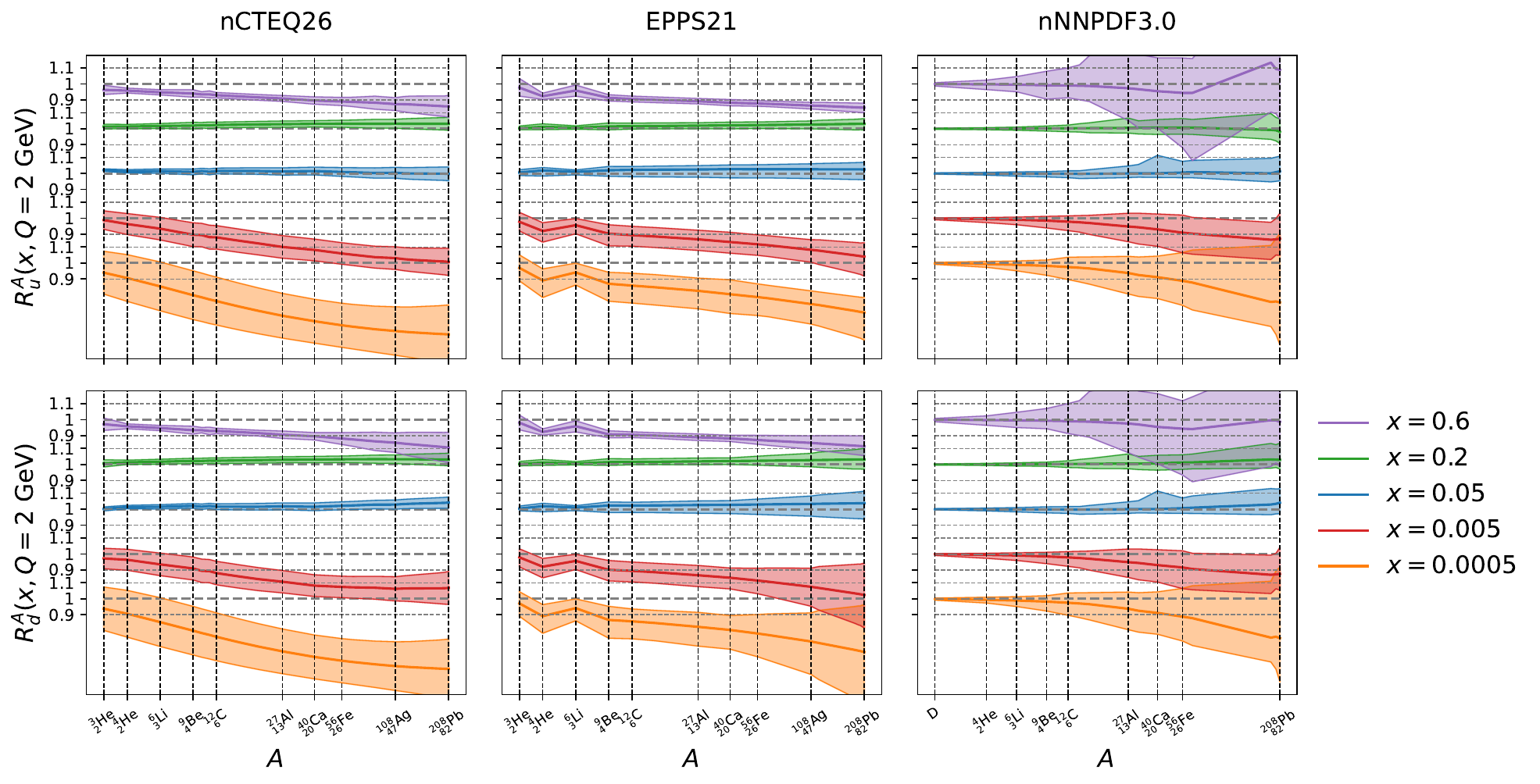}
        \label{subfig:Af-vs-epps-nnpdf-ud}
    }
    \\
    \subfloat[$R^A_i$ for strange and gluon distributions.]{
        \includegraphics[width=0.99\linewidth]{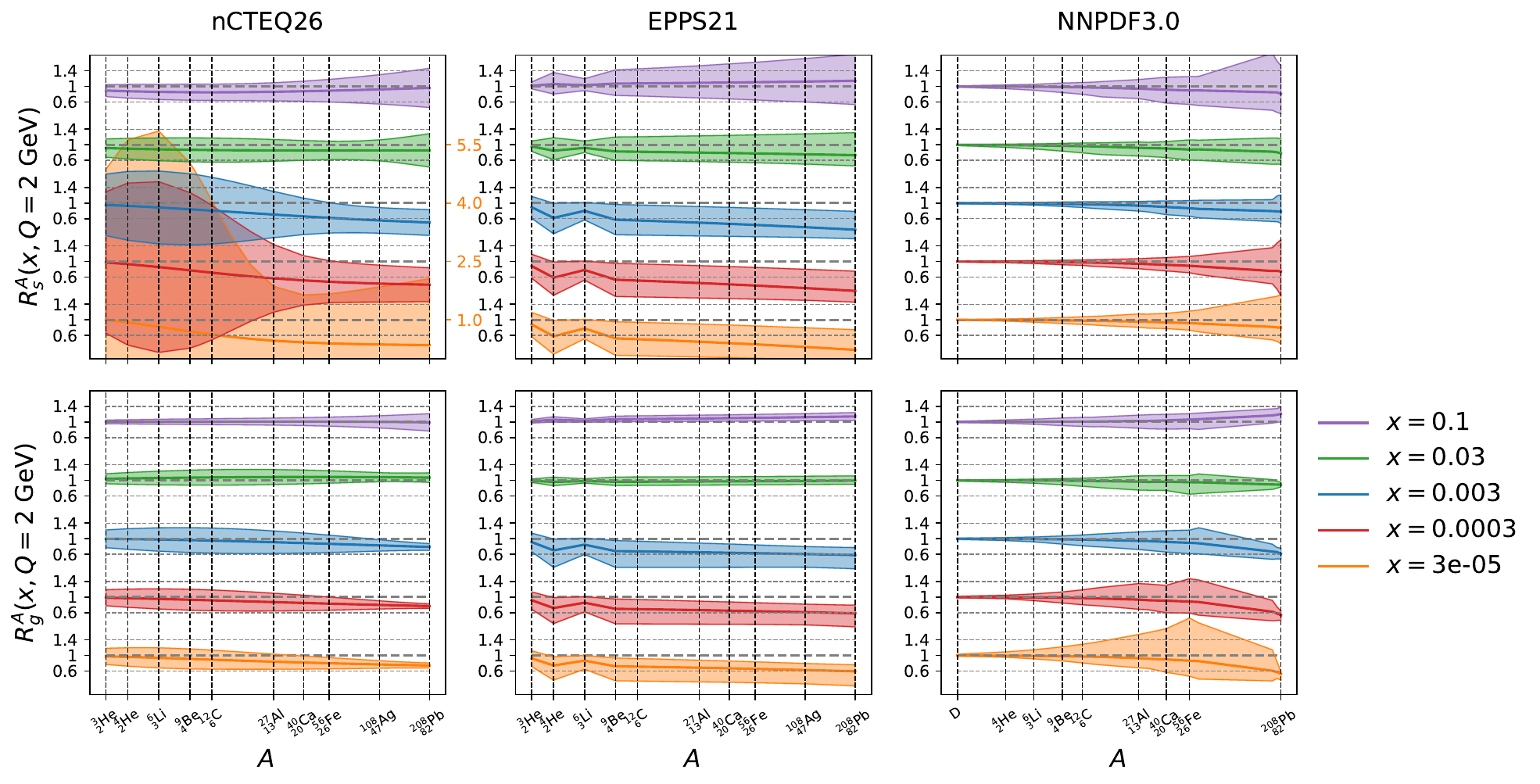}
        \label{subfig:Af-vs-epps-nnpdf-sg}
    }
    \caption{Comparison of nuclear modification ratios $R^A_i$ as defined in \cref{eq:RA} from nCTEQ26, EPPS21 and nNNPDF3.0 at $Q = 2$ GeV as a function of nuclear mass number $A$ and for selected values of $x$. 
    We plot a broad selection of nuclei but the tick labels are only shown for a selected subset.
    For nCTEQ26, we plot all nuclei whose data enter our analysis. 
    For EPPS21, we plot all nuclei published on LHAPDF website except for $^{40}_{18}\mathrm{Ar}$, which has the same mass number as $^{40}_{20}\mathrm{Ca}$. 
    For nNNPDF3.0 we also plot all nuclei published on LHAPDF website with the exception of: $^{108}_{54}\mathrm{Ag}$, $^{119}_{59}\mathrm{Sn}$ and $^{131}_{54}\mathrm{Xe}$, 
    as their uncertainties deviate from the general trend of the $A$-dependence, thereby obstructing the plots.
    }
    \label{fig:Af-vs-epps-nnpdf}
\end{figure*}

In this analysis, we build on this approach, but place greater emphasis on the  $A$-dependence itself. In \cref{fig:Af-vs-epps-nnpdf}, we compare the $A$-dependences obtained in the different global analyses numerically. We first observe that the three fits yield central predictions that, despite of featuring notably different shapes, are largely consistent with each other within their uncertainties throughout most of the $A$-range. The most notable departures from this rule are observed for the up- and down-quark distributions at low $x$ and intermediate $A$. The EPPS21 uncertainty is generally either monotonically increasing or monotonically decreasing in $A$. This can be related to the fact that EPPS has in general a smooth parametrization in $A$ with two extra factors exclusively for ${}^{6}$Li and ${}^{3}$He, where it is not smooth even for strange and gluon. For light nuclei, nNNPDF3.0 has the smallest uncertainties across all shown observables. As discussed in the corresponding paper \cite{AbdulKhalek:2022fyi}, this is due to the large amount of deuterium data included in the analysis as well as the free-proton boundary condition. In contrast, nCTEQ26 and EPPS21 do not fit the deuteron as a separate nucleus, but rather construct it from the proton PDF, eventually applying a deuteron correction. For more details see the discussion in \cref{subsec:dis}. The uncertainties of the gluon in nCTEQ26 and nNNPDF3.0 exhibit maxima approximately at silver (${A=108}$) for nNNPDF3.0 and at carbon (${A = 12}$) for nCTEQ26. The most interesting shape is observed for nCTEQ26 in the strange distribution at $x~\lesssim0.01$ imposed by the dimuon data as discussed in \cref{subsec:Adep}. The nCTEQ26 analysis also has reduced uncertainties in the gluon for heavy nuclei compared to other analyses. \cref{fig:Af-vs-epps-nnpdf} demonstrates that the nCTEQ26 framework allows for larger uncertainties in regions where the data provide limited information, e.g., the strange distribution at low $x$ or gluons and sea quarks in light nuclei, from which we infer that the gluon uncertainty is not constrained by the parametrization itself. Rather, the large body of heavy-quark data included in the fit appears to provide enough sensitivity to meaningfully constrain the gluon distribution. Overall, the $A$-dependence of the central fits obtained by the three analyses is broadly consistent, while considerably greater differences are seen in the corresponding uncertainties. The detailed behavior in $A$ was not a primary target of the previous analyses and, consequently, the features seen in \cref{fig:Af-vs-epps-nnpdf} should not necessarily be regarded as deficiencies of the corresponding fits. Rather, they reflect the role for which these parameterizations were primarily designed and motivate the more detailed approach adopted in the present analysis.

\section{Discussion of selected flavors}
\label{sec:discussion}

In this section we discuss in more detail the strange and gluon PDFs obtained in our analysis. We provide additional context related to the current data constraints on the strange quark and investigate in detail the dependence of our gluon distribution on the use of a data-driven approach for the analysis of the HQ data.

\subsection{Strange-quark distribution}
\label{subsec:strange}

While the proton and nuclear PDFs of the up and down quarks are now well constrained, the strange-quark PDF remains comparatively uncertain. Fixed-target neutrino experiments in the 1990s suggested that the strange-quark PDF was approximately 50\% smaller than the anti-down-quark PDF, which was consistent with theoretical expectations of a suppression due to the larger strange-quark mass~\cite{Olness:2003wz,NuTeV:2003uxv,Foudas:1989rk,Rabinowitz:1993xx,CCFR:1994ikl,CHARMII:1998njb,HERMES:2008pug}. However, the interpretation of these results was complicated by the use of heavy nuclear targets such as iron and lead. It was anticipated that proton-proton measurements at the LHC would help to resolve this issue. Instead, inclusive measurements of $W$- and $Z$-boson production indicated a strange-quark distribution that was comparable to, or even larger than, the anti-down-quark distribution~\cite{ATLAS:2012sjl,ATLAS:2016nqi}, which was a surprising outcome. Similar results, but with larger uncertainties, were also observed in vector-boson production in $p$Pb collisions at the LHC~\cite{Kusina:2020lyz}. At present, no single hypothesis provides a satisfactory explanation for these seemingly discrepant results. However, it should be noted that the strange- to light-sea ratio reported from the LHC gauge-boson production is measured at smaller values of $x$ and higher scales compared to the extractions based on the neutrino DIS data~\cite{Kusina:2012vh}. Furthermore, extractions of the strange- to light-sea ratio from measurements of ${W+c}$ or $W$+jet production at the LHC show that the ratio can be smaller at larger $x$~\cite{CMS:2018dxg,ATLAS:2019ext,Cooper-Sarkar:2018ufj}. In this context, it is important to note that taking into account quark-mass effects in the DGLAP evolution equations may improve the theoretical description. In addition, the NNLO corrections to the neutrino DIS dimuon production cross-section, which is directly sensitive to the strange-quark PDF, are found to be negative at small $x$ and positive at larger values of $x$, which could also partially explain this discrepancy~\cite{Helenius:2026uuz,Bailey:2020ooq}. The recently calculated NNLO correction to inclusive CC DIS may provide further information~\cite{Gao:2017kkx,Gao:2021fle,Risse:2025smp,Caola:2026kvl}.

As can be seen in \cref{subfig:Af-vs-ncteq15-sg,subfig:Af-vs-epps-nnpdf-sg}, the uncertainty of the strange-quark PDF in the nCTEQ26 fit is by far the largest compared to the other flavors and it is particularly large for light nuclei. As already discussed, even small modifications in the parameter $c_1$ can considerably alter the behavior at small $x$, where the term $x^{c_1}$ dominates. Having a detailed look at the different $a_1^k$ parameter scans in \cref{fig:param1d_all}, we find that the natural range of the light-sea and gluon parameters, i.e., the difference of the parameter values at which the tolerance is reached, is of the order of $10^{-2}$. For the valence quarks, this range is even smaller and of the order of \mbox{${\sim}10^{-3}$}. For the strange quark, the variation in the parameter $a_1^{s+\overline{s}}$ is instead about $10^{-1}$, which is drastically more than in every other flavor. When the values of all other parameters are kept fixed, this range correlates with the size of the PDF uncertainty making it much larger for the strange quark at low $x$, which is what we observe.

\begin{figure}[t!]
    \centering
    \includegraphics[width=0.49\textwidth]{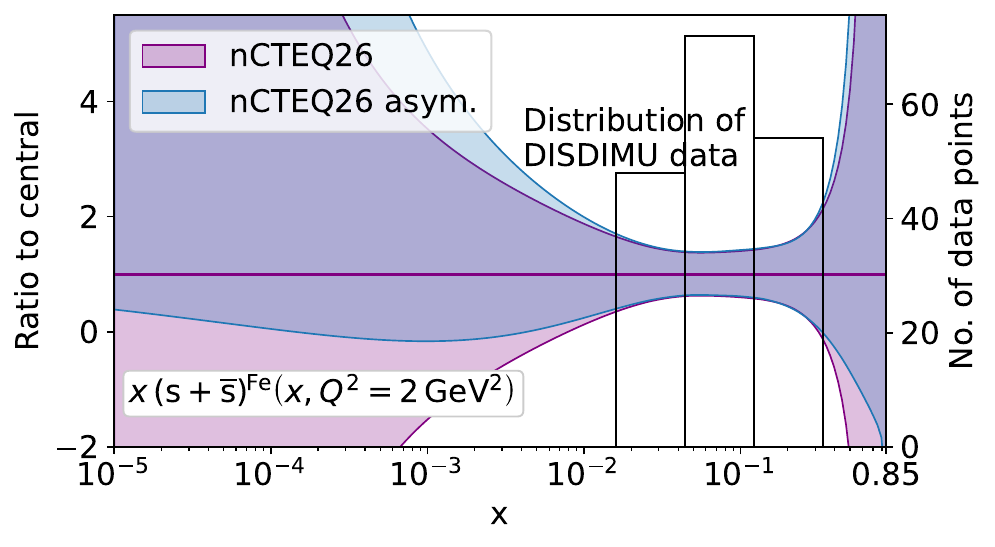}
    \caption{The nCTEQ26 strange-quark PDF in the iron nucleus relative to its central prediction overlaid with the distribution of the number of DIS dimuon data points. We also show the sizes of the uncertainty bands with either a symmetric (\cref{eq:ObservableUncertainty}) or an asymmetric (\cref{eq:ObservableUncertaintyAsym}) definition.}
    \label{fig:strange_unc_sym}
\end{figure}

In \cref{subsec:constraints}, we investigated the different data constraints on the strange-quark PDF by inspecting one-dimensional $\chi^2$-profiles in \cref{subfig:par-ssb_a2,subfig:par-ssb_a0}. What we found was that the dimuon data provided the main constraint for the shape parameters, followed by the inclusive NC and CC DIS data. This illustrated nicely the direct sensitivity of the dimuon data on the strange-quark distribution, since the dimuon data contributes only 174 data points as compared to 1277 and 1068 data points from the inclusive NC and CC DIS datasets, respectively. In \cref{fig:strange_unc_sym}, we now overlay the strange-quark PDF uncertainty band in the iron nucleus with the number of dimuon data points in three $x$ bins. We find that the size of the uncertainty band indeed strongly correlates with the location of the data. In addition, the plot shows uncertainties obtained not only with the symmetric prescription of \cref{eq:ObservableUncertainty}, but also with the asymmetric prescription defined in \cref{eq:ObservableUncertaintyAsym}. We observe that in the region supported by the dimuon data the two prescriptions agree very well, while outside they diverge significantly, indicating weak constraints from other data. The asymmetric uncertainty band shows a much tighter lower boundary for the strange-quark distribution at low~$x$. As in this region the data constraints are limited, we conclude that this boundary is not data-driven and prefer to use the symmetric prescription for calculating the uncertainties, as they account more conservatively for possible biases.

\begin{figure*}[t]
    \centering
    \includegraphics[width=\textwidth]{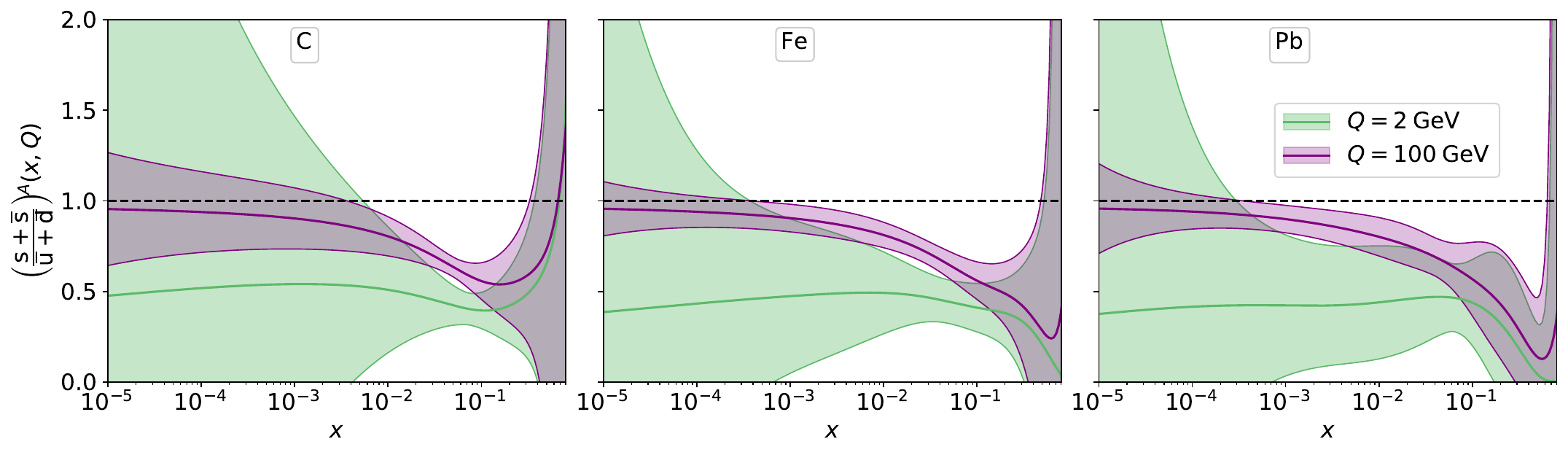}
    \caption{The nCTEQ26 strange- to light-sea ratio \mbox{$r_s=(s+\bar{s})/(\bar{u}+\bar{d})$} at $Q=2$ (green) and $Q=100$ GeV (magenta) as a function of $x$ for C, Fe and Pb nuclei.
    }
    \label{fig:ssbar_over_ubardbar}
\end{figure*}

The situation for the strange-quark normalization, \cref{subfig:par-ssb_a0}, was different. Here, the CC DIS data provided the strongest constraint towards large values of the $a_{\mathrm{sum}}^{s+\bar{s}}$ parameter, while the DIS dimuon, NC DIS and FT DY data constrained it on the low side. The minimum of the total $\chi^2$ was therefore a compromise between those datasets. We complement these results by inspecting the $A$-dependence of the strange normalization parameter $c_{\mathrm{sum}}^{s+\bar{s}}(A)$, which controls the contribution of the strange sea compared to the light sea quarks. In practical terms, this dependence is implemented by calculating the normalization and then rescaling it by the $A$-dependent parameters in $c^{s+\bar{s}}_\text{sum}(A)$. Note that in this case only the linear term of the $A$-dependence enters, since the parameter $b_{\mathrm{sum}}^{s+\bar{s}}$ is fixed to zero due to poor sensitivity. As a consequence, we find a linearly decreasing dependence in $\ln A$ for $c^{s+\bar{s}}_\text{sum}$ with a value of 0.5 for proton ($A=1$), indicating no strange suppression, and a decrease to $0.42$ for iron ($A=56$) and $0.39$ for lead ($A=208$).
In \cref{fig:ssbar_over_ubardbar}, the ratio of the strange to the light sea \mbox{$r_s=(s+\bar{s})/(\bar{u}+\bar{d})$} is shown. A uniform distribution \mbox{($r_s =$ 1)} would represent a ``democratic'' sea and imply that the flavors $u$, $d$ and $s$ contribute equally to the quark sea, whereas a ratio below one would imply that the strange-quark distribution is suppressed compared to $\overline{u}$ and $\overline{d}$. 
We can see that the situation is more complicated and the strange ratio changes with $x$. 
We also examine the dependence of this suppression on the scale by showing predictions at $Q=2$ GeV and \mbox{$Q=100$ GeV} as well as on the nuclear mass number $A$ by showing predictions for carbon, iron and lead nuclei. 
At the low scale, we observe a strong suppression throughout the whole range in $x$ of at least a factor two. However the, uncertainties are large, in particular for light nuclei and at low $x$. For carbon, the ratio is compatible with one throughout a significant region in $x$ with the exception of $x \in [10^{-2}, 10^{-1}]$. For iron and lead, the ratio is incompatible with one in a much larger region of $x \in [5 \times 10^{-4}, 10^{-1}]$. The central value of the suppression $r_s$ has a mild dependence on $A$ in particular at low $x$, where it decreases from 0.5 for carbon to 0.4 for lead. 
At the high scale $Q=100$ GeV, we observe that at low $x\lesssim 10^{-3}$, $r_s$ is compatible with one and the suppression all but disappears for all nuclei. This is consistent with the naive expectation that the strange-quark mass becomes less relevant in the high energy regime. 
This also aligns with the measurements, in which the lower energy experiments prefer a stronger suppression as compared to the LHC measurements from the production of electroweak bosons, which have a mass of about $Q=100$ GeV. At $Q=2$ GeV, $r_s$ is fairly constant up to about $x\sim0.1$, while at $Q=100$ GeV we see a more pronounced shape with a steeply growing suppression starting at around $x\gtrsim 0.01$. 

\begin{figure}
    \includegraphics[width=0.95\linewidth]{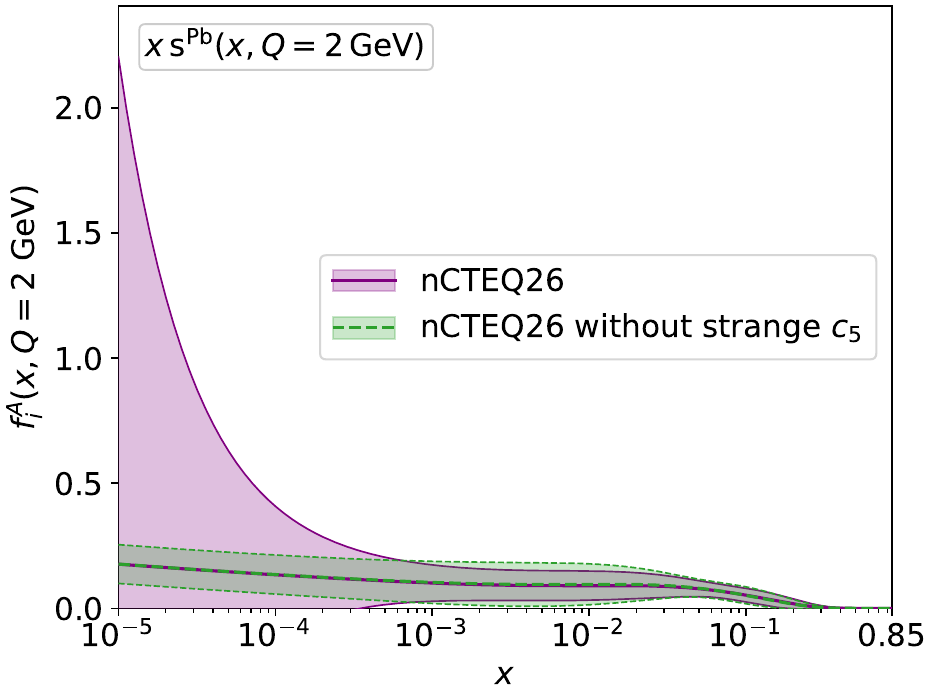}
    \caption{Comparison of strange-quark uncertainties calculated with the nCTEQ26 minimum at $Q=2$ GeV with and without $c_5$ open in the Hessian analysis.}
    \label{fig:c5}
\end{figure}

During our analysis, we found that the obtained strange-quark uncertainty strongly depends on minor details of the fit setup. We therefore introduced the additional parameter $c^{s+\bar{s}}_5 = a^{s+\bar{s}}_5 \ln A + b^{s+\bar{s}}_5 \ln^2 A$, see \cref{eq:parametrization1}, which for all other flavors was kept fixed at zero. We observed in particular, that fits with $c^{s+\bar{s}}_5$ closed typically featured unrealistically small uncertainties in the strange-quark distribution. To illustrate this behavior, we closed the parameters $a^{s+\bar{s}}_5$ and $b^{s+\bar{s}}_5$ in the nCTEQ26 fit keeping their original values and reanalyzed the Hessian uncertainties. This results in a collapse of the strange uncertainty as can be seen in \cref{fig:c5} and therefore suggests an important increase in flexibility of the parametrization through the inclusion of the $c^{s+\bar{s}}_5$ parameter, which then allows for a better exploration of the strange-quark uncertainty.

Another noteworthy feature that we observed is a strong correlation between the $a_1^{s+\bar{s}}$ and $b_1^{s+\bar{s}}$ parameters leading to a $\chi^2$ valley along the diagonal direction. Fits featuring different combinations of the values of these parameters along the diagonal do not differ appreciably in the data quality description as measured by $\chi^2$. However, they yield significantly different uncertainty bands for the strange-quark distribution in regions with weak data constraints, whereas the uncertainties in the region populated by the dimuon data as well as the uncertainty bands for the other flavors remain the same. From this we conclude that the lack of constraints at low values of $x$ creates very flat regions in the strange-quark PDF parameter space. It then depends on the exact shape of the $\chi^2$ function around the minimum for the Hessian eigenvectors to capture this behavior, meaning that slight changes in this shape of the \mbox{$\chi^2$-surface} can produce large differences in the resulting strange PDF uncertainties. For our main fit, we select a combination of $a_1^{s+\bar{s}}$ and $b_1^{s+\bar{s}}$ parameters that yields additional flexibility and allows for a larger strange-quark uncertainty across a broad range of nuclei. We believe that this choice leads to a more faithful propagation of the data constraints into the PDF uncertainty and minimizes the bias in regions with weak constraints. Comparing to the other global analyses, in particular EPPS21 and nNNPDF3.0, the nCTEQ26 analysis yields the largest strange uncertainties in a broad range of $x$ and for a wide selection of nuclei. This, together with the fact that nCTEQ26 features the most comprehensive selection of data sensitive to the strange-quark nPDF, leads us to believe that nCTEQ26 currently provides the most realistic uncertainties. However, the observed instability of this error band implies that the uncertainty on this uncertainty is also very large, so that the true uncertainty of the strange quark at low $x$ could be even larger. 
This is also manifest in the error-PDF decomposition of the strange-quark uncertainty band, where we find that while the bands for light nuclei receive contributions from a multitude of eigenvectors, the bands in heavy nuclei, starting with iron, are driven by a single error PDF.

As already discussed in \cref{subsec:nudis}, there are additional CC DIS measurements from CCFR, NuTeV and CDHSW which provide sensitivity to the strange-quark PDF, but are not included in our analysis because of their incompatibility with the NC DIS data~\cite{Muzakka:2022wey}. Another existing dataset that can provide additional information on the strange-quark PDF is the NOMAD dimuon measurement~\cite{NOMAD:2013hbk}, which will be discussed in \cref{subsec:nomad}. 

For future consideration, important new constraints will become available as on-going neutrino experiments associated with beams from the Forward Physics Facility at the LHC~\cite{Feng:2022inv} continue to accumulate statistics.   Also, precise measurements of charged-current (CC) DIS processes at the future Electron–Ion Collider~\cite{AbdulKhalek:2021gbh,Arratia:2020azl} will be a welcome addition.

\subsection{Gluon distribution}
\label{subsec:gluon}

It is well known that the LHCb open heavy-flavor $D^0$-production data are able to constrain the gluon PDF at low values of $x$, as seen in the lower left section of the leading-order kinematic coverage plot (\cref{fig:x-Q2-plane}) and in the decrease of the gluon uncertainty observed in recent global nPDF analyses \cite{Eskola:2021nhw,Duwentaster:2022kpv,AbdulKhalek:2022fyi}, that include these data. In nCTEQ15HQ \cite{Duwentaster:2022kpv} and in the nCTEQ26 main fit, theory predictions for open heavy-flavor production and quarkonium production are calculated in a data-driven approach with the Crystal-Ball parametrization, which we described in \cref{subsec:hq}. To further examine the consistency of this method, we now consider an alternative fit, referred to as nCTEQ26OHF, in which we replace the data-driven approach with a perturbative QCD calculation for open heavy-flavor (OHF) production and refrain from using the quarkonium data. In this fit, we therefore consider only the $D^0$-production data listed in \cref{tab:HQdatapPb} with 209 out of the 695 data points included in the main fit, which reduces the total number of data points from 3598 to 3112. As in the main fit, 131 points originate from the recent $8.16~\mathrm{GeV}$ LHCb $D^0$ measurement \cite{LHCb:2022dmh}. We do not modify the methodological setup, i.e., we open the same parameters, keep the same kinematic cuts, and retain all data from the other processes. Note that the nCTEQ26OHF fit is only intended to be a consistency check. We still recommend the nCTEQ26 main fit for general use. 

In contrast to the nCTEQ26 fit, the partonic cross section for OHF data in nCTEQ26OHF is computed perturbatively using the NLO SACOT calculation of Refs.~\cite{Kniehl:2004fy,Kniehl:2005mk,Kniehl:2012ti} (therein referred to as GM-VFNS), which we interfaced to PineAPPL \cite{Carrazza:2020gss,Jezo:2026adf} for the production of fast-convolution grids~\cite{Wissmann:2024bca,JezoACOT}. Similarly to the SIH case (\cref{subsec:sih}), the OHF cross section includes, in addition to the two PDFs in the initial state, a fragmentation function for the $D^0$ meson in the final state, for which we choose the KKKS08 fit \cite{Kneesch:2007ey}. Due to the PDFs and the FF, the OHF process depends not only on the renormalization scale ($\mu_\mathrm{r}$) and the factorization scale in the initial state ($\mu_\mathrm{i}$), but also on the factorization scale in the final state ($\mu_\mathrm{f}$). We choose $\mu_j = \xi_j \sqrt{p_T^2 + 4m_c^2}$ with $j \,{\in}\,\{\mathrm{r}, \mathrm{i}, \mathrm{f}\}$, where $p_T$ is the transverse momentum of the $D^0$ meson, $m_c = 1.3~\mathrm{GeV}$ the charm-quark mass, and $\xi_j$ is a constant coefficient. We tune the coefficients $\xi_j$ to the proton-proton $D^0$-production data that is used in the Crystal-Ball proton baseline fit and listed in \cref{tab:HQdatapp}, arriving at $\{\xi_\mathrm{r}, \xi_\mathrm{i}, \xi_\mathrm{f}\} = \{1.0, 0.6, 2.5\}$.

\begin{figure}[t]
    \centering
    \includegraphics[width=\linewidth]{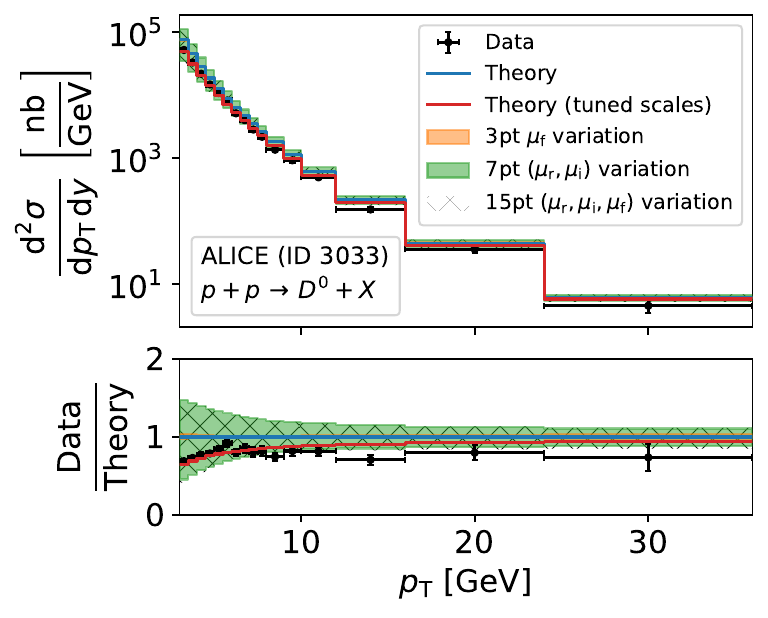}
    \caption{Comparison of nCTEQ26OHF NLO QCD predictions with tuned scales and theory uncertainties to a representative ALICE dataset of $D^0$-meson production in $pp$ collisions at the LHC.}
    \label{fig:tuned-scale}
\end{figure}

In \cref{fig:tuned-scale}, we compare the NLO QCD predictions with this tuned scale choice to a representative $pp$ measurement from ALICE. It is clear that the predictions computed with this scale choice describe the data very well and also lie within the 7-point scale uncertainty band, where only $(\mu_\mathrm{r}, \mu_\mathrm{i})$ are varied by factors of 1/2 and 2, since variations of $\mu_\mathrm{f}$ have significantly smaller impact. Studies of the choice of FFs and the inclusion of additional final states beyond $D^0$ mesons are left for future work. The SACOT calculation exhibits a numerical divergence in the limit $p_\mathrm{T} \to 0$, arising from the choice to use massless quark matrix elements in the subtraction terms for the heavy-mass logarithms. One solution for this (SACOT-$\chi$) is to use a particular choice of $\xi_\mathrm{i}$ and $\xi_\mathrm{f}$, ensuring the subtraction terms vanish in the low-$p_\mathrm{T}$ region \cite{Benzke:2019usl}. Here, such a tuning is not considered, since the low transverse-momentum region $p_T<3$~GeV is excluded as in the main fit. Alternative solutions are given by the SACOT-$m_\mathrm{T}$ \cite{Helenius:2018uul,Helenius:2023wkn} and SACOT-MPS \cite{Xie:2021ycd} schemes, which, however, still suffer from large-scale uncertainties in the low-$p_\mathrm{T}$ region.

\begin{figure}[htbp!]
    \centering
    \subfloat[nPDFs for various flavors as a function of $x$.]{
        \label{subfig:ncteq26-vs-ncteq26ohf}
        \includegraphics[width=\linewidth]{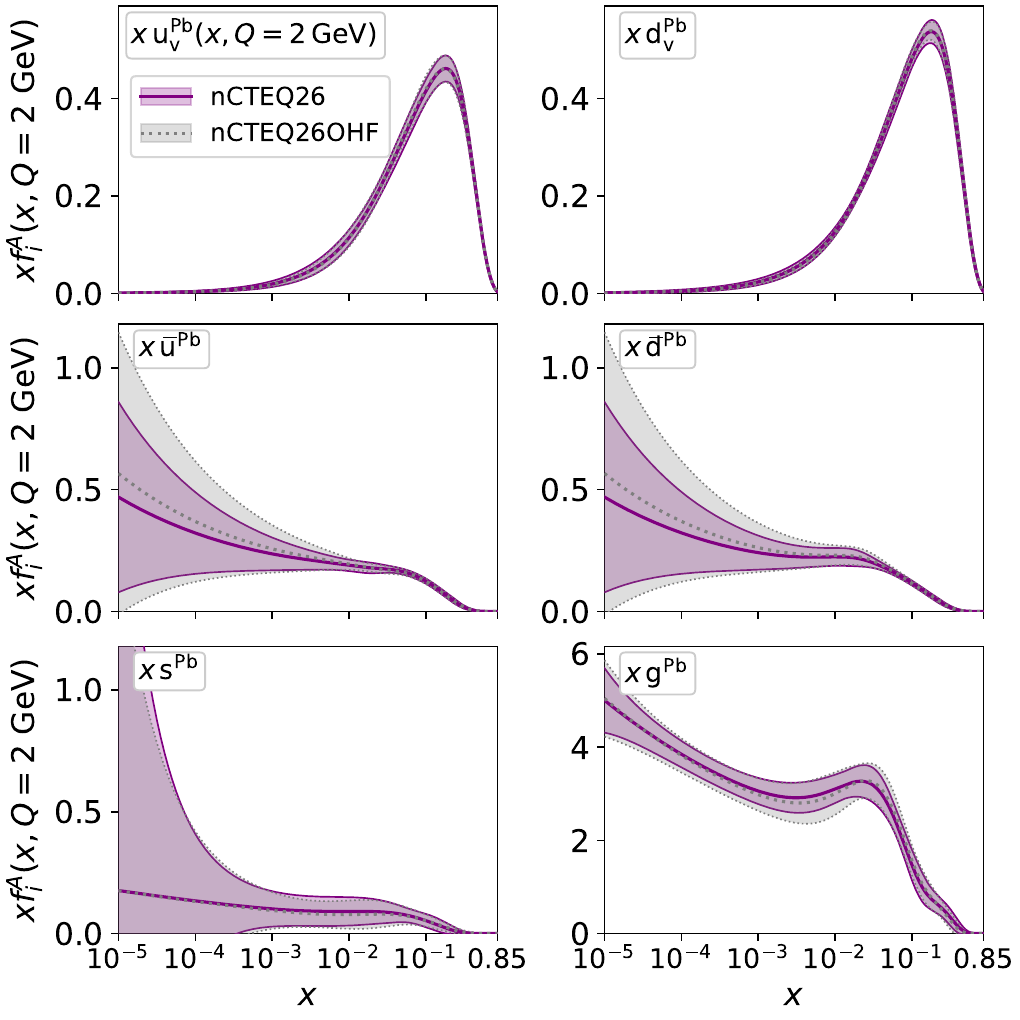}
    }
    \\
    \subfloat[nPDF ratios for various flavors as a function of $x$.]{
        \label{subfig:ncteq26-vs-ncteq26ohf_ratio}
        \includegraphics[width=\linewidth]{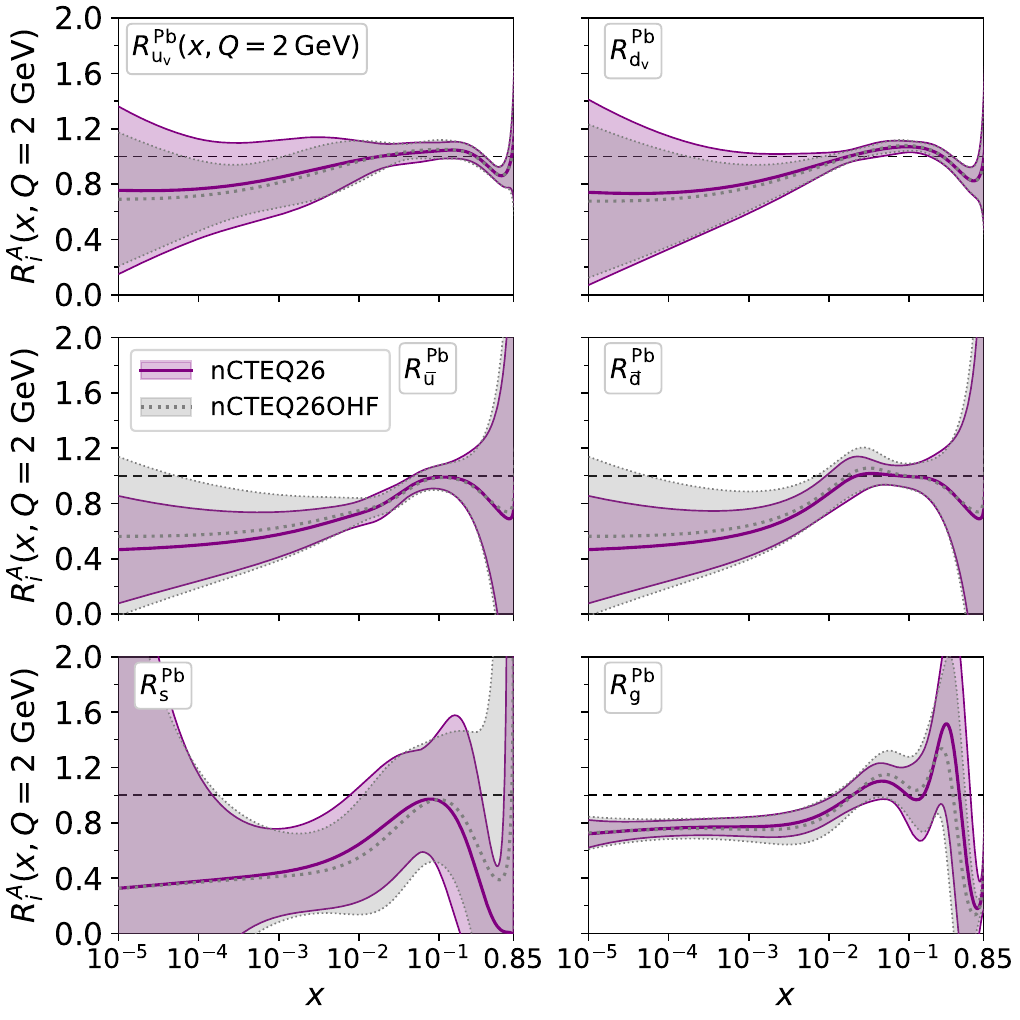}
    }
    \caption{(a) nCTEQ26 and nCTEQ26OHF nPDFs for lead nuclei at the scale $Q=2$ GeV. (b) Corresponding ratios of the full nPDFs for lead nuclei to those constructed from the free-proton baseline using Eq.~\eqref{eq:nPDFA}.}
    \label{fig:ncteq26_ncteq26OHF_Pb_Q2} 
\end{figure}
\begin{figure*}[htp!]
    \centering
    \includegraphics[width=1\linewidth]{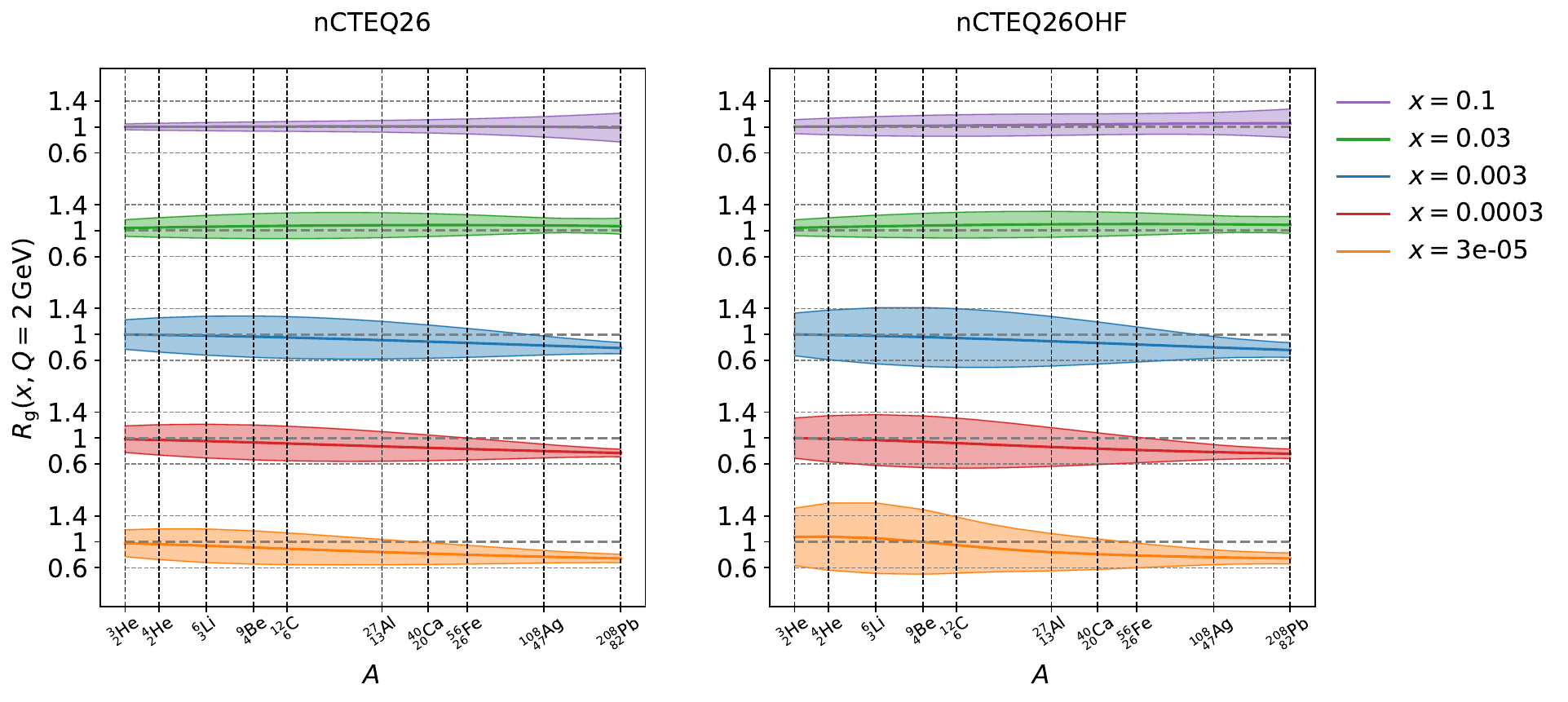}
    \caption{Comparison of the gluon nuclear modification ratios $R^A_g$ as defined in \cref{eq:RA} from nCTEQ26 and nCTEQ26OHF at $Q = 2$ GeV as a function of nuclear mass number $A$ and for selected values of $x$.}
    \label{fig:Af-vs-ncteq26ohf}
\end{figure*}

In \cref{fig:ncteq26_ncteq26OHF_Pb_Q2}, we compare the lead PDFs from the nCTEQ26 fit and the alternative nCTEQ26OHF fit. It is clear that the central PDFs of both fits agree perfectly within uncertainties for all flavors. The differences in the $u_v$, $d_v$, $\bar{u}$ and $\bar{d}$ uncertainties indicate that they are only weakly constrained in the low-$x$ region in both the nCTEQ26 and nCTEQ26OHF fits. In contrast to the data-driven approach, the $gq$ and $q\bar{q}$ channels are taken into account in the SACOT calculation in nCTEQ26OHF. However, since the dominant contribution is still provided by the $gg$ channel, the other channels do not provide strong additional constraints. The size of the uncertainties in the strange PDF is comparable in both fits. The nCTEQ26OHF gluon central value is very close to the main fit, and its uncertainty slightly increases throughout the range in $x$. Since only the gluon-gluon channel is taken into account in the Crystal-Ball matrix element, this indicates that the Crystal-Ball method is consistent with the perturbative QCD approach. It is, however, somewhat counterintuitive that the increase in the gluon uncertainty is not larger, given that the number of quarkonium data points is more than twice as large as the number of OHF points.
\begin{figure}[htb]
    \centering
    \includegraphics[width=0.95\linewidth]{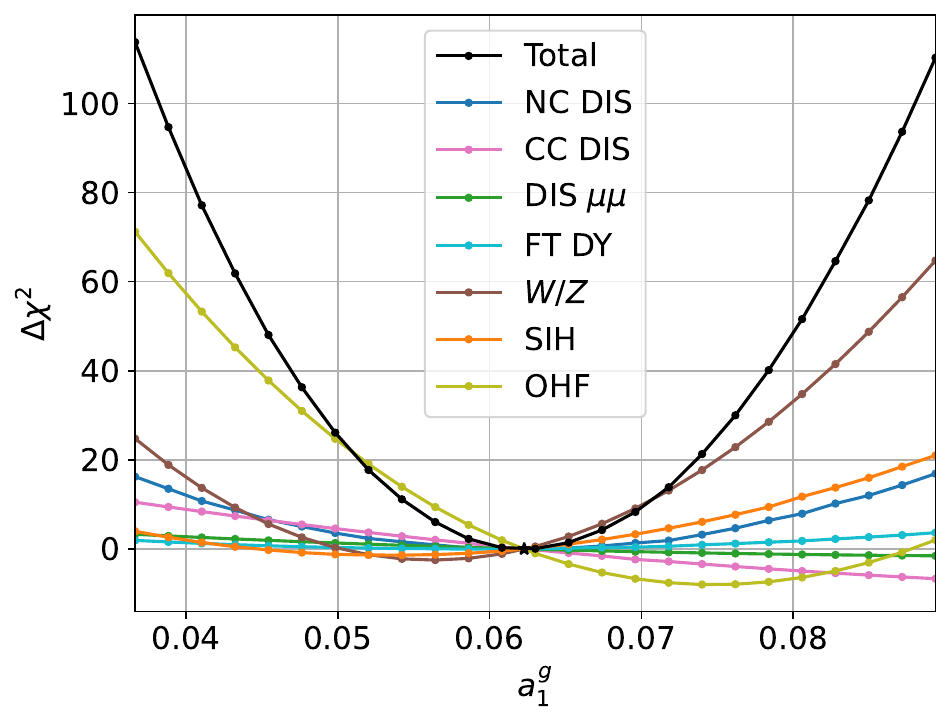}
    \caption{One-dimensional $\chi^2$ scan along the $a_1^g$ parameter in nCTEQ26OHF. Plotted is ${\Delta \chi^2=\chi^2-\chi^2_0}$ together with a decomposition into contributions from different data types.}
    \label{fig:scan-g-a1-ncteq26ohf}
\end{figure}
To investigate this further we compare the $A$-dependence of the main and the alternative fits in \cref{fig:Af-vs-ncteq26ohf}. 
We observe that the gluon uncertainty does indeed increase for the lighter nuclei. This is at first sight again unexpected, since the quarkonium data provide constraints for lead PDFs, so this is where one would expect the biggest increase in uncertainty when removing this constraint. However, the nCTEQ26 central PDFs and their uncertainties are correlated across all values of $A$ due to the analytic \mbox{$A$-dependence} imposed upon the parameters. It is thus not implausible that the change in PDFs, manifest in \cref{fig:Af-vs-ncteq26ohf}, can be explained by a reduction of constraints for lead due to removal of the quarkonium data. For heavier nuclei, we identify two potential reasons for the relatively small change in the gluon uncertainty. First, the nCTEQ26 fit includes theoretical uncertainties associated with the Crystal-Ball fit parameters. These effectively increase the uncertainties in these datasets and, as such, reduce their constraining power. In nCTEQ26OHF, there are, of course, no uncertainties related to the Crystal-Ball parameters, which, in turn, increases the pull of the OHF data.\footnote{In principle, one could add the uncertainties due to the renormalization and factorization scale fit to the data uncertainty, analogously to adding the Crystal-Ball parameter uncertainties in the main fit.} Second, it was shown, e.g., in nNNPDF3.0 \cite{AbdulKhalek:2022fyi}, that the $D^0$ data alone can have a large impact on the gluon and significantly reduce its uncertainties. This can be further investigated in \cref{fig:scan-g-a1-ncteq26ohf}, where we show a $\chi^2$ scan along the gluon parameter $a_1^g$, that dominates the behavior at low $x$. It is apparent that the $W/Z$-boson data have a large impact, constraining this parameter from the right and opposing the constraint from the OHF data, which is still very similar to the one provided by the quarkonium data in the main fit (cf.\ \cref{subfig:par-g_a1}). Together, these constraints lead again to the parabolic shape of the $\chi^2$ profile and decrease the gluon uncertainties.

\begin{figure}
    \centering
    \includegraphics[width=0.95\linewidth]{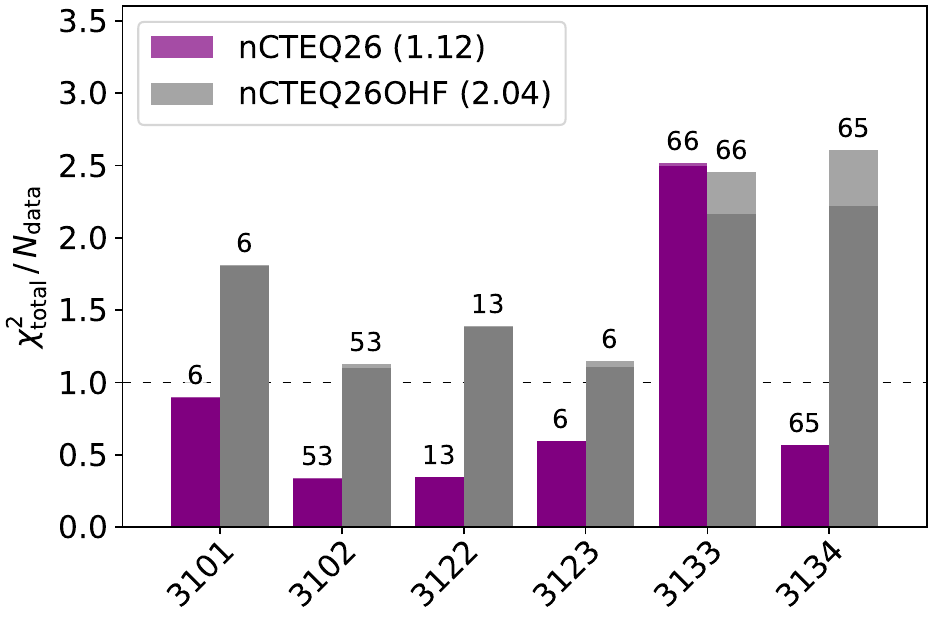}
    \caption{Comparison of the $\chi^2$ values of the open heavy-flavor production data between the nCTEQ26 and nCTEQ26 fits.}
    \label{fig:chi2-breakdown-ohf}
\end{figure}

Even though the nCTEQ26OHF fit yields central values and uncertainties very similar to the main fit, it features appreciable differences in the $\chi^2$ values for the open heavy-flavor production data. The comparison of the $\chi^2$ breakdown per dataset for the OHF category is shown in \cref{fig:chi2-breakdown-ohf} and we refer the reader to \cref{fig:chi2dataset} for the $\chi^2$ values of the other processes. 
We first note the increase in $\chi^2$ of the whole OHF data. It changes from an acceptable value of $1.12$ in nCTEQ26 to a value of $2.04$ in nCTEQ26OHF. In turn, this increases the total $\chi^2$ to $1.004$, compared to the value of $0.943$ in the main fit. Interestingly, dataset 3102 (LHCb $5.02\,\mathrm{TeV}$) is the dataset with the lowest $\chi^2$ in the nCTEQ26OHF fit, whereas datasets 3133 and 3134 (LHCb $8.16\,\mathrm{TeV}$ at forward and backward rapidity) have the highest $\chi^2$, with 3133 also being described poorly in the nCTEQ26 fit. The other OHF datasets are consistently better described in the main fit. Again, one of the reasons for the increase in $\chi^2$ is the additional CB parameter uncertainty considered in the nCTEQ26 fit, which is absent in the nCTEQ26OHF fit.

\begin{figure}[tbhp!]
    \centering
    \subfloat[Forward rapidity.]{
    \label{subfig:data-vs-theory-ohf-3133}
    \includegraphics[width=0.99\linewidth]{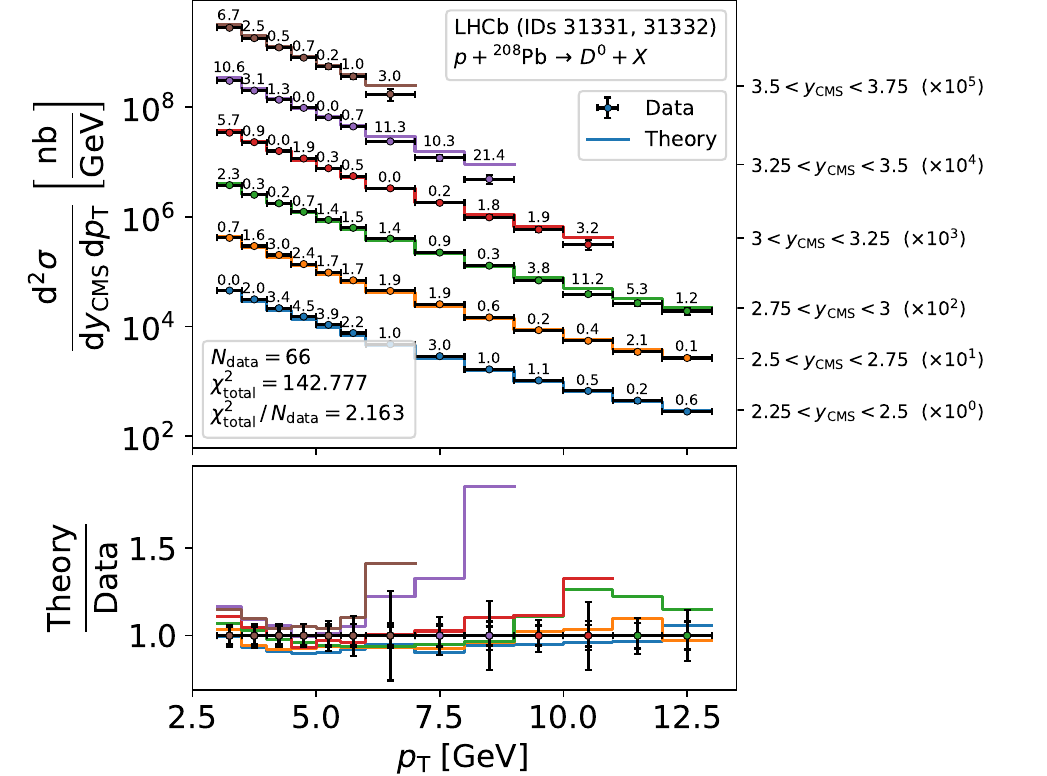}
    }
    \\
    \subfloat[Backward rapidity.]{
    \label{subfig:data-vs-theory-ohf-3134}
    \includegraphics[width=0.99\linewidth]{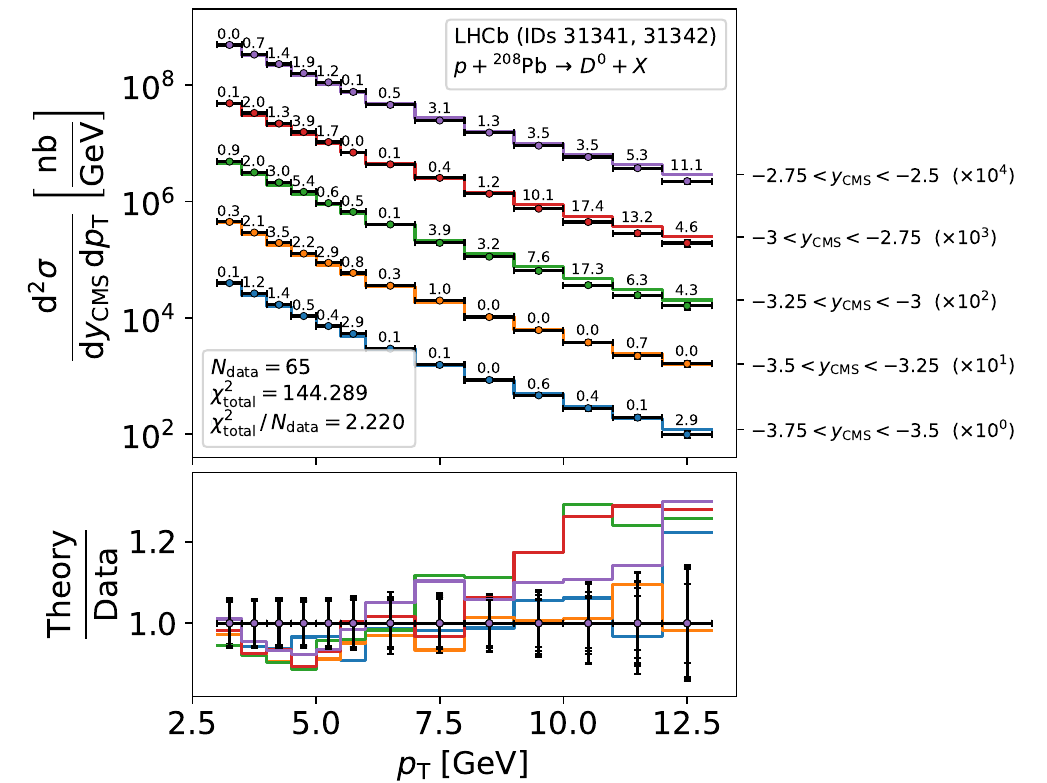}
    }
    \caption{nCTEQ26OHF predictions for LHCb $8\,\mathrm{TeV}$ data at forward and backward rapidity. See \cref{fig:data-vs-theory:HQ3133,fig:data-vs-theory:HQ3134} for the comparison to nCTEQ26.}
    \label{fig:data-vs-theory-ohf-3133-3134}
\end{figure}

Additionally, in \cref{fig:data-vs-theory-ohf-3133-3134}, we compare our nCTEQ26OHF predictions with the LHCb $8\,\mathrm{TeV}$ data (ID 3133 and 3134). The largest deviations are observed, for forward rapidities ($y_\mathrm{CMS} > 2.75$), in the low- and high-$p_T$ regions adjacent to our cuts and, for backward rapidities, in the high-$p_T$ region, but not in the two most backward rapidity bins. 
Interestingly, the CB predictions at forward rapidity \cref{fig:data-vs-theory:HQ3133} show large $\chi^2$ values in all rapidity bins, resulting in a similar $\chi^2$ of dataset 3133 in both fits, whereas the CB predictions at backward rapidity \cref{fig:data-vs-theory:HQ3134} describe almost all $p_T$ and $y$ bins very well. Additionally, the shape of the theory prediction seems to differ from the data in the ratio panel shown in \cref{subfig:data-vs-theory-ohf-3134}. This might be connected to the tuning of the scale coefficients. We plan to investigate this further in future work.

We find the nCTEQ26OHF PDFs, based purely on pQCD theoretical predictions and not including the data-driven Crystal-Ball matrix element, to be exceptionally consistent with the nCTEQ26 fit. There is agreement in both the central and uncertainties of the gluon, showing consistency of the CB OHF and quarkonium predictions with GM-VFNS OHF predictions. Nevertheless, in nCTEQ26OHF, we find the overall OHF data description unsatisfactory due to large $\chi^2$ values. For this reason we consider nCTEQ26OHF not an independent global analysis but a supporting fit, and generally recommend usage of the nCTEQ26 main PDFs for phenomenology. But we would like to further improve the nCTEQ26OHF fit, e.g., by quantifying the uncertainty of the scale coefficients analogous to the CB fit parameter uncertainties, assessing the impact of the choice of FF, etc. Lastly, since the CB approach allows only to include data that have a corresponding proton-proton baseline, there is also potential to expand the data selection by measurements of other charmed-meson final states, for example~$D^+$ \cite{ALICE:2019fhe,LHCb:2023kqs}. We reserve these aspects for future work.

\section{Predictions}
\label{sec:pred}

In this section, we present predictions using our new nCTEQ26 nPDFs for a number of datasets that are not included in our analysis. The decision to present comparisons with these datasets, rather than incorporating them into the fit, is motivated by two broad considerations. 
The first motivation is conceptual, stemming from both the tensions between these datasets and other measurements and the current lack of sufficiently reliable theoretical predictions.
The second reason is practical and encompasses either the limited impact of these datasets on the global analysis or the lack of computationally efficient theoretical calculations. 
In the following, we examine the  datasets under consideration one by one and discuss their potential roles in future global analyses, where they may provide valuable additional constraints on the nPDFs.

\subsection{NOMAD dimuons}
\label{subsec:nomad}

Apart from the CCFR/NuTeV experiments at FNAL, the NOMAD collaboration at CERN has also measured charm dimuon production in neutrino DIS on iron~\cite{NOMAD:2013hbk}. The ratio of the charm (${\mu^+ \mu^-}$) production cross section to the inclusive charged-current DIS cross section was given differentially in three variables: the neutrino energy $E_\nu$, the Bjorken-$x$, and the center-of-mass energy $\sqrt{s}$. These data are not included in the nCTEQ26 analysis, since they require the fast calculation of an extra integral, which is not available within the \ncteqpp{} framework.\footnote{These data have also not been included in any other global analysis of nPDFs. They were, however, used in one analysis of proton PDFs~\cite{Alekhin:2017kpj}.}

\begin{figure}
    \centering
    \subfloat[Distribution in the neutrino energy $E_\nu$.]{
        \includegraphics[width=0.95\linewidth]{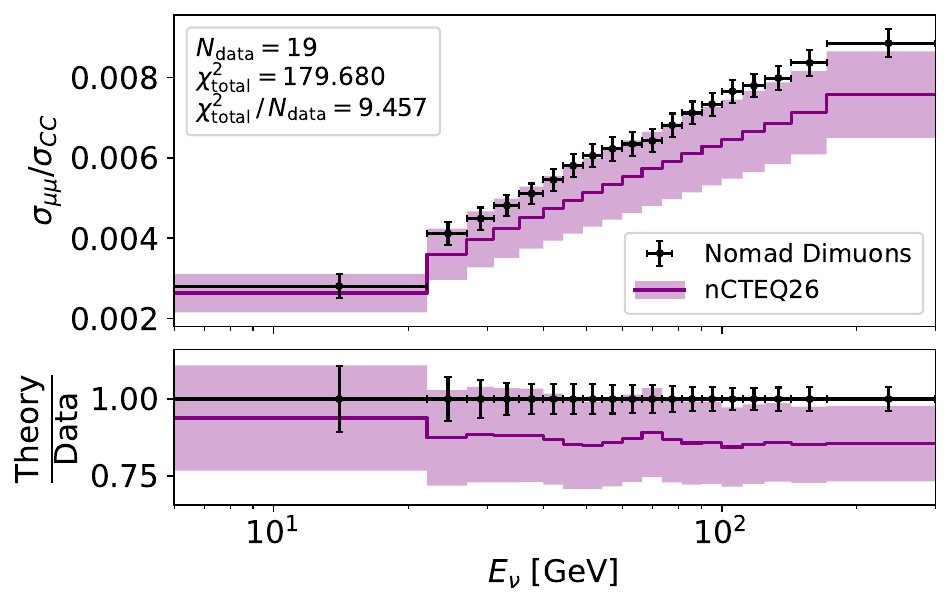}
    }
    \\
    \subfloat[Distribution in Bjorken-$x$.]{
        \includegraphics[width=0.95\linewidth]{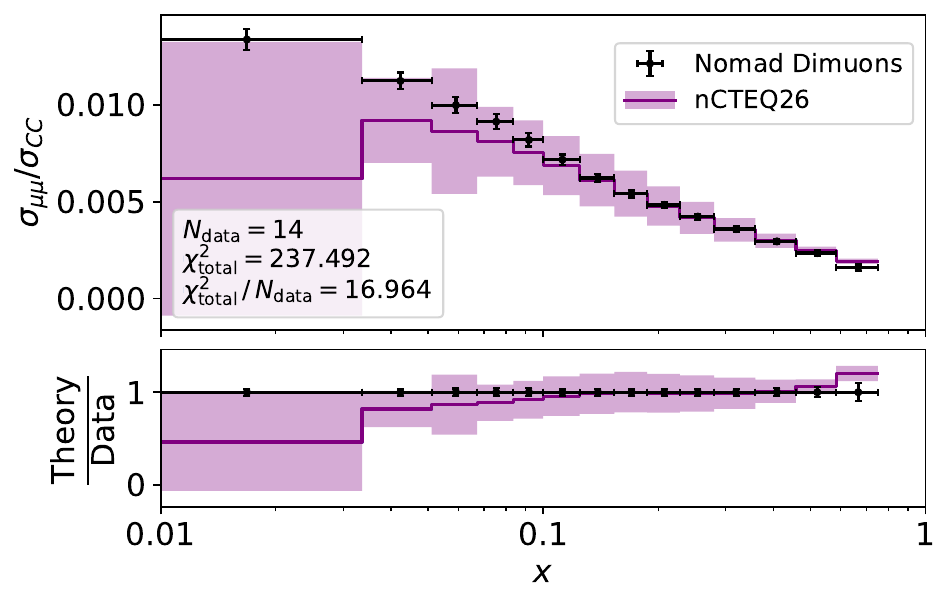}
    }
    \caption{Comparison of nCTEQ26 predictions to NOMAD charm dimuon data in neutrino DIS on iron.}
    \label{fig:Nomad_dimuon}
\end{figure}

In \cref{fig:Nomad_dimuon}, we compare our predictions with nCTEQ26 iron PDFs to the NOMAD distributions in (a) the neutrino energy $E_\nu$ and (b) Bjorken-$x$. For the neutrino-energy distribution, the data are covered by the 90\% nPDF uncertainty band, whose upper end also describes the shape of the data. At the same time, the central value lies clearly below the data over almost the entire range of $E_\nu$ with only the lowest bin falling within the experimental uncertainty as quantified by the $\chi^2$ given for each data point. The situation for the distribution in $x$ is at first sight similar, in particular for the two lowest bins in $x$. However, already for $x>0.1$, where most of the data points lie, the central theory predictions fall within the experimental uncertainty. Since charm dimuon production is mainly induced by strange quarks, one can therefore expect that including these data will increase the strange quark nPDF at small $x$, but most of all considerably reduce its uncertainty at intermediate and large $x$. This will not only affect the iron PDFs, but indirectly, through the continuous $A$-dependence, also all other nuclei.

\subsection{CMS dijets}
\label{subsec:dijets}

\begin{figure*}[htbp!]
    \centering
    \subfloat[CMS dijet data in $pp$ collisions compared to CJ15 predictions.]{
        \includegraphics[page=1, width=\linewidth]{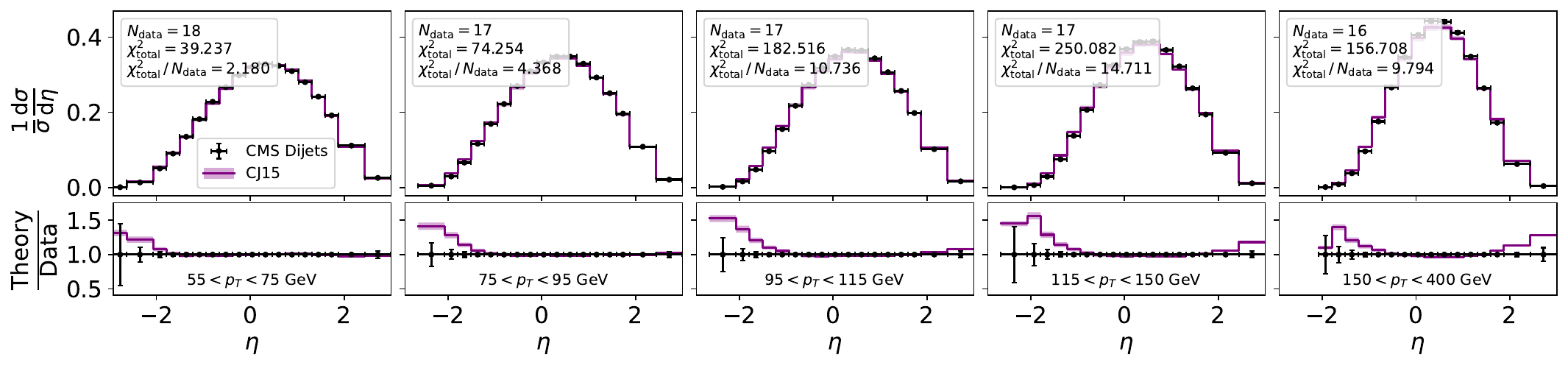}       
        \label{subfig:ncteq26_cms_dijet_pp}
    }
    \\
    \subfloat[CMS dijet data in $p$Pb collisions compared to nCTEQ26 predictions.]{
        \includegraphics[page=1, width=\linewidth]{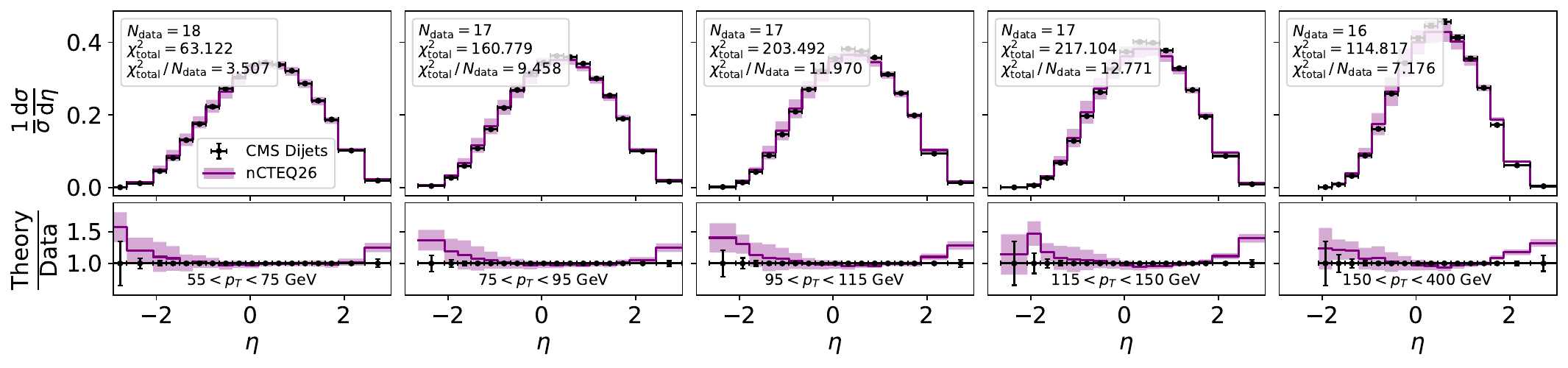}       
        \label{subfig:ncteq26_cms_dijet_pPb}
    }
    \\
    \subfloat[Ratio of $p$Pb to $pp$ CMS dijet data compared to nCTEQ26 predictions.]{
        \includegraphics[page=1, width=\linewidth]{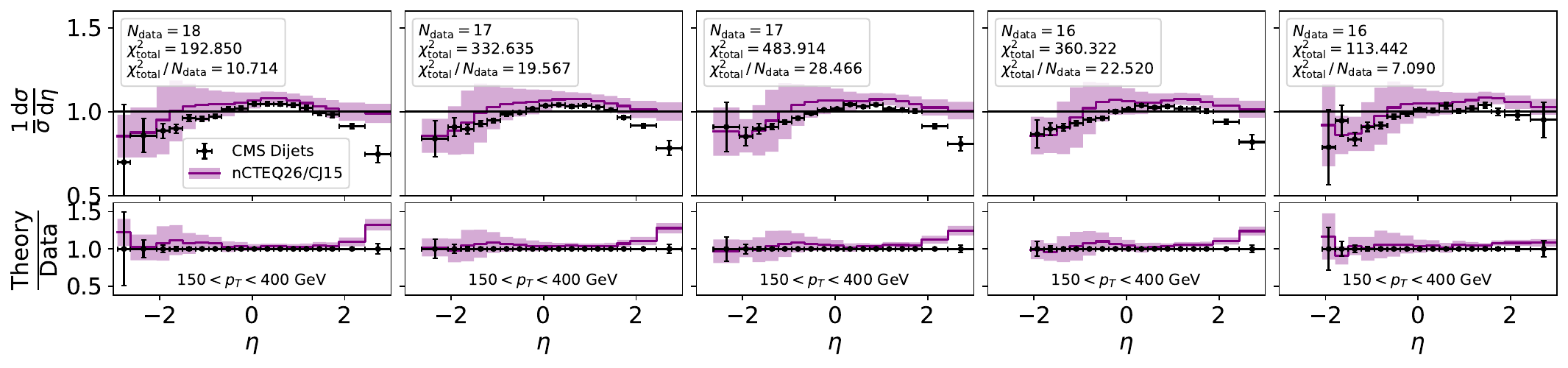}      
        \label{subfig:ncteq26_cms_dijet_ratio}
    }
    \caption{Comparison of CJ15 and nCTEQ26 predictions to CMS dijet data in $pp$ and $p$Pb collisions at LHC Run I.}
    \label{fig:ncteq26_cms_dijet}
\end{figure*}

Dijet production in $p$Pb collisions is induced by all partonic channels, where it probes the intermediate- to large-$x$ regime ($3 \times 10^{-3} \lesssim x \lesssim 0.8$, cf.\ Fig.\ 1 of Ref.\ \cite{CMS:2018jpl}) of nPDFs at large interaction scales. Dijet measurements therefore represent a valuable addition to other data like fixed-target DIS, DY and LHC $W/Z$ production, which are only sensitive to the gluon distribution starting at NLO, or to heavy quarks and quarkonia, which are produced at lower scales. Compared to $pp$ collisions, the interpretation of dijet data in $p$Pb collisions is unfortunately complicated by a significant background from the underlying event and larger hadronization corrections compared to $pp$ collisions.

The most constraining data currently come from CMS Run I measurements of pseudorapidity distributions of dijets in bins of their average transverse momentum at $\sqrt{s}=5.02$ TeV \cite{CMS:2018jpl}. The rapidity distributions in $pp$ and $p$Pb collisions have been normalized to the rapidity-integrated cross sections. Also their ratios, in which systematic uncertainties should largely cancel, were determined, taking into account a shift in rapidity due to the different beam energies. These ratios were found to be broadly consistent with previous nPDF determinations. They have therefore been included in the most recent EPPS21 (83 data points, $\chi^2/N_{\rm data}=1.49$) \cite{Eskola:2021nhw} and nNNPDF3.0 (84 data points, $\chi^2/N_{\rm data}=1.75$) \cite{AbdulKhalek:2022fyi} global analyses, where they have a large impact on the gluon. However, the most forward data points could not be fitted. It is also impossible to describe the $p$Pb and $pp$ distributions individually at NLO (e.g.\ $\chi^2/N_{\rm data}=13.96$ for $p$Pb in nNNPDF3.0). The impact of missing higher-order corrections is currently unclear, but preliminary results of an upcoming study at NNLO do not indicate a sufficient improvement of the description of the individual measurements \cite{ChenDijets}. We have therefore not included the CMS dijet data in our current nCTEQ26 analysis.

We compare our predictions with our baseline CJ15 proton and nCTEQ26 lead PDFs to the CMS $pp$ data in \cref{subfig:ncteq26_cms_dijet_pp}, to the CMS $p$Pb data in \cref{subfig:ncteq26_cms_dijet_pPb}, and to their ratio in \cref{subfig:ncteq26_cms_dijet_ratio}. These predictions were calculated using fastNLO grids from Ref.~\cite{Wobisch:2011ij}. Similarly to previous observations by the EPPS collaboration \cite{Eskola:2019dui}, our CJ15 proton baseline can describe the $pp$ data only for low $p_T$~($<75$~GeV) and positive $\eta$, whereas in all other kinematic regions the predictions do not align with the data. Our $p$Pb predictions obtained with nCTEQ26 PDFs agree with the data within uncertainties for ${|\eta|<1.5}$ for all $p_T$, but not at forward rapidity. This translates into good agreement between data and theory in the ratios in the backward direction ($\eta<0$), but not in the forward direction, which corresponds to the low $x$ momentum fraction in lead. The CMS dijet data and in particular the $p$Pb/$pp$ ratios will therefore only be considered for inclusion in future nCTEQ analyses, when the impact of NNLO corrections \cite{Chen:2022tpk} has been clarified.

It is worth noting that there are also single-inclusive jet data in $p$Pb collisions~\cite{ATLAS:2014cpa,CMS:2016svx,ALICE:2023ama}, but their uncertainties are considerably larger than those of the dijet measurements. Moreover, there are promising new measurements from ATLAS~\cite{ATLAS:2023zfx} and CMS~\cite{CMS:2014qca}, which could provide crucial constraints in the future when no only yields and their ratios, but also absolute cross sections have been published. 

\subsection{ATLAS/ALICE prompt photons}
\label{subsec:photons}

\begin{figure}[tbp!]
    \centering
    \begin{subfigure}[b]{0.49\textwidth}
        \includegraphics[width=\linewidth]{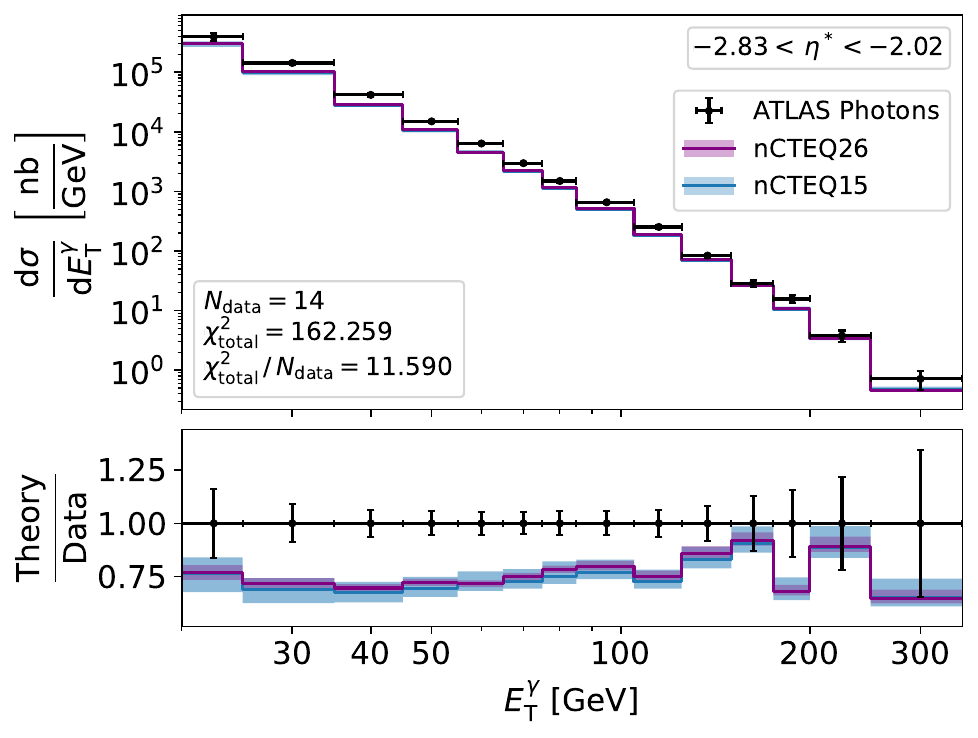}
    \end{subfigure}
    \\
    \begin{subfigure}[b]{0.49\textwidth}
        \includegraphics[width=\linewidth]{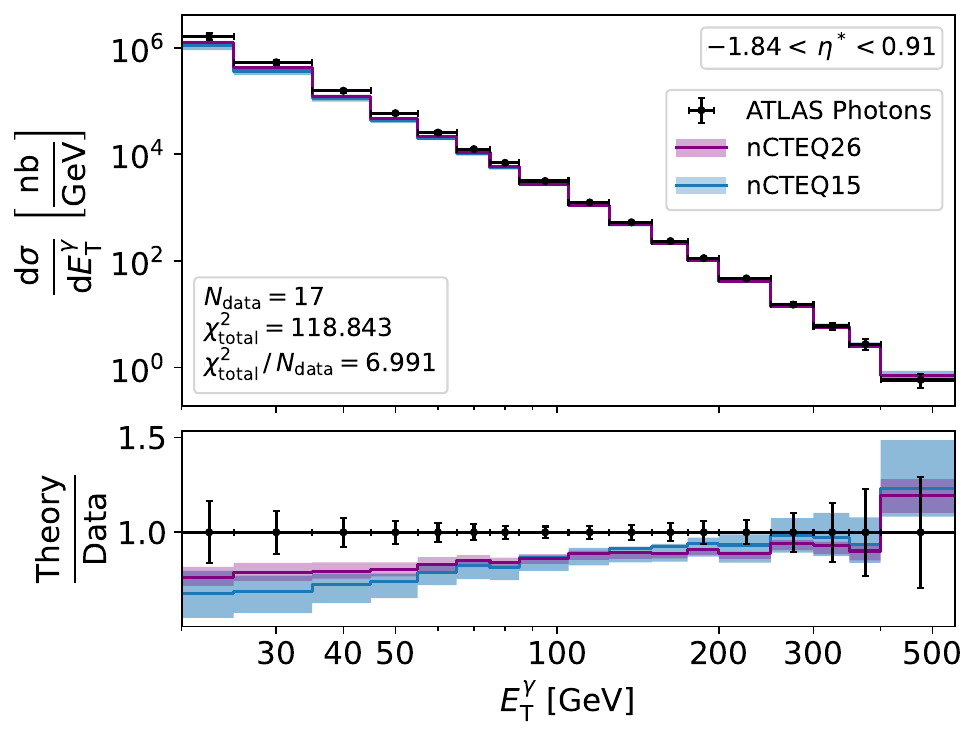}
    \end{subfigure}
    \\
    \begin{subfigure}[b]{0.49\textwidth}
        \includegraphics[width=\linewidth]{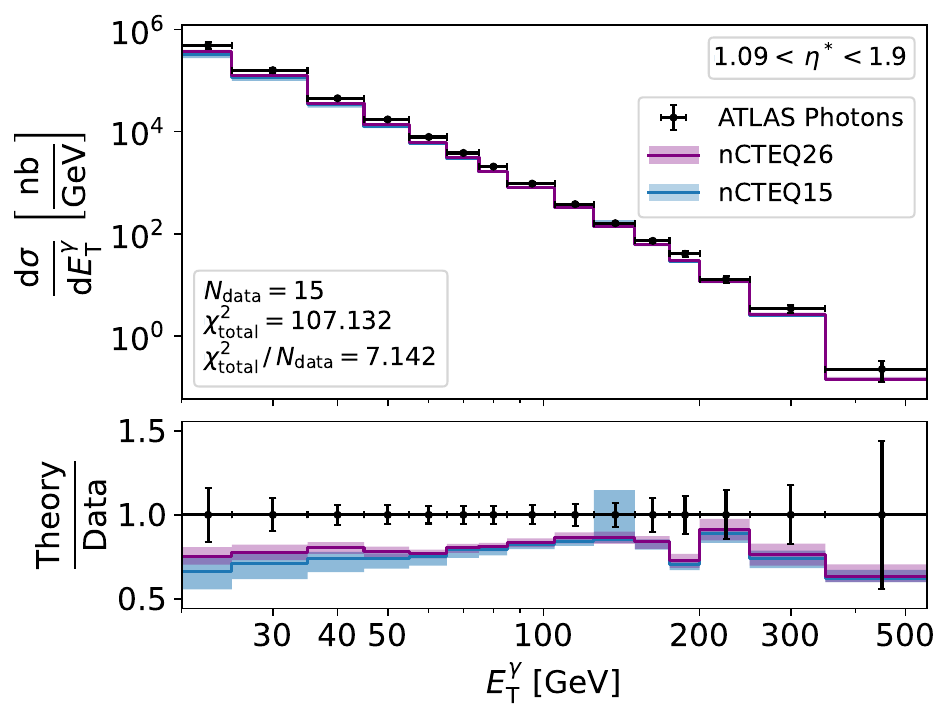}
    \end{subfigure}   
    \caption{Comparison of nCTEQ26 (magenta) and nCTEQ15 (blue) predictions to ATLAS \cite{ATLAS:2019ery} isolated prompt photon data in $p$Pb collisions at LHC Run II in three different center-of-mass rapidity bins. The annotated values display the $\chi^2$ per point between the data and nCTEQ26.}
    \label{fig:photonPPBATLAS}
\end{figure}
\begin{figure}[ptbh!]
    \centering
    \includegraphics[width=0.95\linewidth]{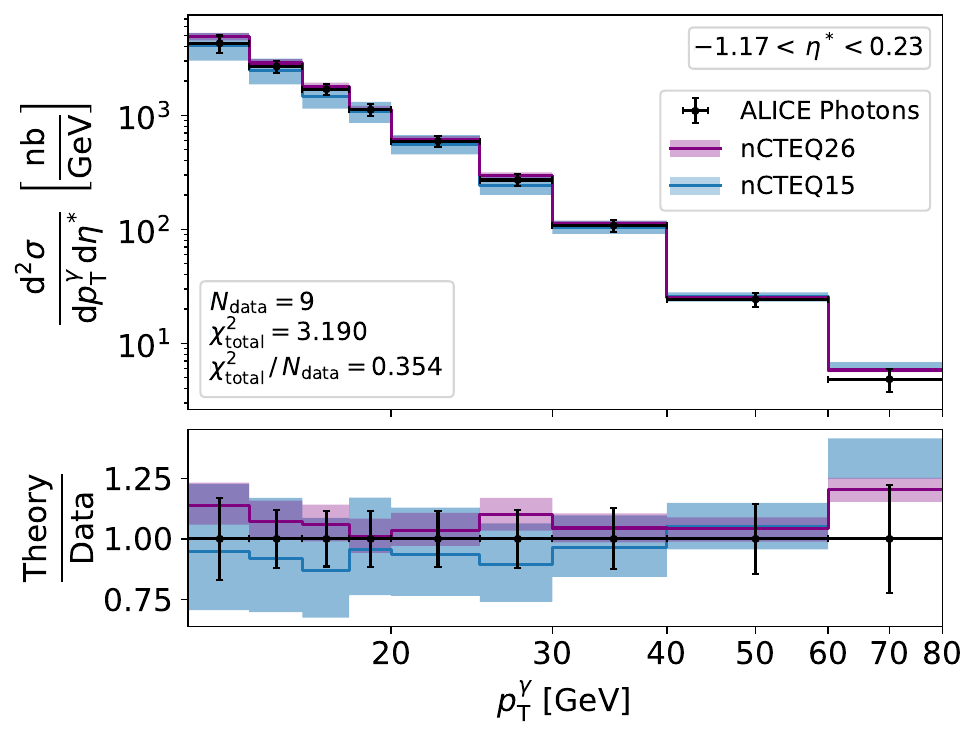}
    \caption{Comparison of nCTEQ26 (magenta) and nCTEQ15 (blue) predictions to ALICE \cite{ALICE:2025bnc} isolated prompt photon data at central rapidity in $p$Pb collisions at LHC Run II. The annotated values display the $\chi^2$ per point between the data and nCTEQ26.}
    \label{fig:photonPPBALICE}
\end{figure}

Prompt photon production proceeds in perturbative QCD through quark-antiquark annihilation and the QCD Compton process, which dominates at large photon transverse momenta ($p_T^\gamma$) and therefore provides direct access to the gluon PDF \cite{dEnterria:2012kvo}. At low $p_T^\gamma$, partonic scatterings and fragmentation into photons become important. Photon isolation criteria, which typically limit the transverse hadronic energy in a cone around the photon, allow to reduce not only the fragmentation contribution, but also the background photons from pion decays \cite{Frixione:1998jh}. Similarly to DY or $W/Z$ boson production, prompt photon production can then serve as a clean probe of nuclear structure.

At the LHC, prompt photon production has been measured by the ATLAS 
\cite{ATLAS:2019ery} and ALICE~\cite{ALICE:2025bnc} collaborations at high (20 GeV $< E_T^\gamma<550$ GeV) and low (12 GeV $<E_T^\gamma<80$ GeV) transverse photon energy, respectively. The ATLAS $E_T^\gamma$-distributions were obtained in $p$Pb collisions at Run II with $\sqrt{s}=8.16$~TeV and in $pp$ collisions at Run I with $\sqrt{s}=8$ TeV for three different center-of-mass rapidity ($\eta^*$) bins together with the $p$Pb/$pp$ ratios, using an extrapolation for the different $pp$ center-of-mass energy and rapidity selection. The ratios were found to be roughly compatible with previously determined nuclear modifications and were therefore included in the nNNPDF3.0 analysis (43 data points, $\chi^2/N_{\rm data}=1.03$) using MCFM calculations \cite{Campbell:2016lzl,Campbell:2018wfu} gridded with APPLgrid \cite{Carli:2010rw} and assuming that the neglected fragmentation contributions are small. The absolute $p$Pb spectra are, however, poorly described ($\chi^2/N_{\rm data}=3.21$). As for the dijet data, missing higher order effects might be part of the explanation. As a consequence, we do not yet include the ATLAS prompt photon data in our global analysis, until the impact of NNLO corrections \cite{Campbell:2016lzl,Chen:2022gpk} has been clarified.

In \cref{fig:photonPPBATLAS} we compare our predictions based on JETPHOX \cite{Catani:2002ny}, which contrary to MCFM \cite{Campbell:2016lzl} includes fragmentation contributions at NLO, to the ATLAS absolute $E_T^\gamma$-distributions \cite{ATLAS:2019ery}. In all three rapidity bins, ATLAS measures softer $p_T$ spectra as compared to the theoretical predictions and a larger fiducial cross section. While the new nCTEQ26 nPDFs (magenta) bring the theoretical predictions closer to the data compared to those obtained with nCTEQ15 (blue), an appreciable difference in the shape remains. Note that the fiducial cross section difference could be partially alleviated by allowing for a floating normalization.

The ALICE collaboration has measured prompt-photon $p_T^\gamma$-spectra and $p$Pb/$pp$ ratios at mid-rapidity ($-1.17<\eta^*<0.23$) combining data from Runs I and II~\cite{ALICE:2025bnc}. 
The obtained ratios are consistent within uncertainties with the corresponding ratio measured by ATLAS in the overlapping kinematic region and considering the different energies and acceptances.
The ALICE data have not been included in any global nPDF analysis, but should not have a significant impact due to their relatively large uncertainties.

In \cref{fig:photonPPBALICE} we compare our nCTEQ26 (magenta) and nCTEQ15 (blue) predictions to the ALICE data. In both cases the description of the data is very good, especially for the nCTEQ26 PDFs, so that these data points could in principle be included in our analysis. However, fast predictions are not yet available for the fragmentation contribution, which is larger at low $p_T^\gamma$, so an inclusion of the ALICE data is left for future work.

Finally, we observe that the nCTEQ26 uncertainties are smaller than the experimental uncertainties for both the ATLAS and ALICE measurements. Consequently, incorporating these data is unlikely to provide significant additional constraints within the global fit. Nevertheless, prompt photons offer a unique color-neutral probe of the gluon PDF. As future measurements become more precise, they could provide an important cross-check on gluon constraints extracted from other processes such as dijet and heavy-flavor production, which might be influenced by final-state effects.

\subsection{ATLAS/CMS top-quark pairs}
\label{subsec:ttbar}

\begin{figure}
    \centering
    \includegraphics[width=0.9\linewidth]{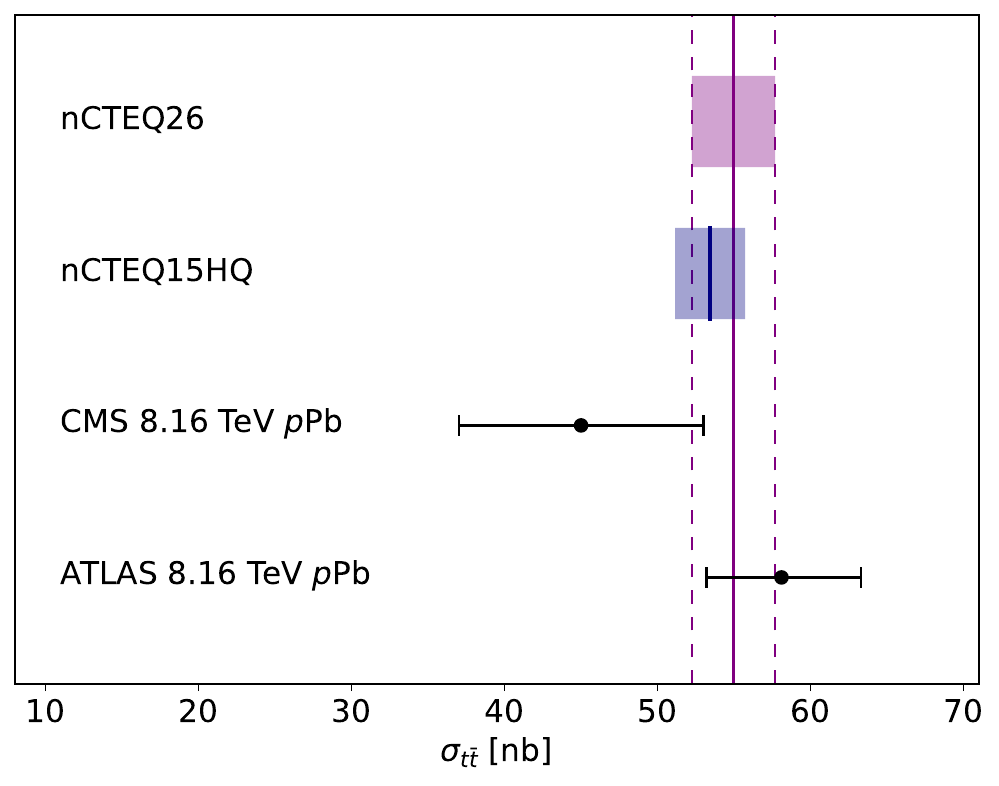}
    \caption{Comparison of nCTEQ26 (magenta) and nCTEQ15HQ \cite{Duwentaster:2022kpv} (blue) predictions to ATLAS \cite{ATLAS:2024qdu} and CMS \cite{CMS:2017hnw} top-quark data in $p$Pb collisions at LHC Run II.}
    \label{fig:ttbar-atlas}
\end{figure}

Finally, we compare in \cref{fig:ttbar-atlas} our nCTEQ26 (magenta) and nCTEQ15HQ~\cite{Duwentaster:2022kpv} (blue) predictions for top-quark pair production to the corresponding total cross sections measured by ATLAS~\cite{ATLAS:2024qdu} and CMS \cite{CMS:2017hnw} in $p$Pb collisions at $\sqrt{s}=8.16$ TeV. Here, we have used calculations with POWHEG/hvq generator~\cite{Nason:2004rx,Frixione:2007nw,Frixione:2007vw,Alioli:2010xd}%
\footnote{Note that there is no code available that natively supports the calculation of the $t\bar{t}$ cross section in asymmetric $p$Pb collisions. In our calculation, we have used POWHEG/hvq, which we have adapted to and thoroughly tested in an asymmetric collision setup. However, we fail to reproduce the theoretical prediction reported by ATLAS \cite{ATLAS:2024qdu}, which was based on MCFM \cite{Campbell:2016dks}.} 
and rescaled them with an NNLO/NLO $K$-factor (${K=1.139}$) originally obtained with Top++ \cite{Czakon:2011xx} and adopted here from Ref.\ \cite{ATLAS:2024qdu}.

We observe that the two data points are mutually compatible at the $1\sigma$ level, with ATLAS preferring a substantially larger value of the cross section and featuring a smaller uncertainty. We also find that within the corresponding PDF uncertainties  both theory predictions agree with both measurements, but slightly favor the ATLAS value. The current nPDF uncertainties are smaller than the uncertainties of the available data. Hence, at the moment it is unlikely that these data would add important constraints to our global nPDF analysis. Nevertheless, top-quark pair production provides access to a kinematic region at relatively large $x$ for the gluon that may otherwise not be available. Therefore, future measurements of kinematic distributions in $pA$ collisions, e.g.~the top-quark rapidity, invariant mass of the top-quark pair, or other distributions of top-quark decay products, could be of great value. Furthermore, the impact of final-state effects in this process (especially in the dilepton channel) could be much smaller than in other colored final states, making it even more important for validating constraints coming from other processes, such as jet or open heavy-quark and quarkonium production.

\section{Conclusions}
\label{sec:sum}

We have presented the nCTEQ26 release of nuclear PDFs from the nCTEQ collaboration. In this new global analysis, we have incorporated a broad set of experimental data and significantly expanded the range of processes and kinematic reach compared to our previous analyses. We also introduced several theoretical and methodological improvements including a new proton PDF baseline and new parametrizations of both the $x$- and $A$-dependence.

In contrast to the nCTEQ15 nPDFs, which were based on 740 data points largely comprising of neutral-current deep-inelastic scattering and the Drell-Yan process at fixed targets, nCTEQ26 includes a much broader set of processes, comprising 3598 data points spanning an extended kinematic range in both $x$ and $Q^2$.
In addition to the traditional NC DIS and DY data sets, we also included data from inclusive CC DIS and CC DIS dimuon production, LHC $W/Z$ production, single-inclusive hadron production, inclusive quarkonium production, and open heavy-flavor meson production. This expanded set of observables provides complementary sensitivity to different partonic flavors, particularly enhancing constraints on the gluon and sea quark distributions.
These data sets also extend the kinematic coverage to significantly smaller values of ${x~\sim10^{-5}}$, which is especially important for constraining the gluon distribution in this region. 
At the same time, we relaxed the kinematic cuts to include lower $Q^2$ and higher $x$ values. This allowed us to incorporate valuable new DIS measurements from JLab that probe the large-$x$ region, and also to include many more data points from older experiments, e.g.~SLAC or Hermes, which were not used in earlier analyses.
This substantial increase in both the volume and diversity of data provides stronger and more comprehensive constraints on the nPDFs, leading to improved precision and greater reliability in phenomenological applications.
It also provides a more complete and balanced coverage of the nuclear kinematic phase space and encourages the ongoing study of neutrino-nucleus interactions in the kinematic range of neutrino experiments, such as DUNE, IceCube or SHiP.

To achieve such extended kinematic coverage, we implemented several theoretical improvements allowing for reliable predictions in this kinematic region. Specifically, we incorporated target-mass corrections and included a dedicated treatment of deuteron, as well as considered higher-twist effects, following the well-established results of the CJ group. 
To this end, we have also updated our proton PDF baseline by adopting the CJ15 PDF set, which ensures that the deuteron corrections are fully consistent. In addition, the CJ15 baseline provides a careful treatment of the high-$x$ region, and it is determined without relying on nuclear data, allowing us to avoid double counting of such data. This ensures a clean separation between proton and nuclear effects in the analysis.
While our nPDF parametrization is modeled on the CJ15 framework and incorporates a detailed treatment of the valence distributions and the $\bar{d}/\bar{u}$ ratio in the high-$x$ region, we further introduced greater flexibility in the strange-quark distribution and extended the CJ15 functional form with an additional term to better accommodate the additional data sets.

Finally, we also revised the nuclear $A$-dependence, replacing the exponential form used in nCTEQ15 with a quadratic logarithmic form. This provides greater flexibility in describing nuclear modifications across different mass numbers while maintaining a stable, well-constrained fit. 
Importantly, the current parametrization reduces the bias associated with imposing overly restrictive constraints from one nucleus onto another. This has been a persistent issue in nPDF fits, where insufficiently flexible $A$-dependence can produce an artificially smooth or monotonic evolution of uncertainties with the nuclear mass number $A$, even in regions where direct experimental constraints are absent. This issue was carefully investigated using the strange-quark distribution and shown to be substantially mitigated by the current parametrization.

There are a number of directions that the presented analysis can be expanded upon in the future. On the theoretical level, one obvious direction is the inclusion of higher-order NNLO corrections in both the evolution and in the matrix elements. The elements needed for this are in principle available for the most important processes (DIS, DY, $W/Z$, dijets), but substantial technical work is required to incorporate them for asymmetric collisions. Another theoretical direction is the study of nuclear higher-twist effects, which is largely unexplored. On the methodological side, next steps could be the incorporation of theoretical uncertainties (as scale variations or theory nuisance parameters~\cite{Tackmann:2024kci}) and uncertainties of proton PDFs. Going further into the future, an interesting development would be to perform a simultaneous fit of proton and nuclear PDFs, removing some of the restrictions of using a fixed proton baseline. Finally, on the experimental side, the inclusion of additional data will be pursued. This includes data that are already available but not well described by the current theoretical calculations, as well as data coming from new experiments or not-yet-analyzed LHC data, such as $p$O data or data from SMOG2.

In summary, the nCTEQ26 nPDF set represents a significant advance, integrating the latest experimental data with important theoretical refinements. 
The improvements relative to nCTEQ15 and our subsequent more specific studies are substantial, enabling more precise and reliable comparisons between experimental measurements and theoretical predictions. Particularly valuable is the currently most precise determination of the gluon nPDF in a broad kinematic range, which in turn supports more stringent tests of our understanding of nuclear structure and sharpens our sensitivity to potential deviations that could signal new or previously unexplored phenomena.

Taken together, nCTEQ26 establishes a new benchmark for global nPDF analyses, substantially advancing both precision and kinematic reach. It provides a robust foundation for precision QCD studies and an essential resource for interpreting current and future measurements, from fixed-target experiments to the High-Luminosity Large Hadron Collider and the Electron–Ion Collider. More broadly, nCTEQ26 will help advance our quantitative understanding of the quark and gluon structure of nuclei.

\addtocontents{toc}{\protect\vspace{+0.8em}}
\addtocontents{toc}{\protect\hspace{-0.4cm}\textcolor{blue}{Appendices} \par} %
\addtocontents{toc}{\protect\vspace{-0.5em}}
\appendix

\section{Usage of nCTEQ26 PDFs}
\label{app:usage}
\addtocontents{toc}{\tocSpace}

The nCTEQ26 nPDFs are provided as full nuclear PDFs, $x f_k^A(x,Q)$, as defined in \cref{eq:nPDFA} and can be downloaded as LHAPDF files~\cite{Buckley:2014ana} from the LHAPDF website \mbox{\url{https://www.lhapdf.org/pdfsets.html}}. Note that these nPDFs are scaled by the nuclear mass number. The resulting cross sections must therefore be interpreted as cross sections per nucleon.
In addition, on the nCTEQ website \mbox{\url{https://ncteq.hepforge.org/ncteq26}} we also provide the bound proton PDFs, $x f_k^{p/A}(x,Q)$, which can be useful in certain applications, e.g.~to construct different isotopes or idealized isoscalar nuclei.

By default we publish LHAPDF grids for all the nuclei that were used in the fit and for a few additional nuclei of interest, e.g.\ oxygen, for which measurements were taken at the LHC. Nevertheless, the analytic $A$-dependent parametrization used in our analysis allows us to provide predictions also for other nuclei. The corresponding nCTEQ26 PDFs will be made available upon request.

When calculating PDF uncertainties of an observable $X$ with nCTEQ26 nPDFs, we recommend using the symmetric formula of \cref{eq:ObservableUncertainty}, i.e.\
\begin{align}
    \Delta X = \frac12 \sqrt{\sum_k^{N_{\mathrm{par}}} (X_k^+-X_k^-)^2}\,.
\end{align}
Note that the nCTEQ26 LHAPDF files use {\tt ErrorType: hessian}. As a result, when computing uncertainties using the LHAPDF library, both symmetric and asymmetric errors are computed. The user should then select the preferred prescription.
Uncertainties for ratios of observables computed for different nuclei should be calculated in a correlated manner, i.e.~the same error PDFs should be used for both numerator and denominator. This, however, does not apply to ratios compared to the proton or deuteron, as those uncertainties are determined independently in the proton baseline fit from the CJ collaboration. The proton PDFs should always be taken from the nCTEQ26 set ({\tt nCTEQ26\_1\_1}), as they align with the CJ PDFs only at the initial scale.

\section{Definition of the tolerance}
\label{app:tolerance}
\addtocontents{toc}{\tocSpace}

The uncertainty of the fitted PDFs depends critically on the definition of the tolerance, i.e.~the maximally allowed increase $\Delta\chi^2$ induced by the variation of the fit parameters. Ideally one would expect $\Delta\chi^2=1$ and 2.71 for one degree of freedom and a 68\% (1$\sigma$) and 90\% ($1.645\sigma$) confidence interval, respectively. However, in realistic global PDF analyses with potentially inconsistent datasets, underestimated experimental (e.g.\ systematic, possibly non-Gaussian) and theoretical (e.g.\ extrapolation and possibly non-Gaussian parameter) uncertainties, or failure of the quadratic approximation in the Hessian method for some parameters, this can lead to unrealistically small uncertainties. Furthermore, the dimensionality of the analysis, which in general does not coincide with the number of open parameters, may not be straightforward to estimate. While a number of prescriptions for a determination of the tolerance has been proposed in the literature \cite{Stump:2001gu,Martin:2009iq,Eskola:2009uj,Kovarik:2015cma,Eskola:2021nhw}, there is currently no consensus. 
We note that in our recent studies~\cite{Risse:2025qlo,Derakhshanian:2026zkx} we investigated whether MCMC methods could be used to measure the tolerance.

In the current analysis we define the tolerance following Wilks' theorem \cite{Wilks:1938dza,Cowan:1998ji} by assuming that each parameter follows a Gaussian distribution with the best-fit value as its mean. We further identify the number of degrees of freedom with the number of open parameters ${N_\text{par}=36}$ and require each parameter to lie within the confidence region constructed by the tolerance with a probability of 90\%. The probability of observing a ${\chi^2 < \chi^2_0 + \Delta\chi^2}$ is given by
\begin{align}
    P(\Delta\chi^2,N_\text{par}) = \int_0^{\Delta\chi^2} p(t, N_\text{par})~dt + \mathcal{O}(\frac{1}{\sqrt{N_\text{data}}}),
    \label{eq:prob}
\end{align}
where ${p(t,N_\text{par})}$ is the $\chi^2$-distribution
\begin{align}
    p(t,N_\text{par})=\frac{e^{-\frac{t}{2}}\left(\frac{t}{2}\right)^{\frac{N_\text{par}}{2}-1}}{2~\Gamma\left(\frac{N_\text{par}}{2}\right)}.
\end{align}
Solving the equation $P(\Delta\chi^2,36)=0.9$, we find a tolerance $\Delta\chi^2$ (at 90\% CL) of $47.2$, which we round up to $T=50$.

\section{Approximate kinematical variables}
\label{app:approximate-kinematical-variables}
\addtocontents{toc}{\tocSpace}

In order to provide an approximate kinematic reach of the fitted data as presented in \cref{fig:x-Q2-plane}, we employ the usual approach \cite{Wang:2018heo} of using LO kinematics to determine the momentum fraction $x$ and the energy scale $Q$ for each data point. In \cref{tab:approximate-x-Q2}, we show the kinematic relations for each of the considered processes. For NC DIS and CC DIS, we use the experimentally measured Bjorken $x$ ($x_B$) and momentum transfer, whereas we use $Q^2 = m_p \, E_\nu \, x \, y$ in DIS $\mu\mu$. For FT DY, we use the $x$ reconstructed and reported by the experiment, and for $Q$ we use the center of bin of the lepton pair mass from the measurement. For $W/Z$, SIH, OHF and ONIA production, we use ${x = Q e^{-\tilde{y}}/\sqrt{s}}$ with $\tilde{y}$ being the rapidity of the tagged final-state, $C$, for SIH, OHF as well as ONIA, and $\eta_\ell$ or $y_{\ell\bar{\ell}}$ for $W$ or $Z$. For the scale in $W/Z$, we use the mass of either the $W$ or $Z$ boson, depending on the measurement. For SIH, we use $p_\mathrm{T}/2$ of the respective final state pion or $\eta$-meson, and for OHF and ONIA we use the transverse mass of the final state $C= \{ D^0, J/\psi, \psi(2S), \Upsilon(1S)\}$ with charm quark mass $m_c = 1.3$ GeV.
\renewcommand\arraystretch{1.3}
\begin{table}[]
    \centering
    \begin{tabular}{|c|c|c|c|}
        \hline
        Process & $x$ & $Q^2$ & $\tilde{y}$ \\
        \hline
        \hline
        NC DIS, CC DIS, DY & $x_\mathrm{B}$ (data) & data & -- \\
        \hline
        DIS $\mu\mu$ & $x_\mathrm{B}$ (data) & $m_p E_\nu x y$ & -- \\
        \hline
        DY W/Z & \multirow{3}*{$Q e^{-\tilde{y}} / \sqrt{s}$} & $m_{W/Z}^2$ & $\eta_{\ell}/y_{\ell\bar{\ell}}$ \\
        \cline{1-1}
        \cline{3-4}
        SIH & & $\max(p_\mathrm{T}^C / 2, m_c)^2$ & $y^C$ \\
        \cline{1-1}
        \cline{3-4}
        OHF, ONIA & & $(p_\mathrm{T}^C)^2 + m_c^2$ & $y^C$ \\
        \hline
    \end{tabular}
    \caption{Approximate relations for $x$ and $Q^2$ used in \cref{fig:x-Q2-plane}.}
    \label{tab:approximate-x-Q2}
\end{table}
\renewcommand\arraystretch{1.0}

\section{Details of the dimuon analysis} 
\label{app:dimu}

Dimuon events from neutrino-induced charged-current charm production can provide strong constraints on the strange sea.
Their signature in a detector is a pair of oppositely charged muons and a hadronic shower.
The second muon is produced in the semileptonic decay of the charmed particle in the reaction
\begin{eqnarray}
    \nu_{\mu} + s &\to& \mu^- + c + X  \to  \mu^- \mu^+ + X \quad . 
    \nonumber
\end{eqnarray}    
With a sign-selected beam, NuTeV measured $s$ and $\bar{s}$ independently. 

The fully differential cross section for the dimuon process is 
\begin{eqnarray}    
 \frac{d\sigma(\nu_{\mu}N\to\mu^- \mu^+ X)}{dx dQ^2 d\eta dz}    
&=&
 \frac{d\sigma(\nu_{\mu}N\to\mu^- c X)}{dx dQ^2 d\eta}  
 \times 
 \nonumber \\
& \times & \ D(z) \ B_c(c\to \mu^+ X) \ ,\quad
\label{eq:dimuon}
\end{eqnarray}
where  $D(z)$ is the fragmentation function for the charm quark, 
and $B_c$ is the semi-muonic branching fraction for charmed hadrons.
Additionally,   $\{x,Q^2\}$ determine the position and energy of the primary muon, 
while  $\{z,\eta\}$ depend on the position and energy of the secondary muon,
where  $\eta$ is the rapidity of the charm quark and 
$z$ is the scaled energy of the charm after fragmentation.

The number of observed events depends on the detector geometry
and the cuts applied to reconstruct the events and filter the backgrounds. 
For example, NuTeV required that the energy of each of the two muons be greater than 5~GeV and that the muons fall within a fiducial region of the detector. 

One can integrate  Eq.~(\ref{eq:dimuon}) over $z$ and $\eta$ in the fiducial region to obtain the observed cross section as a function of $\{x,Q^2\}$,
\begin{eqnarray}
\left[ \frac{d\sigma(\nu_{\mu}N{\to}\mu^- \mu^+)}{dx dQ^2}  
\right]_{obs}  
&=& 
 \int dz \int d\eta \ \frac{d\sigma(\nu_{\mu}N{\to}\mu^- \mu^+)}{dx dQ^2 dz d\eta}      \nonumber \\
&=&
\epsilon(x,Q^2) \
\frac{d\sigma(\nu_{\mu}N\to \mu^- c)}{dx dQ^2}    \quad .
   \nonumber
\end{eqnarray}
In the second line, we have related the observed cross section to the computed differential charm production 
cross section with an acceptance correction $\epsilon(x,Q^2)$. The latter is usually computed via Monte Carlo simulations
to include the proper detector geometry and kinematic cuts. 
While the acceptance corrections are in principle dependent on the input PDFs, the charm fragmentation function, the semi-leptonic charm branching ratios, and other factors (c.f.,~Refs.~\cite{Rabinowitz:1993xx,CCFR:1994ikl,Kretzer:2001tc,Olness:2003wz,Mason:2004yf,Mason:2006qa,NuTeV:2007uwm}), these uncertainties are generally smaller than the overall experimental uncertainties. 

However, the acceptance correction does depend on the order of the partonic process. 
This is because at LO, the partonic process ($sW^+ \to c$)  is a $2\to1$  interaction,  
while at NLO this is a $2\to2$ partonic process  (e.g.,~$sg\to c \bar{s}$).
Thus, in the LO process, the rapidity of the charm is constrained by the $sW^+$ combination, 
whereas at  NLO  the rapidity of the charmed quark maps onto the \mbox{$W$-parton} CMS scattering angle. 
For this reason, the LO and NLO  $\epsilon(x,Q^2)$  values must be computed separately. 

In practice, the values of $\epsilon(x,Q^2)$ for both CCFR and NuTeV range from 20\% to 90\% 
with an average of about 65\% for the LO process and 70\% for the NLO process. 
Thus, for all the charm quarks which decay semi-leptonically to a muon ($B_c\sim 10\%$), 
approximately two thirds satisfy the detector cuts and contribute to the dimuon signal. \\

\input{08acks}

\bibliographystyle{utphys}
\interlinepenalty=10000
\bibliography{extra,refs}
\end{document}

%% file: 0_Front.tex
\preprint{MS-TP-26-23}
\preprint{IFJPAN-IV-2026-16}
\preprint{SMU-HEP-26-06}
\preprint{FERMILAB-PUB-26-0746-PPD}
\preprint{JLAB-NONE-26-4987}

\newcommand{\orcid}[1]{\,\href{https://orcid.org/#1}{\includegraphics[width=9pt]{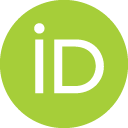}}\,}
\newcommand{\orcidAD}{0009-0009-4902-6408} %
\newcommand{\orcidND}{0000-0003-0962-631X} %
\newcommand{\orcidPD}{0000-0001-7960-7953} %
\newcommand{\orcidOH}{0000-0002-4890-6544} %
\newcommand{\orcidTH}{0000-0002-2729-0015} %
\newcommand{\orcidTJ}{0000-0002-1334-7607} %
\newcommand{\orcidTK}{0000-0002-7516-8292} %
\newcommand{\orcidMK}{0000-0002-4665-3088} %
\newcommand{\orcidKK}{0000-0003-1412-447X} %
\newcommand{\orcidAK}{0000-0002-4090-0084} %
\newcommand{\orcidCL}{0000-0001-7509-5655} %
\newcommand{\orcidJM}{0000-0001-9343-9351} %
\newcommand{\orcidFM}{0000-0002-3888-1697} %
\newcommand{\orcidFO}{0000-0001-6799-2436} %
\newcommand{\orcidEP}{0000-0001-9058-2590} %
\newcommand{\orcidPR}{0000-0002-8570-5506} %
\newcommand{\orcidRR}{0000-0002-3316-2175} %
\newcommand{\orcidIS}{0000-0003-0373-474X} %
\newcommand{\orcidJW}{0009-0009-1136-0849} %
\newcommand{\orcidJY}{0000-0001-8366-0968} %
\newcommand{\orcidKG}{0009-0008-7138-6875} %
\def\mit{\affiliation{Massachusetts Institute of Technology, Cambridge, Massachusetts 02139, USA}}
\def\smu{\affiliation{Department of Physics, Southern Methodist University, Dallas, TX 75275-0175, U.S.A.}}
\def\jlab{\affiliation{Jefferson Lab, Newport News, VA 23606, U.S.A.}}
\def\jlabtheo{\affiliation{Theory Center, Jefferson Lab, Newport News, VA 23606, U.S.A.}}
\def\kit{\affiliation{Institute for Theoretical Physics, KIT, 76131  Karlsruhe, Germany}}
\def\krakow{\affiliation{Institute of Nuclear Physics, Polish Academy of Sciences, 31342 Krakow, Poland}}
\def\muenster{\affiliation{Institute for Theoretical Physics, University of M{\"u}nster, Wilhelm-Klemm-Stra{\ss}e 9, 48149 M{\"u}nster, Germany}}
\def\lpsc{\affiliation{Laboratoire de Physique Subatomique et de Cosmologie, Université Grenoble-Alpes, 
    \\CNRS/IN2P3, 53 avenue des Martyrs, 38026 Grenoble, France}}
\def\hampton{\affiliation{Hampton University, Hampton, VA 23668, USA}}
\def\fnal{\affiliation{Fermi National Accelerator Laboratory, Batavia, Illinois 60510, USA}}
\def\iit{\affiliation{Department of Physics, Illinois Institute of Technology, Chicago, Illinois 60616, USA}}
\def\fsu{\affiliation{Department of Physics, Florida State University, Tallahassee, Florida 32306-4350, USA}}
\def\wales{\affiliation{School of Physics, The University of New South Wales, Sydney NSW 2052, Australia}}
\def\jyv{\affiliation{University of Jyväskylä, Department of Physics, P.O.\ Box 35, 40014 University of Jyväskylä, Finland}}
\def\helsinki{\affiliation{Helsinki Institute of Physics, P.O.\ Box 64, 00014 University of Helsinki, Finland}}
\def\anl{\affiliation{High Energy Physics Division, Argonne National Laboratory, Argonne, Illinois 60439, USA}}
\def\fzj{\affiliation{Institut für Energie- und Klimaforschung, Forschungszentrum Jülich GmbH, 52425 Jülich, Germany}}
\def\telaviv{\affiliation{School of Physics and Astronomy, Tel Aviv University, Tel Aviv 6997845, Israel}}

\title{nCTEQ26 nuclear parton distribution functions}
\author{K.~Greve\orcid{\orcidKG}}
\muenster{}

\author{T.~Je\v{z}o\orcid{\orcidTJ}}
\email{tomas.jezo@uni-muenster.de}
\muenster{}

\author{A.~Kusina\orcid{\orcidAK}} 
\email{aleksander.kusina@ifj.edu.pl}
\krakow{}

\author{J.~Wissmann\orcid{\orcidJW}}
\muenster{}

\author{M.~Klasen\orcid{\orcidMK}}
\muenster{}

\author{K.~Kova\v{r}\'{i}k\orcid{\orcidKK}} 
\muenster{}

\author{C.~L\'eger\orcid{\orcidCL}} 
\lpsc{}

\author{J.G.~Morf\'{i}n\orcid{\orcidJM}} 
\fnal{}

\author{F.I.~Olness\orcid{\orcidFO}} 
\email{olness@smu.edu}
\smu{}

\author{P.~Risse\orcid{\orcidPR}}
\smu{}
\jlabtheo{}

\author{I.~Schienbein\orcid{\orcidIS}} 
\lpsc{}

\author{J.Y.~Yu\orcid{\orcidJY}}
\lpsc{}

%% file: tables/data_new/DIS_NC_AoD.tex
\renewcommand{\arraystretch}{1.25}

\begin{table}[!tbph]
\centering
\begin{tabular}{|>{\rowmacro}p{1.4cm}>{\rowmacro}p{2.5cm}>{\rowmacro}p{0.7cm}>{\rowmacro}C{0.8cm}>{\rowmacro}R{0.5cm}@{\rowmacro /}>{\rowmacro}L{0.5cm}>{\rowmacro}R{1.1cm}<{\clearrowstyle}|}
\hline
\multicolumn{7}{|c|}{NC DIS data $F_{2}^{A}/F_{2}^{\mathrm{D}}$ and $\sigma_{\text{DIS}}^{A}/\sigma_{\text{DIS}}^{\mathrm{D}}$}\\
\hline
\hline
Nucleus & Experiment & ID & Ref. & \multicolumn{2}{c}{$N_{\mathrm{data}}$} & \!\!\!\!$\chi^2/N_\text{data}$ \\
\hline
${}^{3}\mathrm{He}/\mathrm{D}$ & HERMES & 5156 & \cite{HERMES:2002llm} & 182 & 92 & 0.87 \\
\cline{1-7}
\multirow{2}{*}{${}^{4}\mathrm{He}/\mathrm{D}$} & NMC-95,re & 5124 & \cite{NewMuon:1995cua} & 18 & 16 & 1.05 \\
 & SLAC-E139 & 5141 & \cite{Gomez:1993ri} & 18 & 17 & 0.39 \\
\cline{1-7}
${}^{6}\mathrm{Li}/\mathrm{D}$ & NMC-95 & 5115 & \cite{NewMuon:1995tgs} & 24 & 15 & 1.11 \\
\cline{1-7}
${}^{9}\mathrm{Be}/\mathrm{D}$ & SLAC-E139 & 5138 & \cite{Gomez:1993ri} & 17 & 16 & 0.65 \\
\cline{1-7}
\multirow{6}{*}{${}^{12}\mathrm{C}/\mathrm{D}$} & EMC-88 & 5107 & \cite{EuropeanMuon:1988lbf} & 9 & 9 & 0.94 \\
 & EMC-90 & 5110 & \cite{EuropeanMuon:1989mef} & 9 & 2 & 0.05 \\
 & NMC-95 & 5113 & \cite{NewMuon:1995tgs} & 24 & 15 & 0.80 \\
 & NMC-95,re & 5114 & \cite{NewMuon:1995tgs} & 18 & 16 & 1.15 \\
 & FNAL-E665-95 & 5125 & \cite{E665:1995xur} & 11 & 4 & 0.43 \\
 & SLAC-E139 & 5139 & \cite{Gomez:1993ri} & 7 & 7 & 1.25 \\
\cline{1-7}
\multirow{2}{*}{${}^{14}\mathrm{N}/\mathrm{D}$} & BCDMS-85 & 5103 & \cite{BCDMS:1985dor} & 9 & 9 & 0.64 \\
 & HERMES & 5157 & \cite{HERMES:2002llm} & 175 & 92 & 0.53 \\
\cline{1-7}
\multirow{2}{*}{${}^{27}\mathrm{Al}/\mathrm{D}$} & SLAC-E049 & 5134 & \cite{Bodek:1983ec} & 18 & 18 & 1.14 \\
 & SLAC-E139 & 5136 & \cite{Gomez:1993ri} & 17 & 16 & 0.96 \\
\cline{1-7}
\multirow{4}{*}{${}^{40}\mathrm{Ca}/\mathrm{D}$} & EMC-90 & 5109 & \cite{EuropeanMuon:1989mef} & 9 & 2 & 0.53 \\
 & NMC-95,re & 5121 & \cite{NewMuon:1995cua} & 18 & 15 & 1.01 \\
 & FNAL-E665-95 & 5126 & \cite{E665:1995xur} & 11 & 4 & 0.21 \\
 & SLAC-E139 & 5140 & \cite{Gomez:1993ri} & 7 & 7 & 0.41 \\
\cline{1-7}
\multirow{5}{*}{${}^{56}\mathrm{Fe}/\mathrm{D}$} & BCDMS-87 & 5101 & \cite{BCDMS:1987upi} & 10 & 10 & 1.07 \\
 & BCDMS-85 & 5102 & \cite{BCDMS:1985dor} & 6 & 6 & 0.54 \\
 & SLAC-E049 & 5131 & \cite{Bodek:1983qn} & 14 & 14 & 0.99 \\
 & SLAC-E139 & 5132 & \cite{Gomez:1993ri} & 23 & 22 & 1.24 \\
 & SLAC-E140 & 5133 & \cite{Dasu:1993vk} & 10 & 6 & 0.70 \\
\cline{1-7}
\multirow{3}{*}{${}^{64}\mathrm{Cu}/\mathrm{D}$} & EMC-93 & 5104 & \cite{EuropeanMuon:1992pyr} & 10 & 10 & 0.57 \\
 & EMC-93 (chariot) & 5105 & \cite{EuropeanMuon:1992pyr} & 9 & 9 & 0.61 \\
 & EMC-88 & 5106 & \cite{EuropeanMuon:1988lbf} & 9 & 9 & 0.41 \\
\cline{1-7}
${}^{84}\mathrm{Kr}/\mathrm{D}$ & HERMES & 5158 & \cite{HERMES:2002llm} & 167 & 84 & 0.74 \\
\cline{1-7}
${}^{108}\mathrm{Ag}/\mathrm{D}$ & SLAC-E139 & 5135 & \cite{Gomez:1993ri} & 7 & 7 & 0.58 \\
\cline{1-7}
${}^{119}\mathrm{Sn}/\mathrm{D}$ & EMC-88 & 5108 & \cite{EuropeanMuon:1988lbf} & 8 & 8 & 2.18 \\
\cline{1-7}
${}^{131}\mathrm{Xe}/\mathrm{D}$ & FNAL-E665-92 & 5127 & \cite{E665:1992ivw} & 10 & 4 & 0.16 \\
\cline{1-7}
${}^{197}\mathrm{Au}/\mathrm{D}$ & SLAC-E139 & 5137 & \cite{Gomez:1993ri} & 18 & 17 & 0.66 \\
\cline{1-7}
${}^{208}\mathrm{Pb}/\mathrm{D}$ & FNAL-E665-95 & 5129 & \cite{E665:1995xur} & 11 & 4 & 0.37 \\
\hline
\hline
\setrowstyle{\bfseries}
Total &  &  &  & \!\!\!\!\!913 & 582 & 0.79\\
\hline
\end{tabular}
\caption{NC DIS $F_{2}^{A}/F_{2}^{D}$ and $\sigma_{\text{DIS}}^{A}/\sigma_{\text{DIS}}^{D}$ datasets used in this fit. The $N_{\text{data}}$ column indicates the number of data points before/after kinematic cuts.}
\label{tab:dis1data}
\end{table}
\renewcommand{\arraystretch}{1.0}

%% file: tables/data_new/DIS_NC_FAoFAprime.tex
\renewcommand{\arraystretch}{1.25}
\begin{table}[!htbp]
\centering
\begin{tabular}{|>{\rowmacro}p{1.8cm}>{\rowmacro}p{2.0cm}>{\rowmacro}p{0.8cm}>{\rowmacro}C{0.7cm}>{\rowmacro}R{0.6cm}@{\rowmacro /}>{\rowmacro}L{0.6cm}>{\rowmacro}R{1.0cm}<{\clearrowstyle}|}
\hline
\multicolumn{7}{|c|}{NC DIS data $F_{2}^{A}/F_{2}^{A'}$}\\
\hline
\hline
Nucleus & Experiment & ID & Ref. & \multicolumn{2}{c}{$N_{\mathrm{data}}$} & \!\!\!\!$\chi^2/N_\text{data}$ \\
\hline
${}^{12}\mathrm{C}/{}^{6}\mathrm{Li}$ & NMC-95,re & 5123 & \cite{NewMuon:1995cua} & 25 & 20 & 0.88 \\
\cline{1-7}
${}^{40}\mathrm{Ca}/{}^{6}\mathrm{Li}$ & NMC-95,re & 5122 & \cite{NewMuon:1995cua} & 25 & 20 & 0.55 \\
\cline{1-7}
${}^{9}\mathrm{Be}/{}^{12}\mathrm{C}$ & NMC-96 & 5112 & \cite{NewMuon:1996yuf} & 15 & 15 & 0.39 \\
\cline{1-7}
${}^{27}\mathrm{Al}/{}^{12}\mathrm{C}$ & NMC-96 & 5111 & \cite{NewMuon:1996yuf} & 15 & 15 & 0.36 \\
\cline{1-7}
\multirow{2}{*}{${}^{40}\mathrm{Ca}/{}^{12}\mathrm{C}$} & NMC-95,re & 5120 & \cite{NewMuon:1995cua} & 25 & 20 & 0.89 \\
 & NMC-96 & 5119 & \cite{NewMuon:1996yuf} & 15 & 15 & 0.46 \\
\cline{1-7}
${}^{56}\mathrm{Fe}/{}^{12}\mathrm{C}$ & NMC-96 & 5143 & \cite{NewMuon:1996yuf} & 15 & 15 & 0.63 \\
\cline{1-7}
${}^{119}\mathrm{Sn}/{}^{12}\mathrm{C}$ & NMC-96 & 5159 & \cite{NewMuon:1996gam} & 146 & 144 & 0.55 \\
\cline{1-7}
${}^{208}\mathrm{Pb}/{}^{12}\mathrm{C}$ & NMC-96 & 5116 & \cite{NewMuon:1996yuf} & 15 & 15 & 0.22 \\
\hline
\hline
\setrowstyle{\bfseries}
Total &  &  &  & \!296 & 279 & 0.56\\
\hline
\end{tabular}
\caption{NC DIS $F_{2}^{A}/F_{2}^{A'}$ datasets used in this fit. The $N_{\text{data}}$ column indicates the number of data points before/after kinematic cuts.}
\label{tab:dis2data}
\end{table}
\renewcommand{\arraystretch}{1.0}

%% file: tables/data_new/DIS_NC_sigmaAosigmaD_JLAB.tex
\renewcommand{\arraystretch}{1.25}
\begin{table}[!htbp]
\centering
\begin{tabular}{|>{\rowmacro}p{1.7cm}>{\rowmacro}p{2.0cm}>{\rowmacro}p{0.9cm}>{\rowmacro}C{0.7cm}>{\rowmacro}R{0.6cm}@{\rowmacro /}>{\rowmacro}L{0.6cm}>{\rowmacro}R{1.0cm}<{\clearrowstyle}|}
\hline
\multicolumn{7}{|c|}{NC DIS data $\sigma_{\text{DIS}}^{A}/\sigma_{\text{DIS}}^{\mathrm{D}}$  from JLab}\\
\hline
\hline
Nucleus & Experiment & ID & Ref. & \multicolumn{2}{c}{$N_{\mathrm{data}}$} & \!\!\!\!$\chi^2/N_\text{data}$ \\
\hline
\multirow{2}{*}{${}^{3}\mathrm{He}/\mathrm{D}$} & \multirow{2}{*}{Hall C} & 9982 & \multirow{2}{*}{\cite{Seely:2009gt}} & 25 & 17 & 1.25 \\
 &  & 9983 &  & 26 & 16 & 0.70 \\
\cline{1-7}
\multirow{2}{*}{${}^{4}\mathrm{He}/\mathrm{D}$} & \multirow{2}{*}{Hall C} & 9980 & \multirow{2}{*}{\cite{Seely:2009gt}} & 25 & 17 & 0.89 \\
 &  & 9981 &  & 26 & 16 & 0.48 \\
\cline{1-7}
\multirow{3}{*}{${}^{9}\mathrm{Be}/\mathrm{D}$} & \multirow{3}{*}{Hall C} & \textbf{9092} & \cite{HallC:2022utd} & 27 & 20 & 0.74 \\
 &  & 9986 & \multirow{2}{*}{\cite{Seely:2009gt}} & 25 & 17 & 1.01 \\
 &  & 9987 &  & 26 & 16 & 0.25 \\
\cline{1-7}
${}^{10}\mathrm{B}/\mathrm{D}$ & Hall C & \textbf{9093} & \cite{HallC:2022utd} & 27 & 20 & 0.61 \\
\cline{1-7}
${}^{11}\mathrm{B}/\mathrm{D}$ & Hall C & \textbf{9094} & \cite{HallC:2022utd} & 27 & 20 & 0.32 \\
\cline{1-7}
\multirow{12}{*}{${}^{12}\mathrm{C}/\mathrm{D}$} & Hall C & \textbf{9091} & \cite{HallC:2022utd} & 27 & 20 & 0.41 \\
\cline{2-7}
 & CLAS & 9979 & \cite{CLAS:2019vsb} & 25 & 24 & 0.79 \\
\cline{2-7}
 & \multirow{10}{*}{Hall C} & 9990 & \multirow{10}{*}{\cite{Seely:2009gt}} & 25 & 17 & 0.70 \\
 &  & 9991 &  & 17 & 7 & 1.34 \\
 &  & 9992 &  & 26 & 16 & 0.15 \\
 &  & 9993 &  & 18 & 6 & 0.69 \\
 &  & 9994 &  & 17 & 7 & 1.94 \\
 &  & 9995 &  & 15 & 2 & 0.13 \\
 &  & 9996 &  & 19 & 7 & 1.08 \\
 &  & 9997 &  & 16 & 2 & 0.24 \\
 &  & 9998 &  & 20 & 8 & 0.49 \\
 &  & 9999 &  & 17 & 3 & 0.47 \\
\cline{1-7}
${}^{27}\mathrm{Al}/\mathrm{D}$ & CLAS & 9978 & \cite{CLAS:2019vsb} & 25 & 24 & 0.81 \\
\cline{1-7}
${}^{56}\mathrm{Fe}/\mathrm{D}$ & CLAS & 9977 & \cite{CLAS:2019vsb} & 25 & 24 & 0.64 \\
\cline{1-7}
\multirow{2}{*}{${}^{64}\mathrm{Cu}/\mathrm{D}$} & \multirow{2}{*}{Hall C} & 9984 & \multirow{2}{*}{\cite{Seely:2009gt}} & 25 & 17 & 0.53 \\
 &  & 9985 &  & 26 & 16 & 0.58 \\
\cline{1-7}
\multirow{2}{*}{${}^{197}\mathrm{Au}/\mathrm{D}$} & \multirow{2}{*}{Hall C} & 9988 & \multirow{2}{*}{\cite{Seely:2009gt}} & 24 & 17 & 0.21 \\
 &  & 9989 &  & 26 & 16 & 0.33 \\
\cline{1-7}
${}^{208}\mathrm{Pb}/\mathrm{D}$ & CLAS & 9976 & \cite{CLAS:2019vsb} & 25 & 24 & 0.64 \\
\hline
\hline
\setrowstyle{\bfseries}
Total &  &  &  & \!652 & 416 & 0.65\\
\hline
\end{tabular}
\caption{NC DIS $\sigma_{\text{DIS}}^{A}/\sigma_{\text{DIS}}^{\mathrm{D}}$ datasets from JLab used in this fit. The CLAS data from Ref.~\cite{CLAS:2019vsb} and Hall~C data from Ref.~\cite{Seely:2009gt} were already considered in some of our earlier dedicated studies \cite{Segarra:2020gtj,nCTEQ:2023cpo}. Datasets not included in previous analyses are indicated by bold IDs. The $N_{\text{data}}$ column indicates the number of data points before/after kinematic cuts.}
\label{tab:jlabdata}
\end{table}
\renewcommand{\arraystretch}{1.0}

%% file: tables/data_new/neutrino_DIS.tex
\renewcommand{\arraystretch}{1.1}
\begin{table}[!tbp]
\centering
\begin{tabular}{|>{\rowmacro}p{0.9cm}>{\rowmacro}p{1.6cm}>{\rowmacro}p{1.2cm}>{\rowmacro}p{0.8cm}>{\rowmacro}C{0.6cm}>{\rowmacro}R{0.7cm}@{\rowmacro /}>{\rowmacro}L{0.6cm}>{\rowmacro}R{1.0cm}<{\clearrowstyle}|}
\hline
\multicolumn{8}{|c|}{CC DIS and DIS $\mu \mu$ data $d^2 \sigma / dx dy$}\\
\hline
\hline
Nucl. & Exp. & Observ. & ID & Ref. & \multicolumn{2}{c}{$N_{\mathrm{data}}$} & \!\!\!\!$\chi^2/N_\text{data}$ \\
\hline
\multirow{4}{*}{$^{56}\mathrm{Fe}$} & \multirow{2}{*}{CCFR} & $\nu \mu \mu$ & 791 & \cite{NuTeV:2001dfo} & 44 & 44 & 1.17 \\
 &  & $\bar{\nu} \mu \mu$ & 792 & \cite{NuTeV:2001dfo} & 44 & 44 & 0.66 \\
\cline{2-8}
 & \multirow{2}{*}{NuTeV} & $\nu \mu \mu$ & 793 & \cite{NuTeV:2001dfo} & 44 & 44 & 0.61 \\
 &  & $\bar{\nu} \mu \mu$ & 794 & \cite{NuTeV:2001dfo} & 42 & 42 & 0.54 \\
\cline{1-8}
\multirow{2}{*}{$^{208}\mathrm{Pb}$} & \multirow{2}{*}{CHORUS} & $\nu$ & 5946 & \cite{CHORUS:2005cpn} & 607 & 534 & 1.36 \\
 &  & $\bar{\nu}$ & 5947 & \cite{CHORUS:2005cpn} & 607 & 534 & 1.15 \\
\hline
\hline
\setrowstyle{\bfseries}
Total &  &  &  &  & \!1388 & 1242 & 1.18\\
\hline
\end{tabular}
\caption{Neutrino DIS and dimuon production datasets used in this analysis. The $N_{\text{data}}$ column indicates the number of data points before/after kinematic cuts.}
\label{tab:nudata}
\end{table}
\renewcommand{\arraystretch}{1.0}

%% file: tables/data_new/Drell_Yan.tex
\renewcommand{\arraystretch}{1.25}
\begin{table}[!tbp]
\centering
\begin{tabular}{|>{\rowmacro}p{1.6cm}>{\rowmacro}p{2.4cm}>{\rowmacro}p{0.8cm}>{\rowmacro}C{0.6cm}>{\rowmacro}R{0.5cm}@{\rowmacro /}>{\rowmacro}L{0.5cm}>{\rowmacro}R{1.1cm}<{\clearrowstyle}|}
\hline
\multicolumn{7}{|c|}{Fixed-target (FT) Drell-Yan (DY) data $\sigma_{\text{DY}}^{pA}/\sigma_{\text{DY}}^{pA'}$}\\
\hline
\hline
Nucleus & Experiment & ID & Ref. & \multicolumn{2}{c}{$N_{\mathrm{data}}$} & \!\!$\chi^2/N_\text{data}$ \\
\hline
$^{12}\mathrm{C}/\mathrm{D}$ & FNAL-E772-90 & 5203 & \cite{Alde:1990im} & 9 & 9 & 1.38 \\
\cline{1-7}
$^{40}\mathrm{Ca}/\mathrm{D}$ & FNAL-E772-90 & 5204 & \cite{Alde:1990im} & 9 & 9 & 0.25 \\
\cline{1-7}
$^{56}\mathrm{Fe}/\mathrm{D}$ & FNAL-E772-90 & 5205 & \cite{Alde:1990im} & 9 & 9 & 0.31 \\
\cline{1-7}
$^{184}\mathrm{W}/\mathrm{D}$ & FNAL-E772-90 & 5206 & \cite{Alde:1990im} & 9 & 9 & 1.03 \\
\cline{1-7}
$^{56}\mathrm{Fe}/^{9}\mathrm{Be}$ & FNAL-E866-99 & 5201 & \cite{NuSea:1999egr} & 28 & 28 & 0.72 \\
\cline{1-7}
$^{184}\mathrm{W}/^{9}\mathrm{Be}$ & FNAL-E866-99 & 5202 & \cite{NuSea:1999egr} & 28 & 28 & 0.90 \\
\hline
\hline
\setrowstyle{\bfseries}
Total &  &  &  & \!92 & 92 & 0.78\\
\hline
\end{tabular}
\caption{Fixed-target Drell-Yan $\sigma_{\text{DY}}^{pA}/\sigma_{\text{DY}}^{pA'}$ datasets used in this fit. The $N_{\text{data}}$ column indicates the number of data points before/after kinematic cuts.}
\label{tab:DYdata}
\end{table}
\renewcommand{\arraystretch}{1.0}

%% file: tables/data_new/WZ.tex
\renewcommand{\arraystretch}{1.1}
\begin{table}[!tbp]
\centering
\begin{tabular}{|>{\rowmacro}p{1.2cm}>{\rowmacro}p{1.7cm}>{\rowmacro}C{1.0cm}>{\rowmacro}p{0.8cm}>{\rowmacro}C{0.6cm}>{\rowmacro}R{0.55cm}@{\rowmacro /}>{\rowmacro}L{0.55cm}>{\rowmacro}R{1.1cm}<{\clearrowstyle}|}
\hline
\multicolumn{8}{|c|}{LHC $W/Z$-boson and DY data $d\sigma/dy$ in pPb collisions}\\
\hline
\hline
Exp. & Observ. & \!\!\!\!\!\!\!\!$\sqrt{s}$~[TeV] & ID & Ref. & \multicolumn{2}{c}{$N_{\mathrm{data}}$} & \!\!\!\!$\chi^2/N_\text{data}$ \\
\hline
\multirow{4}{*}{ALICE} & $W^-$ & 5.02 & 6251 & \cite{ALICE:2016rzo} & 2 & 2 & 0.89 \\
 & $W^+$ & 5.02 & 6253 & \cite{ALICE:2016rzo} & 2 & 2 & 0.20 \\
 & $Z~\text{forw.}$ & 5.02 & \textbf{6255} & \cite{ALICE:2016rzo} & 1 & 1 & 0.50 \\
 & $Z~\text{back.}$ & 5.02 & \textbf{6256} & \cite{ALICE:2016rzo} & 1 & 1 & 0.00 \\
\cline{1-8}
\multirow{6}{*}{ALICE} & $Z~\text{forw.}$ & 8.16 & \textbf{6257} & \cite{ALICE:2020jff} & 1 & 1 & 0.18 \\
 & $Z~\text{back.}$ & 8.16 & \textbf{6258} & \cite{ALICE:2020jff} & 1 & 1 & 0.02 \\
 & $W^-~\text{forw.}$ & 8.16 & \textbf{6261} & \cite{ALICE:2022cxs} & 3 & 3 & 1.74 \\
 & $W^-~\text{back.}$ & 8.16 & \textbf{6262} & \cite{ALICE:2022cxs} & 3 & 3 & 4.76 \\
 & $W^+~\text{forw.}$ & 8.16 & \textbf{6263} & \cite{ALICE:2022cxs} & 3 & 3 & 2.62 \\
 & $W^+~\text{back.}$ & 8.16 & \textbf{6264} & \cite{ALICE:2022cxs} & 3 & 3 & 0.80 \\
\cline{1-8}
\multirow{3}{*}{ATLAS} & $W^-$ & 5.02 & 6211 & \cite{TheATLAScollaboration:2015lnm} & 10 & 10 & 0.46 \\
 & $W^+$ & 5.02 & 6213 & \cite{TheATLAScollaboration:2015lnm} & 10 & 10 & 0.20 \\
 & $Z$ & 5.02 & 6215 & \cite{ATLAS:2015mwq} & 14 & 14 & 1.48 \\
\cline{1-8}
\multirow{3}{*}{CMS} & $W^-$ & 5.02 & 6231 & \cite{CMS:2015ehw} & 10 & 10 & 1.09 \\
 & $W^+$ & 5.02 & 6233 & \cite{CMS:2015ehw} & 10 & 10 & 1.27 \\
 & $Z$ & 5.02 & 6235 & \cite{CMS:2015zlj} & 12 & 12 & 0.02 \\
\cline{1-8}
\multirow{4}{*}{CMS} & $W^-$ & 8.16 & 6232 & \cite{CMS:2019leu} & 24 & 24 & 0.88 \\
 & $W^+$ & 8.16 & 6234 & \cite{CMS:2019leu} & 24 & 24 & 0.74 \\
 & $Z~\text{low mass}$ & 8.16 & \textbf{6236} & \cite{CMS:2021ynu} & 12 & 12 & 1.46 \\
 & $Z~\text{high mass}$ & 8.16 & \textbf{6237} & \cite{CMS:2021ynu} & 24 & 24 & 1.86 \\
\cline{1-8}
LHCb & $Z$ & 5.02 & 6275 & \cite{LHCb:2014jgh} & 2 & 2 & 0.63 \\
\cline{1-8}
\multirow{2}{*}{LHCb} & $Z~\text{forw.}$ & 8.16 & \textbf{6276} & \cite{LHCb:2022kph} & 5 & 5 & 0.57 \\
 & $Z~\text{back.}$ & 8.16 & \textbf{6277} & \cite{LHCb:2022kph} & 3 & 3 & 2.79 \\
\hline
\hline
\setrowstyle{\bfseries}
Total &  &  &  &  & \!\!180 & 180 & 1.10\\
\hline
\end{tabular}
\caption{$W/Z$-boson and Drell-Yan $d\sigma/dy$ datasets from LHC experiments used in this fit.  Datasets not included in previous analyses are indicated by bold IDs. The $N_{\text{data}}$ column indicates the number of data points before/after kinematic cuts.}
\label{tab:WZdata}
\end{table}
\renewcommand{\arraystretch}{1.0}

%% file: tables/data_new/SIH.tex
\renewcommand{\arraystretch}{1.1}
\begin{table}[!tbp]
\centering
\begin{tabular}{|>{\rowmacro}p{0.7cm}>{\rowmacro}p{1.4cm}>{\rowmacro}p{0.8cm}>{\rowmacro}C{1.1cm}>{\rowmacro}p{0.8cm}>{\rowmacro}C{0.7cm}>{\rowmacro}R{0.4cm}@{\rowmacro /}>{\rowmacro}L{0.4cm}>{\rowmacro}R{1.1cm}<{\clearrowstyle}|}
\hline
\multicolumn{9}{|c|}{Single-inclusive hadron (SIH) production data $d\sigma/dp_\mathrm{T}$}\\
\hline
\hline
Nucl. & Exp. & \!\!\!\!\!Obs. & \!\!\!\!\!\!\!\!$\sqrt{s}$~[TeV] & ID & Ref. & \multicolumn{2}{c}{$N_{\mathrm{data}}$} & \!\!$\chi^2/N_\text{data}$ \\
\hline
\multirow{5}{*}{\;\;\rotatebox{90}{$R_{\vspace{0.07cm}\mathrm{D}\,\mathrm{Au}}$}} & \multirow{3}{*}{PHENIX} & $\pi^0$ & 0.2 & 4003 & \cite{PHENIX:2006mhb} & 21 & 17 & 0.37 \\
 &  & $\pi^\pm$ & 0.2 & 4103 & \cite{PHENIX:2006mhb} & 20 & 0 & --\;\; \\
 &  & $K^\pm$ & 0.2 & 4203 & \cite{PHENIX:2013kod} & 15 & 0 & --\;\; \\
\cline{2-9}
 & \multirow{2}{*}{STAR} & $\pi^0$ & 0.2 & 4002 & \cite{STAR:2009qzv} & 13 & 9 & 0.17 \\
 &  & $\pi^\pm$ & 0.2 & 4102 & \cite{STAR:2006xud} & 23 & 8 & 0.82 \\
\cline{1-9}
\multirow{4}{*}{\;\;\rotatebox{90}{$R_{\vspace{0.07cm}\mathrm{p}\,\mathrm{Pb}}$}} & \multirow{4}{*}{ALICE} & $\pi^0$ & 5.02 & 4001 & \cite{ALICE:2018vhm} & 31 & 15 & 0.46 \\
 &  & $\pi^0$ & 8.16 & 4004 & \cite{ALICE:2021est} & 30 & 19 & 0.77 \\
 &  & $\pi^\pm$ & 5.02 & 4101 & \cite{ALICE:2016dei} & 58 & 22 & 0.12 \\
 &  & $K^\pm$ & 5.02 & 4201 & \cite{ALICE:2016dei} & 58 & 22 & 0.28 \\
\hline
\hline
\setrowstyle{\bfseries}
Total &  &  &  &  &  & \!\!269 & 112 & 0.40\\
\hline
\end{tabular}
\caption{Single-inclusive light-meson production $d\sigma/dp_\mathrm{T}$ datasets from RHIC and LHC experiments used in this fit. The $N_{\text{data}}$ column indicates the number of data points before/after kinematic cuts.}
\label{tab:SIHdata}
\end{table}
\renewcommand{\arraystretch}{1.0}

%% file: tables/data_new/HQ_pPb.tex
\renewcommand{\arraystretch}{1.1}
\begin{table}[!tbp]
\centering
\begin{tabular}{|>{\rowmacro}p{1.1cm}>{\rowmacro}p{2.2cm}>{\rowmacro}C{0.8cm}>{\rowmacro}p{0.8cm}>{\rowmacro}C{0.6cm}>{\rowmacro}R{0.5cm}@{\rowmacro /}>{\rowmacro}L{0.5cm}>{\rowmacro}R{1.0cm}<{\clearrowstyle}|}
\hline
\multicolumn{8}{|c|}{OHF and quarkonium data $d^2\sigma/dp_\mathrm{T} dy$ in pPb collisions}\\
\hline
\hline
Exp. & Observ. & \!\!\!\!\!\!\!\!$\sqrt{s}$~[TeV] & ID & Ref. & \multicolumn{2}{c}{$N_{\mathrm{data}}$} & \!\!\!\!$\chi^2/N_\text{data}$ \\
\hline
\multirow{14}{*}{ALICE} & $\text{prompt}~D^0$ & 5.02 & 3101 & \cite{ALICE:2014xjz} & 10 & 6 & 0.90 \\
 & $\text{incl.}~J/\psi$ & 5.02 & 3103 & \cite{ALICE:2013snh} & 12 & 0 & --\;\; \\
 & $\text{incl.}~J/\psi$ & 5.02 & 3104 & \cite{ALICE:2015sru} & 25 & 10 & 0.99 \\
 & $\text{incl.}~\Upsilon(1S)$ & 5.02 & 3110 & \cite{ALICE:2014ict} & 4 & 0 & --\;\; \\
 & $\text{incl.}~J/\psi$ & 8.16 & 3112 & \cite{ALICE:2018mml} & 24 & 9 & 1.35 \\
 & $\text{incl.}~\Upsilon(1S)$ & 8.16 & 3114 & \cite{ALICE:2019qie} & 10 & 3 & 0.08 \\
 & $\text{prompt}~D^0$ & 5.02 & 3122 & \cite{ALICE:2019fhe} & 21 & 13 & 0.35 \\
 & $\text{prompt}~D^0$ & 5.02 & 3123 & \cite{ALICE:2016yta} & 11 & 6 & 0.60 \\
 & $\text{incl.}~\psi(2S)$ & 8.16 & 3126 & \cite{ALICE:2020vjy} & 10 & 3 & 0.67 \\
 & $\text{incl.}~\psi(2S)$ & 5.02 & 3127 & \cite{ALICE:2014cgk} & 8 & 2 & 3.23 \\
 & $\text{prompt}~J/\psi$ & 5.02 & \textbf{3128} & \cite{ALICE:2021lmn} & 5 & 4 & 0.62 \\
 & $\text{non-pr.}~J/\psi$ & 5.02 & \textbf{3129} & \cite{ALICE:2021lmn} & 5 & 4 & 0.60 \\
 & $\text{prompt}~J/\psi$ & 8.16 & \textbf{3130} & \cite{ALICE:2022zig} & 5 & 4 & 0.41 \\
 & $\text{non-pr.}~J/\psi$ & 8.16 & \textbf{3131} & \cite{ALICE:2022zig} & 5 & 4 & 0.93 \\
\cline{1-8}
\multirow{7}{*}{ATLAS} & $\text{incl.}~\Upsilon(1S)$ & 5.02 & 3109 & \cite{ATLAS:2017prf} & 8 & 6 & 0.39 \\
 & $\text{non-pr.}~J/\psi$ & 5.02 & 3116 & \cite{ATLAS:2017prf} & 8 & 8 & 0.03 \\
 & $\text{prompt}~J/\psi$ & 5.02 & 3117 & \cite{ATLAS:2017prf} & 8 & 8 & 0.29 \\
 & $\text{prompt}~J/\psi$ & 5.02 & 3118 & \cite{ATLAS:2015mpz} & 10 & 10 & 0.47 \\
 & $\text{non-pr.}~J/\psi$ & 5.02 & 3119 & \cite{ATLAS:2015mpz} & 10 & 10 & 0.42 \\
 & $\text{prompt}~\psi(2S)$ & 5.02 & 3124 & \cite{ATLAS:2017prf} & 8 & 8 & 0.71 \\
 & $\text{non-pr.}~\psi(2S)$ & 5.02 & 3125 & \cite{ATLAS:2017prf} & 8 & 8 & 1.47 \\
\cline{1-8}
\multirow{4}{*}{CMS} & $\text{prompt}~\psi(2S)$ & 5.02 & 3115 & \cite{CMS:2018gbb} & 17 & 17 & 1.13 \\
 & $\text{prompt}~J/\psi$ & 5.02 & 3120 & \cite{CMS:2017exb} & 53 & 51 & 1.09 \\
 & $\text{non-pr.}~J/\psi$ & 5.02 & 3121 & \cite{CMS:2017exb} & 53 & 51 & 0.68 \\
 & $\text{incl.}~\Upsilon(1S)$ & 5.02 & \textbf{3132} & \cite{CMS:2022wfi} & 6 & 4 & 0.93 \\
\cline{1-8}
\multirow{9}{*}{LHCb} & $\text{prompt}~D^0$ & 5.02 & 3102 & \cite{LHCb:2017yua} & 92 & 53 & 0.34 \\
 & $\text{prompt}~J/\psi$ & 8.16 & 3105 & \cite{LHCb:2017ygo} & 140 & 88 & 1.17 \\
 & $\text{non-pr.}~J/\psi$ & 8.16 & 3106 & \cite{LHCb:2017ygo} & 140 & 88 & 0.75 \\
 & $\text{non-pr.}~J/\psi$ & 5.02 & 3107 & \cite{LHCb:2013gmv} & 40 & 25 & 0.77 \\
 & $\text{prompt}~J/\psi$ & 5.02 & 3108 & \cite{LHCb:2013gmv} & 40 & 25 & 1.68 \\
 & $\text{incl.}~\Upsilon(1S)$ & 5.02 & 3111 & \cite{LHCb:2014rku} & 2 & 0 & --\;\; \\
 & $\text{incl.}~\Upsilon(1S)$ & 8.16 & 3113 & \cite{LHCb:2018psc} & 64 & 36 & 0.99 \\
 & $\text{prompt}~D^0~\text{forw.}$ & 8.16 & \textbf{3133} & \cite{LHCb:2022dmh} & 170 & 66 & 2.52 \\
 & $\text{prompt}~D^0~\text{back.}$ & 8.16 & \textbf{3134} & \cite{LHCb:2022dmh} & 163 & 65 & 0.57 \\
\hline
\hline
\setrowstyle{\bfseries}
Total &  &  &  &  & \!\!\!\!\!1195 & 695 & 0.99\\
\hline
\end{tabular}
\caption{Open heavy-flavor (OHF) and quarkonium (ONIA) data from LHC pPb experiments used in this fit. Datasets not included in the nCTEQ15HQ analysis are indicated by bold IDs. The $N_{\text{data}}$ column indicates the number of data points before/after kinematic cuts. }
\label{tab:HQdatapPb}
\end{table}
\renewcommand{\arraystretch}{1.0}

%% file: tables/data/HQ_pp.tex
\renewcommand{\arraystretch}{1.05}
\begin{table}[htbp]
  \centering
  \begin{tabular}{|p{0cm}p{1.2cm}p{2.1cm}p{1.3cm}p{0.7cm}p{0.6cm}p{1.2cm}|}
    \hline
    \multicolumn{7}{|c|}{OHF and quarkonium data $d^2\sigma/dp_\mathrm{T} dy$ in pp collisions}\\
    \hline
    \hline
    & Exp. & Observable & $\sqrt{s}$ [TeV] &  \;ID &  Ref. & $N_{\text{data}}$ \\
    \hline 
    & \multirow{17}{*}[-0.0em]{ALICE} 
    & prompt $D^0$ & \;\;\;7 & 3008 & \cite{ALICE:2012mhc} &  9/7 \\
    & & incl. $\Upsilon(1S)$ & \;\;\;7 & 3009 & \cite{ALICE:2014uja} &  5/3 \\
    & & incl. $J/\psi$ & \;\;\;5.02 & 3016 & \cite{ALICE:2019pid} &  7/4 \\
    & & prompt $D^0$ & \;\;\;7 & 3021 & \cite{ALICE:2016yta} &  10/7 \\
    & & incl. $J/\psi$ & \;\;\;8 & 3022 & \cite{ALICE:2015pgg} &  13/10 \\
    & & incl. $J/\psi$ & \;\;\;5.02 & 3023 & \cite{ALICE:2017leg} &  11/8 \\
    & & incl. $J/\psi$ & \;\;\;13 & 3024 & \cite{ALICE:2017leg} &  18/14 \\
    & & incl. $\psi(2S)$ & \;\;\;8 & 3028 & \cite{ALICE:2015pgg} &  9/6 \\
    & & incl. $\psi(2S)$ & \;\;\;13 & 3030 & \cite{ALICE:2017leg} &  12/9 \\
    & & incl. $\Upsilon(1S)$ & \;\;\;8 & 3031 & \cite{ALICE:2015pgg} &  5/3 \\
    & & prompt $D^0$ & \;\;\;7 & \textbf{3032} & \cite{ALICE:2017olh} &  13/10 \\
    & & prompt $D^0$ & \;\;\;5.02 & \textbf{3033} & \cite{ALICE:2019nxm} &  22/16 \\
    & & prompt $D^0$ & \;\;\;13 & \textbf{3035} & \cite{ALICE:2021edd} &  9/7 \\
    & & non-pr. $J/\psi$ & \;\;\;5.02 & \textbf{3036} & \cite{ALICE:2021edd} & 3/2 \\
    & & non-pr. $J/\psi$ & \;\;\;13 & \textbf{3037} & \cite{ALICE:2021edd} & 6/4 \\
    & & prompt $J/\psi$ & \;\;\;5.02 & \textbf{3038} & \cite{ALICE:2021edd} & 3/2 \\
    & & prompt $J/\psi$ & \;\;\;13 & \textbf{3039} & \cite{ALICE:2021edd} & 6/4 \\
    \cline{2-7}
    & \multirow{8}{*}[-0.0em]{ATLAS}  
    & prompt $J/\psi$ & \;\;\;7 & 3003 &\cite{ATLAS:2011aqv} & 64/63 \\
    & & non-pr. $J/\psi$ & \;\;\;7 & 3006 &\cite{ATLAS:2011aqv} & 64/63 \\
    & & incl. $\Upsilon(1S)$ & \;\;\;7 & 3012 &\cite{ATLAS:2012lmu} & 100/88 \\
    & & prompt $J/\psi$ & \;\;\;5.02 & 3017 &\cite{ATLAS:2017prf} & 33/33 \\
    & & non-pr. $J/\psi$ & \;\;\;5.02 & 3018 &\cite{ATLAS:2017prf} & 33/33 \\
    & & incl. $\Upsilon(1S)$ & \;\;\;5.02 & 3025 &\cite{ATLAS:2017prf} & 24/18 \\
    & & prompt $\psi(2S)$ & \;\;\;5.02 & 3026 &\cite{ATLAS:2017prf} & 15/15 \\
    & & non-pr. $\psi(2S)$ & \;\;\;5.02 & 3027 &\cite{ATLAS:2017prf} & 15/15 \\
    \cline{2-7}
    &  \multirow{8}{*}[-0.0em]{CMS} 
    &   prompt $J/\psi$ & \;\;\;7 & 3001 &\cite{CMS:2011rxs} & 44/39 \\
    & & non-pr. $J/\psi$ & \;\;\;7 & 3005 &\cite{CMS:2011rxs} & 44/39 \\
    & & non-pr. $J/\psi$ & \;\;\;7 & 3007 &\cite{CMS:2010nis} & 14/9 \\
    & & incl. $\Upsilon(1S)$ & \;\;\;7 & 3013 &\cite{CMS:2013qur} & 42/30 \\
    & & non-pr. $J/\psi$ & \;\;\;5.02 & 3019 &\cite{CMS:2017exb} & 52/50 \\
    & & prompt $J/\psi$ & \;\;\;5.02 & 3020 &\cite{CMS:2018gbb} & 52/50 \\
    & & prompt $\psi(2S)$ & \;\;\;5.02 & 3029 &\cite{CMS:2022wfi} & 10/10 \\
    & & incl. $\Upsilon(1S)$ & \;\;\;5.02 & \textbf{3040} &\cite{CMS:2022wfi} & 6/4 \\
    \cline{2-7}
    & \multirow{15}{*}[-0.0em]{LHCb} 
    &   prompt $J/\psi$ & \;\;\;7 & 3002 & \cite{LHCb:2011zfl} & 66/43 \\
    & & non-pr. $J/\psi$ & \;\;\;7 & 3004 & \cite{LHCb:2011zfl} & 66/30 \\
    & & incl. $\Upsilon(1S)$ & \;\;\;7 & 3010 & \cite{LHCb:2012aa} & 75/48 \\
    & & incl. $\Upsilon(1S)$ & \;\;\;7 & 3011 & \cite{LHCb:2015log} & 109/82 \\
    & & prompt $D^0$ & \;\;\;7 & 3014 & \cite{LHCb:2013xam} & 38/20 \\
    & & incl. $\Upsilon(1S)$ & \;\;\;8 & 3015 & \cite{LHCb:2015log} & 109/82 \\
    & & prompt $D^0$ & \;\;\;5.02 & \textbf{3034} & \cite{LHCb:2016ikn} & 45/27 \\
    & & prompt $D^0$ & \;\;\;5.02 & \textbf{3034} & \cite{LHCb:2016ikn} & 45/27 \\
    & & incl. $\Upsilon(1S)$ & \;\;\;13 & \textbf{3041} & \cite{LHCb:2018yzj} &  111/83 \\
    & & non-pr. $J/\psi$ & \;\;\;5.02 & \textbf{3042} & \cite{LHCb:2021pyk} &  50/28 \\
    & & prompt $J/\psi$ & \;\;\;5.02 & \textbf{3043} & \cite{LHCb:2021pyk} &  50/28 \\
    & & prompt $\psi(2S)$ & \;\;\;7 & \textbf{3044} & \cite{LHCb:2019eaj} &  55/44 \\
    & & non-pr. $\psi(2S)$ & \;\;\;7 & \textbf{3045} & \cite{LHCb:2019eaj} &  55/44 \\
    & & prompt $\psi(2S)$ & \;\;\;13 & \textbf{3046} & \cite{LHCb:2019eaj} &  79/61 \\
    & & non-pr. $\psi(2S)$ & \;\;\;13 & \textbf{3047} & \cite{LHCb:2019eaj} &  79/61 \\
    & & prompt $D^0$ & \;\;\;13 & \textbf{3048} & \cite{LHCb:2015swx} &  78/48 \\
    \hline 
    \multicolumn{3}{|l}{\textbf{Total}} & &  & & \hspace*{-0.5cm}\textbf{1838/1436} \\
    \hline
  \end{tabular}
  \caption{Open heavy-flavor (OHF) and quarkonium (ONIA) data from LHC pp experiments used in the Crystal Ball fit. Datasets not included in the nCTEQ15HQ analysis are indicated by bold IDs. The $N_{\text{data}}$ column indicates the number of data points before/after kinematic cuts.}
  \label{tab:HQdatapp}
\end{table}
\renewcommand{\arraystretch}{1.0}

%% file: tables/crystal_ball.tex
\begin{table}[htb!]
    \renewcommand{\arraystretch}{1.2}
    \setlength\tabcolsep{5pt}
	\centering
	\vspace{0.2cm}
	\begin{tabular}{|c||c|c|c|}
		\hline 
		     & $D^0$ & $J/\psi$ & $B \rightarrow J/\psi$  \\
		\hline
		\hline
		 $\kappa$               & 0.68223 & 0.51608 & 0.23748  \\
		\hline
		 $\lambda$              & 1.65235 & 0.31361 & 0.12402 \\
		\hline
		 $\left<p_T \right>$    &-0.40783 & 4.95117 & 5.58369 \\
		\hline
		 $n$                    & 1.70145 & 2.21832 & 1.53569  \\
		\hline 
		 $a$                    &-0.01489 & 0.02800 &-0.10225 \\
		\hline
		\hline 
		$N_{\mathrm{points}}$ & 135 & \multicolumn{2}{c|}{569}   \\
		\hline 
		$\chidof$ & 0.59 & \multicolumn{2}{c|}{1.02}   \\
		\hline 
        \addlinespace[7pt] 
        \hline
		      & $\Upsilon(1S)$  & $\psi(2S)$ & $B \rightarrow \psi(2S)$ \\
		\hline
		\hline
		 $\kappa$                & 1.95673 & 0.49509 & 0.22663 \\
		\hline
		 $\lambda$               & 0.06099 & 0.13345 & 0.08186 \\
		\hline
		 $\left<p_T \right>$     &-3.75593 & 5.23376 & 7.3664 \\
		\hline
		 $n$                     & 1.91625 & 2.20183 & 1.43383 \\
		\hline 
		 $a$                    &-0.36407 & 0.06620 & -0.07833 \\
		\hline 
		\hline 
		$N_{\mathrm{points}}$   & 420  & \multicolumn{2}{c|}{265} \\
		\hline 
		$\chidof$   & 0.90  & \multicolumn{2}{c|}{0.87} \\
		\hline 
	\end{tabular}     	
	\caption{Crystal Ball parameters and $\chidof$ values for the HQ production processes considered in this fit.}
	\label{tab:CrystalBallBaseline}
\end{table}

%% file: tables/parameter_nCTEQ26.tex
\begin{table}[t] 
 \centering 
 \setlength{\extrarowheight}{0pt}
\begin{subtable}[t]{0.50\textwidth}
 \centering 
\begin{tabular}{|C{0.7cm}|R{1.8cm}|R{1.8cm}|R{1.8cm}|} 
\hline
 & $p$ & $a$ & $b$ \\
\hline
\end{tabular}
\end{subtable}%
\vspace{1em}
\begin{subtable}[t]{0.50\textwidth}
 \centering 
\begin{tabular}{|C{0.7cm}|R{1.8cm}|R{1.8cm}|R{1.8cm}|} 
\hline
$c^g_{\text{sum}}$ & $0.38870$ & $\mathbf{0.00501}$ & $\mathbf{0.00003}$ \\
$c^g_1$ & $0.60307$ & $\mathbf{0.06195}$ & $\mathbf{0.01033}$ \\
$c^g_2$ & $6.48120$ & $\mathbf{0.99826}$ & $\mathbf{-0.00425}$ \\
$c^g_3$ & $-3.30640$ & $\mathbf{-0.23433}$ & $0$ \\
$c^g_4$ & $3.17210$ & $\mathbf{0.47323}$ & $0$ \\
$c^g_5$ & $0$ & $0$ & $0$ \\
\hline
\end{tabular}
\end{subtable}%
\vspace{1em}
\begin{subtable}[t]{0.50\textwidth}
 \centering 
\begin{tabular}{|C{0.7cm}|R{1.8cm}|R{1.8cm}|R{1.8cm}|} 
\hline
$c^{u_v}_0$ & \multicolumn{3}{c|}{sum rule}  \\
\cline{2-4}
$c^{u_v}_1$ & $0.61537$ & $\mathbf{0.01034}$ & $\mathbf{-0.00356}$ \\
$c^{u_v}_2$ & $3.54330$ & $\mathbf{-0.18494}$ & $0$ \\
$c^{u_v}_3$ & $0$ & $\mathbf{1.75286}$ & $0$ \\
$c^{u_v}_4$ & $3.46090$ & $\mathbf{-2.49003}$ & $0$ \\
$c^{u_v}_5$ & $0$ & $0$ & $0$ \\
\hline
\end{tabular}
\end{subtable}%
\vspace{1em}
\begin{subtable}[t]{0.50\textwidth}
 \centering 
\begin{tabular}{|C{0.7cm}|R{1.8cm}|R{1.8cm}|R{1.8cm}|} 
\hline
\rowpad $c^{d_v}_0$ & \multicolumn{3}{c|}{sum rule}  \\
\cline{2-4}
$c^{d_v}_1$ & $1.15950$ & $\mathbf{-0.00126}$ & $\mathbf{0.00016}$ \\
$c^{d_v}_2$ & $6.55140$ & $\mathbf{0.19671}$ & $\mathbf{-0.00708}$ \\
$c^{d_v}_3$ & $-3.50300$ & $\mathbf{-0.08267}$ & $0$ \\
$c^{d_v}_4$ & $4.67870$ & $\mathbf{0.23513}$ & $0$ \\
$c^{d_v}_5$ & $0$ & $0$ & $0$ \\
$c^{d_v}_6$ & $0.00360$ & $\mathbf{0.00183}$ & $0$ \\
$c^{d_v}_7$ & $2.00000$ & $0$ & $0$ \\
\hline
\end{tabular}
\end{subtable}%
\vspace{1em}
\begin{subtable}[t]{0.50\textwidth}
 \centering 
\begin{tabular}{|C{0.7cm}|R{1.8cm}|R{1.8cm}|R{1.8cm}|} 
\hline
\rowpad $c^{\bar{u}+\bar{d}}_0$ & \multicolumn{3}{c|}{sum rule}  \\
\cline{2-4}
$c^{\bar{u}+\bar{d}}_1$ & $-0.20775$ & $\mathbf{0.02483}$ & $\mathbf{-0.00546}$ \\
$c^{\bar{u}+\bar{d}}_2$ & $8.32860$ & $\mathbf{0.03115}$ & $0$ \\
$c^{\bar{u}+\bar{d}}_3$ & $0$ & $0$ & $\mathbf{0.47427}$ \\
$c^{\bar{u}+\bar{d}}_4$ & $14.60600$ & $\mathbf{-3.82396}$ & $\mathbf{1.32808}$ \\
$c^{\bar{u}+\bar{d}}_5$ & $0$ & $0$ & $0$ \\
\hline
\end{tabular}
\end{subtable}%
\vspace{1em}
\begin{subtable}[t]{0.50\textwidth}
 \centering 
\begin{tabular}{|C{0.7cm}|R{1.8cm}|R{1.8cm}|R{1.8cm}|} 
\hline
$c^{s+\bar{s}}_{\text{sum}}$ & $0.50002$ & $\mathbf{-0.02023}$ & $0$ \\
$c^{s+\bar{s}}_1$ & $-0.20775$ & $\mathbf{-0.07522}$ & $\mathbf{0.04795}$ \\
$c^{s+\bar{s}}_2$ & $8.32860$ & $\mathbf{-1.66412}$ & $\mathbf{0.54728}$ \\
$c^{s+\bar{s}}_3$ & $0$ & $0$ & $0$ \\
$c^{s+\bar{s}}_4$ & $14.60600$ & $\mathbf{-2.31066}$ & $0$ \\
$c^{s+\bar{s}}_5$ & $0$ & $\mathbf{0.06969}$ & $\mathbf{0.00071}$ \\
\hline
\end{tabular}
\end{subtable}%
\vspace{1em}
\begin{subtable}[t]{0.50\textwidth}
 \centering 
\begin{tabular}{|C{0.7cm}|R{1.8cm}|R{1.8cm}|R{1.8cm}|} 
\hline
\rowpad $c^{\bar{d}/\bar{u}}_0$ & $35712.00000$ & $0$ & $0$ \\
$c^{\bar{d}/\bar{u}}_1$ & $4.02490$ & $0$ & $0$ \\
$c^{\bar{d}/\bar{u}}_2$ & $20.15400$ & $0$ & $0$ \\
$c^{\bar{d}/\bar{u}}_3$ & $17.00000$ & $\mathbf{-27.27820}$ & $0$ \\
$c^{\bar{d}/\bar{u}}_4$ & $51.15600$ & $\mathbf{1.10192}$ & $0$ \\
\hline
\end{tabular}
\end{subtable}%
\caption{Values of the PDF parameters for the nCTEQ26 fit at the initial scale $Q_0=1.3$ GeV. Bold entries denote free parameters, non-bold entries are fixed. Normalizations not listed are determined by number ($c_0^{u_v},c_0^{d_v}$) and momentum ($c_0^{\bar{u}+\bar{d}}$) sum rules (cf.\ \cref{subsec:param}).}
\label{tab:params}
\end{table}

%% file: 08acks.tex
\section*{Acknowledgments}

We thank P.\ Duwentäster and K.~F.\ Muzakka for their contributions during the early stages of this project, as well as C.~Keppel, J.~Niehues, E.~R.~Nocera and J.~F.~Owens for valuable discussions.
We would also like to thank \mbox{I.~Grabowska-Bołd} for help with understanding the $t\bar{t}$ results from ATLAS.

Work at the University of Münster has been supported by the BMFTR under contract 05P24PMA. Numerical calculations for this publication were performed on the HPC cluster PALMA II of the University of Münster, subsidized by the DFG (INST 211/667-1). 
AK~is grateful for the support of Narodowe Centrum Nauki under the SONATA Bis grant no.\ 2019/34/E/ST2/00186.
The work at SMU was supported by 
the U.S.\ DoE  Grant No.~DE-SC0010129,
and 
by the  U.S.\ DoE, Office of Science, Office of Nuclear Physics, 
within the framework of the Saturated Glue (SURGE) Topical Theory Collaboration.
The work of FO was performed in part at the Aspen Center for Physics, which is supported by National Science Foundation grant PHY-2210452.
The work of PR was work supported by the U.S.\ Department of Energy, Office of Science, Office of Nuclear Physics under Contract No.\ 89243126CSC000213.
The work of JGM was supported by FermiForward Discovery Group, LLC under Contract No. 89243024CSC000002 with the U.S. Department of Energy, Office of Science, Office of High Energy Physics.